\documentclass{aa}  
\usepackage{natbib}
\bibpunct{(}{)}{;}{a}{}{,} 
\usepackage[colorinlistoftodos]{todonotes}
\usepackage{afterpage}
\usepackage{placeins}
\usepackage{float}
\usepackage{graphicx}
\usepackage{units}
\usepackage{txfonts}
\usepackage[colorlinks=true,linkcolor=blue,citecolor=blue]{hyperref}

\newcommand{\lsun}{\mbox{L$_\odot$}\xspace}
\newcommand{\msun}{\mbox{M$_\odot$}\xspace}
\newcommand{\chcn}{\mbox{CH$_3$CN}\xspace}
\newcommand{\SgrB}{Sgr\,B2}
\newcommand{\isochcn}{\mbox{$^{13}$CH$_3$CN}\xspace}
\newcommand{\hii}{\mbox{\textsc{H\,ii}}\xspace}
\newcommand{\scm}{\mbox{cm$^{-2}$}\xspace}
\newcommand{\qcm}{\mbox{cm$^{-3}$}\xspace}

\newcommand{\phn}{\phantom{0}}

\newcommand{\pandora}{Pandora}

\def\ee#1{\ensuremath{10^{#1}}}
\def\tee#1{\ensuremath{\times10^{#1}}}

\begin{document} 

\title{The physical and chemical structure of Sagittarius~B2}

\subtitle{III. Radiative transfer simulations of the hot core Sgr\,B2(M) for methyl cyanide}

\author{S.~Pols 				 \inst{1}  	 \and
        A.~Schw\"{o}rer 	 	 \inst{1} 	 \and
        P.~Schilke				 \inst{1}    \and
        A.~Schmiedeke 			 \inst{1, 2} \and
        \'{A}.~S\'{a}nchez-Monge \inst{1}    \and          
        Th.~M\"{o}ller 			 \inst{1}}

\institute{I. Physikalisches Institut, Universit\"at zu K\"oln,
		   Z\"ulpicher Str. 77, D-50937 K\"oln, Germany\\
           \email{pols, schwoerer, schilke, sanchez, moeller @ph1.uni-koeln.de}
           \and
           Max-Planck Institute for Extraterrestrial Physics,
           Giessenbachstrasse 1, D-85748 Garching, Germany\\
           \email{schmiedeke @mpe.mpg.de}
}
           
\date{Received; accepted}

 
\abstract
   {We model the emission of methyl cyanide (\chcn) lines towards the massive hot molecular core \SgrB(M).}
   {We aim at reconstructing the \chcn\ abundance field and investigating the gas temperature distribution as well as the velocity field.}
   {\SgrB(M) was observed with the Atacama Large Millimeter/sub-millimeter Array (ALMA) in a spectral line survey from \unit[211 to 275]{GHz}. This frequency range includes several transitions of \chcn\ (including isotopologues and vibrationally excited states). We employ the three-dimensional radiative transfer toolbox \pandora\ in order to retrieve the velocity and abundance field by modeling different \chcn\ lines. For this purpose, we base our model on the results of a previous study that determined the physical structure of \SgrB(M), i.e.\ the distribution of dust dense cores, ionized regions and heating sources.}
   {The morphology of the \chcn\ emission can be reproduced by a molecular density field that consists of a superposition of cores with modified Plummer-like density profiles. The averaged relative abundance of \chcn\ with respect to H$_2$ ranges from 4\tee{-11} to 2\tee{-8} in the northern part of \SgrB(M) and from 2\tee{-10} to 5\tee{-7} in the southern part. In general, we find that the relative abundance of \chcn\ is lower at the center of the very dense and hot cores, causing the general morphology of the \chcn\ emission to be shifted with respect to the dust continuum emission. The dust temperature calculated by the radiative transfer simulation based on the available luminosity reaches values up to \unit[900]{K}. However, in some regions  vibrationally excited transitions of \chcn\ are underestimated by the model, indicating that the predicted gas temperature, which is assumed to be equal to the dust temperature, is partly underestimated. The determination of the velocity component along the line of sight reveals that a velocity gradient from the north to the south exists in \SgrB(M).}
   {}


\maketitle

\section{Introduction}\label{sec:Introduction}

Almost all information we have about the interstellar medium (ISM) is obtained from the analysis of electromagnetic radiation received by telescopes. Hence, in astrophysics it is crucial to predict the radiative properties of a physical model when interpreting astronomical observations. For this purpose, three-dimensional radiative transfer codes have been developed, which can calculate the radiation signature associated with a three-dimensional physical model of a specific source. The results can subsequently be compared to observations to study the physical and chemical properties of astrophysical objects. 

In the struggle to understand how the evolution of high-mass stars proceeds from the fragmentation of large giant molecular clouds down to dense hot molecular cores \citep[e.g.\ ][]{tan2014, schilke2015}, it is important to study the kinematics, density structure, temperature distribution and chemical variations of the dense gas associated with these objects, as these parameters determine their dynamical evolution. The study of the emission of lines of certain molecular species can provide this crucial information as the intensity and the line profile are determined by the physical and excitation conditions of the gas.

There are several molecules in the ISM that can be used to study the dense cores where star formation takes place. However, hot molecular cores \citep[e.g.\ ][]{kurtz2000, cesaroni2005} associated with high temperatures ($T$\unit[$>$100]{K}) and high densities ($n_\mathrm{H_2}$>$10^{6}$~cm$^{-3}$) are traced best by complex molecules (i.e.\ molecules with 6 or more atoms) that are mainly formed on dust grains and released into the gas-phase during the warm-up through the heating produced by the newly-born star. Among these complex molecules, \chcn has turned out to be an excellent tool to determine the gas temperature due to its symmetry structure \citep[e.g.][]{araya2005}, as well as the distribution and velocity field of the high-density gas from where high-mass stars form \citep[see e.g.\ ][]{remijan2004, sanchez-monge2013, hernandez-hernandez2014}. 

Sagittarius~B2 (hereafter \SgrB) is one of the most massive molecular clouds in the Galaxy. Located at a projected distance of $\sim$100~pc from the Galactic center, which has a distance of 8.34$\pm$0.16~kpc from the Sun \citep{reid2014}, \SgrB\ is an excellent source for studying the formation of high-mass clusters and astrochemistry.  It has been shown to harbor active star formation sites \citep{gordon1993} and contains a wide range of different molecules including complex organic species, many of which were found for the first time in space toward \SgrB\ (e.g.\ acetic acid \citep[CH$_3$COOH,][]{mehringer1997} or ethyl formate \citep[C$_2$H$_5$OCHO,][]{belloche2009}). The \SgrB\ complex contains a total mass of \unit[$\sim$\ee{7}]{\msun}, and a total luminosity of about \unit[\ee{7}]{\lsun} \citep{goldsmith1990}. Based on its density structure, \SgrB\ was proposed to be composed of three different parts \citep{huettemeister1993}: a low density envelope with $n_\mathrm{H_2}$\unit[$\sim$\ee{3}]{\qcm}, a moderate density region with $n_\mathrm{H_2}$\unit[$\sim$\ee{5}]{\qcm} extending around local hot molecular cores, which are the most compact and densest regions with $n_\mathrm{H_2}$\unit[$\sim$\ee{7}]{\qcm}. The two main sites of star formation are the central hot molecular cores \SgrB(N) and \SgrB(M), which have sizes of \unit[$\sim$0.5]{pc} and contain numerous ultracompact \hii\ regions \citep[see e.g.][]{mehringer1993,gaume1995}. A more detailed description of \SgrB\ can be found in \cite[][hereafter Paper~I]{schmiedeke2016} and \cite[][herafter Paper~II]{sanchez-monge2017a}.

This is the third in a series of papers investigating the star-forming region \SgrB. In this work, we model line transitions of methyl cyanide, \chcn, by employing radiative transfer codes. This modeling is based on a physical model of \SgrB, which has been developed in the first paper of this series (Paper~I). The modeling of the \chcn\ transitions is based on high-resolution observations of \SgrB\ obtained with the Atacama Large Millimeter/sub-millimeter Array (ALMA, \citealt{ALMApartnership2015}). The continuum emission of these observations was presented and analyzed in the second paper of this series (Paper~II). In the current paper, we focus our analysis on the hot molecular core \SgrB(M) which exhibits a very complex and fragmented structure, and appears to be in a more evolved stage of evolution than \SgrB(N). Earlier observations using the Submillimeter Array (SMA) by \citet{qin2011} revealed at least 12 dense cores at a frequency of 345~GHz. Recent ALMA observations at 242~GHz (Paper~II) increased this number to 27 sources including dust dense cores and some \hii\ regions, and suggested a more complex structure with hints of the presence of filaments. 

The current paper is organized as follows. In Sect.~\ref{sec:Observations} we give a brief overview of the ALMA observations. In Sect.~\ref{sec:Results}, we present the \chcn\ data and compare the distribution of the dense gas to the continuum emission. The modeling procedure is described in Sect.~\ref{sec:ModelingProcedure}. In Sect.~\ref{sec:PhysicalModel} we describe the physical model used to model the \chcn\ emission towards \SgrB(M). This is followed by a comparison of the simulations with the observational data in Sect.~\ref{sec:Comparison}. Finally, Sect.~\ref{sec:Conclusion} summarizes the main results of the paper. The appendices contain a description of the parameters of the physical model as well as several spectra comparing the observational and simulated data. 

\section{Observations}\label{sec:Observations}

The observations of \SgrB(M) were conducted using ALMA during its Cycle~2 in June 2014 and June 2015 (project number 2013.1.00332.S). The configuration used 34 and 36 antennas, with extended baselines in the range between 30~m and 650~m. At the frequency range of our observations (from 211~GHz to 275~GHz) this results in an angular resolution of $0\farcs4$--$0\farcs8$. \SgrB(M) was observed in track-sharing mode, together with \SgrB(N), with the phase center at $\alpha(J2000)=17^\textrm{h}47^\textrm{m}20^\textrm{s}.157$, $\delta(J2000)=-28^\circ 23' 04\farcs53$. The observations were done in the spectral scan mode surveying the whole ALMA band~6 from 211 to 275~GHz. The achieved spectral resolution is 0.5--0.7~km~s$^{-1}$ across the full frequency band. Calibration and imaging were performed in CASA\footnote{The Common Astronomy Software Applications (CASA; \citealt{mcmullin2007}) software can be downloaded at http://casa.nrao.edu} version 4.4.0. The resulting data cubes used in this project were restored with a circular Gaussian beam of $0\farcs7$. Further information on the calibration, observational details and imaging can be found in Paper~II.

\begin{table}
\caption{\label{tab:transitions} \chcn\ transitions analyzed in this paper}
\centering
\begin{tabular}{c c c}
\hline\hline
Transition
&Frequency~(GHz)
&$E_\mathrm{lower}$~(K)
\\
\hline
\multicolumn{3}{c}{Main isotopologue, ground state} \\
\hline
$12_0 - 11_0$  & 220.747261 & \phn57.87 \\
$12_1 - 11_1$  & 220.743011 & \phn64.96 \\
$12_2 - 11_2$  & 220.730261 & \phn86.25 \\
$12_3 - 11_3$  & 220.709017 & 121.71 \\
$12_4 - 11_4$  & 220.679287 & 171.36 \\
$12_5 - 11_5$  & 220.641084 & 235.17 \\
$12_6 - 11_6$  & 220.594423 & 313.13 \\
$13_0 - 12_0$  & 239.137916 & \phn68.39 \\
$13_1 - 12_1$  & 239.133313 & \phn75.48 \\
$13_2 - 12_2$  & 239.119504 & \phn96.77 \\
$13_3 - 12_3$  & 239.096497 & 132.23 \\
$13_4 - 12_4$  & 239.064299 & 181.88 \\
$13_5 - 12_5$  & 239.022924 & 245.68 \\
$13_6 - 12_6$  & 238.972390 & 323.64 \\
$13_7 - 12_7$  & 238.912715 & 415.73 \\
$13_8 - 12_8$  & 238.843926 & 521.93 \\
$13_9 - 12_9$  & 238.766049 & 642.21 \\
$14_0 - 13_0$  & 257.527384 & \phn79.79 \\
$14_1 - 13_1$  & 257.522428 & \phn86.88 \\
$14_2 - 13_2$  & 257.507561 & 108.16 \\
$14_3 - 13_3$  & 257.482792 & 143.63 \\
$14_4 - 13_4$  & 257.448128 & 193.27 \\
$14_5 - 13_5$  & 257.403584 & 257.07 \\
$14_6 - 13_6$  & 257.349179 & 335.03 \\
$14_7 - 13_7$  & 257.284935 & 427.11 \\
$14_8 - 13_8$  & 257.210877 & 533.31 \\
$14_9 - 13_9$  & 257.127035 & 653.59 \\
\hline
\multicolumn{3}{c}{Vibrationally excited $v_8=1$} \\
\hline
$12_1 - 11_{-1}$  & 221.625828 & 577.13 \\    
\hline
\multicolumn{3}{c}{$^{13}$\chcn\ isotopologue, ground state} \\
\hline
$13_0 - 12_0$  & 232.234188 & \phn66.38 \\
$13_1 - 12_1$  & 232.229822 & \phn73.48 \\
$13_2 - 12_2$  & 232.216726 & \phn94.79 \\
$13_3 - 12_3$  & 232.194906 & 130.30 \\
$13_4 - 12_4$  & 232.164369 & 180.01 \\
$13_5 - 12_5$  & 232.125130 & 243.89 \\
$13_6 - 12_6$  & 232.077203 & 321.94 \\
$13_7 - 12_7$  & 232.020609 & 414.15 \\
\hline
\end{tabular}
\tablefoot{Parameters of the transitions obtained from the molecular databases CDMS \citep{mueller2005,endres2016} and JPL \citep{picket1998}. In more detail, the \chcn\ and \isochcn\ entries are based on \citet{mueller2015} and \citet{mueller2009}, respectively, with important additional data especially from \citet{cazzoli2006} and \citet{pearson1996}. The dipole moment of \chcn\, used to derive the Einstein $A_{ul}$ values, is taken from \citet{gadhi1995}; the one of \isochcn\ is assumed to agree with that of the main isotopologue, as is frequently done.}
\end{table}

\begin{figure*}[t!]
\begin{center}
\begin{tabular}[b]{c}
        \includegraphics[width=0.98\textwidth]{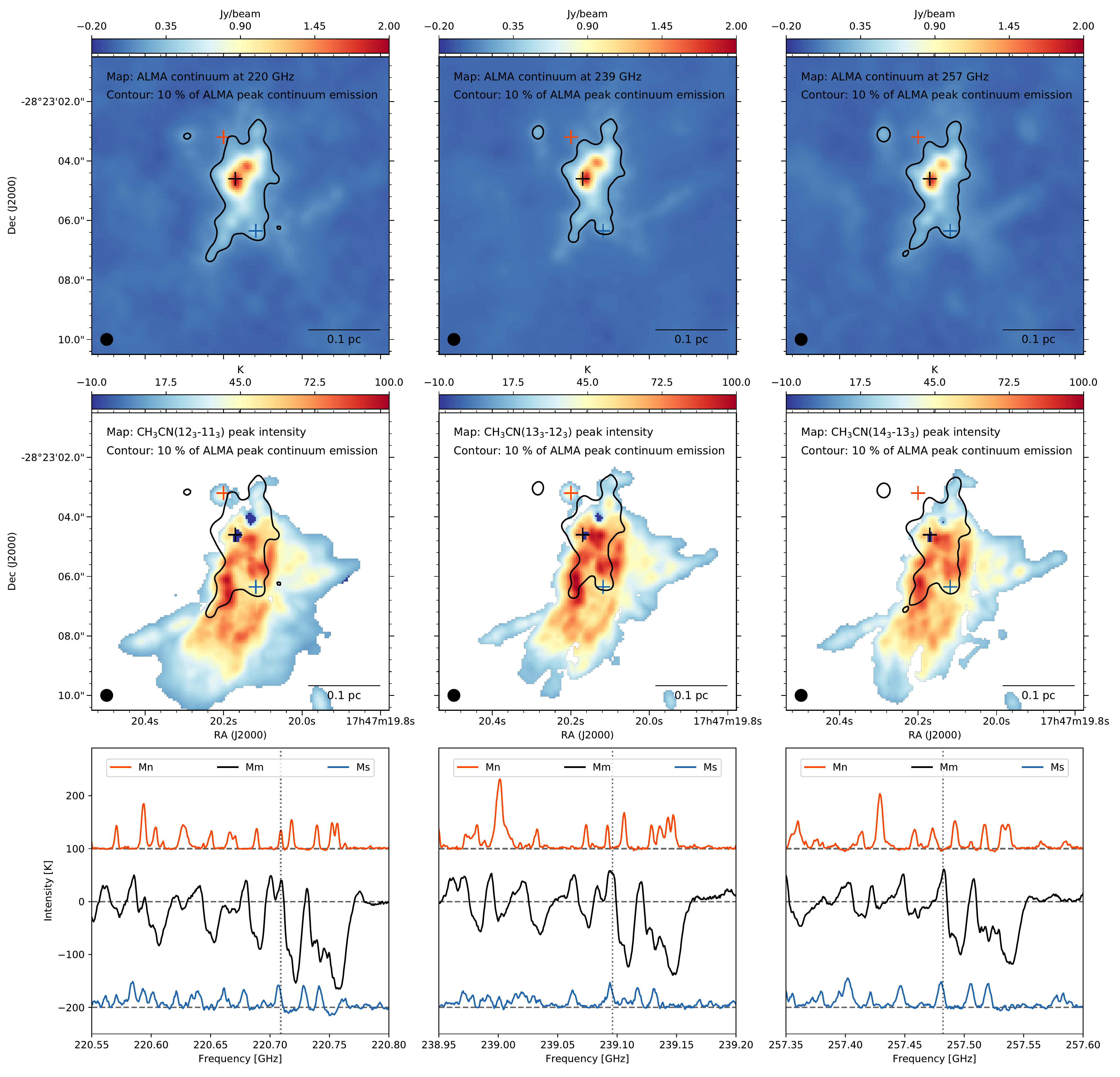} \\
\end{tabular}
\caption{\textit{Top panels}: Continuum maps towards \SgrB(M) determined using STATCONT \citep{sanchez-monge2018} at the frequencies of the spectral windows that cover the three \chcn\ ground state transitions: \unit[220]{GHz} in the left, \unit[239]{GHz} in the center, and \unit[257]{GHz} in the right column. The black contour shows the 10\% level of the continuum emission peak. \textit{Middle panels}: Intensity peak maps of different \chcn\ transitions: $J$=12--11,~$K$=3 in the left, $J$=13--12,~$K$=3 in the center, and $J$=14--13,~$K$=3 in the right panel.The intensity peak maps are obtained from a narrow frequency range around the frequency indicated with a vertical dotted line in the bottom panels. The black contour depicts the 10\% level of the continuum emission as in the top panels. \textit{Bottom panels}: Spectra extracted at three different positions marked with crosses in the top and middle panels. The select positions are located in the north (Mn, in red), in the center of the region, associated with the brightest continuum peak (Mm, in black), and towards the south (Ms, in blue) To enhance clarity in the plot, we shifted the red spectrum by +100 K and the blue spectrum by -200 K. The relative continuum levels are indicated by the dashed gray lines.}
\label{fig:observations}
\end{center}
\end{figure*}

\begin{figure}[t!]
	\centering
	\includegraphics[width=0.98\columnwidth]{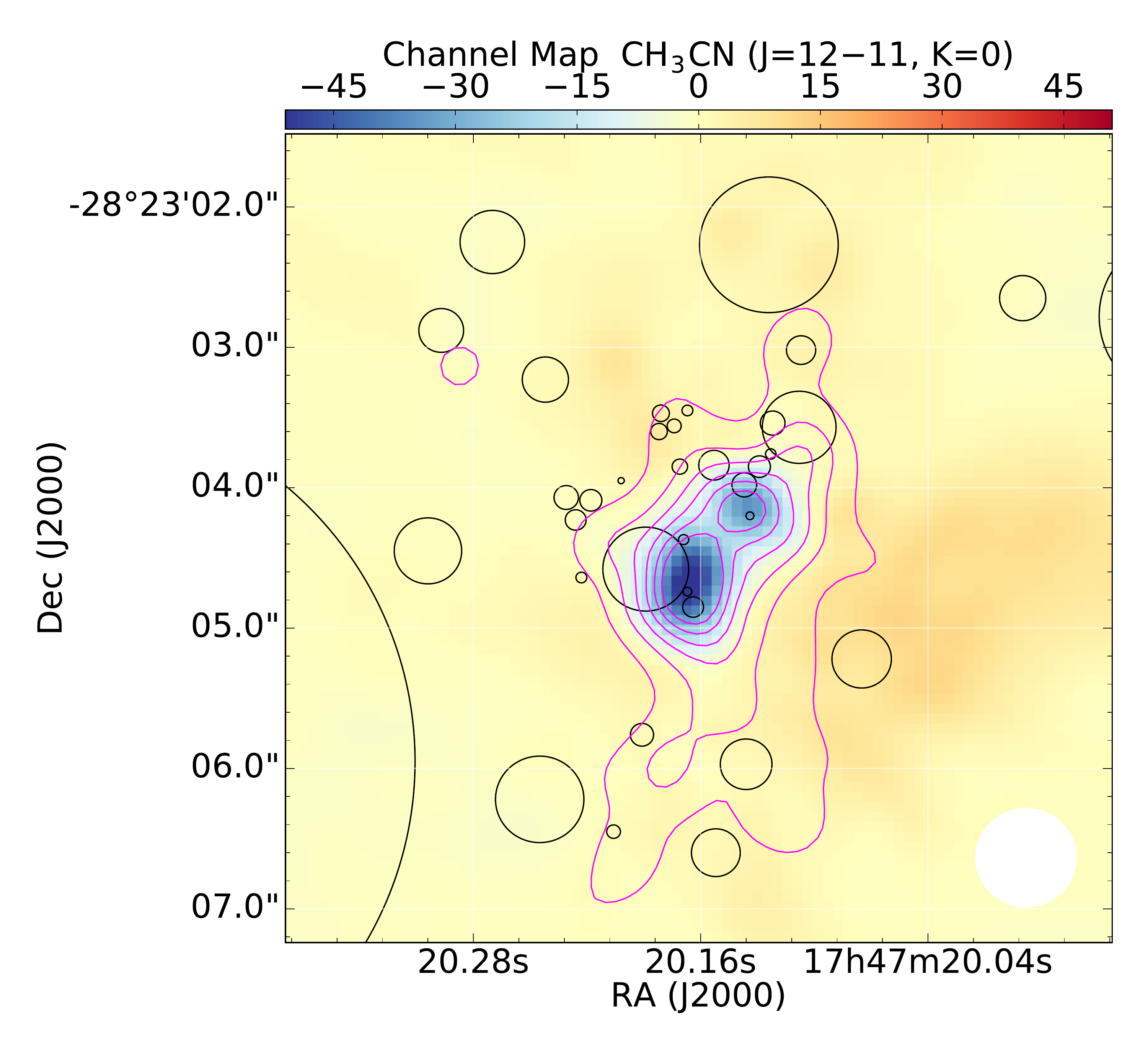}
\caption{Channel map at a frequency of \unit[220.765]{GHz} corresponding to the line of \chcn\ ($J$=12-11, $K$=0). Clear absorption is observed in the center of the map. Magenta contours show the ALMA \unit[242]{GHz} continuum emission from Paper~II. Black circles indicate the position and extend of the \hii\ regions (see e.g.\ Paper~I).}
\label{fig:absorption}
\end{figure}

\section{Observational results} \label{sec:Results}

The frequency range of the ALMA observations covers several rotational transitions of \chcn\ in the vibrational ground state (e.g. $J$=12--11 at \unit[220.7]{GHz}, $J$=13--12 at \unit[239.1]{GHz}, $J$=14--13 at \unit[257.5]{GHz}) as well as in the vibrationally excited state $v_8=1$. An overview of the \chcn\ transitions analyzed in this paper is given in Table~\ref{tab:transitions}. 

The middle panels of Fig.~\ref{fig:observations} show the \SgrB(M) peak intensity maps of three \chcn\ lines compared with the ALMA continuum emission (black contour; from Paper~II). The image maps of the continuum emission for the frequencies of each one of the selected \chcn\ transitions are shown in the top panels. A first comparison of the distribution of the dense gas and the continuum emission reveals that the \chcn\ emission is much more extended than the continuum emission, specially towards the south. Moreover, the local maxima in the line emission do not necessarily coincide with the position of the peaks in the continuum maps. This is clearly seen in the northern part of \SgrB(M) where a compact \chcn\ condensation appears shifted towards the north-east with respect to the continuum. The southern region exhibits a significantly brighter and more extended \chcn\ emission than towards the northern part.

In the bottom panels of Fig.~\ref{fig:observations}, we present spectra extracted towards selected positions indicated with crosses in the top and middle panels of the Figure. The spectra show the emission of the $K$ lines of the ground state \chcn\ transitions $J$=12--11 (left panel), $J$=13--12 (central panel) and $J$=14--13 (right panel). The spectra show the characteristic $K$-ladder of a symmetric top molecule, but also lines of other molecules that are partly blending the \chcn\ lines. For a better identification of which lines actually belong to \chcn, the reader is referred to Fig.~\ref{fig:spectra1} and \ref{fig:spectra2}, where the observational spectra are compared to the results of our simulations. Depending on the location within \SgrB(M), the regions show different spectral characteristics. The spectrum extracted from the northern region (Mn and red cross in Fig.~\ref{fig:observations}) has clear and well-defined lines all of them in emission. The southern position (Ms and blue cross in the Figure) has slightly broader lines shifted towards lower frequencies (i.e.\ red-shifted velocities). Finally, the spectra extracted towards the brightest continuum source (Mm and black cross in the Figure) is completely different to those obtained in the south and north. The central spectra show a complex combination of emission and absorption features belonging to \chcn. The absorption features are related to the strong continuum emission present in the central region, which at the depicted frequency is about 180~K. The emission of a channel map of the strongest absorption feature of \chcn\ ($J$=12--11, $K$=0) is shown in Fig.~\ref{fig:absorption}. The possible origin of the strong continuum emission that causes the absorption features is the presence of the numerous \hii\ regions in \SgrB(M). However, a comparison of the position and extent of the known \hii\ regions with the area of absorption shows that they only partly coincide (see black circles in Fig.~\ref{fig:absorption}). Moreover, the absorption feature is also present at regions with no \hii\ regions. On the contrary, the ALMA 242~GHz continuum emission, magenta contours in Fig.~\ref{fig:absorption}, clearly coincides with the area of absorption, indicating that bright dust continuum emission is the most probable origin of the absorption features found in the central region of \SgrB(M). A similar absorption produced against bright dust continuum is also found in the massive hot molecular core G31.41+0.31 \citep{cesaroni2017}.

\begin{figure}[t!]
	\centering
    \includegraphics[width=0.98\columnwidth]{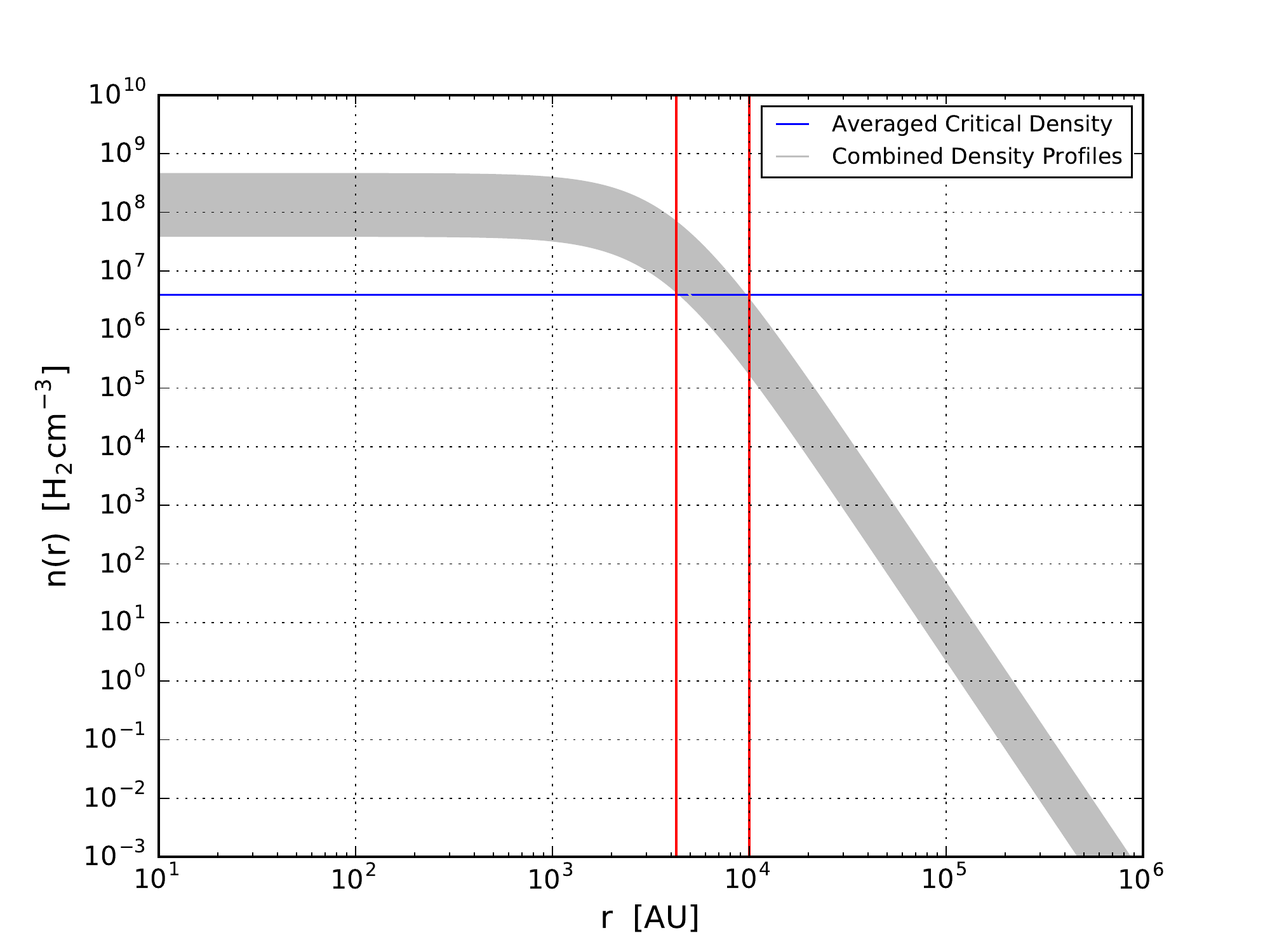}
\caption{Combined density profile (grey area) of all the dense cores included in the model of \SgrB(M). The average critical density of the transitions listed in Table~\ref{tab:transitions} is shown as an horizontal blue line. The vertical red lines at \unit[4.25\tee{3} and \ee{4}]{au} indicate up to which distance from the center of the cores the LTE approximation is valid.}
\label{fig:LTEapproximation}
\end{figure}

\section{Modeling procedure}\label{sec:ModelingProcedure}

We use the radiative transfer tool box \pandora\ (presented in Paper~I), which in one of its implementations can use the radiative transfer code RADMC-3D to model the \chcn\ line transitions in \SgrB(M). A detailed description of the framework \pandora\ and the RADMC-3D code can be found in Paper~I and \citet{dullemond2012}. In this section we focus on the new routines implemented in \pandora\ that are necessary for the line modeling and that were therefore not described in the previous paper.  

\subsection{Modeling \chcn\ with RADMC-3D}\label{ssec:RADMC3Dlines}

In order to solve the radiative transfer problem, RADMC-3D samples the density structure defined by the user by generating a grid via the method of adaptive mesh refinement. This means that parent cells are refined into children cells if the density gradient within the parent cell exceeds a certain value (details can be found in Paper~I). For modeling the \chcn\ line transitions towards \SgrB(M), we used a cubic grid with a side length of 1~pc centered around \SgrB(M). We set the allowed maximum density difference between the cells to 10\%. This configuration results into a minimum cell size of 100~au and roughly $10^{7}$ final cells. After the grid is generated the dust temperature is calculated self-consistently based on the luminosity sources by using a Monte Carlo approach \citep{bjorkman2001}.

In the last step of the radiative transfer simulations, we then use RADMC-3D to calculate the \chcn\ line emission. For this purpose, RADMC-3D first calculates the level populations for each cell in dependence of the gas temperature. We make two assumptions: (i) local thermodynamic equilibrium (LTE) conditions, i.e.\ the energy levels are populated according to the Boltzmann distribution and (ii) the gas temperature equals the dust temperature, which is a valid assumption in high-density regions like those found towards \SgrB(M).  

Due to the high densities present in the small scale structures of \SgrB(M), the LTE approximation is a valid assumption. This can be seen by calculating the critical density $n_\mathrm{crit}$, which is defined by the ratio of the Einstein $A_{ul}$ coefficient and the collisional rate coefficient $C_{ul}$ 
\begin{equation}
n_\mathrm{crit}=\frac{A_{ul}}{C_{ul}}.
\end{equation}

The collisional terms involve the square of the density, whereas the radiative terms only increase linearly with density. Thus, for densities much larger than $n_\mathrm{crit}$, the collisional terms become dominant and the level populations are Boltzmann distributed, meaning that the LTE approximation is valid. Because $A_{ul}\propto \nu^3$, transitions with higher frequencies need higher densities to be thermally excited. For the transitions within one K-ladder of \chcn this does not really matter because the transitions are closely spaced. We calculated the critical density of all $J$=12-11 transitions of the main isotopologue based on the values of the Einstein $A_{ul}$ coefficient and the collisional rate coefficients at 100~K from the LAMDA (Leiden Atomic and Molecular DAtabase; \citealt{schoeier2005}) database. For all transitions the critical density is on the order of $\sim$$10^{6}$~cm$^{-3}$. As the collisional rate coefficient depends on the temperature, also the critical density depends on the temperature. However, in the considered range of temperatures, the order of magnitude remains the same. If we compare now the critical densities with the densities in \SgrB, we find that the LTE approximation is valid within the small scale structure of \SgrB(M), for which the densities range from $4\times10^{7}$ to $5\times10^{8}$~cm$^{-3}$. The density profiles of the cores of \SgrB(M) are described by modified Plummer density profiles (see Paper~I and Sect.~\ref{sec:PhysicalModel}). In Fig.~\ref{fig:LTEapproximation} the modified Plummer density profiles used in the model are plotted together with the average of the critical densities. Depending on the exact parameters of the core, the LTE approximation is valid within $\sim$$10^{4}$~au from the center of the core. We note that this is only a lower limit, since the overall density is obtained as the superposition of the density profiles of all the cores, thus increasing the value of the density.

For calculating the line transitions in the LTE approximation, RADMC-3D needs information about the molecular energy levels and the transition parameters. The energy levels, transition frequencies and Einstein $A_{ul}$ coefficients for the ground state transitions were taken from the CDMS (Cologne Database for Molecular Spectroscopy) catalog \citep{mueller2005,endres2016}, while the data of the vibrationally excited state $v_8=1$ were taken from the JPL (Jet Propulsion Laboratory) database \citep{picket1998}. All these data are available via the Virtual Atomic and Molecular Data Centre (VAMDC).

\subsection{Post-processing with CASA}\label{ssec:CASApostprocessing}

The final step in the modeling procedure aims at conducting a proper comparison of the radiative transfer images to the observational maps obtained with ALMA. For this, we developed a post-processing routine that uses the software package CASA\footnote{In the original version of \pandora\ the post-processing is done with the software package MIRIAD. In order to compare with observational data obtained with ALMA, and processed with CASA, we developed this second alternative within \pandora\ for the post-processing.}. The routine uses the \textit{uv}-coverage of the ALMA observations to sample the simulated data cubes. This ensures that the post-processed files filter out and recover the same emission and spatial scales detectable in the observations. In a second step, we image the synthetic data cubes with the same imaging parameters used for the observational ALMA data. Finally, the post-processed synthetic data can be (visually) compared to the observational data searching for deviations that require a modification in the initial model. If so, the physical structure of the model (i.e.\ molecular density, velocity field; see Sect.~\ref{sec:PhysicalModel}) is modified. This cycle is iterated to increase the consistency of the simulated images with the observational maps.

\begin{figure*}
\centering
    \includegraphics[width=0.98\textwidth]{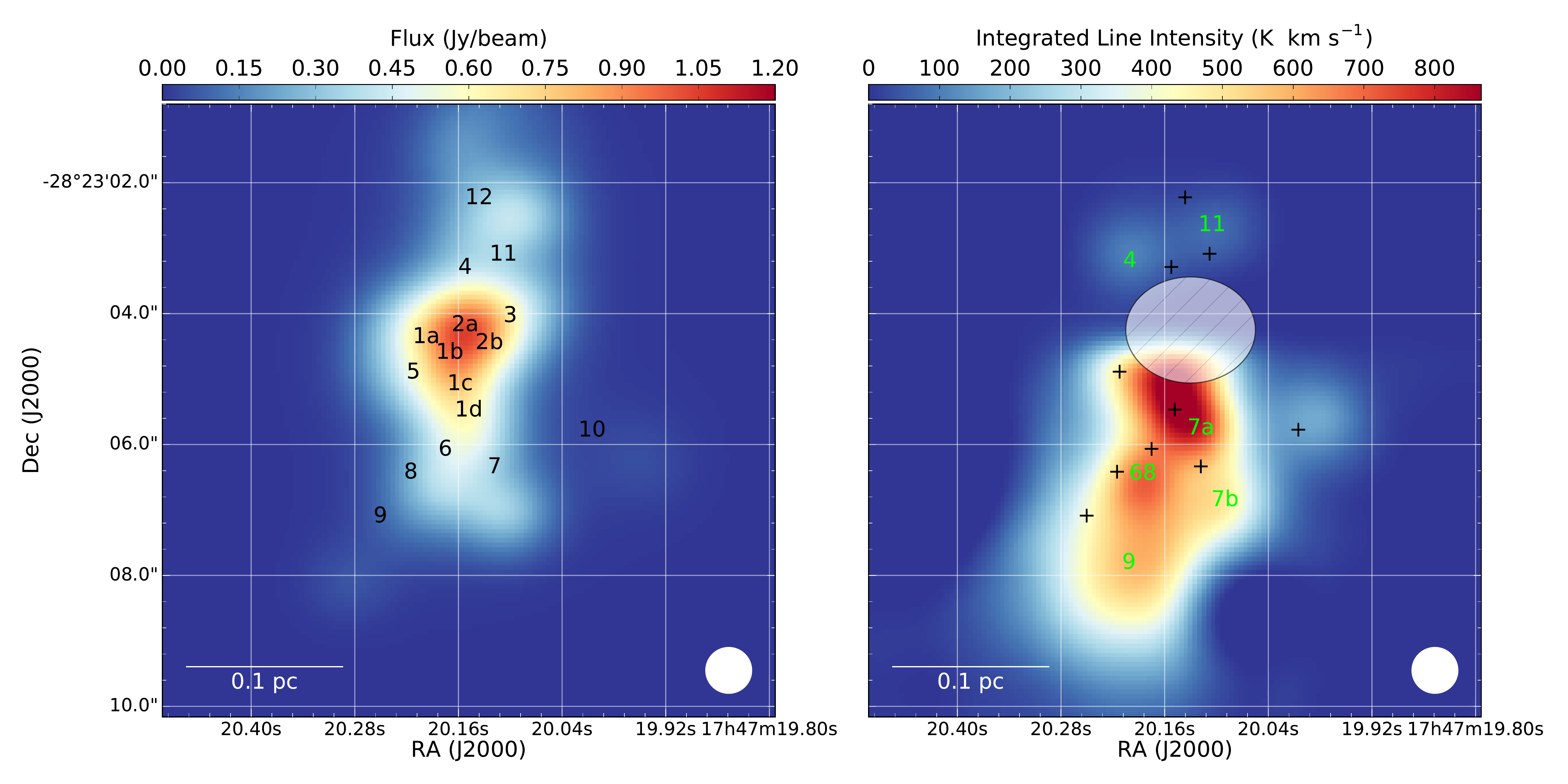}
\caption{\textit{Left panel}: Synthetic map of the continuum emission of \SgrB(M) at \unit[220]{GHz}. The black numbers indicate the positions of the dust cores from the physical model introduced in Paper~I. \textit{Right panel}: Integrated line intensity map of the $J$=12--11, $K$=3 transition predicted by the model. The black crosses indicate the positions of the dust cores as shown in the right panel. The green numbers indicate the position of the newly introduced molecular centers (see Sect.~\ref{ssec:ParametrizationMolecularDensity}). The central region is masked out by the transparent hatched area as the \chcn\ emission in this region could not be reproduced by the model (see Sect.~\ref{ssec:CentralRegion}).
}
\label{fig:orientation}
\end{figure*}

\section{The physical model}\label{sec:PhysicalModel}

The model used in this paper is based on the three-dimensional physical model of \SgrB\ developed in Paper~I. The model was created by modeling the dust and free-free continuum emission of \SgrB\ by employing the radiative transfer toolbox \pandora. The density distribution of the cloud complex was reconstructed by comparing the emission of the model to available observational maps, including both single-dish (e.g.\ from \textit{Herschel}/HiGAL, \citealt{molinari2010}; \textit{Herschel}/HIFI, \citealt{bergin2010}) and high-resolution interferometric images (e.g.\ VLA radio-continuum maps, \citealt{depree1998, rolffs2011}; SMA 345~GHz maps, \citealt{qin2011}) covering a frequency range from 1~GHz to 4.5~THz.

The continuum maps obtained with ALMA (see Paper~II) are compared with predictions from the model developed in Paper~I and presented in the Fig.~12 of Paper~II. We found a good correspondence between the predicted emission and the observed, with deviations of less than a factor of 2. Continuum maps at different frequencies together with spectral index\footnote{The spectral index $\alpha$ is determined as $S_\nu\propto\nu^\alpha$ where $S_\nu$ is the flux at a given frequency $\nu$. A comparison of the observed and synthetic spectral index maps is found in Fig.~13 of Paper~II} maps were also compared to individual continuum synthetic maps at the same frequencies and to spectral index synthetic maps showing a good agreement between them (see Paper~II). The main discrepancies were found to occur towards the central and brightest cores. The discrepancies can be due to different dust opacity coefficients throughout the \SgrB(M) region. In a following paper, we will present an updated model for the continuum emission of \SgrB, obtained by using high-angular resolution ALMA observations at frequencies from \unit[80]{GHz} up to \unit[275]{GHz}. In the current paper, we based our analysis on the physical model setup presented in Paper~I, and we model several \chcn\ line transitions observed with ALMA.

\subsection{The physical model of \cite{schmiedeke2016}}\label{ssec:PhysicalModel}

The physical model of \SgrB\ developed in Paper~I consists of dense (gas and dust) cores, stars and \hii\ regions. The model covers spatial scales from 45~pc down to 100~au. The density profiles of the individual cores are assumed to be modified Plummer functions given by
\begin{equation}
n_i(r)=\frac{n_c}{\left(1+|\vec{r}|^2\right)^{\eta/2}},
\end{equation}
where $n_c$ is the central density given in H$_2$ (cm$^{-3}$), and $|\vec{r}|$ is the Euclidean norm including scaling factors
\begin{equation}
\label{eq:SpatialScale}
|\vec{r}|=\sqrt{\left(\frac{r_x}{r_{0,x}}\right)^2+\left(\frac{r_y}{r_{0,y}}\right)^2+\left(\frac{r_z}{r_{0,z}}\right)^2},
\end{equation}
where $r_{x,y,z}$ are the coordinates and $r_{0,x}$, $r_{0,y}$ and $r_{0,z}$ set the size of the core in direction of each principal axis. The overall density structure is obtained by the superposition of the density profiles of all cores. The density in a cell j is then given by
\begin{equation}
n_j=\sum_{i=1}^{N} n_{i}(\vec{r}),
\end{equation}
where $i$ is the index of the dust core and $N$ the total number of dust cores. The parameters of the cores are described in Table~\ref{tab:cores}. The left panel of Fig.~\ref{fig:orientation} shows a synthetic map of the continuum emission at \unit[242]{GHz}. The black numbers indicate the central position of the dust cores included in the model (see Table~\ref{tab:cores}). The model presented in Paper~I, and describing all the scales from 100 au to 45 pc, also includes large-scale components to reproduce the extended envelope visible in the single-dish (and low-resolution) observations. The ALMA observations filter out the extended emission and are only sensitive to the small-scale dense cores.

All stars are assumed to be point sources, i.e.\ their radius is not taken into account. The model includes early-type high-mass stars and their \hii\ regions, which have been observed \citep[e.g.\ ][]{depree1998, rolffs2011}. The stars included in the model, based on the existence of \hii\ regions, account only for stars down to the B0 spectral type. All later spectral types with masses between 0.01~\msun\ and $\sim$19~\msun, i.e.\ stars which cannot produce \hii\ regions detectable with current observations, were placed by an algorithm that randomly draws stars from the initial mass function (IMF) of \cite{kroupa2001} and distributes the stars randomly throughout the computational domain based on the gravitational potential. The observed \hii\ regions created by massive stars are included in the model as Str\"omgen spheres, i.e.\ spherical structures with uniform electron density with no dust. A more detailed description of the model can be found in Paper~I. In the following section we describe how the molecular density and velocity fields (necessary for line modeling) are set up in \pandora. All other constituents of the model, i.e.\ stars, \hii\ regions, dense cores, introduced in Paper~I remain unchanged.

\subsection{Molecular density and velocity field}\label{ssec:AbundanceField}
  
With the aim of modeling the \chcn\ line emission, we implemented a molecular density and velocity field within \pandora\ that are added to the physical model retrieved from the continuum modeling.

The molecular density is set up based on the Plummer density cores described before. Within \pandora\ it is possible to define a global molecular abundance factor $\Omega$ for each species. Based on this global abundance factor the molecular density in a cell $j$ is obtained by multiplying the density in the cell $n_j$ by $\Omega$. However, due to the fact that the molecular abundance can vary significantly from position to position due to e.g.\ different temperatures, presence of a star, shocks, it is necessary to be able to modify the molecular density on a local level. For this reason, we implemented an option to define a local abundance factor $\omega_i$ for each Plummer density core. In summary, the molecular density $n(\chcn)_j$ in a cell $j$ is given by
\begin{equation}
n(\chcn)_j=\Omega\sum_{i=1}^{N} n_{i}(\vec{r})\omega_i,
\end{equation}
where $i$ is the index of the dust core, $N$ the total number of dust cores, $\Omega$ the global abundance factor and $\omega_i$ the local abundance factor of the i-th dust core. The abundance for each cell is then calculated by dividing the total molecular density by the total H$_2$ density of the cell.

Besides the molecular density we also include a velocity field in order to describe the line profiles. This velocity field includes the velocity along the line of sight as well as the line width due to turbulent motions of the gas.
Within \pandora\, the velocity along the line of sight for each position is determined by defining the velocity $v_i$ for each individual Plummer core. This enables the possibility of having different velocities at different positions. In order to ensure a smooth velocity field, the velocity of the gas $v_j$ in a cell $j$ is determined by weighting the contribution of each core by its density at this specific cell as
\begin{equation}
v_{j}=\frac{\sum_{i=1}^{N} v_i n_{i}(\vec{r})}{\sum_{i=1}^{N} n_{i}(\vec{r})}.
\end{equation}

This means that the resulting velocity is dominated by the value $v_i$ of the core with the highest density at the considered cell. The turbulent line width $\sigma_\textrm{linewidth}$ is set up equivalently. Besides the turbulent line width, the resulting line width is determined also by the thermal broadening. 

\begin{figure*}[t!]
\centering
\includegraphics[width=0.98\textwidth]{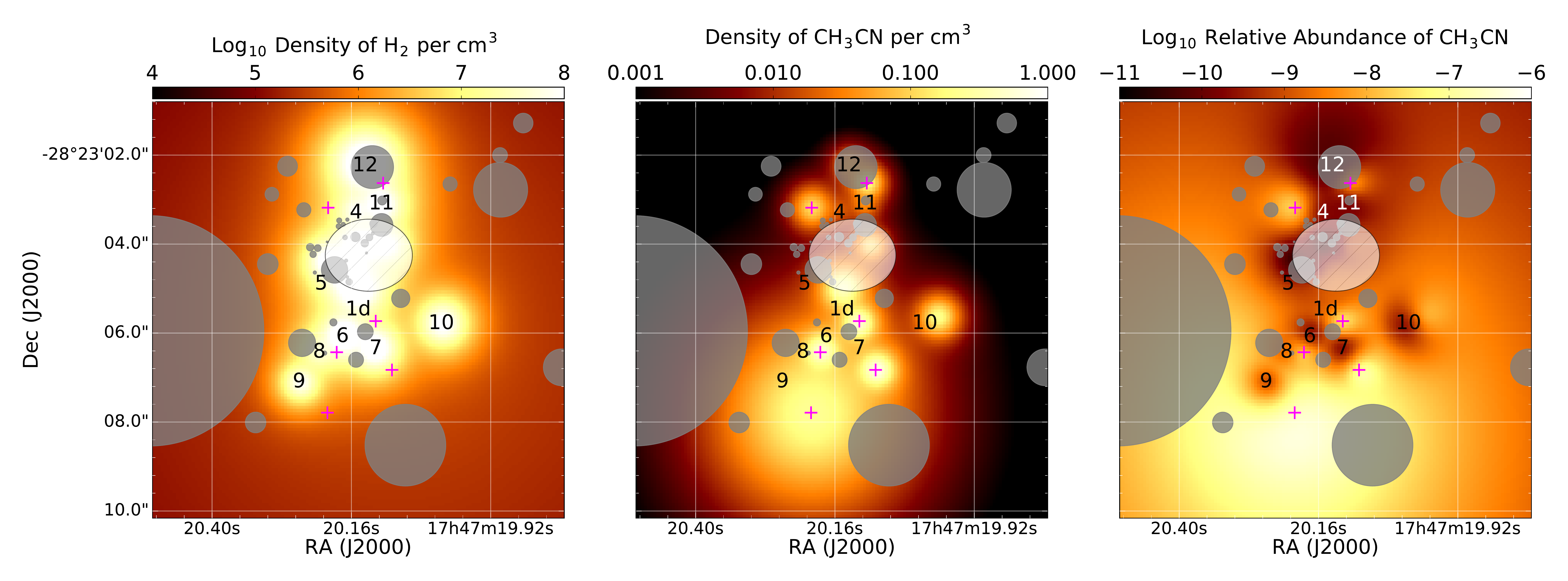}
\caption{Maps of the H$_2$ (\textit{left panel}) and \chcn\ (\textit{middle panel}) density and relative abundance of \chcn\ with respect to H$_2$ (\textit{right panel}). The black numbers show the position of the dust cores, and the magenta crosses indicate the position of the newly introduced molecular centers, as shown in Fig.~\ref{fig:orientation} (see also Table~\ref{tab:cores}). The gray circles indicate the position and extent of \hii\ regions, where the H$_2$ as well as the \chcn\ density is zero. The central region is masked out (transparent hatched area) as the \chcn\ transitions in this region can not be well reproduced with the model (see Sect.~\ref{ssec:CentralRegion}).}
\label{fig:density_map}
\end{figure*}

\subsection{Parametrization of the molecular density field}\label{ssec:ParametrizationMolecularDensity}

As described in Sect.~\ref{ssec:AbundanceField} the line transitions of \chcn\ were calculated by introducing an abundance field consisting of individual cores with Plummer-like density profiles. Although we based our \chcn\ abundance field on the density structure retrieved from the continuum emission by Paper~I, we had to introduce modifications. These modifications are necessary because the morphology of the emission of \chcn\ does not completely coincide with the continuum emission (see Fig.~\ref{fig:observations} and Sect.~\ref{sec:Results}) and therefore also not with the position of the dust cores.

Moreover, the emission of \chcn\ is much more extended than the continuum emission. We inspected the emission of the less abundant \isochcn\ isotopologue to determine if the shift between line and continuum emission might be produced by opacity effects. The same morphology is retrieved for the isotopologue than for the main species. This indicates that the \chcn\ abundance is in fact enhanced in the regions with local maxima and the morphology of the emission is not caused by opacity effects.

For this reason it is not possible to simply use the dust density structure retrieved from the continuum and define abundance factors in order to model the \chcn\ emission. Instead we also introduced new structures, which we call molecular centers (MC). These structures are located at the local maxima of the molecular line emission as seen in the CH3CN lines\footnote{We note that shifting the position of the MC by some pixels do not significantly vary the results of the model compared to the observations. This gives us some freedom in locating the MCs, but also reveals that there is no unique set of MCs that can reproduce the observations. Follow-up papers aimed at modeling other molecular species can help to set more constraints in the location and properties (both physical and chemical) of these structures.}.These molecular centers are diffuse in dust (H$_2$ density), in order to make sure that the continuum remains unchanged, but they possess a high \chcn\ abundance in order to reconstruct the local maxima of the line emission. The location of the molecular centers is shown in the right panel of Fig.~\ref{fig:orientation}. It is important to note that these newly introduced molecular centers are not necessarily real entities, but are a consequence of the current implementation of the abundance field parametrization in \pandora. By introducing these molecular centers, we are able to model the complex structure of the \chcn\ emission which does not coincide with the continuum emission. Hence, only the abundance field resulting from the superposition of the cores and molecular centers as depicted in Fig.~\ref{fig:density_map} has to be considered as a real existent physical structure.

The parameters from the continuum cores as well as the parameters of the newly introduced molecular centers are given in Table~\ref{tab:cores}. Each core and molecular center is described by 10 parameters: right ascension, declination, the displacement along the line of sight, the spatial scales as defined in Eq.~\ref{eq:SpatialScale}, the exponent of the Plummer function, the central density of H$_2$, the central density of \chcn\ and the velocity along the line of sight. For the cores which are located at the central region of \SgrB(M) (M-SMA-1a, M-SMA-1b, M-SMA-1c, M-SMA-1d, M-SMA-2a, M-SMA-2b) no central density of \chcn\ is given because the line emission of this region was not modeled (see Sect.~\ref{ssec:CentralRegion}).

In Fig.~\ref{fig:density_map} density maps of H$_2$ and \chcn as well as of the relative abundance of \chcn with respect to H$_2$ are depicted.

The relative abundance is considerably lower at the position of the continuum cores, where the H$_2$ density is the highest. The correlation between the low abundance of \chcn and the position of the continuum cores indicates that the chemistry at these very dense and hot regions is different and leads to the destruction of \chcn (see Sect.~\ref{ssec:Abundance}). 

\section{Analysis of simulated and observational data}\label{sec:Comparison}

In the following sections we compare the simulated data to the observational data. For this purpose, we first (Sects.~\ref{ssec:GeneralMorphology} and \ref{ssec:Spectra}) analyze the general morphology of the \chcn\ emission by comparing integrated intensity maps, and spectra taken from the center of the cores. In Sect.~\ref{ssec:TemperatureDistribution} we discuss the validity of the temperature distribution determined by RADMC-3D. The abundance and velocity field are discussed in Sects.~\ref{ssec:Abundance} and \ref{ssec:Velocities}. Afterwards, we discuss the observational data from the central region, and the limitations of the current model in this region (Sect.~\ref{ssec:CentralRegion}). In Sect.~\ref{ssec:Isotopologue} we present the results from the model compared to the observations for the isotopologue \isochcn.

\begin{figure*}[t!]
\begin{center}
   \includegraphics[width=0.98\textwidth]{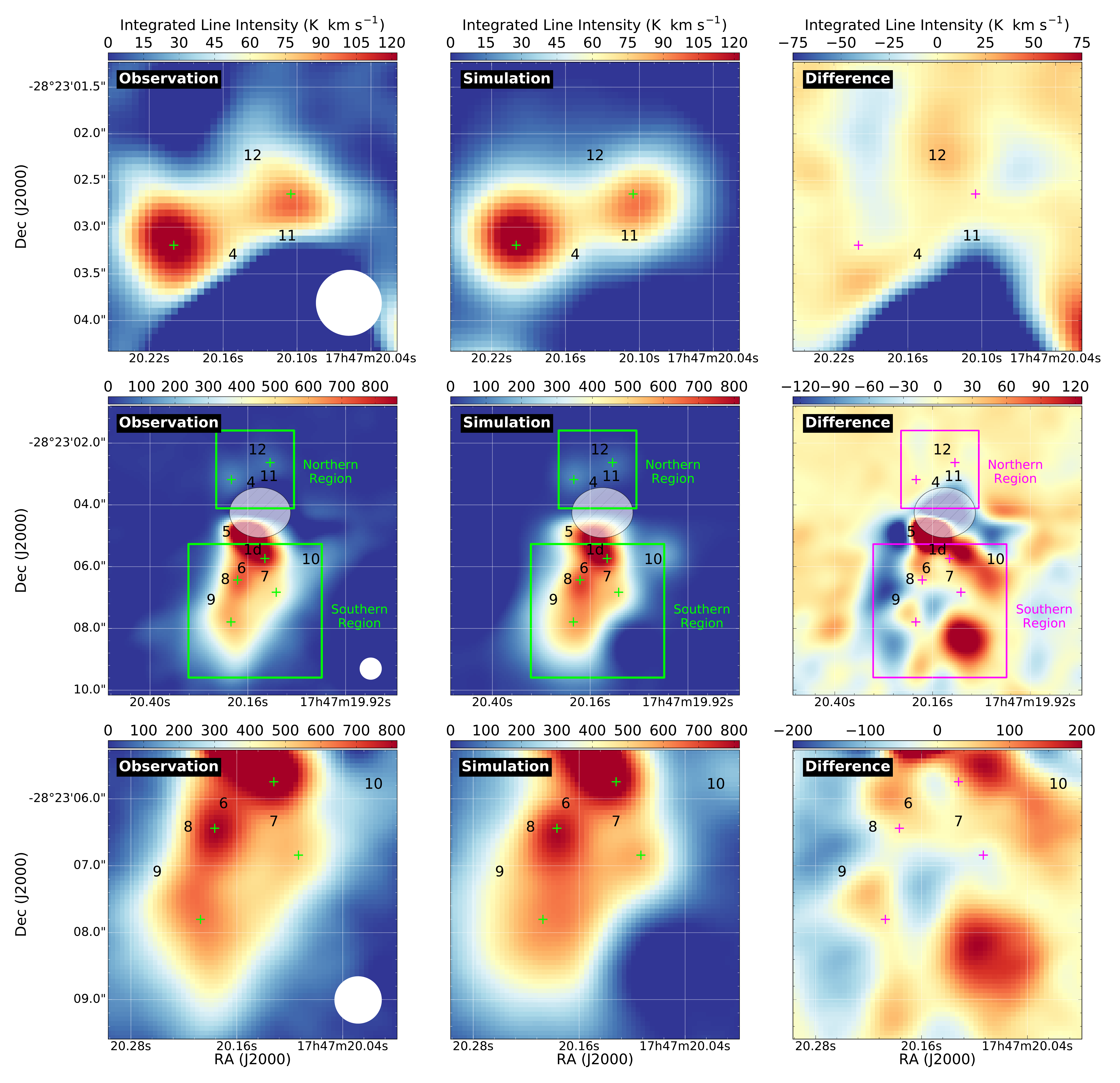}
\caption{Integrated intensity maps (or zeroth-order moment maps) of the observational (\textit{left column}) and simulated (\textit{middle column}) data for the \chcn\ $J$=12-11, $K$=3 transition. The \textit{right column} panels show the difference observation $-$ simulated data. The \textit{central row} shows the entire \SgrB(M) complex except for the central region, which is masked out with a transparent hatched area. The green  and magenta boxes indicate the zoomed-in regions that are depicted in the \textit{top row} (northern part of \SgrB(M)) and the \textit{bottom row} (southern part of \SgrB(M)). As in previous figures, the black numbers indicate the position of the dust cores, whereas the green and magenta crosses depict the position of the molecular centers (see Table~\ref{tab:cores}).}
\label{fig:observation_simulation}
\end{center}
\end{figure*}

\subsection{Distribution of the \chcn\ emission}\label{ssec:GeneralMorphology}

As already discussed in Sect.~\ref{ssec:AbundanceField}, the morphology of the \chcn\ emission does not completely coincide with the continuum emission. For this reason, it was necessary to introduce molecular centers with a high \chcn abundance in order to reconstruct the \chcn emission from the observations. The general morphology of the \chcn emission can be investigated by producing integrated intensity maps or zeroth-order moment maps following the expression
\begin{equation}
\mu(x,y)=\int \limits_{-\infty}^{\infty}  (F(x,y)-F_{\mathrm{\scriptsize cont}}(x,y)) d\nu,
\end{equation}
where $F(x,y)$ is the flux at position $(x,y)$ and $F_{\mathrm{\scriptsize cont}}$ the corresponding continuum flux. The zeroth-order moment maps of the $J$=12--11, $K$=3 transition obtained from the observations and the simulated model are shown in Fig.~\ref{fig:observation_simulation}. This line was used to calculate the zeroth-order moment maps because it is in most cases not affected by significant line blending of other species and it is one of the brightest lines. The comparison of the observed and simulated map shows that the emission obtained from the physical model exhibits in general the same morphology as the observed emission. It is evident that the emission indeed peaks at the position of the newly introduced molecular centers and not at the continuum cores. In the northern part we find two regions with a strong emission that do not coincide with the continuum cores. For this reason it was necessary to introduce two molecular centers MC-4 and MC-11 in order to account for the strong emission in those regions. The integrated line intensity in the northern part reaches values up to \unit[120]{K km s$^{-1}$}. This is significantly lower than in the southern part, where \chcn is more abundant and the emission peaks at around \unit[800]{K km s$^{-1}$}. The central part of \SgrB(M) exhibits a complex combination of \chcn\ emission and absorption features (cf.\ Fig.~\ref{fig:observations}) and shows at the considered frequency range of the depicted maps only absorption. In Sect.~\ref{ssec:CentralRegion} we discuss in more detail the modeling of the central region.
   
In the southern part four molecular centers were introduced to reconstruct the morphology of the \chcn emission and an additional molecular center was placed near the continuum core M-SMA-10. The molecular center MC-9 was introduced in order to account for the strong emission in the most southern part, where the \chcn emission is much more extended than the continuum. Furthermore the molecular center MC-68 was placed between the continuum cores M-SMA-6 and M-SMA-8 and two further molecular centers MC-7a and MC-7b were placed north and south of the continuum core M-SMA-7, respectively.

\begin{figure*}[!t]
   \resizebox{\hsize}{!}
   {\includegraphics[scale=0.05]{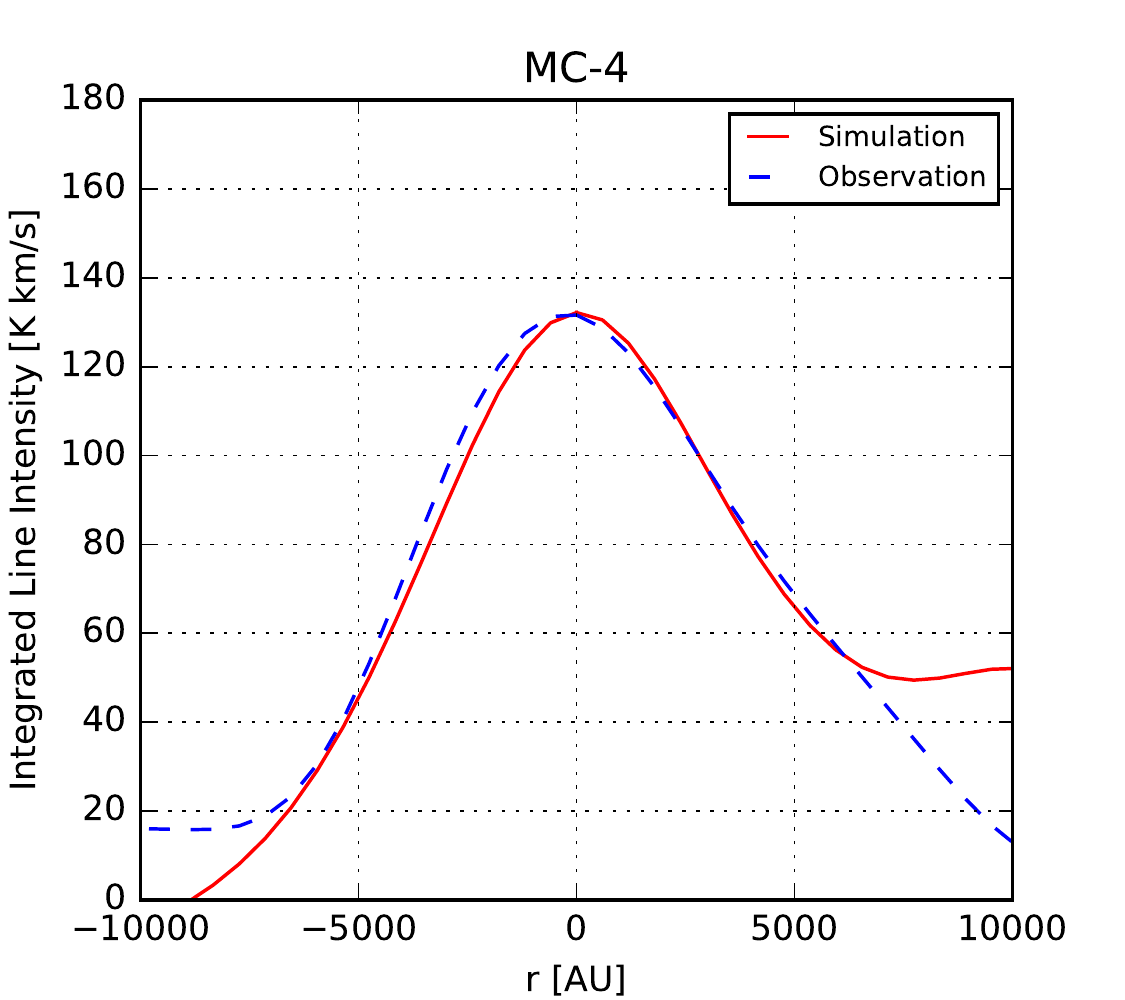}
    \includegraphics[scale=0.05]{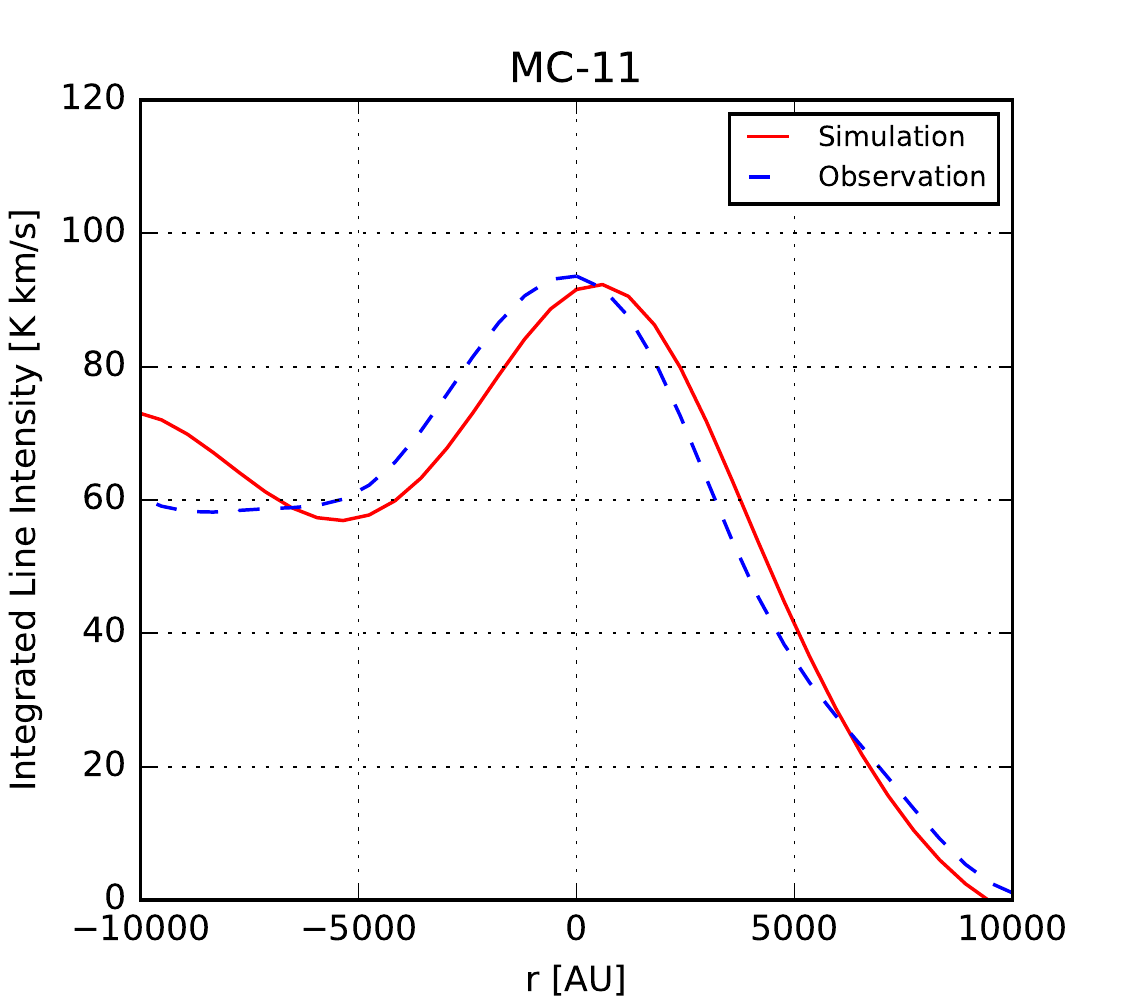}}
   \resizebox{\hsize}{!}
   {\includegraphics[scale=0.05]{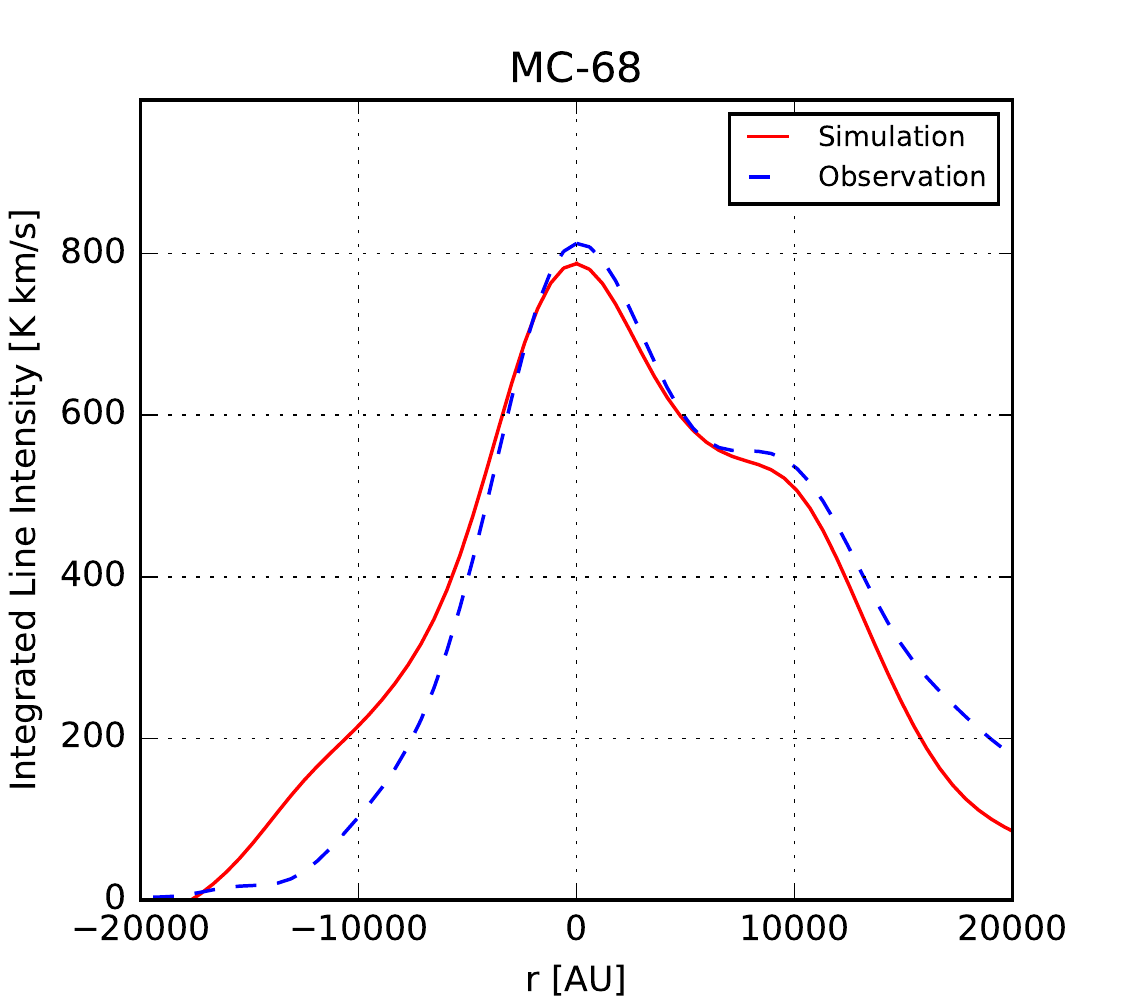}
    \includegraphics[scale=0.05]{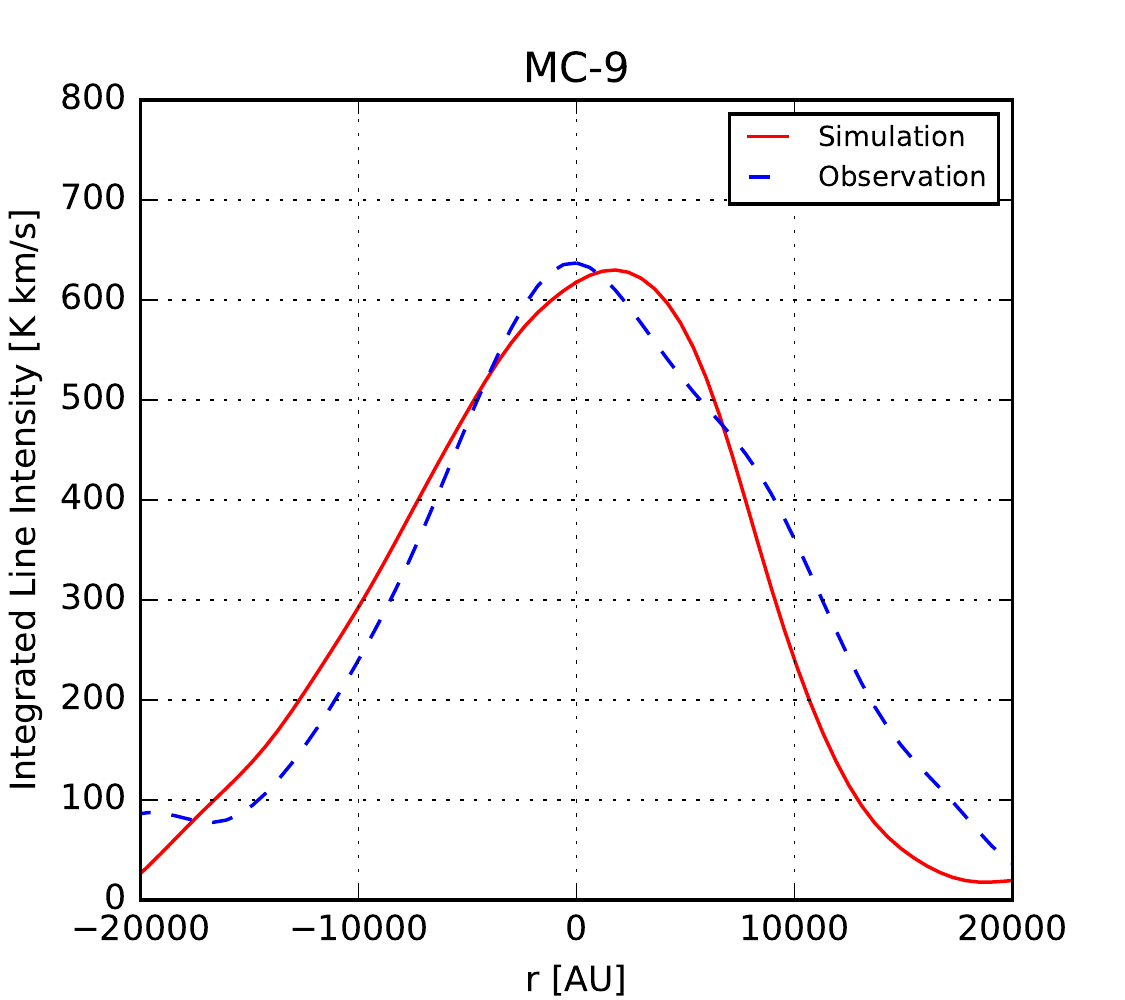}}
\caption{Radial profiles of the integrated line intensity for the \chcn\ $J$=12--11, $K$=3 transition. The radial profiles proceed from west to east and $r=0$ corresponds to the center of the considered molecular centers. The first two profiles (MC-4 and MC-11) are taken from the newly introduced molecular centers in the northern region (cf.\ Fig.~\ref{fig:orientation}). The third profile (MC-68) is taken from the molecular center located between the continuum cores M-SMA-6 and M-SMA-8. The last profile (MC-9) is centered around the molecular center south to the continuum core M-SMA-9. In all panels the blue dashed line shows the observational data, whereas the red solid line corresponds to the simulated data.}
\label{fig:radial_profiles}
\end{figure*}

From the difference map of the observed and simulated data, we find a good agreement in regions where the \chcn\ emission is strong and decreases in regions with weaker emission. This behavior can be more clearly seen by investigating radial profiles of the integrated line intensity. Radial profiles taken from the center of different molecular centers are shown in Fig.~\ref{fig:radial_profiles}. The radial profiles proceed in all cases from west to east. These cuts illustrate that the morphology of the \chcn emission can be described considerably well with modified Plummer-like functions. However, due to the modeling of the cores as spherical structures deviations from the real structures with a certain degree of asymmetry are inevitable. Additionally radial profiles are useful tools to identify local maxima in the emission, e.g.\ the radial profile centered around molecular center MC-68 reveals that there is an additional peak in the integrated intensity further to the east, which is associated with the molecular center MC-7b. 

\subsection{Spectra of the main isotopologue}\label{ssec:Spectra}

In this section we compare the spectra of the observational and simulated data taken from the center of the different cores and molecular centers. On the basis of the transitions presented in this section, we reconstructed the abundance field of \chcn (see Sect~\ref{ssec:Abundance}). The rotational spectrum of a symmetric-top molecule like CH$_3$CN consists of a series of components (determined by the quantum number $K$) for each rotational transitions (determined by the quantum number $J$). This $K$-ladder structure results in multiple lines with different energies that are closely spaced in frequency. Several spectra taken from the center of different cores and molecular centers are shown in Figs.~\ref{fig:spectra1} and \ref{fig:spectra2}. The intensity of the spectra is given in brightness temperatures. The displayed spectra are solely taken from cores and molecular centers located at the northern or southern regions of \SgrB(M) as shown in Fig.~\ref{fig:observation_simulation}, while the spectra of the central region are discussed in Sect.~\ref{ssec:CentralRegion}.

The spectra show the frequency range from \unit[220.55 to 220.80]{GHz}, which includes the first seven $K$-components of the $J$=12--11 transition of the main isotopologue. The $K=0$ and $K=1$ are spaced very closely in frequency and therefore often appear as just one line. The observational data also contain higher $K$-components, but they are weak and sometimes blended with other species. Spectra of the $J$=13--12 and $J$=14--13 transitions can be found in Figs.~\ref{fig:spectra_J=13-12_1} and \ref{fig:spectra_J=14-13_2}. The $K=3$ component, i.e.\ the line with the fourth highest transition frequency, was used to determine the zeroth-order moment maps discussed in Sect.~\ref{ssec:GeneralMorphology} and presented in Fig.~\ref{fig:observation_simulation}. The intensity  of this line is relatively high because the $K=3$ state belongs to the ortho-configuration of \chcn\ with a higher statistical weight. 

The intensity of the simulated lines is in  agreement with the observational spectra. The agreement is very good for the first $K$-components up to $K=3$ and decreases for higher components. However, it is important to point out that the deviations between observational and simulated data are partly caused by line blending. Line blending describes the possibility that spectral lines are interfered by lines of other molecular species or unidentified features and poses a major difficulty in analyzing and comparing the intensities and widths of observed spectral lines. \SgrB\ exhibits an extraordinarily rich chemistry and is composed of various kinds of molecules. As a consequence, the number of detectable species and spectral lines is large and some of the \chcn\ lines are indeed blended by other lines. For example, the $K=5$ component is strongly blended by two spectral features, one of them being brighter than the \chcn\ transition itself. For this reason the intensity of this component in the simulated data is usually smaller when compared to the observational data. The $K=4$ transition is also blended with at least another strong line, but not as drastic as the $K=5$ component. The significance of the line blending effects depend also on the position of the map. At locations close to the center of \SgrB(M), e.g.\ the center of the molecular center MC-7a, the line blending effects are very strong due to the high line density. The spectrum of the molecular center MC-7a shows that at this position the intensity even does not return to the continuum level between the individual lines. In contrast, at positions further away from the dense center, e.g.\ core M-SMA-12, the individual lines are more isolated and line blending effects are less significant. Taking the blending into account would require detailed radiative transfer modeling of all species, which is way beyond the scope of this paper.

Molecules that blend with the $J$=12--11 transitions are among others HNCO and SO$_2$. The $J$=10--9, $K_a$=1, $K_c$=9--8 transition of HNCO has a rest frequency of \unit[220.58475]{GHz}. This strong line is located very close to the $J$=12--11, $K=6$ transition with \unit[220.593987]{GHz} and is causing significant line blending due to its higher intensity. SO$_2$ has transitions at \unit[220.59714]{GHz} and \unit[220.618499]{GHz}. The later could be the origin of the line detected between the $K=6$ and $K=5$ transitions of \chcn. CH$_3$NH$_2$ has three transitions, which are detected, at rest frequencies of 220.76060, 220.78077 and \unit[220.80552]{GHz}. The blending lines were identified using the software package XCLASS \citep{moeller2017} that makes use of the CDMS and JPL catalogues. A full identification of the molecular content in the \SgrB(M) will be presented in a forthcoming paper (M\"oller et al.\ in prep).

Nonetheless, there are also $K$-components like $K=4$ which are not affected significantly by line blending, but the line intensity is still underestimated by the model. This indicates that the gas temperature, which is assumed to be equal to the dust temperature, is too low and therefore the higher $K$-transitions, which exhibit higher energies, are underpopulated. A detailed discussion of the temperature distribution and the possible reasons for the discrepancy is presented in Sect.~\ref{ssec:TemperatureDistribution}.

Moreover, the spectra emphasize that although the continuum emission is brighter at the position of the continuum cores than at the newly introduced molecular centers, the continuum-subtracted line emission is much stronger at the position of the molecular centers. This can be seen by comparing the spectra towards the molecular center MC-9 and the close dust core M-SMA-9. Also noticeable are the spectra from M-SMA-10 and MC-10, which are very close in space. These regions have relatively small line widths, but the lines exhibit extended wings, which could be related to the presence of a high-velocity outflow.

Finally, there is a qualitative difference between spectra from the northern and the southern regions of \SgrB(M). The spectra from the southern region only show emission features, whereas the northern regions have a combination of emission and absorption features. The absorptions, which are overestimated by the model, are likely to be produced by the presence of \hii\ regions (numerous in the northern and central regions of \SgrB(M), see Fig.~\ref{fig:absorption}) that act as bright background continuum sources.

\begin{figure}[t!]
\begin{center}
\begin{tabular}[b]{c}
        \includegraphics[width=0.98\columnwidth]{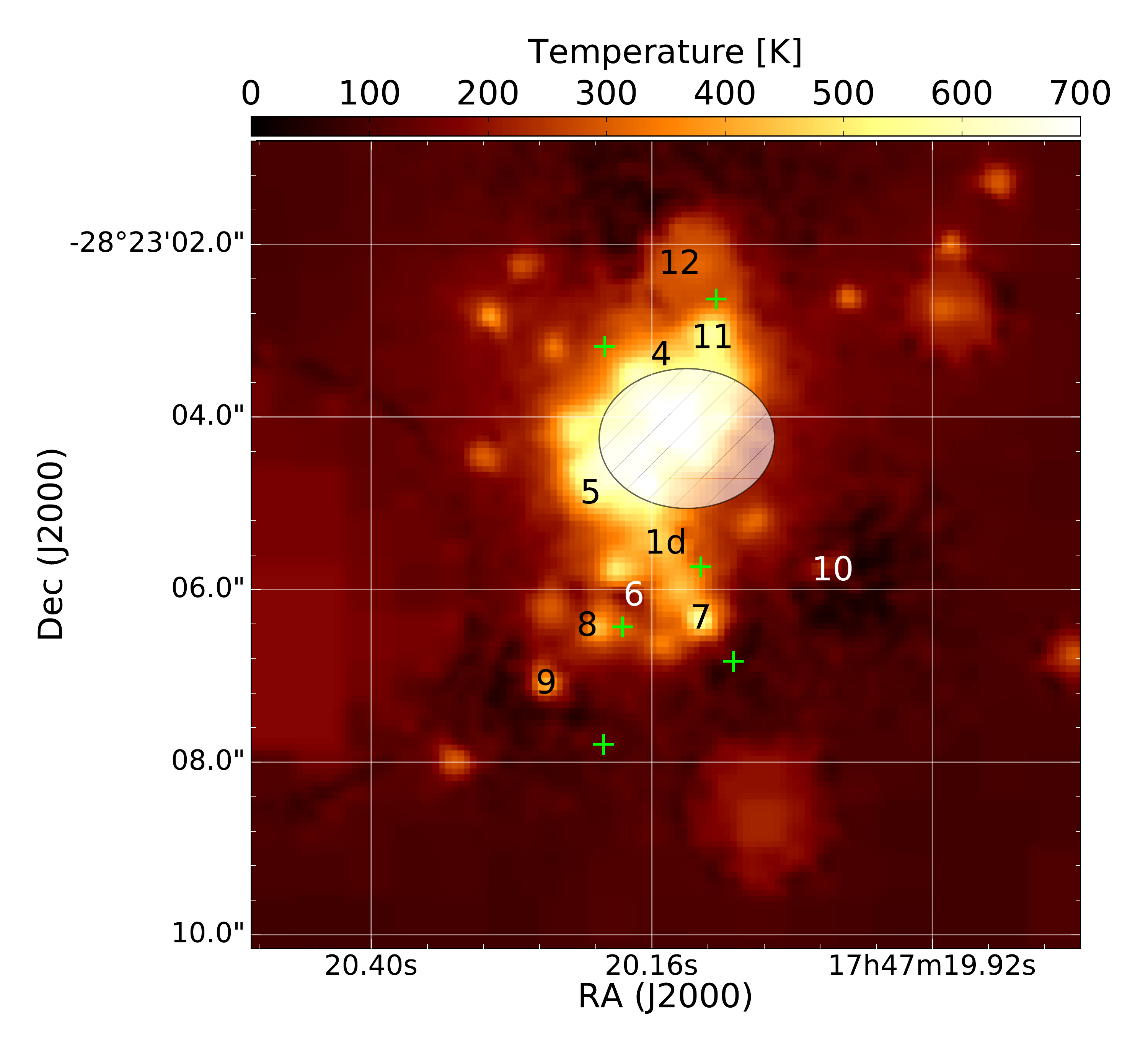} \\
\end{tabular}
\caption{Temperature distribution at the $z=0$ plane for \SgrB(M) as produced by the model presented in Sect.~\ref{sec:PhysicalModel}.The central region, which is discussed in Sect~\ref{ssec:CentralRegion}, is shown with a transparent, hatched area.}
\label{fig:temp_distribution}
\end{center}
\end{figure}

\subsection{Temperature distribution}\label{ssec:TemperatureDistribution}

The temperature of the gas is a crucial ingredient in shaping the spectral features of the line emission of a molecule. In the LTE approximation the temperature solely determines the level populations and thereby affects the intensity of transitions. In our model the dust temperature is calculated self-consistently by RADMC-3D based on the available luminosity produced by the stars of the cluster. We assumed the gas temperature to be equal to the dust temperature, since at these high densities one expects them to be coupled \citep[e.g.\ ][]{goldsmith2001}.  

The temperature distribution determined by RADMC-3D for \SgrB(M) at the $z=0$ plane is shown in Fig.~\ref{fig:temp_distribution}. The temperature ranges from around 50 to \unit[900]{K}, with the hottest regions located at the position of the continuum cores. The spherical structure in the right lower corner of the southern part is associated with an \hii\ region. A comparison of Fig.~\ref{fig:temp_distribution} and \ref{fig:density_map} shows that the hot temperature regions coincide with the continuum cores, where the relative abundance of \chcn is relatively low.  

In Sect.~\ref{ssec:Spectra}, we compared the observational and simulated results of the $J$=12--11 transition. Although, for most of the lines there is a good agreement between simulation and observation, the higher $K$-transitions are underestimated in some cases. This is partly caused by line blending, so that the line intensities appear to be stronger. However, there are also $K$-transitions that are apparently unaffected by line blending, but nonetheless show an underestimated intensity. This may suggest that the gas temperature in the model is too low. As described in Sect.~\ref{sec:ModelingProcedure}, the temperature is calculated based on the stellar luminosity, and it does not include a possible contribution from accretion events. Furthermore, the gas temperature can also be increased by other heating mechanisms, for example with a mechanical origin like turbulent dissipation or outflow shocks that are not included in the model, or local X-rays which are not considered at all. For this reason, the temperature distribution should be regarded as a lower limit for the gas temperature. Another possibility for the underestimation of the excitation is that it is affected by radiative pumping from external dust radiation, which might involve the vibrationally excited level \citep[e.g.\ ][]{hauschildt93}.  This effect can produce deviations from LTE, if the radiation temperatures differs from the local gas kinetic temperature, even if that is coupled to the local dust temperature.

\begin{figure*}
   \resizebox{\hsize}{!}
   {\includegraphics[scale=0.36]{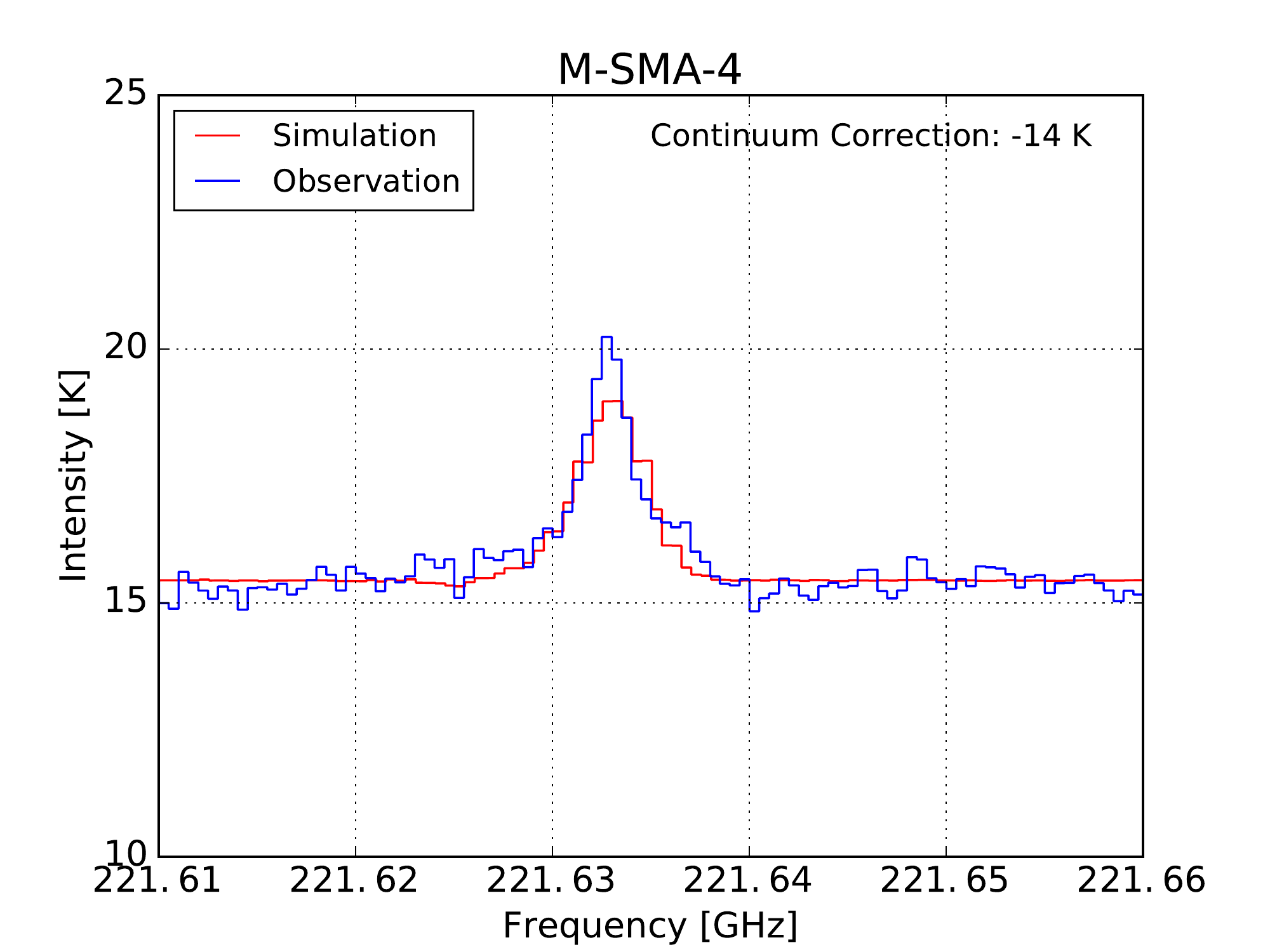}
    \includegraphics[scale=0.36]{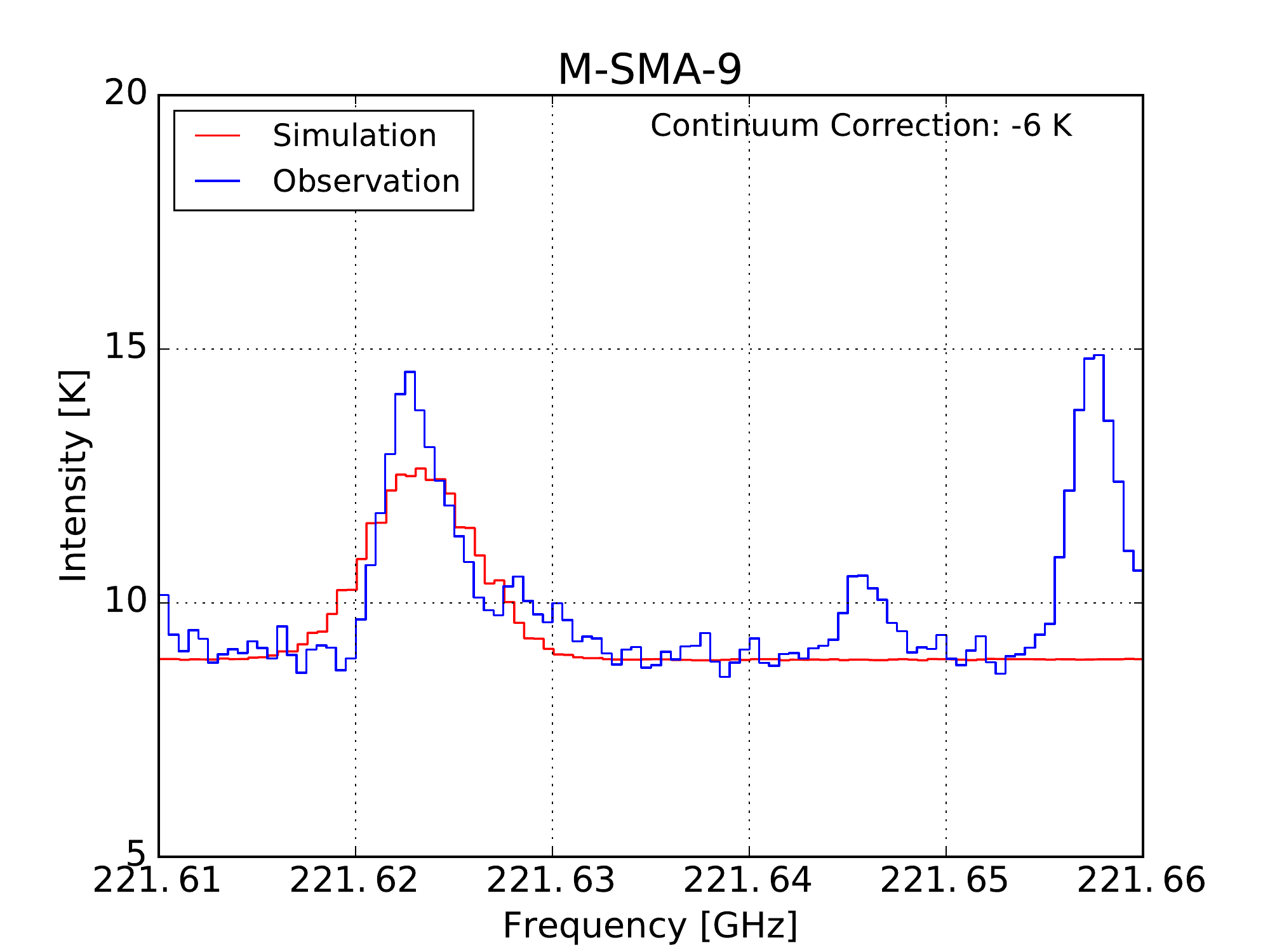}
    }
   \resizebox{\hsize}{!}
   {\includegraphics[scale=0.36]{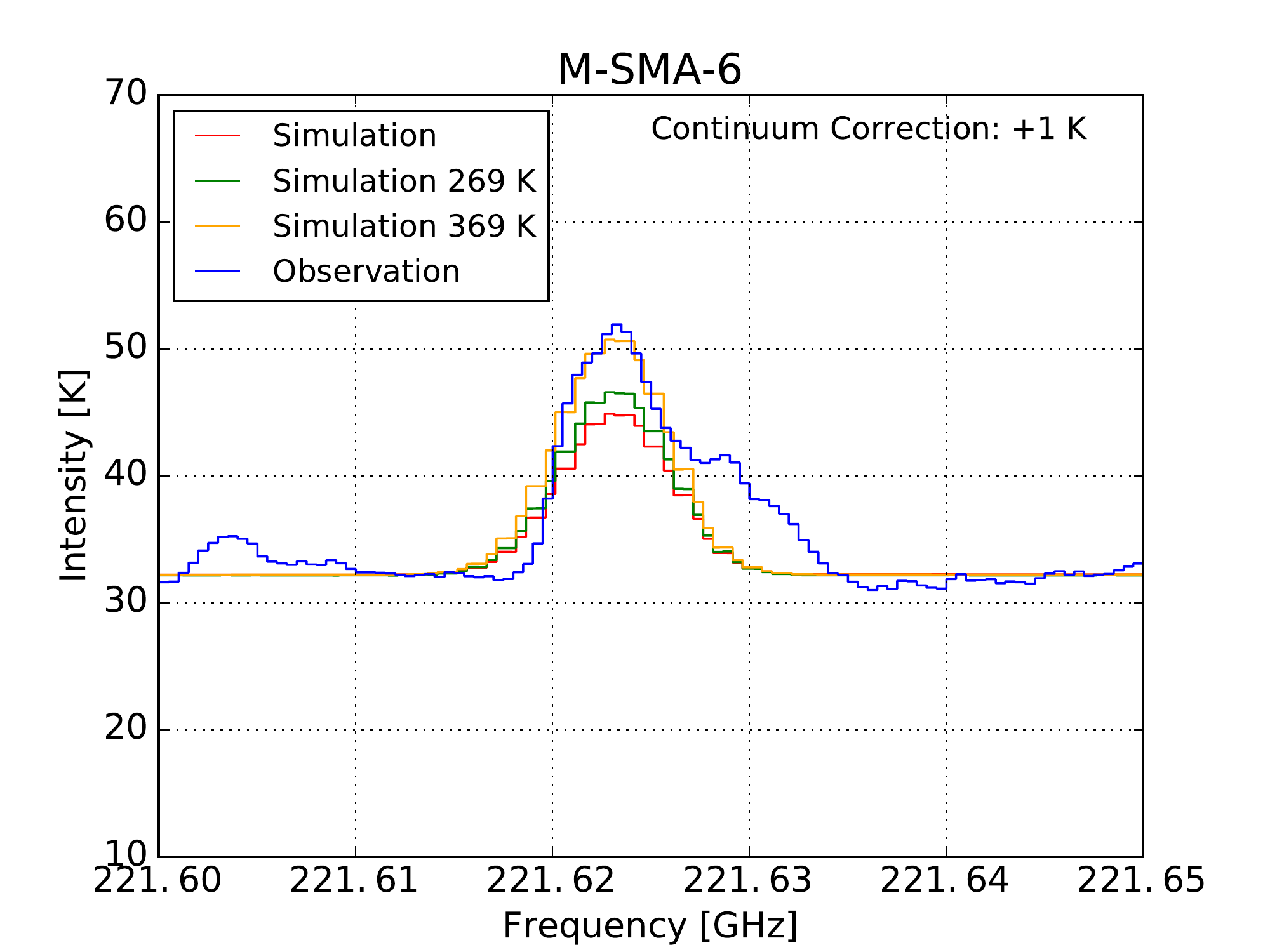}
   \includegraphics[scale=0.36]{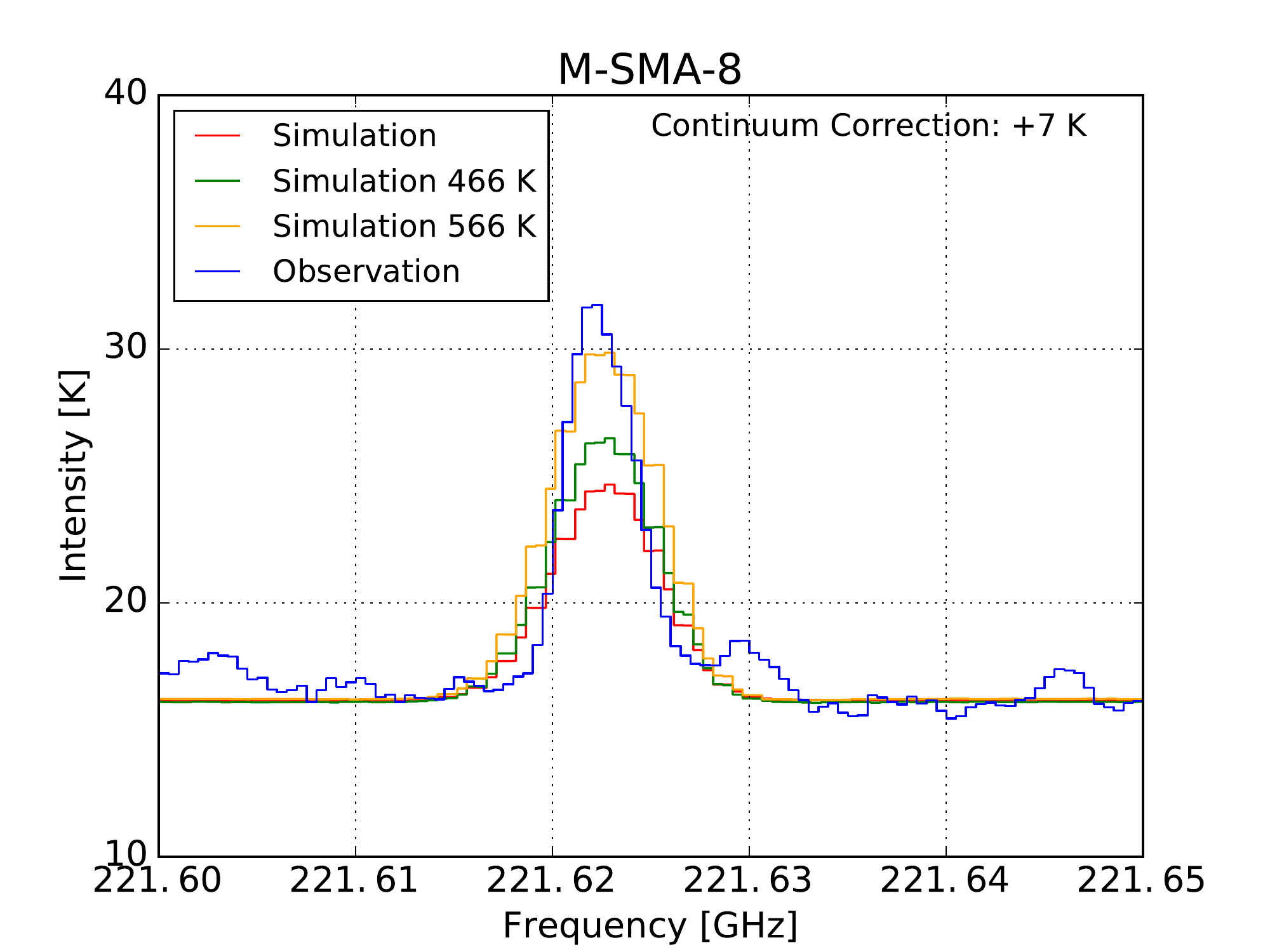}
   }
\caption{Spectra of the vibrationally excited transition $\nu_8=1$, $J$=12--11, $K=1-(-1)$ taken from the center of different cores. The blue solid line shows the observational data, while the red solid lines show the simulated data when the gas temperature equals the dust temperature. The bottom panels additionally show the simulated spectra after modifying the gas temperature by \unit[50]{K} (green) and \unit[150]{GHz} (yellow), as discussed in Sect.~\ref{ssec:TemperatureDistribution}. The continuum correction necessary to match the continuum level of the observation and simulation is indicated in the top-right side of each panel.}
\label{fig:vib_spectra}
\end{figure*}

The states involved in the transitions considered in Sect.~\ref{ssec:Spectra} have energies in the range from approximately 60 up to \unit[325]{K}, so that most of the states have relatively low energies. In order to investigate the validity of the temperature distribution, we have additionally analyzed higher energy transitions from the \chcn\ vibrationally excited state $v_8=1$. In particular, the transition $\nu_8=1$, $J$=12--11 and $K=1-(-1)$, with a rest frequency of \unit[221.62582]{GHz}. The involved states have energies of \unit[577]{K} and \unit[588]{K} and hence primarily trace the hot gas. Fig.~\ref{fig:vib_spectra} shows observational spectra as well as the prediction by our model for the vibrationally excited transition. Further spectra are shown in Figs.~\ref{fig:spectra_vib_appendix_1} to \ref{fig:spectra_vib_appendix_2}. At the center of most cores and molecular centers the vibrationally excited transition is clearly detected. However, there are also regions, where the transition is hardly visible, e.g.\ core M-SMA-12. In most cases the intensity of the transition is only slightly underestimated by our model (see for example cores M-SMA-4 and M-SMA-9), indicating that the calculated temperature is consistent with the observations. However, other regions like cores M-SMA-6 and M-SMA-8 have observed lines for the vibrationally excited transition brighter than the prediction of the model.
   
In order to investigate by how much the gas temperature is underestimated, we produce different models in which the gas temperature is increased compared to the dust temperature. The results of these tests are shown in Fig.~\ref{fig:vib_spectra} for cores M-SMA-6 and M-SMA-8. The figures show the spectra of the vibrationally excited transition for the temperature calculated by RADMC-3D and for two runs where the gas temperature was increased by \unit[50]{GHz} and \unit[150]{K}. While the simulated intensities using the dust temperature as the gas temperature are below the observed intensities, the intensity of the lines in the simulated data is in good agreement for both cores when considering an increase of \unit[150]{K} for the gas temperature. This suggests that the gas temperature is not coupled to the dust temperature and should be higher in some regions, although it is not drastically underestimated.
 
\begin{figure}[t!]
\begin{center}
\begin{tabular}[b]{c}
        \includegraphics[width=\hsize]{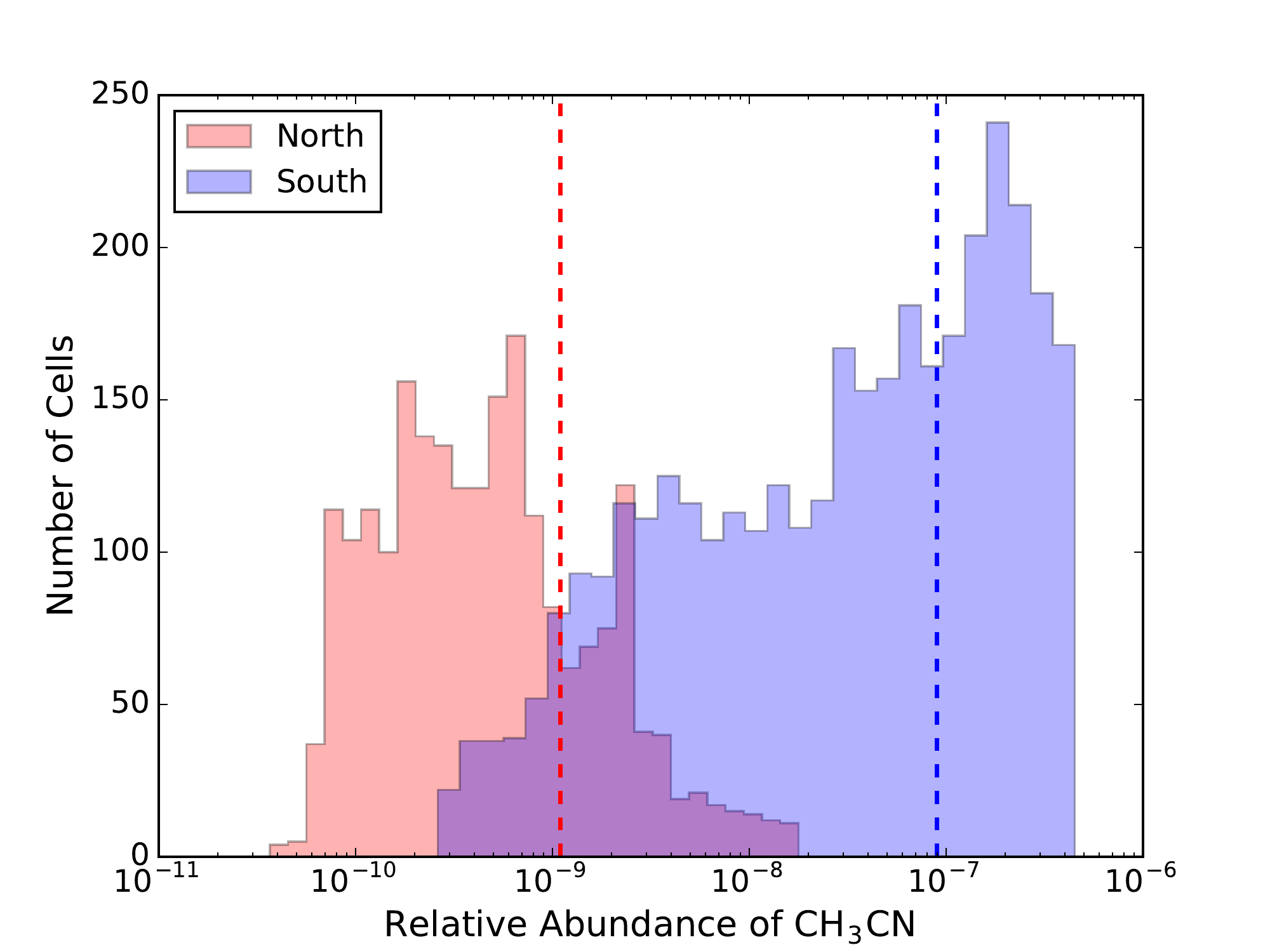} \\
\end{tabular}
\caption{Histogram of the distribution of the relative abundance of \chcn for the northern (red) and southern (blue) region of Sgr B2(M). For this purpose cubes were placed around the northern and southern part, which were composed of cells with the side length of 695 AU. The vertical lines indicate the averaged values.}
\label{fig:histogram_abundance}
\end{center}
\end{figure}
   
\begin{figure}[t!]
\begin{center}
\begin{tabular}[b]{c}
        \includegraphics[width=\hsize]{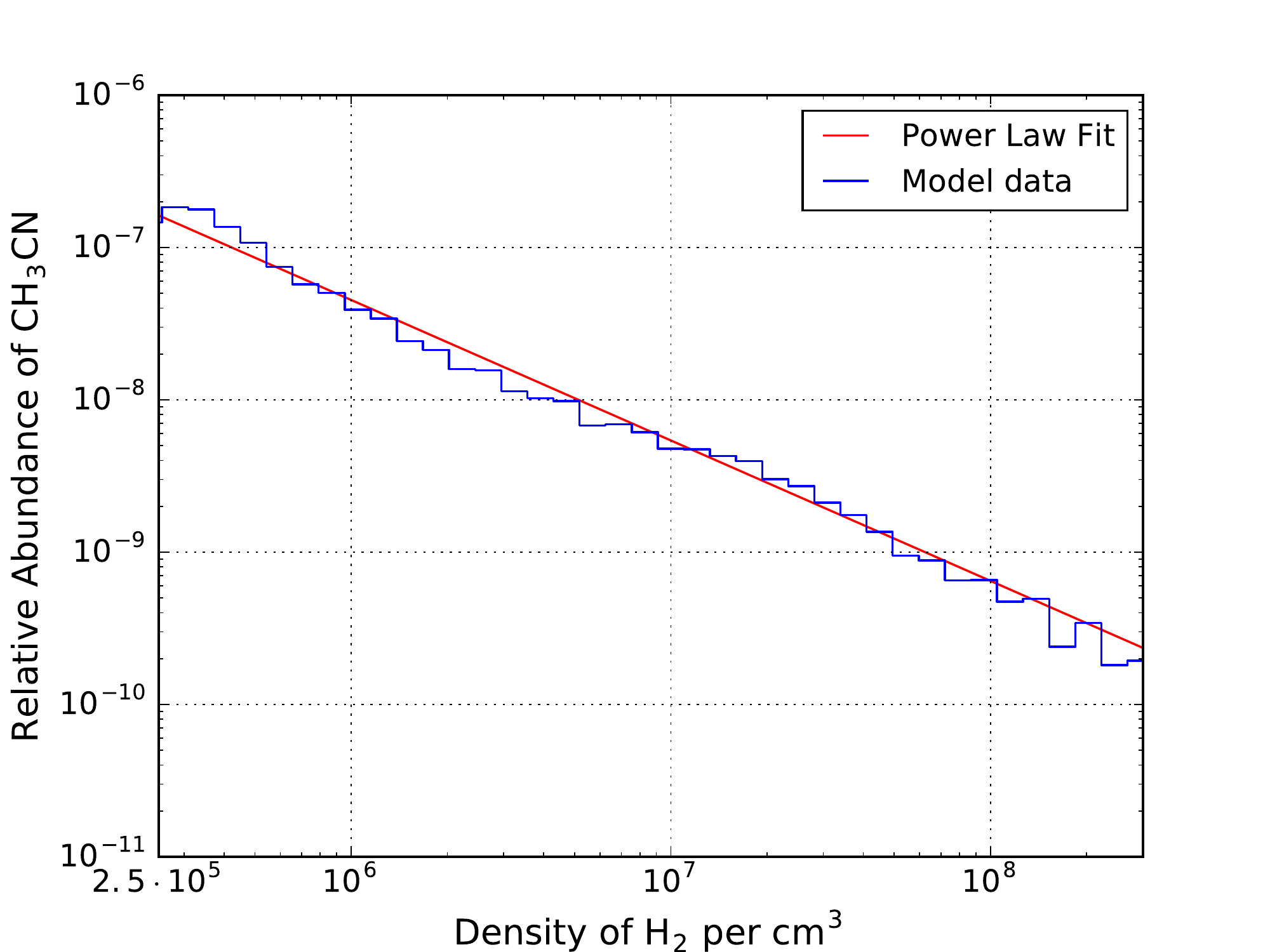} \\
\end{tabular}
\caption{Histogram of the relative abundance of \chcn\ as function of the H$_2$ density based on the developed physical model. The \chcn\ abundances decrease with the H$_2$ density with a function that can be described as a power lay with an exponent of $-0.92$.}
\label{fig:density_abundance}
\end{center}
\end{figure}
   
\begin{figure}[t!]
\begin{center}
\begin{tabular}[b]{c}
        \includegraphics[width=\hsize]{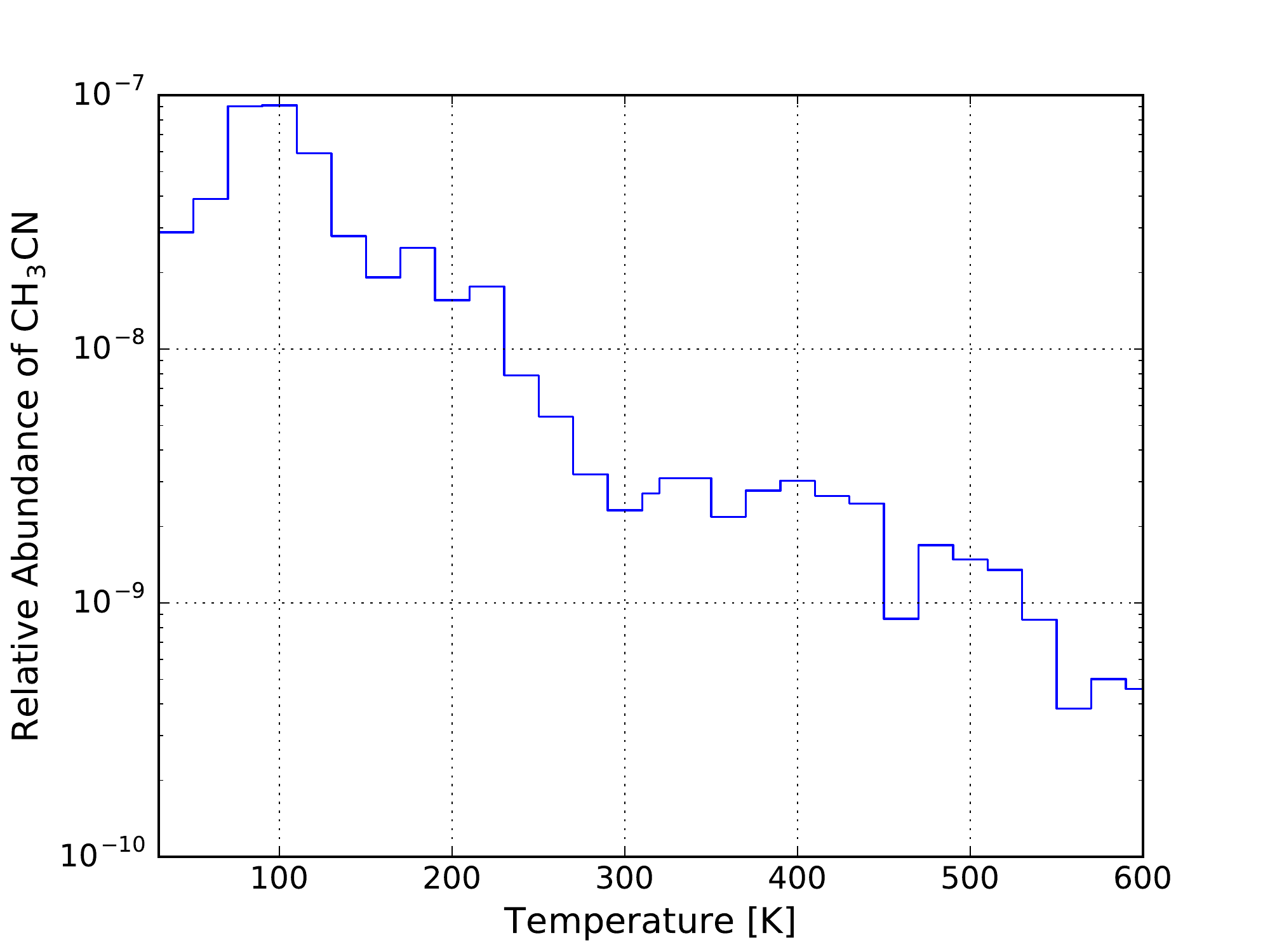} \\
\end{tabular}
\caption{Histogram of the relative abundance of \chcn as function of the temperature.}
\label{fig:temp_abundance}
\end{center}
\end{figure}

\subsection{\chcn\ abundance}\label{ssec:Abundance}   
   
In this section we discuss the abundance distribution of \chcn\ in \SgrB(M). The relative abundance of \chcn\ varies by three orders of magnitude within the northern and the southern regions of \SgrB(M). A histogram of the pixel distribution of the relative abundance (\chcn\ with respect to H$_2$) can be seen in Fig.~\ref{fig:histogram_abundance}. For the northern region the relative abundance ranges from 4\tee{-11} to 2\tee{-8} with an averaged value of 1.1\tee{-9}. In the southern region the relative abundance ranges from 2\tee{-10} to 5\tee{-7} with an averaged value of 9.0\tee{-8}. The histogram illustrates that the abundance of \chcn\ is significantly higher in the southern region compared to the northern region. This may be related to the stronger feedback destroying \chcn\ in the northern regions, where more \hii\ regions are present. A previous study of the \chcn\ emission in \SgrB\ derive an average abundance for the hot cores of $\sim$ 7\tee{-11} \citep{devicente1997}, which is much lower than our estimated value. However, the angular resolution of the observations by \citet{devicente1997} ($\sim$10--30\arcsec) is coarser and most likely smears out the \chcn\ emission associated with small scale structures that are resolved in our data (with a resolution of $0\farcs7$). 

In Fig.~\ref{fig:density_map} wee can see a anti-correlation between the \chcn\ abundance and the position of the dense cores with large H$_2$ densities. The anti-correlation between H$_2$ and \chcn\ is better shown in Fig.~\ref{fig:density_abundance}, where we show the relation between the relative abundance of \chcn\ as a function of the H$_2$ density. The relative abundance of \chcn decreases with increasing density according to a power law with an exponent of $-0.92$. We find a similar behavior for the \chcn\ abundance as a function of the temperature (see Fig.~\ref{fig:temp_abundance}). In the figure we show how the relative \chcn\ abundance decreases with increasing temperature. The relative \chcn\ abundance is high at regions with temperatures around 150 to \unit[250]{K}, and decreases by one or two orders of magnitude for temperatures in the range 300--500~K.

All this together suggest the existence of chemical processes in the dense regions that lead to the destruction of \chcn. Although this needs further investigation that will be done in forthcoming papers using chemical models, we indicate some aspects that may help to understand the anticorrelations shown in Figs.~\ref{fig:density_abundance} and \ref{fig:temp_abundance}. First, the temperature itself has an important impact on the chemistry because it determines the efficiency of chemical reactions especially on the dust grains. Second, the radiation field of the stars, which are evolving inside the dense cores, affects the chemistry by photo-dissociation reactions and ionizations and might reduce the abundance of \chcn\ at the central part of these cores. Chemical models that investigated the chemistry of nitrogen-bearing molecules including \chcn\ are conducted by \citet{rodgers2001}. In their models they assumed a density of \unit[\ee{7}]{H$_2$~\qcm} and constant temperatures in the range of 100--300~K. Additionally they considered different ice compositions. Depending on the time evolution, they find abundance values of \chcn\ ranging from \ee{-9} to \ee{-7} for a temperature of \unit[300]{K}. We find an averaged abundance of roughly 3\tee{-9} at \unit[300]{K}, which in their model would correspond to a time of \unit[2\tee{4}]{yrs}. For a temperature of \unit[100]{K} they find significantly lower abundance values ranging from \ee{-9} to only \ee{-8}. In contrast to our results which show a decreasing \chcn\ abundance with increasing temperature, they seem to find an opposite behavior. However, in these models they do not take into account photo-dissociation reactions. If these reactions are included, they point out that the abundance of \chcn drops by a factor of 100 after \unit[5\tee{5}]{yrs}, even if only cosmic-ray-induced photo-dissociation is included. Their results show that \chcn\ can be destroyed effectively by photo-dissociation reactions. Hence, as many continuum cores in \SgrB(M) already contain evolving stars that have a radiation field, this could be the reason for the low abundance of \chcn\ at the center of these cores. In the future it will be interesting to investigate if chemical models of hot cores with the parameters (e.g.\ density, radiation field) as those found towards the cores of \SgrB(M) can reproduce the radial distribution of \chcn. 

\begin{figure*}
   	\resizebox{\hsize}{!}
   	{\includegraphics[scale=0.36]{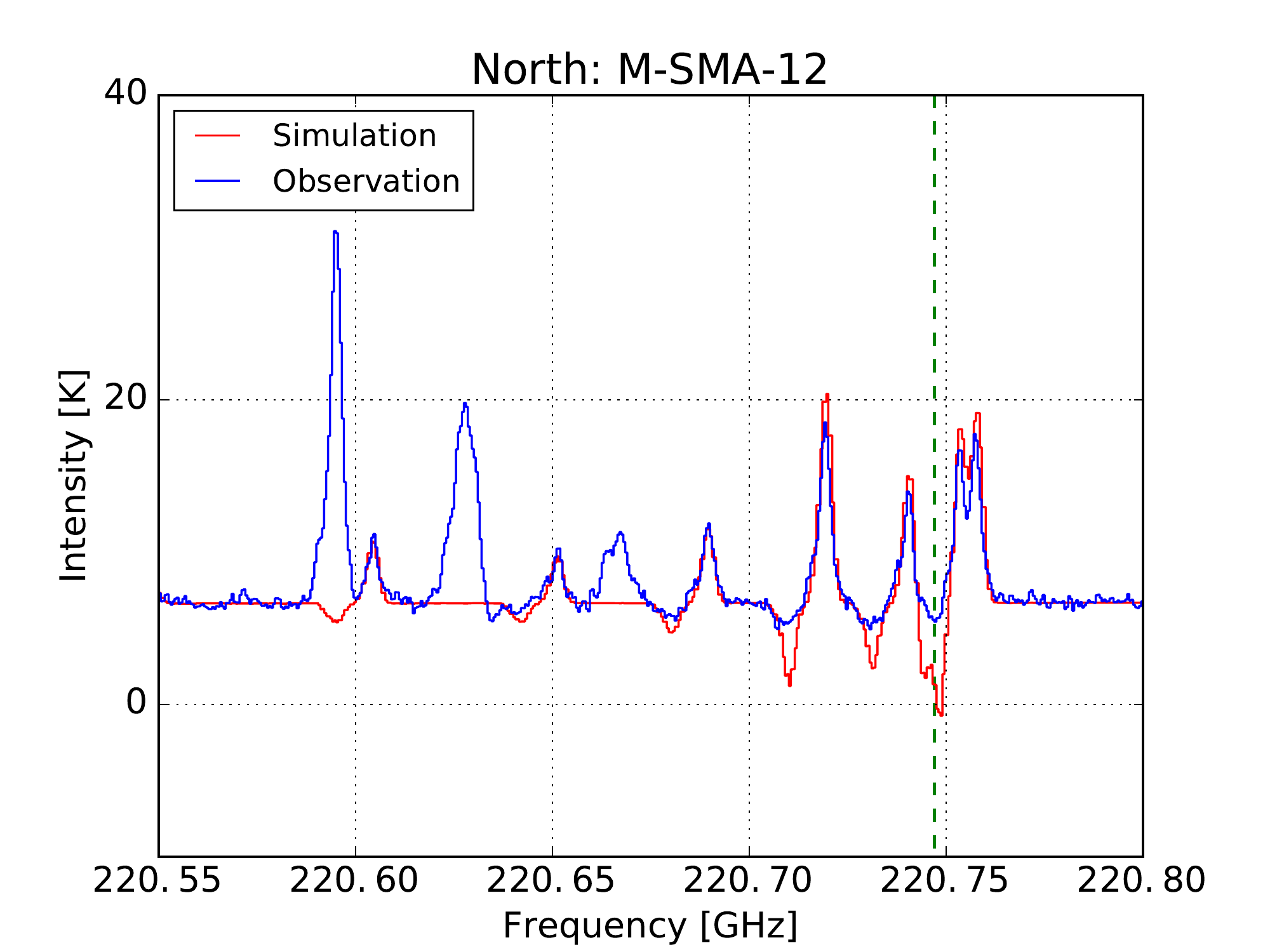}
     \includegraphics[scale=0.36]{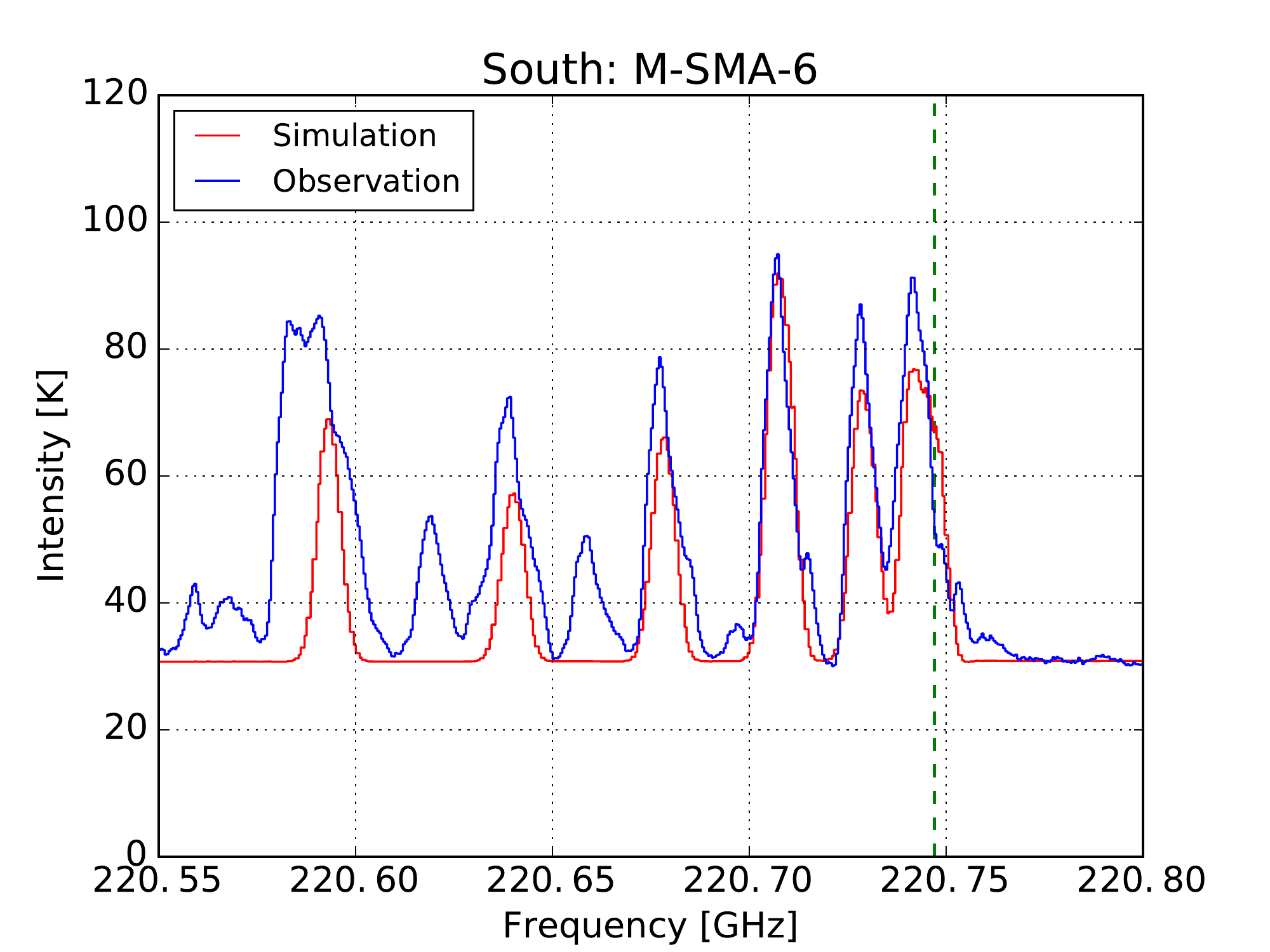}}
\caption{Spectra taken from a northern (M-SMA-12) and a southern (M-SMA-6) core to illustrate the shift in frequency due to the velocity component along the line of sight. The green, dashed vertical lines indicate the rest frequency of the $J$=12--11, $K=0$ transition. The core M-SMA-12 exhibits a blue-shifted velocity along the line of sight of \unit[50]{km~s$^{-1}$}, while core M-SMA-6 has a red-shifted velocity of \unit[67]{km~s$^{-1}$}. The systemic velocity of \SgrB(M) is about \unit[64]{km~s$^{-1}$}.}
\label{fig:velocity_spectra}
\end{figure*} 

\begin{figure}[t!]
\begin{center}
\begin{tabular}[b]{c}
        \includegraphics[width=0.98\columnwidth]{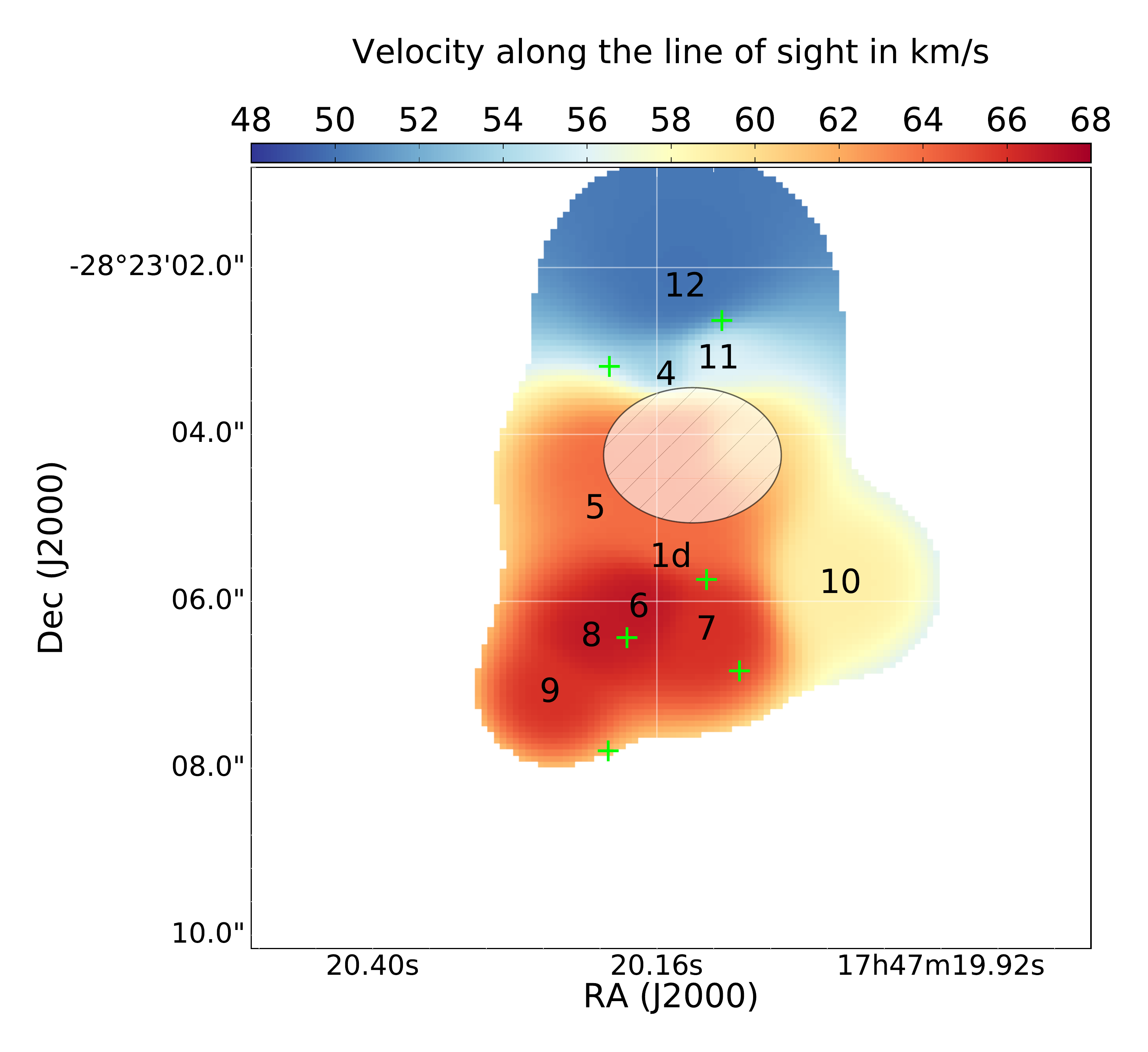} \\
\end{tabular}
\caption{Distribution of the velocity along the line of sight at the $z=0$ plane as derived from the model of \SgrB(M). The central region is masked out by a transparent, hatched area. The black numbers and green crosses indicate the dense cores and molecular centers, respectively, as in Fig.~\ref{fig:orientation}.}
\label{fig:velocity_map}
\end{center}
\end{figure}

\subsection{Core velocities and turbulent line width}\label{ssec:Velocities}
 
The analysis of spectral lines does not only provide information about the temperature and the density of certain molecular species, but also provides information about the velocity structure of a cloud and its internal turbulent motion. The component of the velocity along the line of sight of individual cores can be obtained by the shift of the spectral line away from the rest frequency due to the Doppler effect. Information about the internal turbulent motion can be obtained by analyzing the line shape of spectral lines. 

The spectra shown in Fig.~\ref{fig:velocity_spectra} show the difference in velocity for the emission in different cores. For a northern core like N-SMA-12 (left panel in the figure) the lines are clearly shifted to high frequencies (i.e.\ blue-shifted) with respect to the rest frequency. On the contrary, a southern core like M-SMA-6 has the lines shifted towards lower frequencies (i.e.\ red-shifted). The different velocities of the cores can be seen in the values listed in Table~\ref{tab:cores}. Figure~\ref{fig:velocity_map} shows a velocity (along the line of sight) map of the \chcn\ emission in \SgrB(M) as obtained from the model. Southern regions are clearly red-shifted with respect to the northern regions. This suggests an overall velocity gradient from north (blue-shifted velocities around \unit[50]{km~s$^{-1}$}) to the south (red-shifted velocities around \unit[70]{km~s$^{-1}$}).

The width of a spectral line is determined by the thermal broadening, together with non-thermal motions like e.g.\ turbulence and global scale motions unresolved in our beam size. The line width $\sigma_{\textrm{\scriptsize linewidth}}$ as calculated by RADMC-3D in \pandora\ includes a turbulent and a thermal component as
\begin{equation}
\sigma_{\textrm{\scriptsize linewidth}}^2=\sigma_{\textrm{\scriptsize turb}}^2+\frac{2kT_{\textrm{\scriptsize gas}}}{\mu m_{\textrm{\scriptsize H}}},
\end{equation}
where $\sigma_{\textrm{\scriptsize turb}}$ is the (micro)turbulent line width, $T_{\textrm{\scriptsize gas}}$ the gas temperature, $k$ the Boltzmann constant, $\mu$ the mean molecular weight and $m_{\textrm{\scriptsize H}}$ the mass of an hydrogen atom. In our model of \SgrB(M) we find differences for the turbulent line width between the northern and southern regions. From the depicted spectra in Fig.~\ref{fig:velocity_spectra} (see also Fig.~\ref{fig:observations}) the northern cores (e.g.\ core N-SMA-12) have a narrower line width compared to the southern cores (e.g.\ core N-SMA-6). All southern cores and molecular centers are modeled with a turbulent line width of \unit[5]{km~s$^{-1}$}, except for M-SMA-10 and MC-10 for which we used \unit[2.5]{km~s$^{-1}$}. In contrast, all the northern cores and molecular centers (MC-4, MC-11, M-SMA-4, M-SMA-11, M-SMA-12) exhibit a turbulent line width of only \unit[2.5]{km~s$^{-1}$}. The central region exhibits broad line widths similar to the southern region. This indicates that the internal turbulent motions are larger in the southern and central regions of \SgrB(M) compared to the northern region.

\subsection{\chcn features at the central region}\label{ssec:CentralRegion}

So far we have only presented the simulated and observational data from the northern and southern regions of \SgrB(M) as selected in Fig.~\ref{fig:observation_simulation}, which except for a few cases only show \chcn\ spectral features in emission. The central region exhibits much more complex features of \chcn. Spectra taken from the center of cores, which are located in the central region of \SgrB(M), are presented in Fig.~\ref{fig:center_spectra} and exhibit a combination of emission and absorption features.

The reason for these complex spectra is that the continuum emission is very strong at the center of \SgrB(M). As a consequence, the \chcn\ between the observer and the sources leads to deep absorption features in the spectra. Due to the fact that many \hii\ regions are present in the center of \SgrB(M), one might think that they are the reason for the deep absorption features. However, the positions of the known \hii\ regions do not entirely coincide with the spatial extent of the observed absorption lines, indicating that rather the dust emission is the cause (see Fig.~\ref{fig:absorption}).

As already mentioned, the physical model presented in this paper is not able to reproduce the complex spectra at the center of \SgrB(M). We investigated various modifications to the model in order to retrieve the behavior at the center, but could not find a model that reproduces all features of the \chcn\ transitions. Nonetheless, we give an overview of these models and briefly discuss where they failed to reproduce the observed features. 

\begin{figure*}
   	\resizebox{\hsize}{!}
    {\includegraphics[scale=0.36]{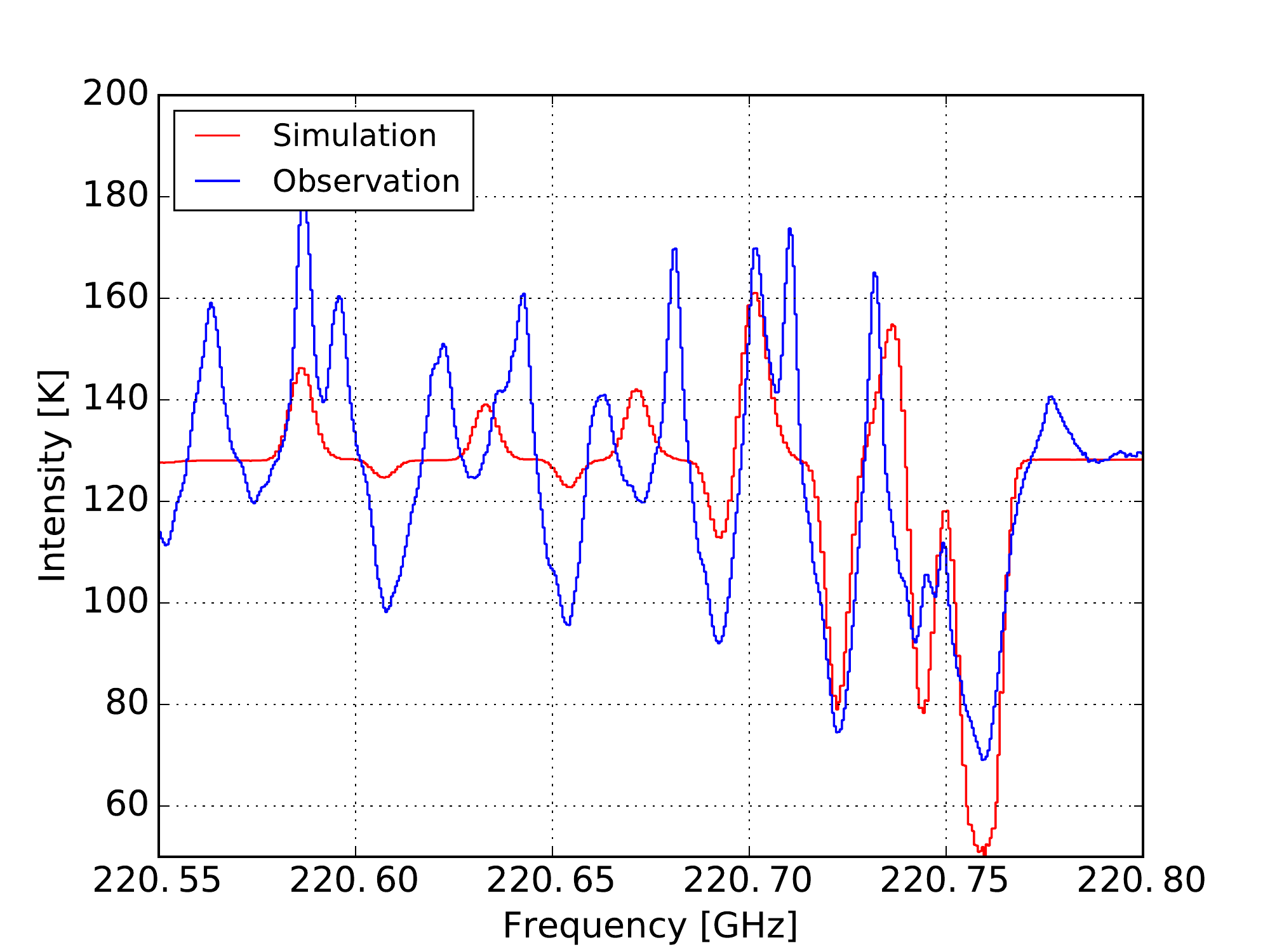}
	 \includegraphics[scale=0.36]{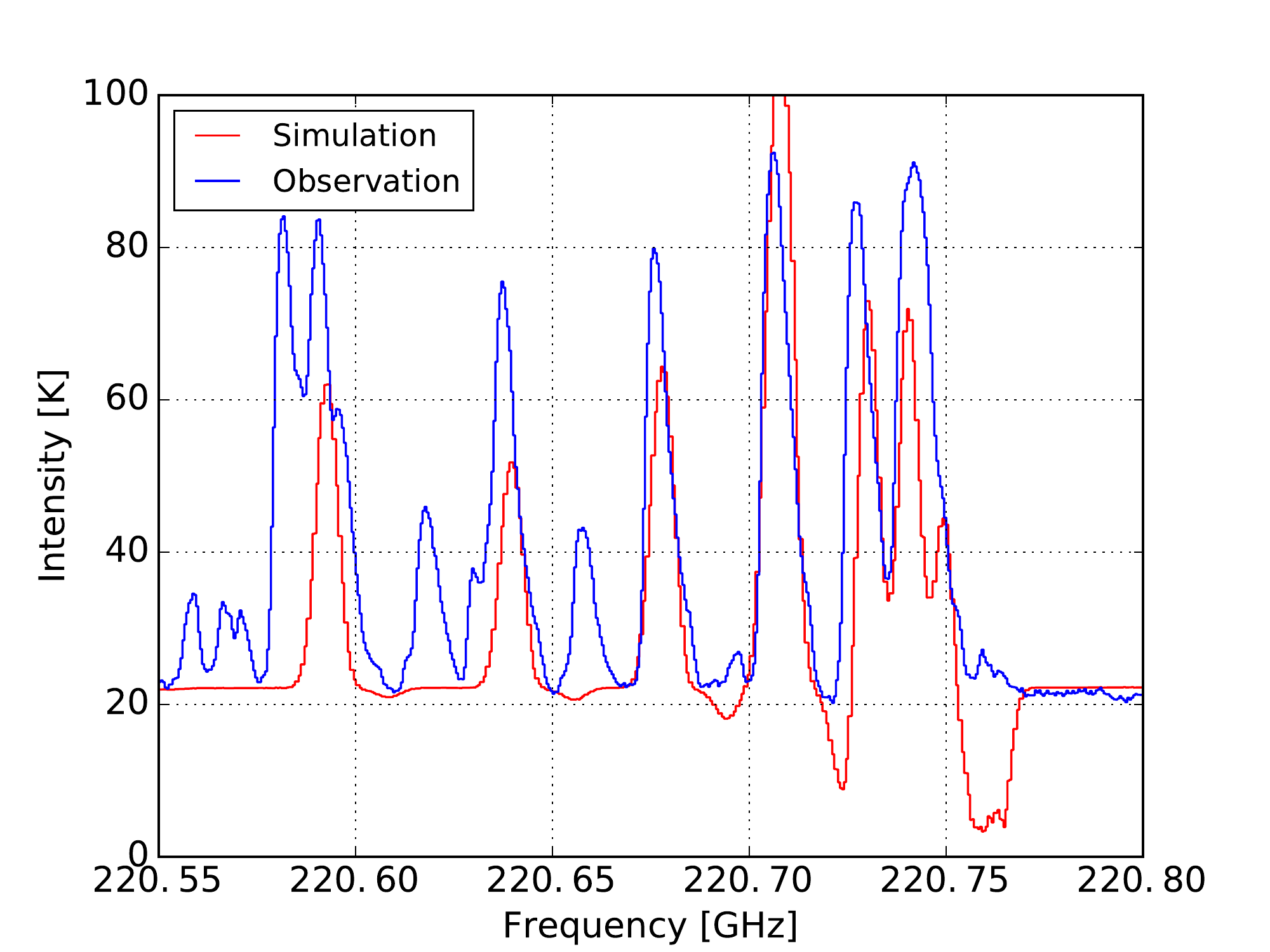}}
\caption{Spectra of the model after including the large-scale envelope with a \chcn\ abundance of  \ee{-7}. The \textit{left panel} shows a spectrum towards the center where absorption features associated with the low-$K$ transitions are reproduced. The \textit{right panel} shows a spectrum towards the southern location MC-68. This model predicts absorptions for the low-$K$ transitions that are not visible in the observational data.}
\label{fig:envelope}
\end{figure*} 

\begin{figure}
\centering
	\includegraphics[width=\hsize]{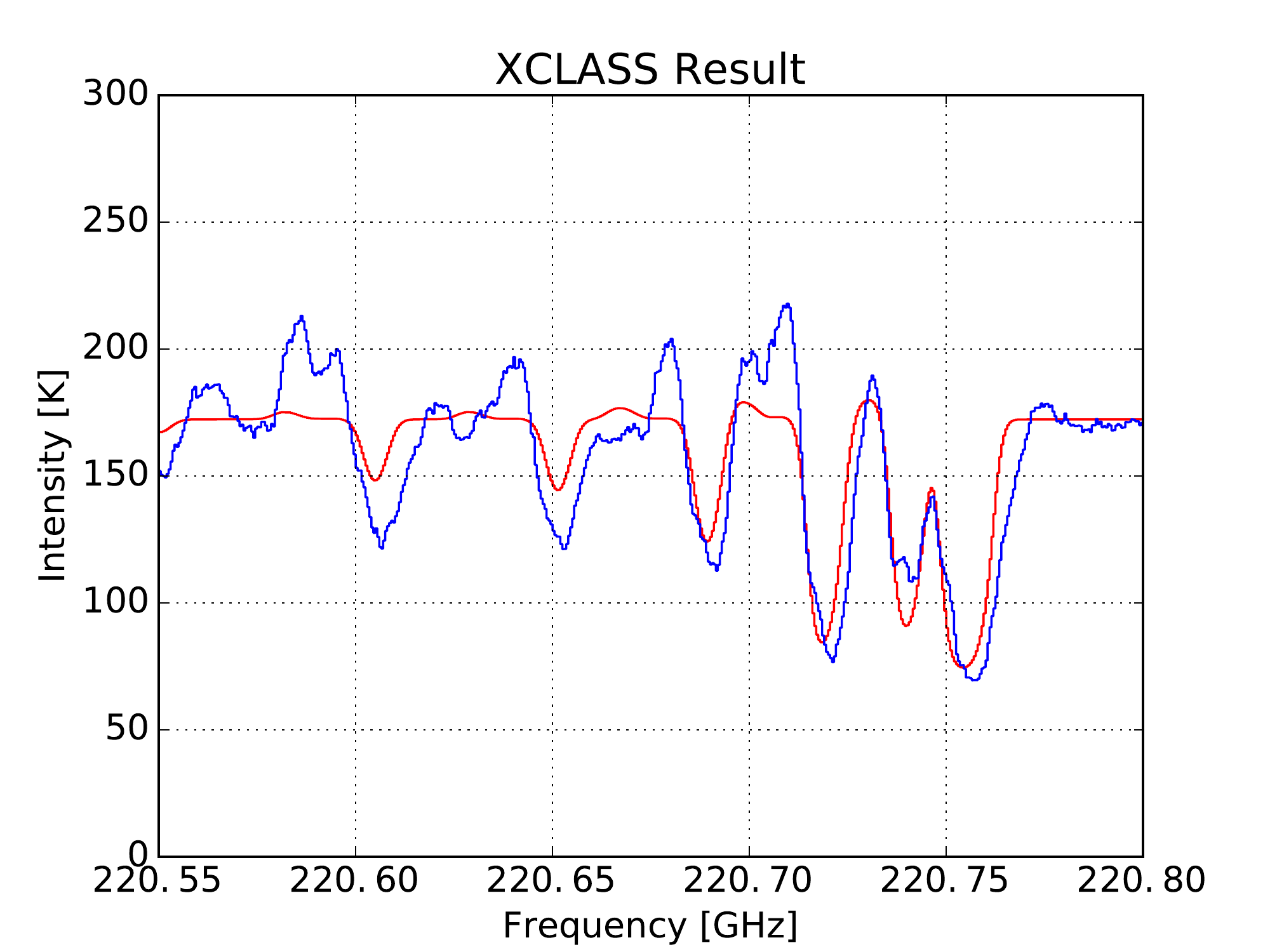}
\caption{Comparison of the \chcn\ $J$=12--11 observational spectrum (in black) towards a central position in \SgrB(M) and a fit (in red) obtained with the software package XCLASS. The best fit is obtained for a temperature range \unit[$\sim$ 100 -- 150]{K} and a column density of \unit[2.9\tee{14}]{\scm}.}
\label{fig:xclass_center}
\end{figure}

\begin{figure}
\centering
	\includegraphics[width=\hsize]{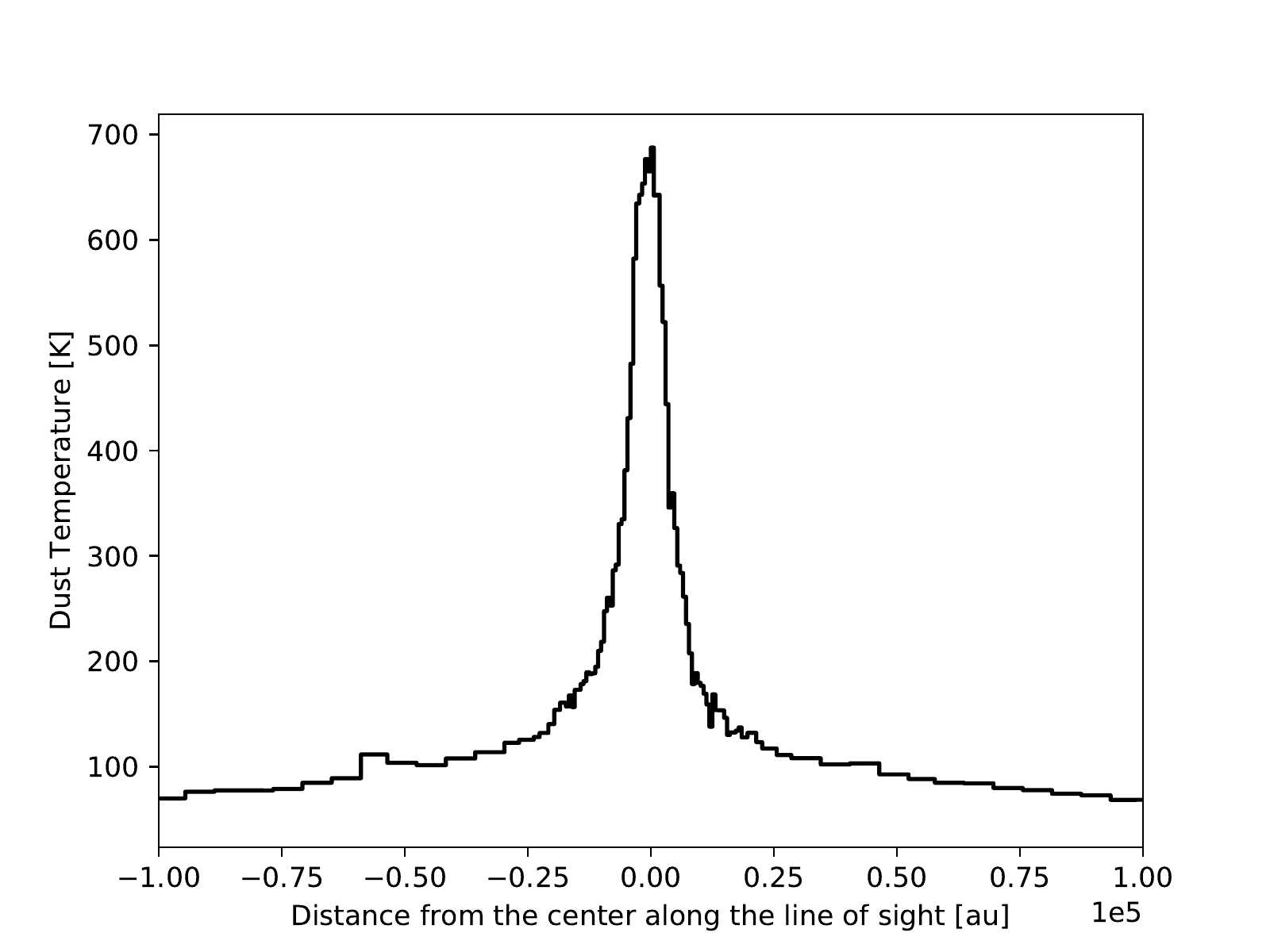}
\caption{Distribution of the dust temperature relative to the center of the model along the line of sight. The plane $z=0$ has a temperature about \unit[700]{K}, while temperatures of about \unit[$\sim$100 -- 150]{K} are found at a distance of \unit[$\sim$ 20,000]{au}.}
\label{fig:temp_dis_central}
\end{figure}

\begin{figure*}[t!]
\centering
\includegraphics[width=0.98\textwidth]{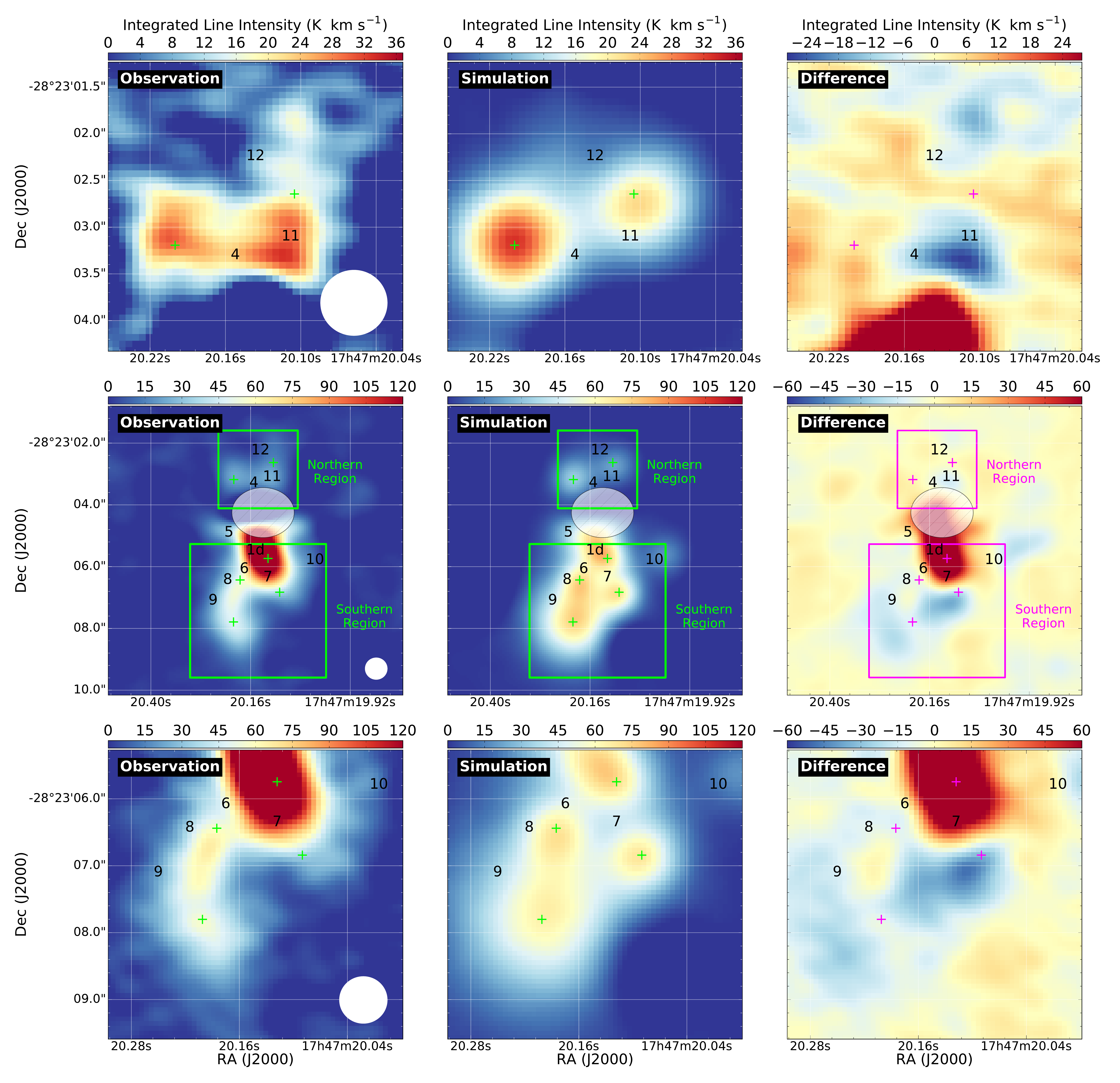}
\caption{Integrated intensity maps (or zeroth-order moment maps) of the observational (\textit{left column}) and simulated (\textit{middle column}) data for the \chcn\ $J$=13-12, $K$=0,1 transitions of the isotopologue \isochcn. The \textit{right column} panels show the difference observation $-$ simulated data. The \textit{central row} shows the entire \SgrB(M) complex except for the central region, which is masked out with a transparent, hatched area. The green boxes and magenta indicate the zoomed-in regions that are depicted in the \textit{top row} (northern part of \SgrB(M)) and the \textit{bottom row} (southern part of \SgrB(M)). As in previous figures, the black numbers indicate the position of the dust cores, whereas the green and magenta crosses depict the position of the molecular centers (see Table~\ref{tab:cores}).}
\label{fig:isotopologue_observation_simulation}
\end{figure*}

Firstly, we tried to reproduce the absorption features by including the large scale envelope surrounding \SgrB(M). Besides the small scale structure of \SgrB(M), the physical model developed by Paper~I also includes an envelope surrounding \SgrB(M) with an extent of \unit[$\sim$ \ee{4} -- \ee{5}]{au} and densities of \unit[$\sim$ \ee{5} -- \ee{6}]{\qcm}. One indication that the absorption is caused by \chcn\ located in the larger scale structure, where temperatures are much lower, is that in the vibrationally excited transition which have very high energies, we see almost no absorption features (see Fig.~\ref{fig:center_spectra}). By defining an abundance factor of \chcn\ for the envelope of \ee{-7} and a velocity along the line of sight of \unit[44]{km/s}, we were able to reproduce the deep absorption features at the center (see Fig.~\ref{fig:envelope}). However, the model with a \chcn\ abundance of \ee{-7} in the envelope produces also absorption features in the northern and southern regions of \SgrB(M), where no deep absorption features are observed. This is likely due to the spatial extent of the envelope, which  does not cover only the central region of \SgrB(M) but the entire complex. As a consequence, the \chcn\ in the envelope is not only causing absorption features towards the center, but also in the northern and southern regions, despite the continuum emission in those regions is much weaker (see Fig.~\ref{fig:envelope}). 

With the knowledge that the absorption is not caused by the gas in the small scale structure, but by the gas which is located in the more outer parts, we investigate a model with a shell of enhanced \chcn\ density surrounding the central part of \SgrB(M). We consider a shell with an extent of \unit[5000]{au}, which only covers the central part but not the northern and southern regions of \SgrB(M). This prevents the model from producing absorption features at these regions. However, the spectra towards the central positions show emission and no clear sign of absorption. In order to evaluate the origin of this effect we used the software package XCLASS \citep{moeller2017} to fit the spectrum towards the center of \SgrB(M) and determine the temperature of the absorption layer. The best fit of the absorbing \chcn\ is obtained for a temperature in the range \unit[$\sim$ 100 -- 150]{K}. In XLASS a model with a Gaussian like density distribution centered around \unit[5000]{au}, a temperature between \unit[$\sim$100 -- 150]{K} and a column density of \unit[2.9\tee{14}]{\scm} can reproduce the deep absorption features towards the central region (see Fig.~\ref{fig:xclass_center}). The temperature distribution calculated in our model (obtained self-consistently by RADMC-3D) along the line of sight at the center (see Fig.~\ref{fig:temp_dis_central}) shows that the temperature range of \unit[$\sim$ 100 -- 150]{K} is reached at a distance of \unit[$\sim$ 20,000]{au}, whereas a shell at \unit[5000]{au} would have a temperature of \unit[$\sim$ 300 -- 400]{K}, and therefore would still lead to emission features rather than absorption. For this reason the model with a shell of increased \chcn\ abundance does not lead to absorption when using the temperature distribution of RADMC-3D.

All this suggest that the gas temperature distribution predicted by the model may differ in the central part of \SgrB(M) with respect to the real distribution. In this sense, the analysis of the ALMA continuum images (see Paper~II) revealed a number of continuum sources not included in the original physical model (Paper~I) which might have an impact on the temperature distribution. An update of the continuum model using the ALMA data presented in Paper~II and new ALMA data at lower frequencies (80~GHz to 150~GHz) is planned in a forthcoming paper. Moreover, we also have to be aware of the fact that in all the described models the \chcn\ density field is constructed based on spherically symmetric Plummer-like clumps. Hence, effects of deviations from this spherical symmetry, like elongated structures along the line of sight, may help to explain the observed absorption features, since higher column densities can be reached at regions where the temperature is favorable for absorption.

\subsection{Isotopologue \isochcn}\label{ssec:Isotopologue}

The spectral line survey contains transitions of the isotoplogues of \chcn. We consider the isotopologue in which one of the carbon atoms $^{12}$C is exchanged by the isotope $^{13}$C. Due to the increased mass of the isotopologue, it also possess a larger moment of inertia, which results in a shift of the energy levels. For this reason, the isotopologues exhibit  transition frequencies which are different from the main isotopologue. Due to the fact that the moment of inertia depends also on the position of the $^{13}$C atom, also the two isotopologues \isochcn\ and CH$_{3}{}^{13}$CN have different transition frequencies. The isotopologue CH$_{3}{}^{13}$CN exhibits transitions which are very close to the main isotopologue and therefore they are strongly blended, whereas the isotopologue \isochcn\ emits at frequency ranges different from the main isotopologue. For this reason, we considered \isochcn\ in order to study the isotope ratio. 

Spectra of \isochcn\ taken from the center of the cores and molecular centers of \SgrB(M) are shown in Figs.~\ref{fig:spectra13CH3CN_1} to \ref{fig:spectra13CH3CN_2}. Zeroth-order moment maps comparing the observational data and simulated data of the isotopologue are shown in Fig.~\ref{fig:isotopologue_observation_simulation}. From the observational maps it is evident that although the emission of the isotopologue does not show exactly the same morphology as the main isotopologue, the emission at least coincides with the positions of the newly introduced molecular centers. This indicates that the abundance is in fact enhanced in those regions and the morphology is not caused by opacity effects.

We used an isotopologue ratio of $^{12}$\chcn/\isochcn=20 in order to obtain a fit of the isotopologue transitions. The images in Fig.~\ref{fig:isotopologue_observation_simulation} illustrate that the model reproduces the general morphology of the emission. Nonetheless, there are some differences. In the northern region the strong emission peak in the east is underestimated by the model and it is slightly shifted to the north. In the southern part the strong emission towards MC-7a and M-SMA-7 is considerably underestimated by the model, whereas the emission at core MC-7b is overestimated. Moreover, the spectra depicted in Fig.~\ref{fig:spectra13CH3CN_1} and \ref{fig:spectra13CH3CN_2} in the appendix show that at some cores and molecular centers (e.g.\ MC-68, M-SMA-7, M-SMA-6, M-SMA-7a) the emission is slightly underestimated. In order to reproduce the strong emission at these regions, it is necessary to decrease the isotopologue ratio for these cores considerably. However, studies of the isotope ratio $^{12}$C/$^{13}$C at the galactic center determined values of $\sim$20 \citep{wilson1994}. Furthermore, the isotopologue ratio of 20 used in the model gives good results when comparing the emission in the other regions of \SgrB(M), and therefore it would be necessary to introduce an isotopologue ratio which varies within \SgrB(M) in order to fit all cores and molecular centers.

\section{Summary and outlook}\label{sec:Conclusion}

In this paper we present ALMA observations of different \chcn\ transitions in the range 211--275~GHz (see details on the observations in Paper~II) towards the hot molecular core \SgrB(M). The achieved angular resolution of $0\farcs7$ (or $\sim$6000~au) permits us to study the structure of the \chcn\ emission at scales not previously studied before. We model the \chcn\ emission using a three-dimensional radiative transfer model (presented in Paper~I) that contains the physical structure (i.e.\ distribution of dense cores, \hii\ regions, stars) of \SgrB(M). We reconstruct the abundance field of \chcn\ towards \SgrB(M), and study the velocity and line width variations in the region. Our main results are listed in the following.

\renewcommand{\labelitemi}{$\bullet$}

\begin{itemize}
\item We derive an average relative abundance of \chcn\ (with respect to H$_2$) of 1.1\tee{-9}, covering a range from 4\tee{-11} to 2e\tee{-8}, towards the north of \SgrB(M) and 9\tee{-8} , covering a range from 2\tee{-10} to 5\tee{-7}, towards the south.
\item We find that the \chcn\ relative abundance is lower at the center of the very dense and hot continuum cores, causing the general morphology of the \chcn\ emission to not completely coincide with the dust continuum emission. 
\item The spectral features of the \chcn\ transitions (e.g.\ line intensity, line width) are reproduced well by the model, however and especially in the southern region of \SgrB(M), the higher $K$-transitions and the vibrationally excited transition are underestimated. This indicates that the gas temperature is higher than the dust temperature or radiative pumping plays a role.  
\item The velocity component along the line of sight varies from \unit[67]{km~s$^{-1}$} (towards the south) to \unit[50]{km~s$^{-1}$} (towards the north), where the systemic velocity of the cloud is about \unit[64]{km~s$^{-1}$}.
\item The central region of \SgrB(M) exhibits complex \chcn\ spectra with a combination of absorption and emission features. We study if the absorption features are related to the presence of \hii\ regions but conclude that their origin is the presence of bright dust, which at this high angular resolution reaches a bright brightness temperature (about 180~K) and produces the absorption features. Despite different attempts in the modeling of this region, our current model does not produce accurate results for all the \chcn\ transitions. We suggest that the calculated temperature distribution deviates from the real distribution towards the center, or that the assumed spherical symmetry of the dense cores included in the model may be not accurate enough. 
\end{itemize}

Modeling the emission of \chcn\ for \SgrB(M) raised some interesting issues, which deserve further investigation in the future. Some of them are listed in the following.

\begin{itemize}
\item Can chemical models of hot molecular cores reproduce the low relative \chcn\ abundance found towards the densest and hottest regions of the cores in \SgrB(M)? In this sense, it is worth studying in more detail which are the chemical processes that are crucial in the destruction of \chcn.
\item The velocity field determined from \chcn\ needs to be compared with the velocity field of other molecular tracers (e.g.\ SiO which traces outflow and shocked gas) and recombination lines (which trace the motions of the ionized gas), in order to study the origin of the velocity gradient from north to south observed in the dense gas.
\item Can a better determination of the continuum model or the inclusion of non-spherical symmetric structure reproduce the absorption and emission spectral line features observed towards the central region of \SgrB(M)? Further investigations and an on-going improvement of the model may help to better constrain the physical properties of this central region.
\item Extending the model of \chcn\ to the northern hot core of \SgrB, i.e.\ \SgrB(N) will be more challenging (due to the higher density of spectral line features) but will allow to compare the properties derived in two distinct but close high-mass star forming sites: \SgrB(M) being more evolved and less dense than \SgrB(N).
\end{itemize}

\begin{acknowledgements}
This work was supported by the Deutsche For\-schungs\-ge\-mein\-schaft (DFG) through grant SFB 956 (subproject A6) and from BMBF/Verbundforschung through the Projects ALMA-ARC 05A11PK3 and 05A14PK1. This paper makes use of the following ALMA data: ADS/JAO.ALMA\#2013.1.00332.S. ALMA is a partnership of ESO (representing its member states), NSF (USA) and NINS (Japan), together with NRC (Canada) and NSC and ASIAA (Taiwan), in cooperation with the Republic of Chile. The Joint ALMA Observatory is operated by ESO, AUI/NRAO and NAOJ.
\end{acknowledgements}

%
%

\bibliography{references}

\begin{thebibliography}{45}
\expandafter\ifx\csname natexlab\endcsname\relax\def\natexlab#1{#1}\fi

\bibitem[{{ALMA Partnership} {et~al.}(2015){ALMA Partnership}, {Fomalont},
  {Vlahakis}, {Corder}, {Remijan}, {Barkats}, {Lucas}, {Hunter}, {Brogan},
  {Asaki}, \& et~al.}]{ALMApartnership2015}
{ALMA Partnership}, {Fomalont}, E.~B., {Vlahakis}, C., {et~al.} 2015, \apjl,
  808, L1

\bibitem[{{Araya} {et~al.}(2005){Araya}, {Hofner}, {Kurtz}, {Bronfman}, \&
  {DeDeo}}]{araya2005}
{Araya}, E., {Hofner}, P., {Kurtz}, S., {Bronfman}, L., \& {DeDeo}, S. 2005,
  \apjs, 157, 279

\bibitem[{{Belloche} {et~al.}(2009){Belloche}, {Garrod}, {M{\"u}ller},
  {Menten}, {Comito}, \& {Schilke}}]{belloche2009}
{Belloche}, A., {Garrod}, R.~T., {M{\"u}ller}, H.~S.~P., {et~al.} 2009, \aap,
  499, 215

\bibitem[{{Bergin} {et~al.}(2010){Bergin}, {Phillips}, {Comito}, {Crockett},
  {Lis}, {Schilke}, {Wang}, {Bell}, {Blake}, {Bumble}, {Caux}, {Cabrit},
  {Ceccarelli}, {Cernicharo}, {Daniel}, {de Graauw}, {Dubernet},
  {Emprechtinger}, {Encrenaz}, {Falgarone}, {Gerin}, {Giesen}, {Goicoechea},
  {Goldsmith}, {Gupta}, {Hartogh}, {Helmich}, {Herbst}, {Joblin}, {Johnstone},
  {Kawamura}, {Langer}, {Latter}, {Lord}, {Maret}, {Martin}, {Melnick},
  {Menten}, {Morris}, {M{\"u}ller}, {Murphy}, {Neufeld}, {Ossenkopf}, {Pagani},
  {Pearson}, {P{\'e}rault}, {Plume}, {Roelfsema}, {Qin}, {Salez}, {Schlemmer},
  {Stutzki}, {Tielens}, {Trappe}, {van der Tak}, {Vastel}, {Yorke}, {Yu}, \&
  {Zmuidzinas}}]{bergin2010}
{Bergin}, E.~A., {Phillips}, T.~G., {Comito}, C., {et~al.} 2010, \aap, 521, L20

\bibitem[{{Bjorkman} \& {Wood}(2001)}]{bjorkman2001}
{Bjorkman}, J.~E. \& {Wood}, K. 2001, \apj, 554, 615

\bibitem[{{Cazzoli} \& {Puzzarini}(2006)}]{cazzoli2006}
{Cazzoli}, G. \& {Puzzarini}, C. 2006, Journal of Molecular Spectroscopy, 240,
  153

\bibitem[{{Cesaroni}(2005)}]{cesaroni2005}
{Cesaroni}, R. 2005, in IAU Symposium, Vol. 227, Massive Star Birth: A
  Crossroads of Astrophysics, ed. R.~{Cesaroni}, M.~{Felli}, E.~{Churchwell},
  \& M.~{Walmsley}, 59--69

\bibitem[{{Cesaroni} {et~al.}(2017){Cesaroni}, {S{\'a}nchez-Monge},
  {Beltr{\'a}n}, {Johnston}, {Maud}, {Moscadelli}, {Mottram}, {Ahmadi},
  {Allen}, {Beuther}, {Csengeri}, {Etoka}, {Fuller}, {Galli},
  {Galv{\'a}n-Madrid}, {Goddi}, {Henning}, {Hoare}, {Klaassen}, {Kuiper},
  {Kumar}, {Lumsden}, {Peters}, {Rivilla}, {Schilke}, {Testi}, {van der Tak},
  {Vig}, {Walmsley}, \& {Zinnecker}}]{cesaroni2017}
{Cesaroni}, R., {S{\'a}nchez-Monge}, {\'A}., {Beltr{\'a}n}, M.~T., {et~al.}
  2017, \aap, 602, A59

\bibitem[{{De Pree} {et~al.}(1998){De Pree}, {Goss}, \& {Gaume}}]{depree1998}
{De Pree}, C.~G., {Goss}, W.~M., \& {Gaume}, R.~A. 1998, \apj, 500, 847

\bibitem[{{de Vicente} {et~al.}(1997){de Vicente}, {Martin-Pintado}, \&
  {Wilson}}]{devicente1997}
{de Vicente}, P., {Martin-Pintado}, J., \& {Wilson}, T.~L. 1997, \aap, 320, 957

\bibitem[{{Dullemond}(2012)}]{dullemond2012}
{Dullemond}, C.~P. 2012, {RADMC-3D: A multi-purpose radiative transfer tool},
  Astrophysics Source Code Library

\bibitem[{{Endres} {et~al.}(2016){Endres}, {Schlemmer}, {Schilke}, {Stutzki},
  \& {M{\"u}ller}}]{endres2016}
{Endres}, C.~P., {Schlemmer}, S., {Schilke}, P., {Stutzki}, J., \&
  {M{\"u}ller}, H.~S.~P. 2016, Journal of Molecular Spectroscopy, 327, 95

\bibitem[{{Gadhi} {et~al.}(1995){Gadhi}, {Lahrouni}, {Legrand}, \&
  {Demaison}}]{gadhi1995}
{Gadhi}, J., {Lahrouni}, A., {Legrand}, J., \& {Demaison}, J. 1995, Journal de
  chimie physique et de physico-chimie biologique, 92, 1984

\bibitem[{{Gaume} {et~al.}(1995){Gaume}, {Claussen}, {de Pree}, {Goss}, \&
  {Mehringer}}]{gaume1995}
{Gaume}, R.~A., {Claussen}, M.~J., {de Pree}, C.~G., {Goss}, W.~M., \&
  {Mehringer}, D.~M. 1995, \apj, 449, 663

\bibitem[{{Goldsmith}(2001)}]{goldsmith2001}
{Goldsmith}, P.~F. 2001, \apj, 557, 736

\bibitem[{{Goldsmith} {et~al.}(1990){Goldsmith}, {Lis}, {Hills}, \&
  {Lasenby}}]{goldsmith1990}
{Goldsmith}, P.~F., {Lis}, D.~C., {Hills}, R., \& {Lasenby}, J. 1990, \apj,
  350, 186

\bibitem[{{Gordon} {et~al.}(1993){Gordon}, {Berkermann}, {Mezger}, {Zylka},
  {Haslam}, {Kreysa}, {Sievers}, \& {Lemke}}]{gordon1993}
{Gordon}, M.~A., {Berkermann}, U., {Mezger}, P.~G., {et~al.} 1993, \aap, 280,
  208

\bibitem[{{Hauschildt} {et~al.}(1993){Hauschildt}, {Gusten}, {Phillips},
  {Schilke}, {Serabyn}, \& {Walker}}]{hauschildt93}
{Hauschildt}, H., {Gusten}, R., {Phillips}, T.~G., {et~al.} 1993, \aap, 273,
  L23

\bibitem[{{Hern{\'a}ndez-Hern{\'a}ndez}
  {et~al.}(2014){Hern{\'a}ndez-Hern{\'a}ndez}, {Zapata}, {Kurtz}, \&
  {Garay}}]{hernandez-hernandez2014}
{Hern{\'a}ndez-Hern{\'a}ndez}, V., {Zapata}, L., {Kurtz}, S., \& {Garay}, G.
  2014, \apj, 786, 38

\bibitem[{{H{\"u}ttemeister} {et~al.}(1993){H{\"u}ttemeister}, {Wilson},
  {Henkel}, \& {Mauersberger}}]{huettemeister1993}
{H{\"u}ttemeister}, S., {Wilson}, T.~L., {Henkel}, C., \& {Mauersberger}, R.
  1993, \aap, 276, 445

\bibitem[{{Kroupa}(2001)}]{kroupa2001}
{Kroupa}, P. 2001, \mnras, 322, 231

\bibitem[{{Kurtz} {et~al.}(2000){Kurtz}, {Cesaroni}, {Churchwell}, {Hofner}, \&
  {Walmsley}}]{kurtz2000}
{Kurtz}, S., {Cesaroni}, R., {Churchwell}, E., {Hofner}, P., \& {Walmsley},
  C.~M. 2000, Protostars and Planets IV, 299

\bibitem[{{McMullin} {et~al.}(2007){McMullin}, {Waters}, {Schiebel}, {Young},
  \& {Golap}}]{mcmullin2007}
{McMullin}, J.~P., {Waters}, B., {Schiebel}, D., {Young}, W., \& {Golap}, K.
  2007, in Astronomical Society of the Pacific Conference Series, Vol. 376,
  Astronomical Data Analysis Software and Systems XVI, ed. R.~A. {Shaw},
  F.~{Hill}, \& D.~J. {Bell}, 127

\bibitem[{{Mehringer} {et~al.}(1993){Mehringer}, {Palmer}, {Goss}, \&
  {Yusef-Zadeh}}]{mehringer1993}
{Mehringer}, D.~M., {Palmer}, P., {Goss}, W.~M., \& {Yusef-Zadeh}, F. 1993,
  \apj, 412, 684

\bibitem[{{Mehringer} {et~al.}(1997){Mehringer}, {Snyder}, {Miao}, \&
  {Lovas}}]{mehringer1997}
{Mehringer}, D.~M., {Snyder}, L.~E., {Miao}, Y., \& {Lovas}, F.~J. 1997, \apjl,
  480, L71

\bibitem[{{Molinari} {et~al.}(2010){Molinari}, {Swinyard}, {Bally}, {Barlow},
  {Bernard}, {Martin}, {Moore}, {Noriega-Crespo}, {Plume}, {Testi}, {Zavagno},
  {Abergel}, {Ali}, {Anderson}, {Andr{\'e}}, {Baluteau}, {Battersby},
  {Beltr{\'a}n}, {Benedettini}, {Billot}, {Blommaert}, {Bontemps}, {Boulanger},
  {Brand}, {Brunt}, {Burton}, {Calzoletti}, {Carey}, {Caselli}, {Cesaroni},
  {Cernicharo}, {Chakrabarti}, {Chrysostomou}, {Cohen}, {Compiegne}, {de
  Bernardis}, {de Gasperis}, {di Giorgio}, {Elia}, {Faustini}, {Flagey},
  {Fukui}, {Fuller}, {Ganga}, {Garcia-Lario}, {Glenn}, {Goldsmith}, {Griffin},
  {Hoare}, {Huang}, {Ikhenaode}, {Joblin}, {Joncas}, {Juvela}, {Kirk},
  {Lagache}, {Li}, {Lim}, {Lord}, {Marengo}, {Marshall}, {Masi}, {Massi},
  {Matsuura}, {Minier}, {Miville-Desch{\^e}nes}, {Montier}, {Morgan}, {Motte},
  {Mottram}, {M{\"u}ller}, {Natoli}, {Neves}, {Olmi}, {Paladini}, {Paradis},
  {Parsons}, {Peretto}, {Pestalozzi}, {Pezzuto}, {Piacentini}, {Piazzo},
  {Polychroni}, {Pomar{\`e}s}, {Popescu}, {Reach}, {Ristorcelli}, {Robitaille},
  {Robitaille}, {Rod{\'o}n}, {Roy}, {Royer}, {Russeil}, {Saraceno}, {Sauvage},
  {Schilke}, {Schisano}, {Schneider}, {Schuller}, {Schulz}, {Sibthorpe},
  {Smith}, {Smith}, {Spinoglio}, {Stamatellos}, {Strafella}, {Stringfellow},
  {Sturm}, {Taylor}, {Thompson}, {Traficante}, {Tuffs}, {Umana}, {Valenziano},
  {Vavrek}, {Veneziani}, {Viti}, {Waelkens}, {Ward-Thompson}, {White},
  {Wilcock}, {Wyrowski}, {Yorke}, \& {Zhang}}]{molinari2010}
{Molinari}, S., {Swinyard}, B., {Bally}, J., {et~al.} 2010, \aap, 518, L100+

\bibitem[{{M{\"o}ller} {et~al.}(2017){M{\"o}ller}, {Endres}, \&
  {Schilke}}]{moeller2017}
{M{\"o}ller}, T., {Endres}, C., \& {Schilke}, P. 2017, \aap, 598, A7

\bibitem[{{M{\"u}ller} {et~al.}(2015){M{\"u}ller}, {Brown}, {Drouin},
  {Pearson}, {Kleiner}, {Sams}, {Sung}, {Ordu}, \& {Lewen}}]{mueller2015}
{M{\"u}ller}, H.~S.~P., {Brown}, L.~R., {Drouin}, B.~J., {et~al.} 2015, Journal
  of Molecular Spectroscopy, 312, 22

\bibitem[{{M{\"u}ller} {et~al.}(2009){M{\"u}ller}, {Drouin}, \&
  {Pearson}}]{mueller2009}
{M{\"u}ller}, H.~S.~P., {Drouin}, B.~J., \& {Pearson}, J.~C. 2009, \aap, 506,
  1487

\bibitem[{{M{\"u}ller} {et~al.}(2005){M{\"u}ller}, {Schl{\"o}der}, {Stutzki},
  \& {Winnewisser}}]{mueller2005}
{M{\"u}ller}, H.~S.~P., {Schl{\"o}der}, F., {Stutzki}, J., \& {Winnewisser}, G.
  2005, Journal of Molecular Structure, 742, 215

\bibitem[{{Pearson} \& {M\"uller}(1996)}]{pearson1996}
{Pearson}, J.~C. \& {M\"uller}, H.~S.~P. 1996, \apj, 471, 1067

\bibitem[{{Pickett} {et~al.}(1998){Pickett}, {Poynter}, {Cohen}, {Delitsky},
  {Pearson}, \& {M{\"u}ller}}]{picket1998}
{Pickett}, H.~M., {Poynter}, R.~L., {Cohen}, E.~A., {et~al.} 1998, JQSRT, 60,
  883

\bibitem[{{Qin} {et~al.}(2011){Qin}, {Schilke}, {Rolffs}, {Comito}, {Lis}, \&
  {Zhang}}]{qin2011}
{Qin}, S.-L., {Schilke}, P., {Rolffs}, R., {et~al.} 2011, \aap, 530, L9

\bibitem[{{Reid} {et~al.}(2014){Reid}, {Menten}, {Brunthaler}, {Zheng}, {Dame},
  {Xu}, {Wu}, {Zhang}, {Sanna}, {Sato}, {Hachisuka}, {Choi}, {Immer},
  {Moscadelli}, {Rygl}, \& {Bartkiewicz}}]{reid2014}
{Reid}, M.~J., {Menten}, K.~M., {Brunthaler}, A., {et~al.} 2014, \apj, 783, 130

\bibitem[{{Remijan} {et~al.}(2004){Remijan}, {Sutton}, {Snyder}, {Friedel},
  {Liu}, \& {Pei}}]{remijan2004}
{Remijan}, A., {Sutton}, E.~C., {Snyder}, L.~E., {et~al.} 2004, \apj, 606, 917

\bibitem[{{Rodgers} \& {Charnley}(2001)}]{rodgers2001}
{Rodgers}, S.~D. \& {Charnley}, S.~B. 2001, \apj, 546, 324

\bibitem[{{Rolffs} {et~al.}(2011){Rolffs}, {Schilke}, {Wyrowski}, {Dullemond},
  {Menten}, {Thorwirth}, \& {Belloche}}]{rolffs2011}
{Rolffs}, R., {Schilke}, P., {Wyrowski}, F., {et~al.} 2011, \aap, 529, A76

\bibitem[{{S{\'a}nchez-Monge} {et~al.}(2013){S{\'a}nchez-Monge}, {Cesaroni},
  {Beltr{\'a}n}, {Kumar}, {Stanke}, {Zinnecker}, {Etoka}, {Galli}, {Hummel},
  {Moscadelli}, {Preibisch}, {Ratzka}, {van der Tak}, {Vig}, {Walmsley}, \&
  {Wang}}]{sanchez-monge2013}
{S{\'a}nchez-Monge}, {\'A}., {Cesaroni}, R., {Beltr{\'a}n}, M.~T., {et~al.}
  2013, \aap, 552, L10

\bibitem[{{S{\'a}nchez-Monge} {et~al.}(2018){S{\'a}nchez-Monge}, {Schilke},
  {Ginsburg}, {Cesaroni}, \& {Schmiedeke}}]{sanchez-monge2018}
{S{\'a}nchez-Monge}, {\'A}., {Schilke}, P., {Ginsburg}, A., {Cesaroni}, R., \&
  {Schmiedeke}, A. 2018, \aap, 609, A101

\bibitem[{{S{\'a}nchez-Monge} {et~al.}(2017){S{\'a}nchez-Monge}, {Schilke},
  {Schmiedeke}, {Ginsburg}, {Cesaroni}, {Lis}, {Qin}, {M{\"u}ller}, {Bergin},
  {Comito}, \& {M{\"o}ller}}]{sanchez-monge2017a}
{S{\'a}nchez-Monge}, {\'A}., {Schilke}, P., {Schmiedeke}, A., {et~al.} 2017,
  \aap, 604, A6, (Paper~II)

\bibitem[{{Schilke}(2015)}]{schilke2015}
{Schilke}, P. 2015, in EAS Publications Series, Vol.~75, EAS Publications
  Series, 227--235

\bibitem[{{Schmiedeke} {et~al.}(2016){Schmiedeke}, {Schilke}, {M{\"o}ller},
  {S{\'a}nchez-Monge}, {Bergin}, {Comito}, {Csengeri}, {Lis}, {Molinari},
  {Qin}, \& {Rolffs}}]{schmiedeke2016}
{Schmiedeke}, A., {Schilke}, P., {M{\"o}ller}, T., {et~al.} 2016, \aap, 588,
  A143, (Paper~I)

\bibitem[{{Sch{\"o}ier} {et~al.}(2005){Sch{\"o}ier}, {van der Tak}, {van
  Dishoeck}, \& {Black}}]{schoeier2005}
{Sch{\"o}ier}, F.~L., {van der Tak}, F.~F.~S., {van Dishoeck}, E.~F., \&
  {Black}, J.~H. 2005, \aap, 432, 369

\bibitem[{{Tan} {et~al.}(2014){Tan}, {Beltr{\'a}n}, {Caselli}, {Fontani},
  {Fuente}, {Krumholz}, {McKee}, \& {Stolte}}]{tan2014}
{Tan}, J.~C., {Beltr{\'a}n}, M.~T., {Caselli}, P., {et~al.} 2014, Protostars
  and Planets VI, 149

\bibitem[{{Wilson} \& {Rood}(1994)}]{wilson1994}
{Wilson}, T.~L. \& {Rood}, R. 1994, \araa, 32, 191

\end{thebibliography}

\onecolumn
\begin{appendix}
\section{Parameters of the cores and molecular centers of the model of \SgrB(M)}\label{sec:TableCores}

In Table~\ref{tab:cores} we list the cores included in the model of \SgrB(M). The first cores reproduce the dust continuum emission are correspond to the cores introduced in the physical model presented in Paper~I. The second list of cores correspond to the molecular centers introduced to reproduce the extension of the \chcn\ emission.

\begin{table*}[h]
\caption{\label{tab:cores} Parameters of the cores and molecular centers of the model of \SgrB(M)}
\centering
\begin{tabular}{l c c c c c c c c c}
\hline\hline

&R.A.
&Dec.
&dz
&$r_0$
&$\eta$
&$n_1$
&$n_2$
&$v_z$
&
\\
ID
&(J2000)
&(J2000)
&($10^5$~au)
&($10^3$~au)
&
&($10^7$~H$_2$~cm$^{-3}$)
&(\chcn\ cm$^{-3}$)
&(km~s$^{-1}$)
&star
\\
\hline
\multicolumn{10}{c}{Dense cores included in the model, from Paper~I} \\
\hline
M-SMA-1a & 17:47:20.197 & $-$28:23:04.36 & 1     & 3.0   & 5     & 20     & \ldots     & \ldots   & B0   \\
M-SMA-1b & 17:47:20.170 & $-$28:23:04.60 & 1     & 3.1   & 5     & 35     & \ldots     & \ldots   & B0.5 \\
M-SMA-1c & 17:47:20.158 & $-$28:23:05.08 & 1     & 3.6   & 5     & 10     & \ldots     & \ldots   & B0.5 \\
M-SMA-1d & 17:47:20.148 & $-$28:23:05.48 & 1     & 3.6   & 5     & 20     & 0.4        & $+$64.0  & B0.5 \\
M-SMA-2a & 17:47:20.152 & $-$28:23:04.18 & 0     & 3.2   & 5     & 19     & \ldots     & \ldots   & B0.5 \\
M-SMA-2b & 17:47:20.124 & $-$28:23:04.45 & 0     & 3.3   & 5     & 14     & \ldots     & \ldots   & B0.5 \\
M-SMA-3  & 17:47:20.100 & $-$28:23:04.04 & 0     & 2.8   & 5     & 14     & 0.112      & $+$60.0  & B0.5 \\
M-SMA-4  & 17:47:20.152 & $-$28:23:03.30 & 0     & 3.4   & 5     & 7      & 0.0098     & $+$53.0  & none \\
M-SMA-5  & 17:47:20.212 & $-$28:23:04.90 & 0     & 3.2   & 5     & 7      & 0.14       & $+$64.0  & B0.5 \\
M-SMA-6  & 17:47:20.175 & $-$28:23:06.08 & 0     & 3.0   & 5     & 35     & 0.00245    & $+$67.0  & none \\
M-SMA-7  & 17:47:20.118 & $-$28:23:06.35 & 0     & 2.9   & 5     & 30     & 0.0054     & $+$66.0  & B0.5 \\
M-SMA-8  & 17:47:20.215 & $-$28:23:06.43 & 0     & 3.6   & 5     & 4      & 0.0006     & $+$67.0  & none \\
M-SMA-9  & 17:47:20.250 & $-$28:23:07.10 & 0     & 2.8   & 5     & 9      & 0.000234   & $+$66.0  & B0.5 \\
M-SMA-10 & 17:47:20.005 & $-$28:23:05.79 & 0     & 3.3   & 5     & 15     & 0.0021     & $+$59.0  & none \\
M-SMA-11 & 17:47:20.108 & $-$28:23:03.10 & 1     & 3.2   & 5     & 20     & 0.028      & $+$56.0  & B0.5 \\
M-SMA-12 & 17:47:20.136 & $-$28:23:02.24 & 1     & 3.8   & 5     & 50     & 0.03       & $+$50.0  & B0   \\
\hline
\multicolumn{10}{c}{Molecular centers included in the model, this work} \\
\hline
MC-4     & 17:47:20.200 & $-$28:23:03.20 & 0     & 3.6   & 5     & 0.0005  & 0.11      & $+$49.0  & none \\
MC-68    & 17:47:20.185 & $-$28:23:06.45 & 0     & 3.0   & 5     & 0.00035 & 0.525     & $+$67.0  & none \\
MC-7a    & 17:47:20.118 & $-$28:23:05.79 & 0     & 2.9   & 5     & 0.0003  & 0.54      & $+$66.0  & none \\
MC-7b    & 17:47:20.090 & $-$28:23:06.85 & 0     & 2.9   & 5     & 0.0003  & 0.72      & $+$66.0  & none \\
MC-9     & 17:47:20.201 & $-$28:23:07.81 & 0     & $r_{0,x,y}=14$ & 5 & 0.0009 & 0.234 & $+$66.0  & none \\  
         &              &                &       & $r_{0,z}=2.8$  &   &        &       &          &      \\
MC-10    & 17:47:19.981 & $-$28:23:05.63 & 0     & 3.3   & 5     & 0.00015 & 0.21      & $+$59.0  & none \\
MC-11    & 17:47:20.105 & $-$28:23:02.65 & 0     & 2.7   & 5     & 0.0002  & 0.192     & $+$58.5  & none \\
\hline
\end{tabular}
\tablefoot{\textbf{ID}: Identifier of the cores used in the model.
\textbf{R.A.}: Right ascension of the density component given in units of hours:minutes:seconds in the equatorial coordinate system.
\textbf{Dec.}: Declination of the density component given in units of degrees:arcminutes:arcseconds in the equatorial coordinate system.
\textbf{dz}: Displacement along the line of sight with respect to the model center. The z-axis is oriented such that it points towards the observer.
$\mathbf{r_0}$: Radius defining the density profile as described in Eq.~\ref{eq:SpatialScale}.
$\boldsymbol{\eta}$: Exponent of the density profile.
$\mathbf{n_1}$: Central density of H$_2$.
$\mathbf{n_2}$: Central density of \chcn.
$\mathbf{v_z}$: Velocity component along the line of sight.
\textbf{star}: This columns indicates whether an additional star was included at the center of the core. If this is the case, the spectral type of the star is given}
\end{table*}

\section{Observed and simulated spectra}\label{sec:AdditionalSpectra}

In Figures~\ref{fig:spectra1} to \ref{fig:spectra13CH3CN_2} we show the observed (in blue) and the simulated (in red) spectra of the different \chcn, \isochcn, and vibrationally excited states towards the different dense cores and molecular centers included in the model (see Table~\ref{tab:cores}).

\begin{figure*}[!h]
\centering
    \includegraphics[scale=0.34]{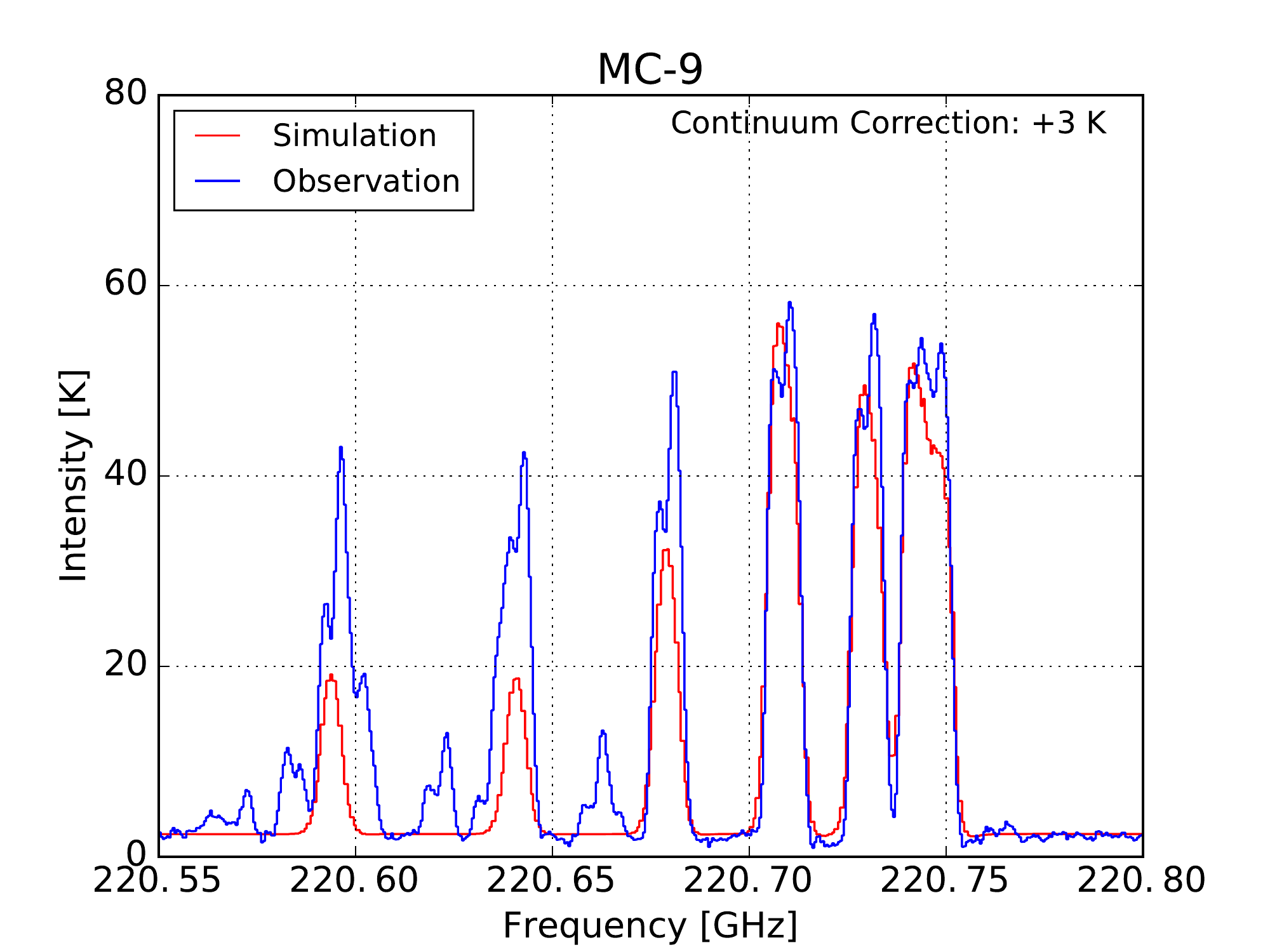}
    \includegraphics[scale=0.34]{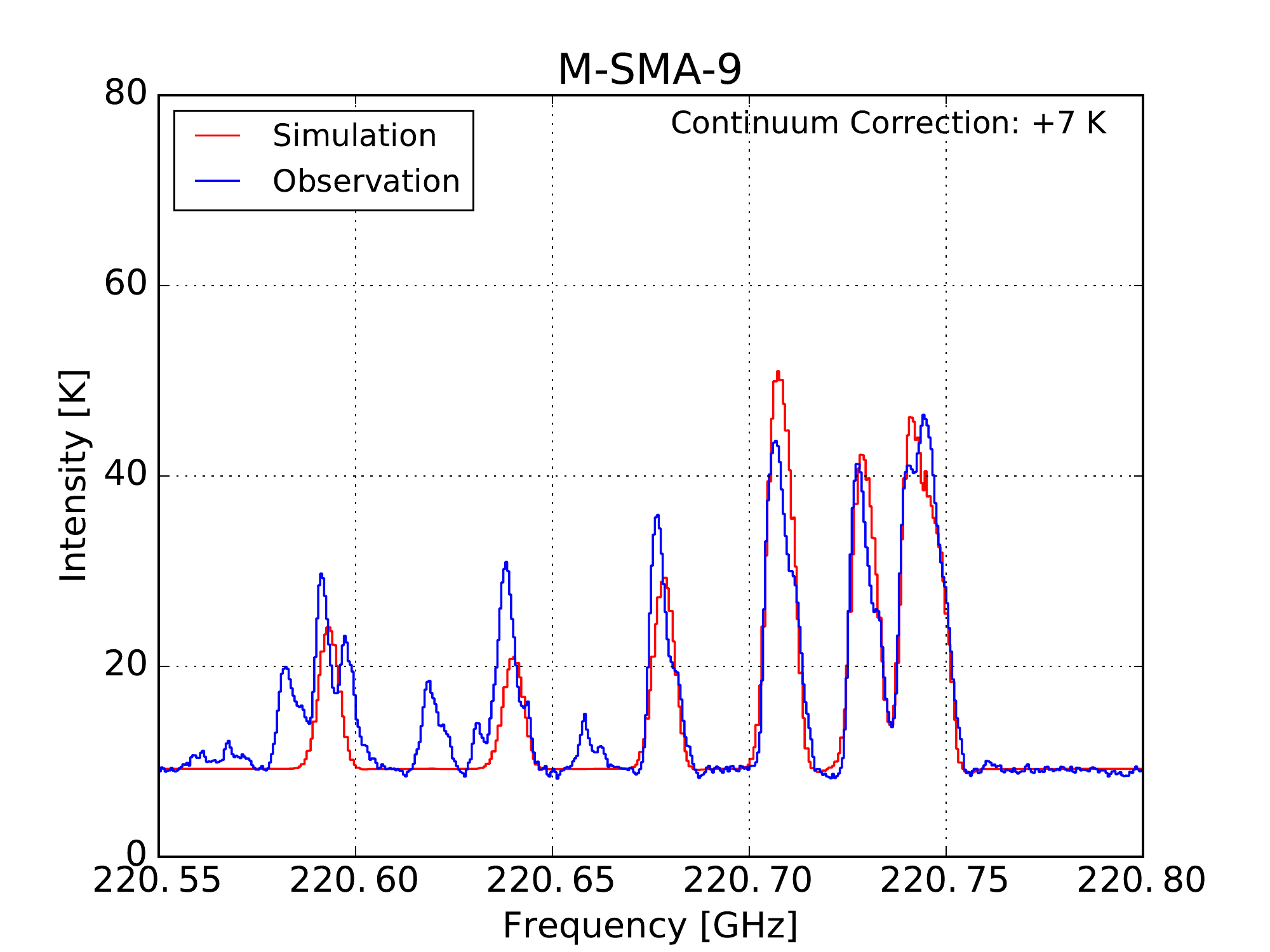}
    \includegraphics[scale=0.34]{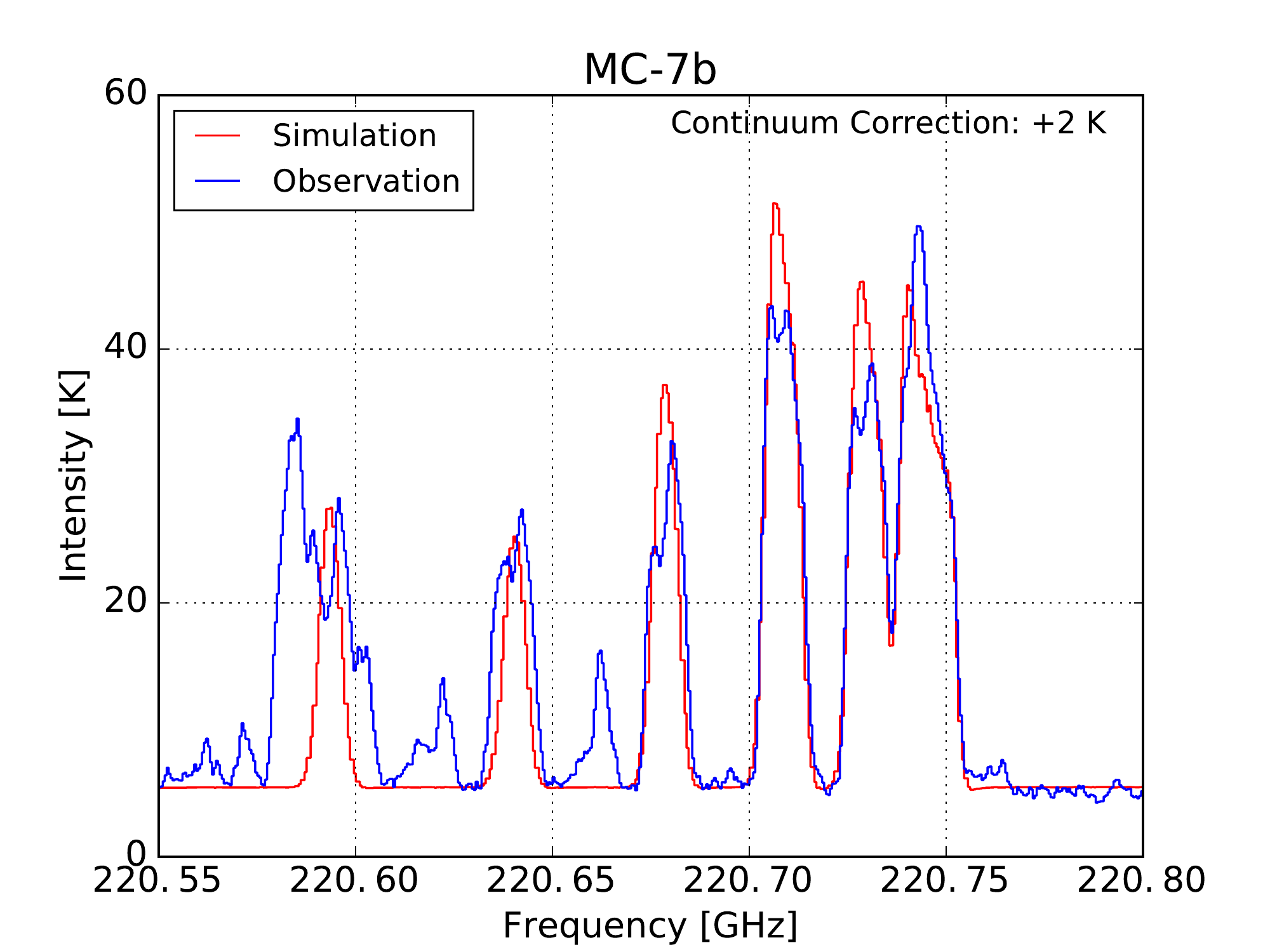}
    \includegraphics[scale=0.34]{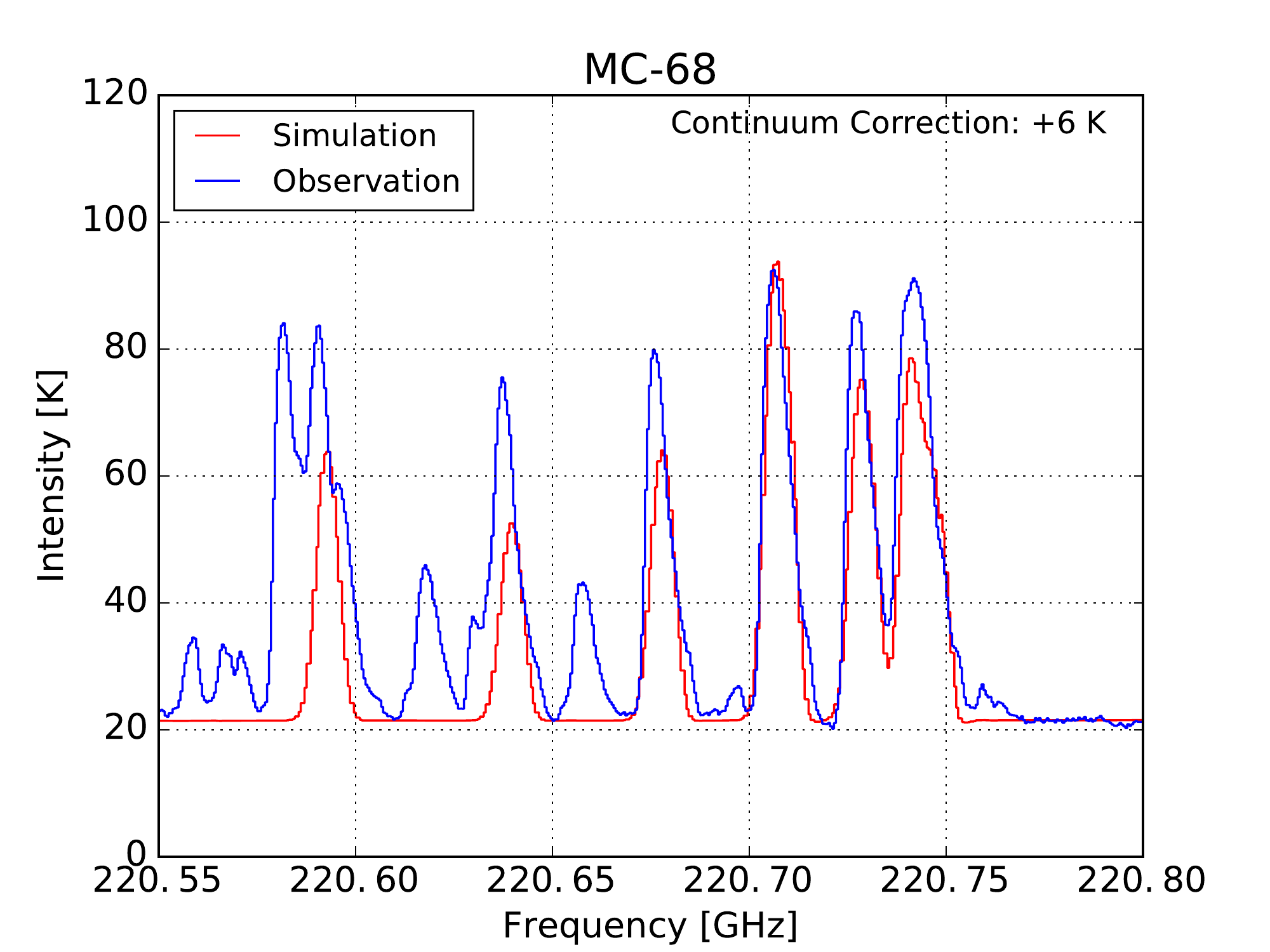}
    \includegraphics[scale=0.34]{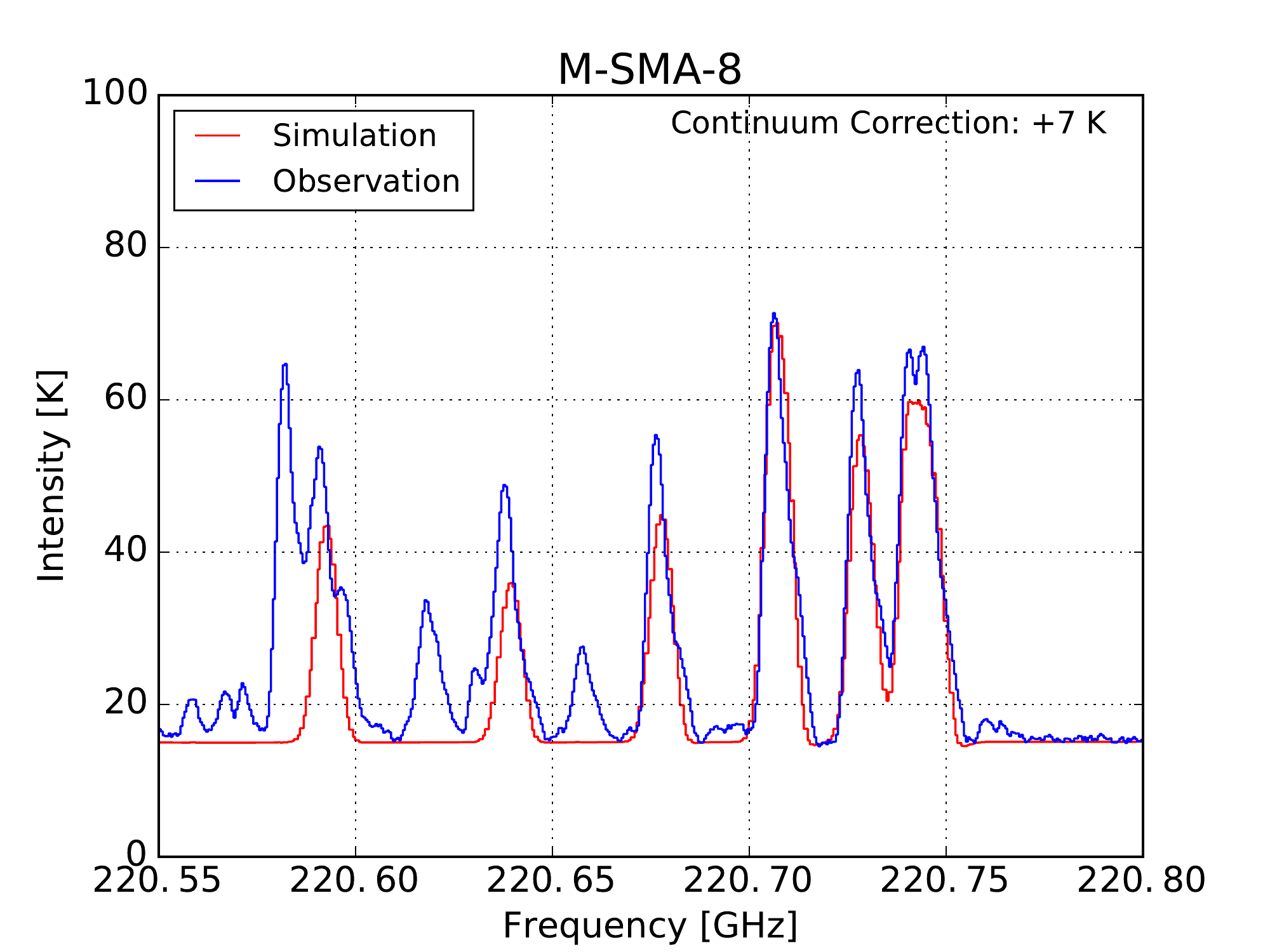}
    \includegraphics[scale=0.34]{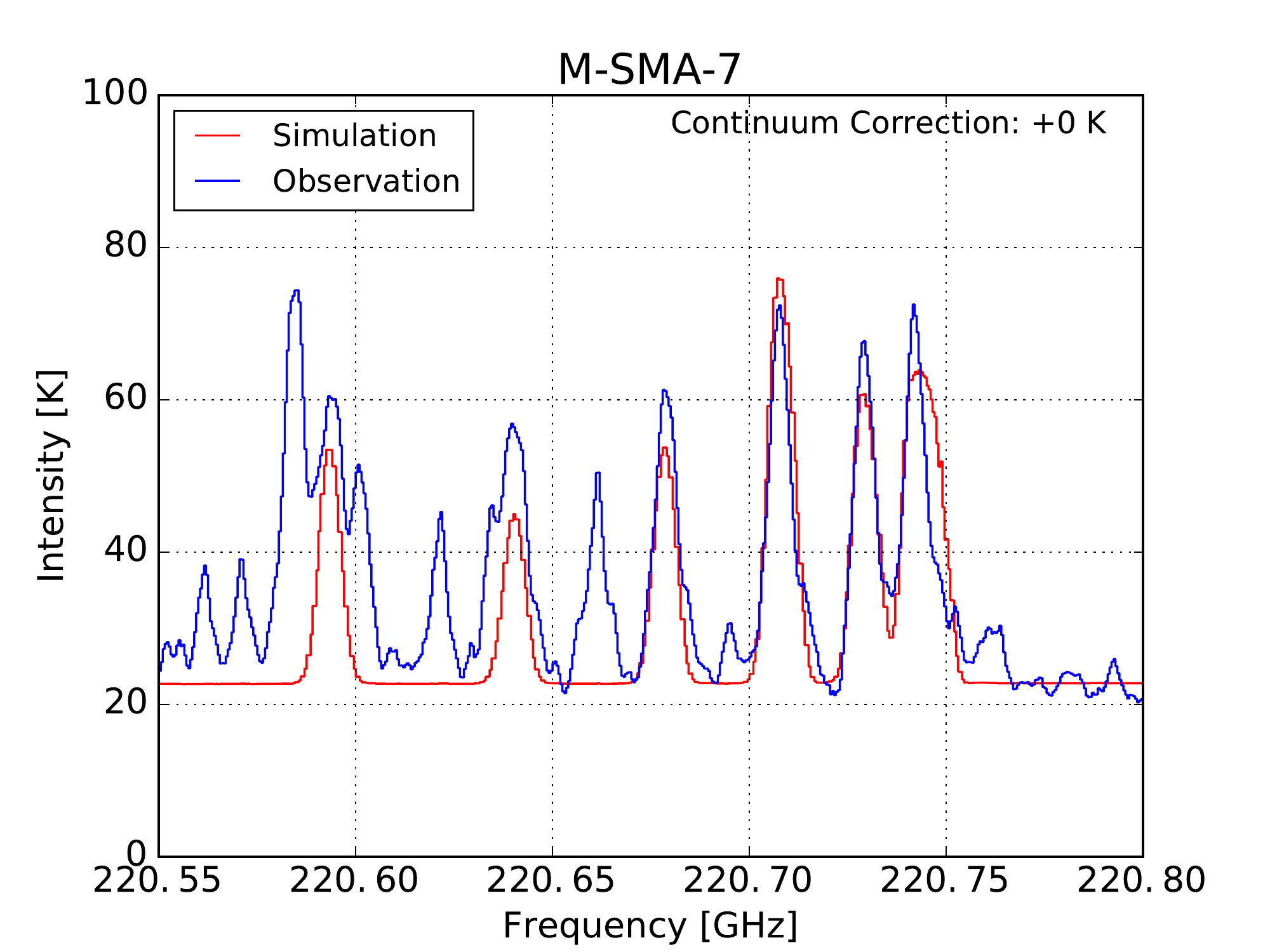}
    \includegraphics[scale=0.34]{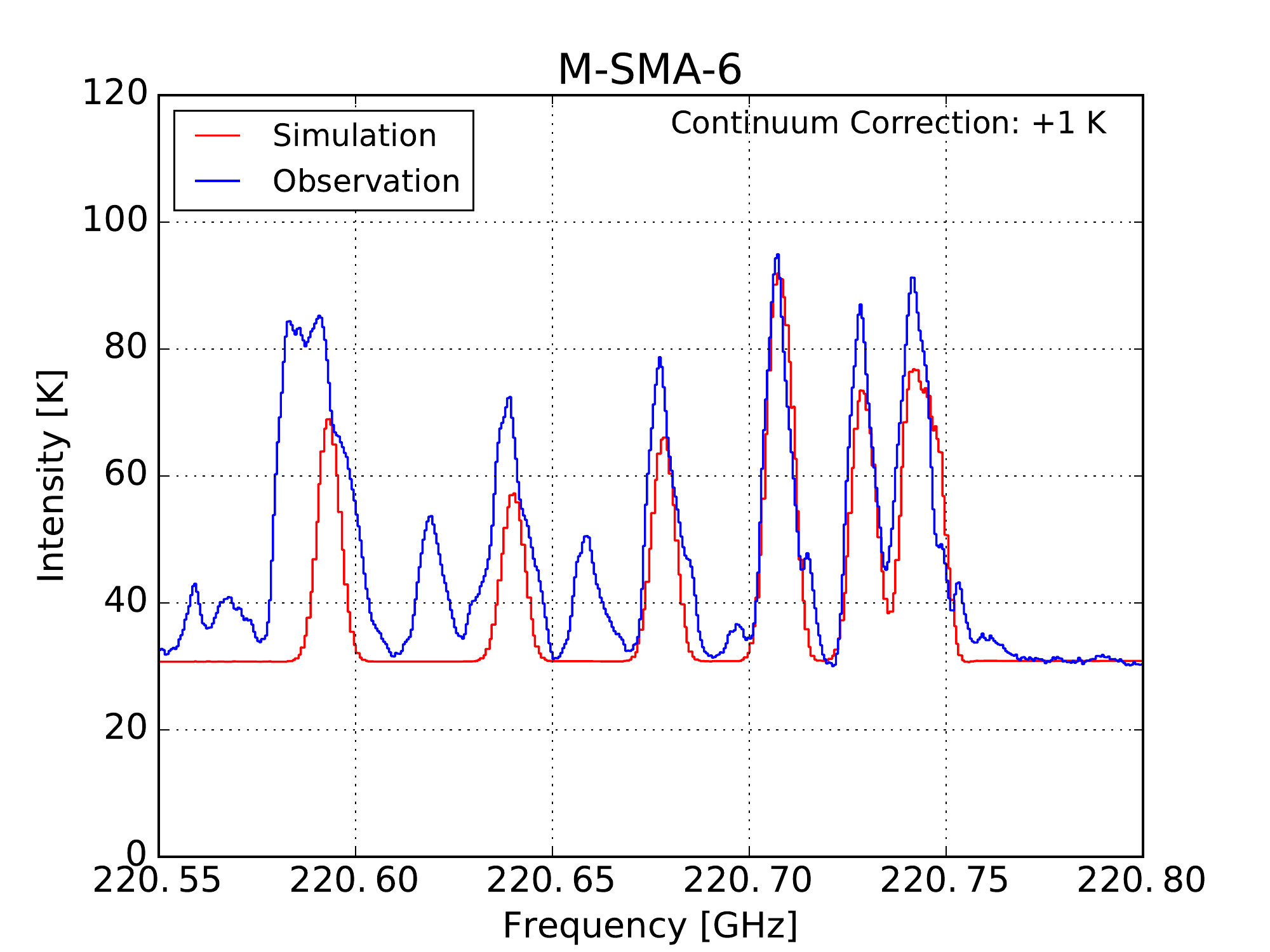}
    \includegraphics[scale=0.34]{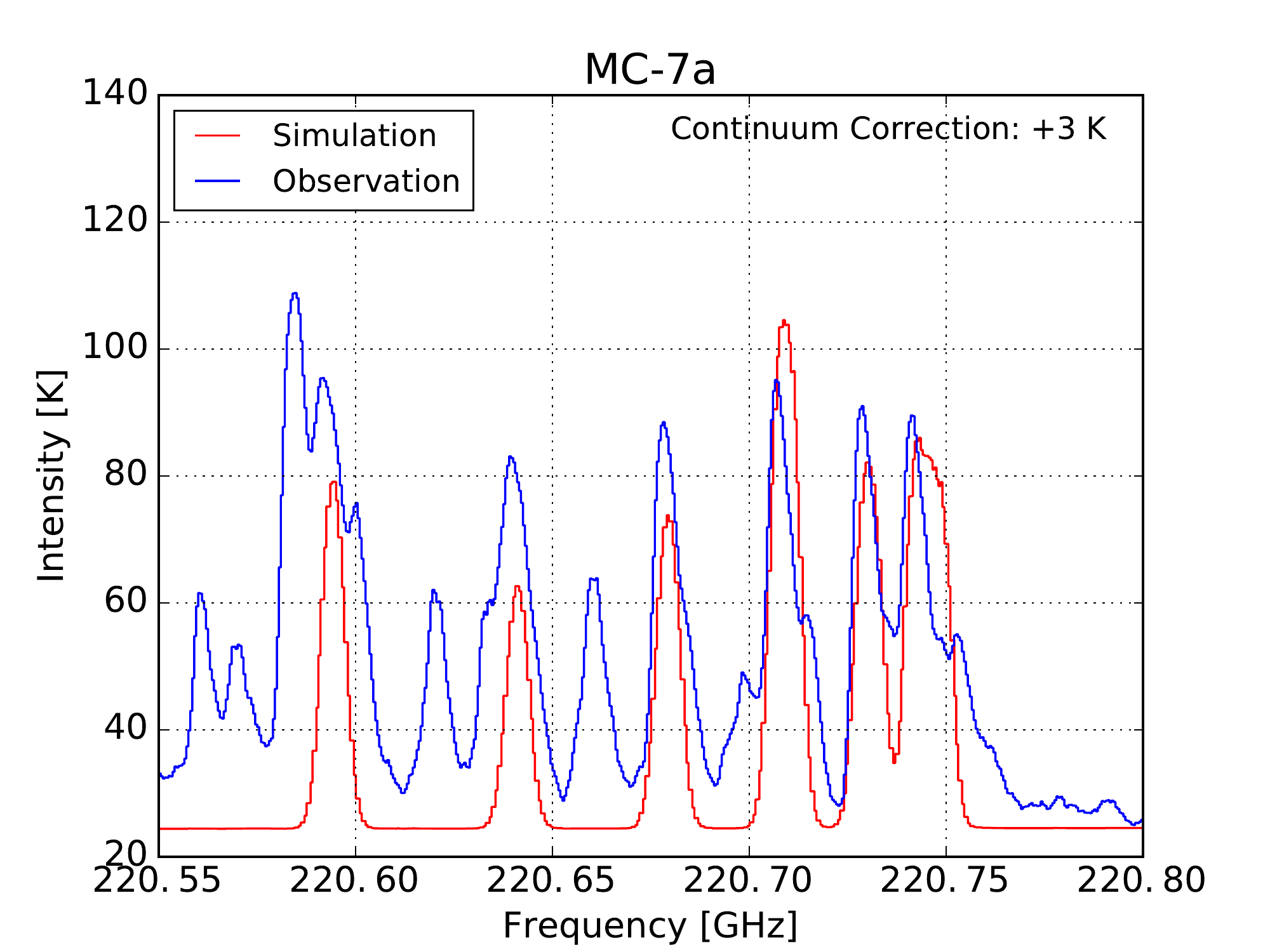}
\caption{Spectra of the observational (in blue) and simulated (in red) data extracted from the position of different cores and molecular centers for the \chcn\ $J$=12--11 transition. The spectra are arranged according to the position of the corresponding cores and molecular centers along a south-north direction, starting from the most southern core. The name of the core and molecular center is shown for each panel. The continuum correction necessary to match the continuum level of the observation and simulation is indicated in the top-right side of each panel.}
\label{fig:spectra1}
\end{figure*}
  
\clearpage  
\begin{figure*}
\centering
    \includegraphics[scale=0.36]{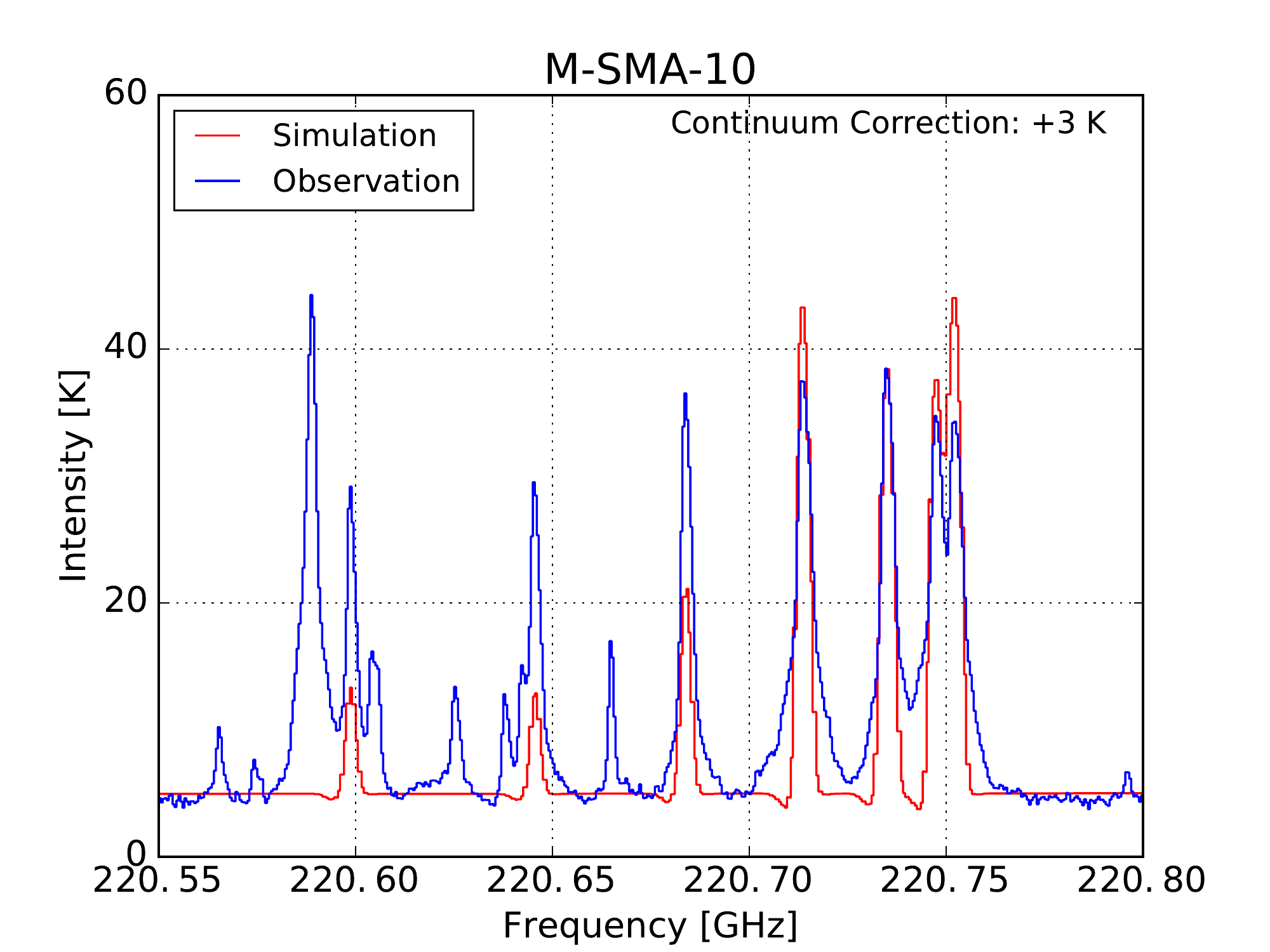}
    \includegraphics[scale=0.36]{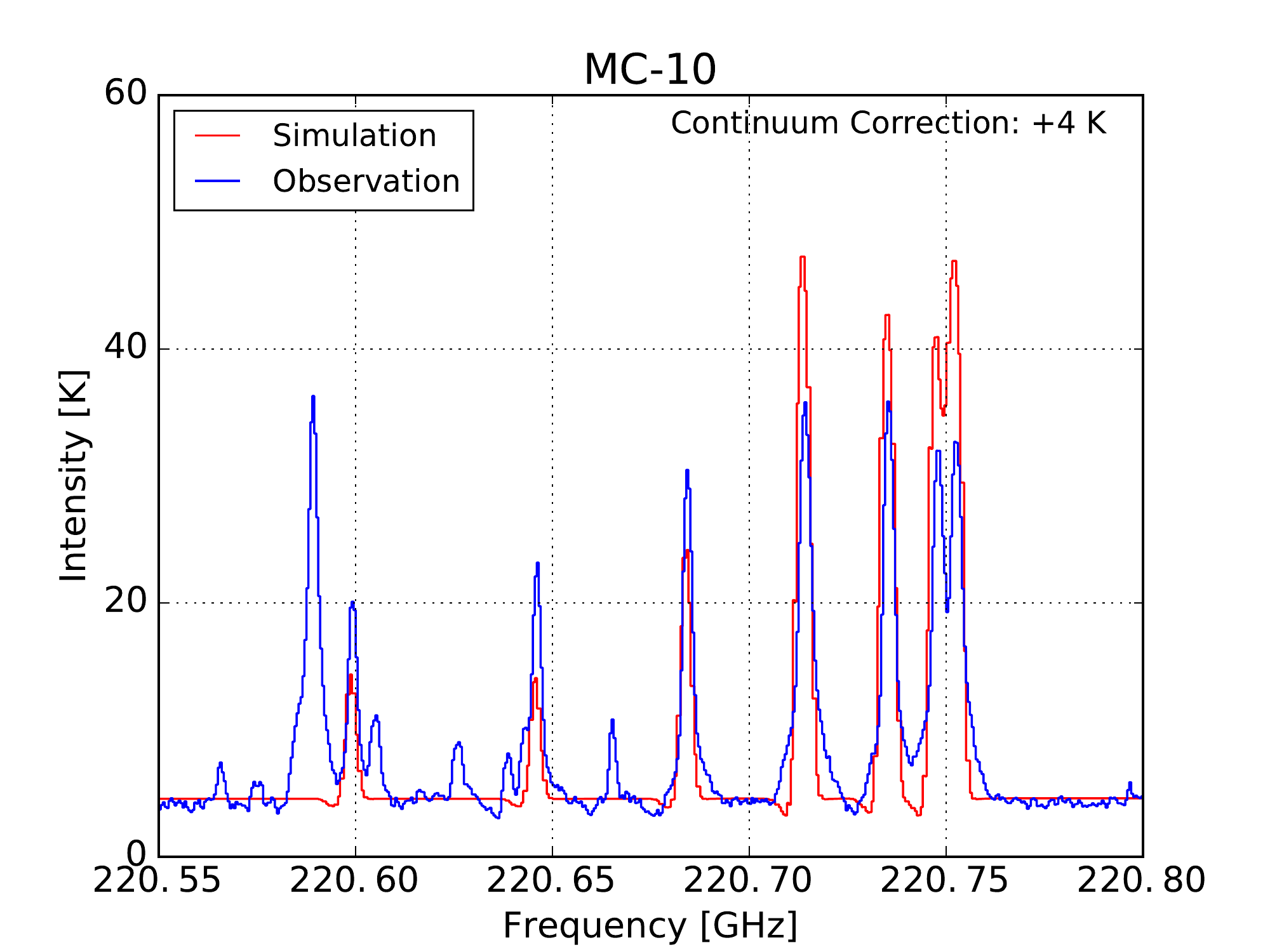}
    \includegraphics[scale=0.36]{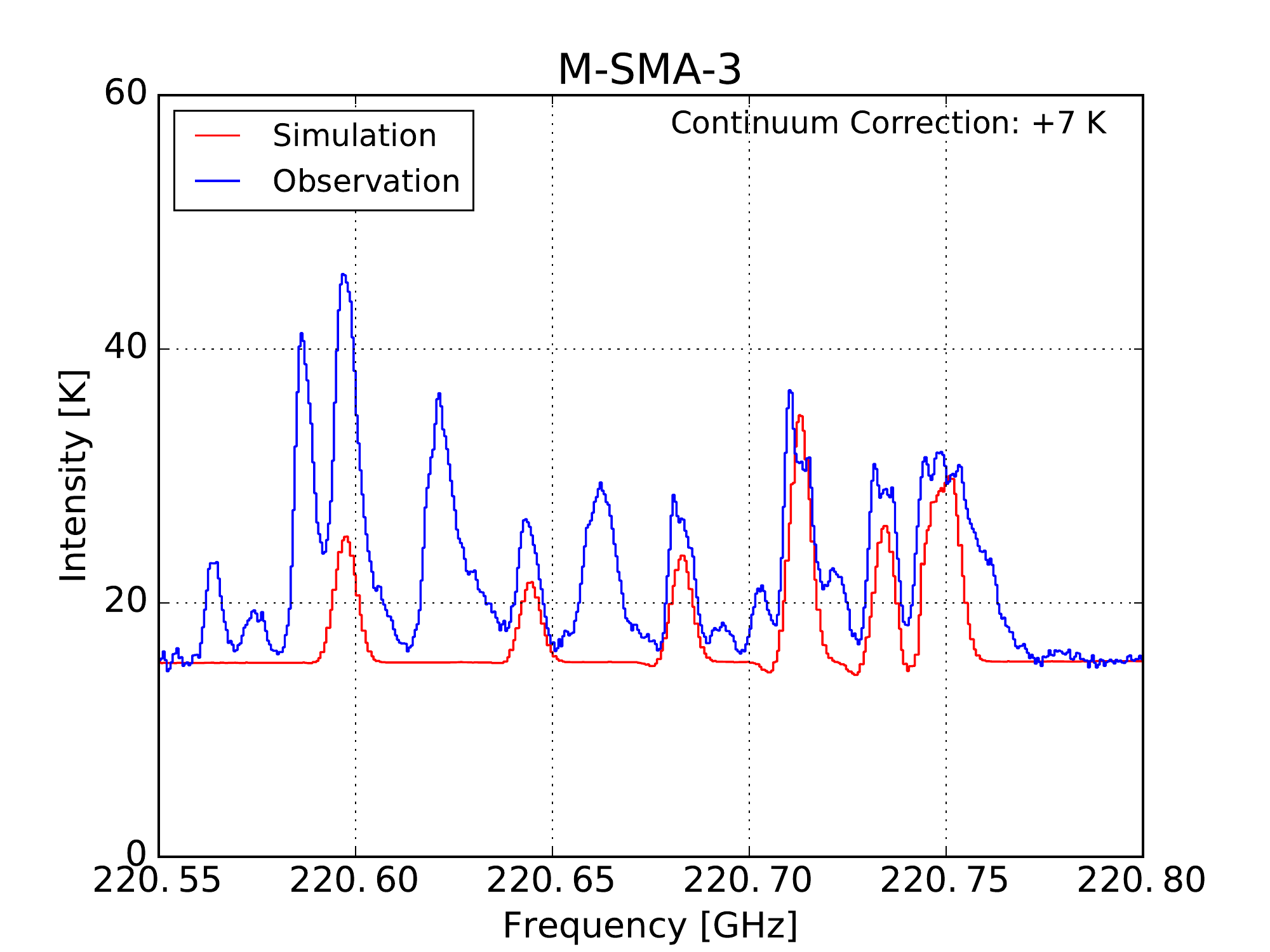}
    \includegraphics[scale=0.36]{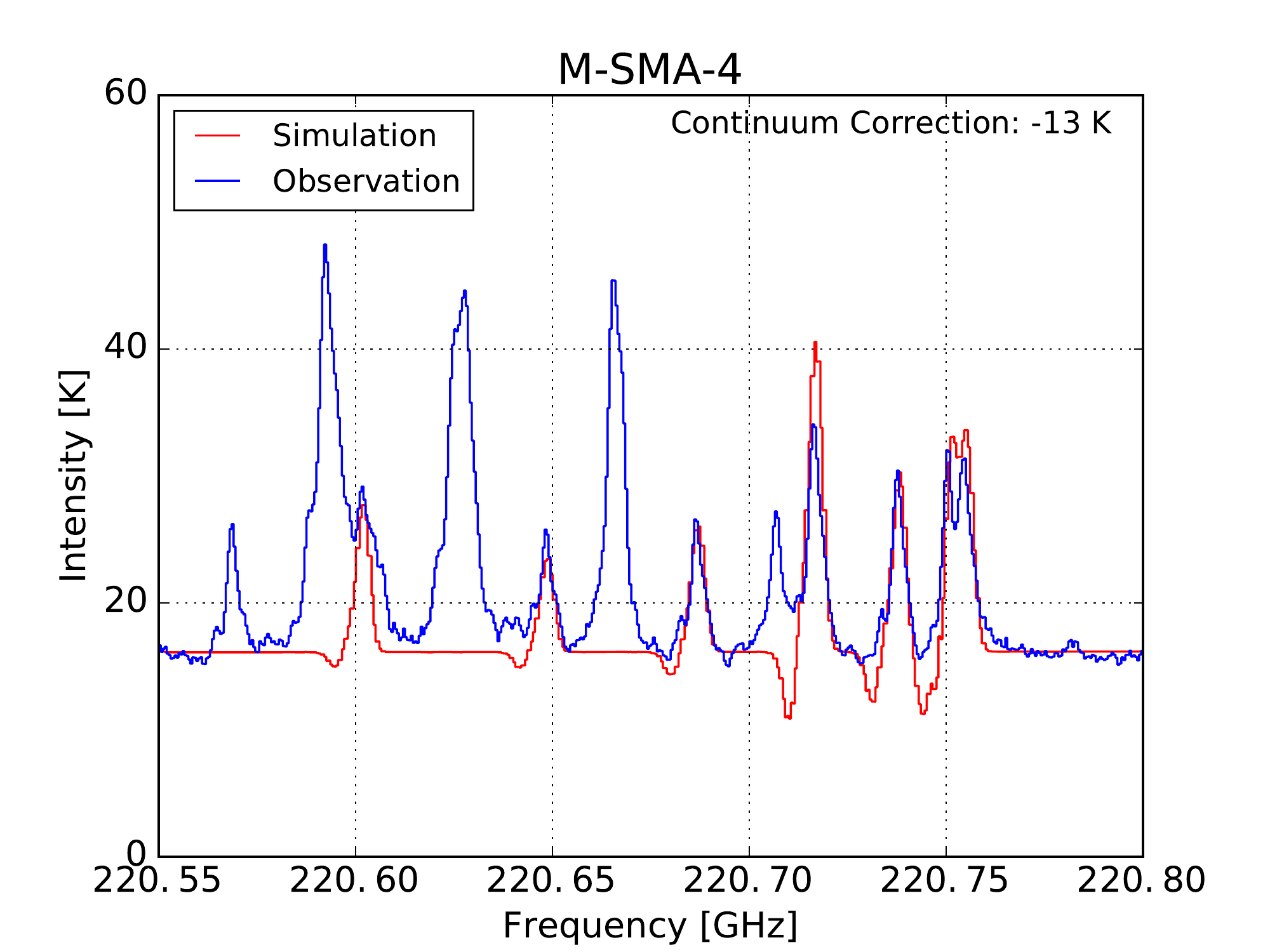}
    \includegraphics[scale=0.36]{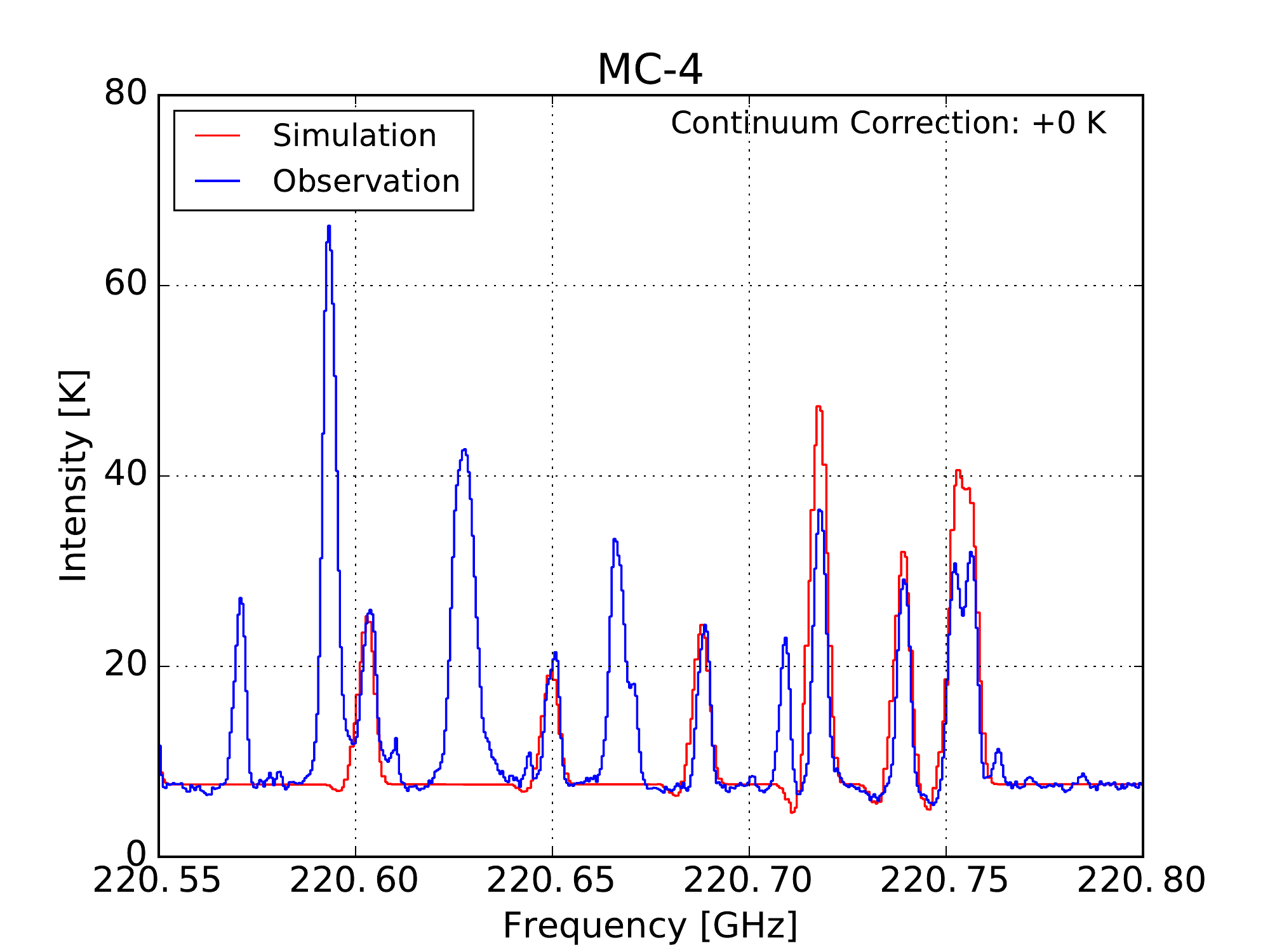}
    \includegraphics[scale=0.36]{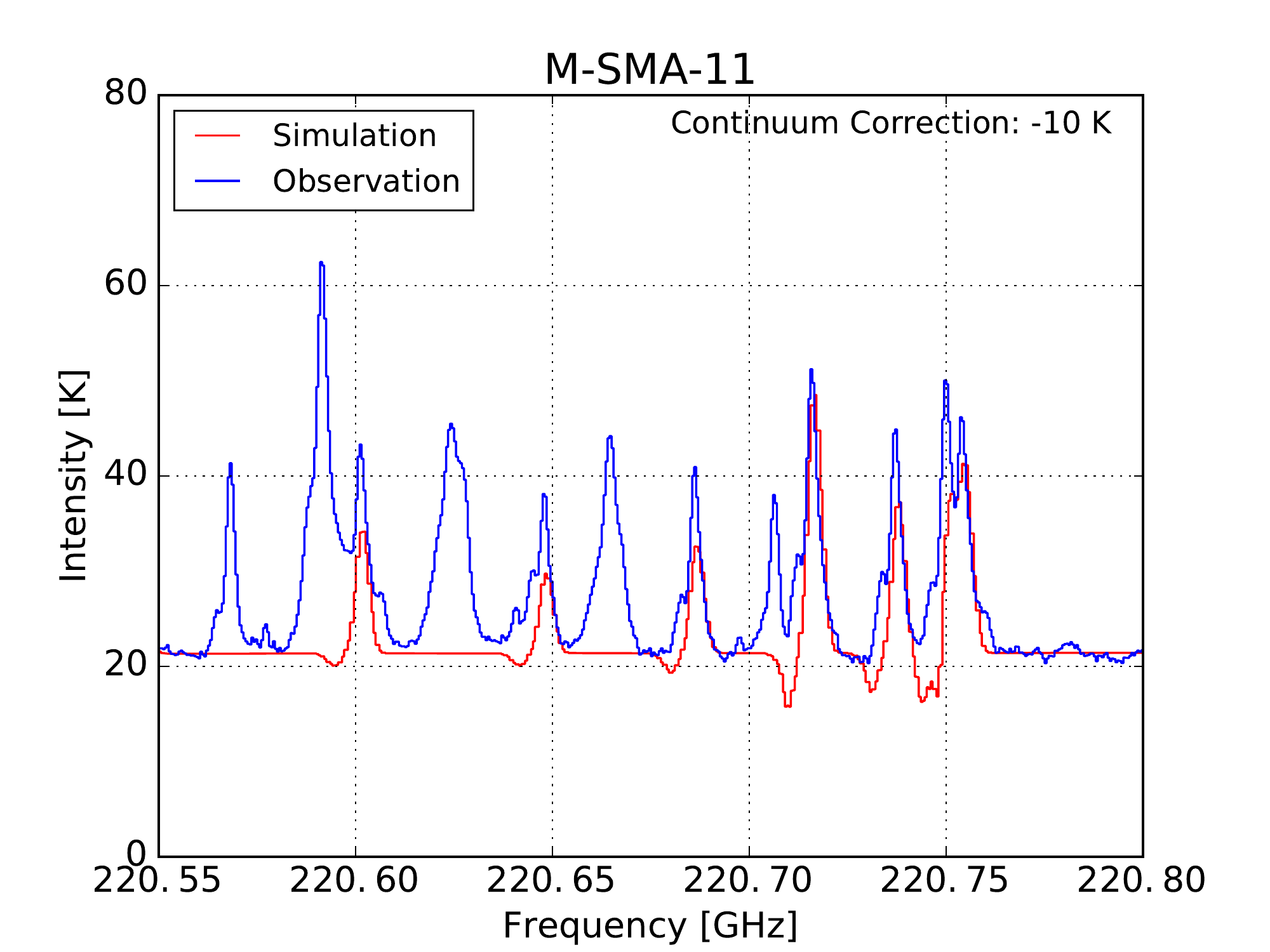}
    \includegraphics[scale=0.36]{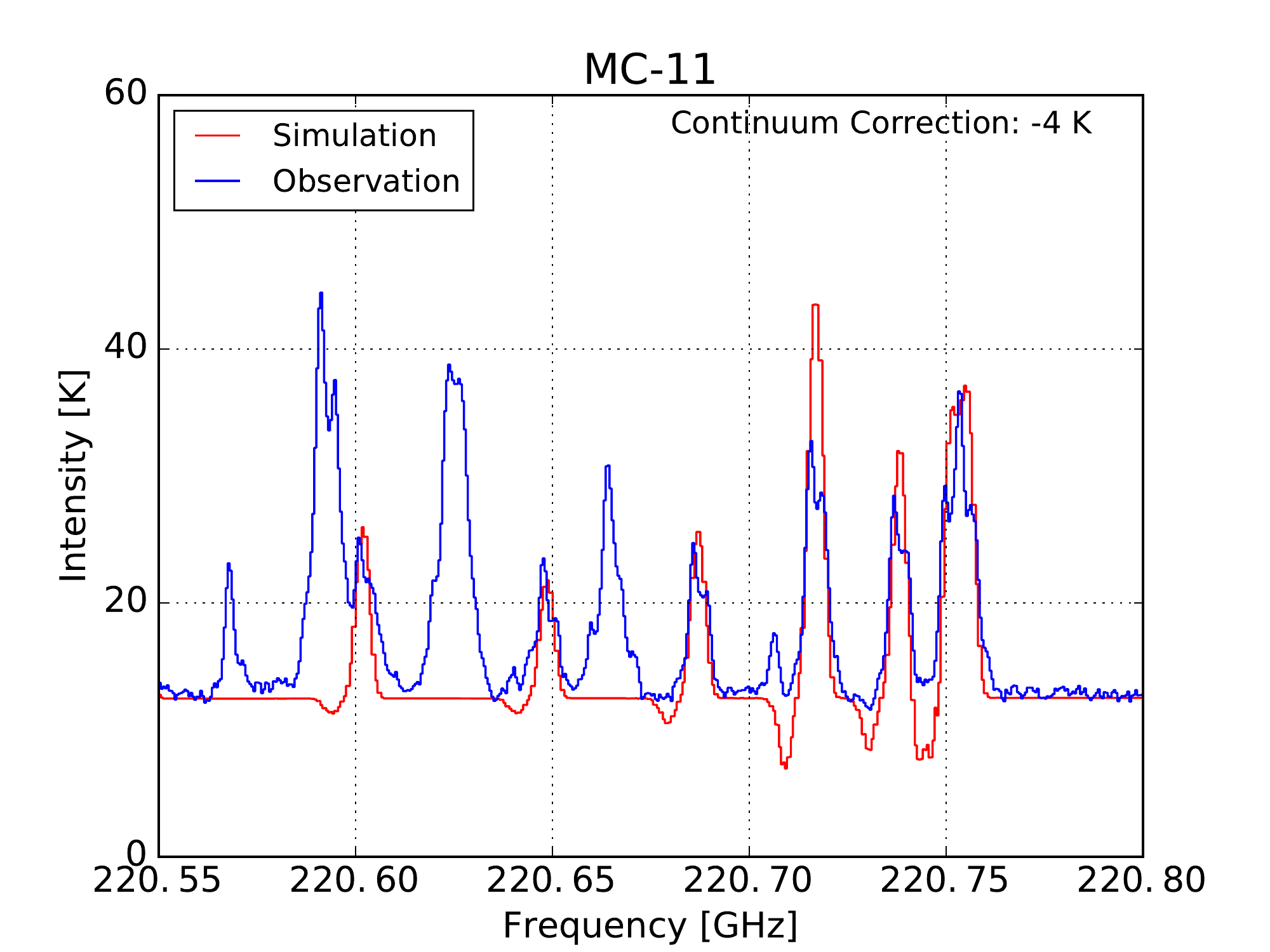}
    \includegraphics[scale=0.36]{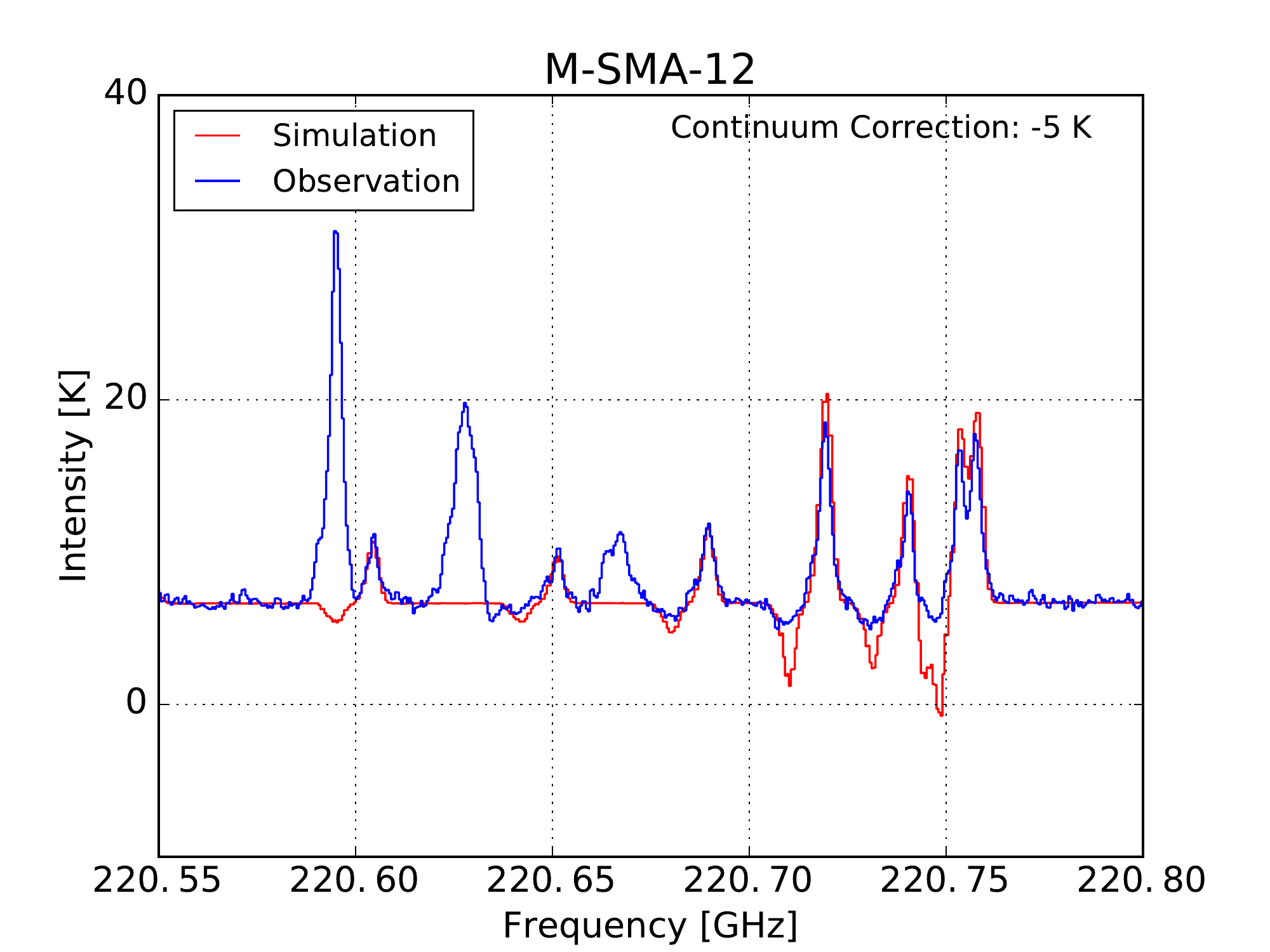}
\caption{Continuation of Fig.~\ref{fig:spectra1} for the remaining cores and molecular centers.}
\label{fig:spectra2}
\end{figure*}

\clearpage
\begin{figure*}
\centering
    \includegraphics[scale=0.36]{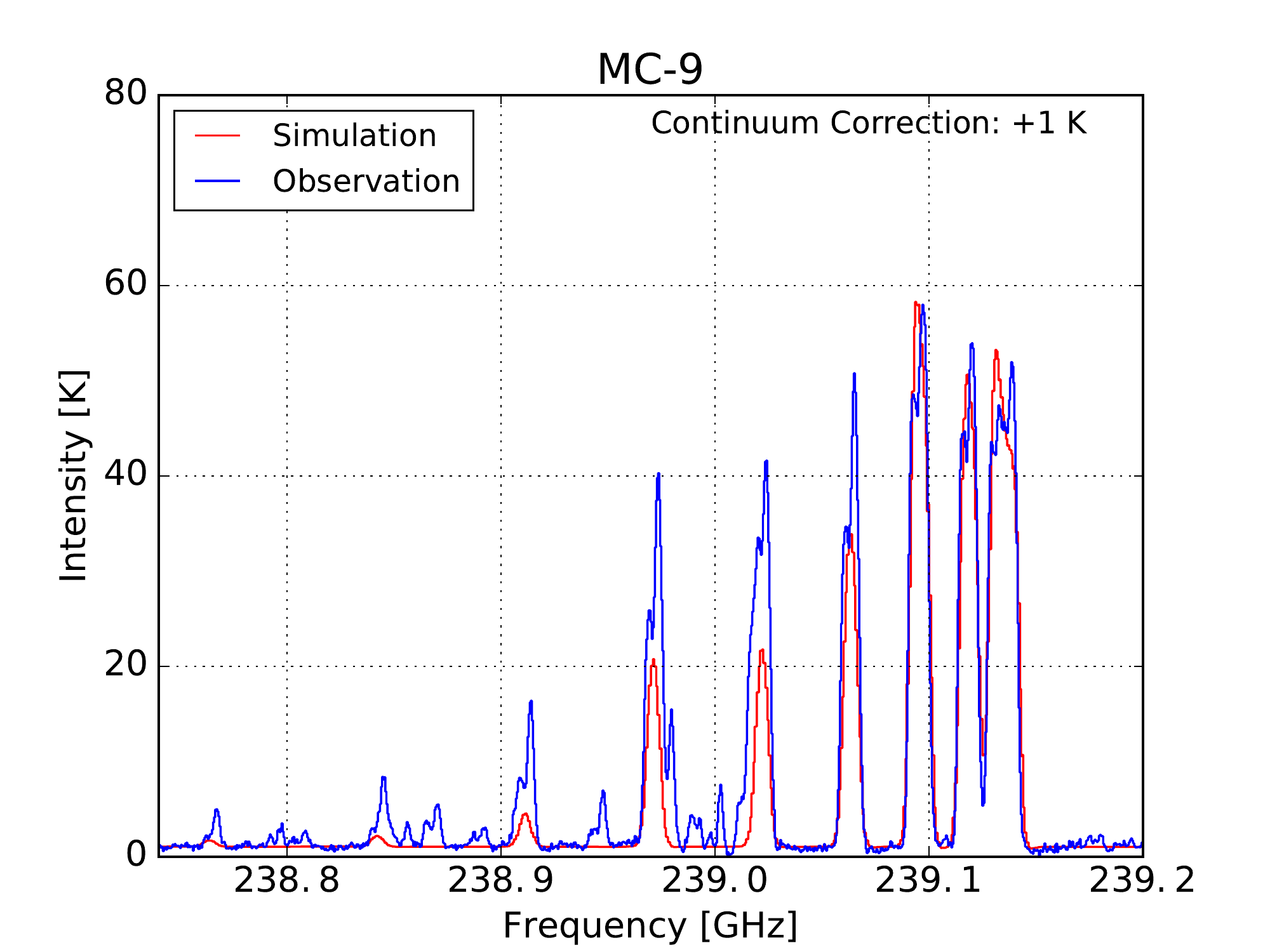}
    \includegraphics[scale=0.36]{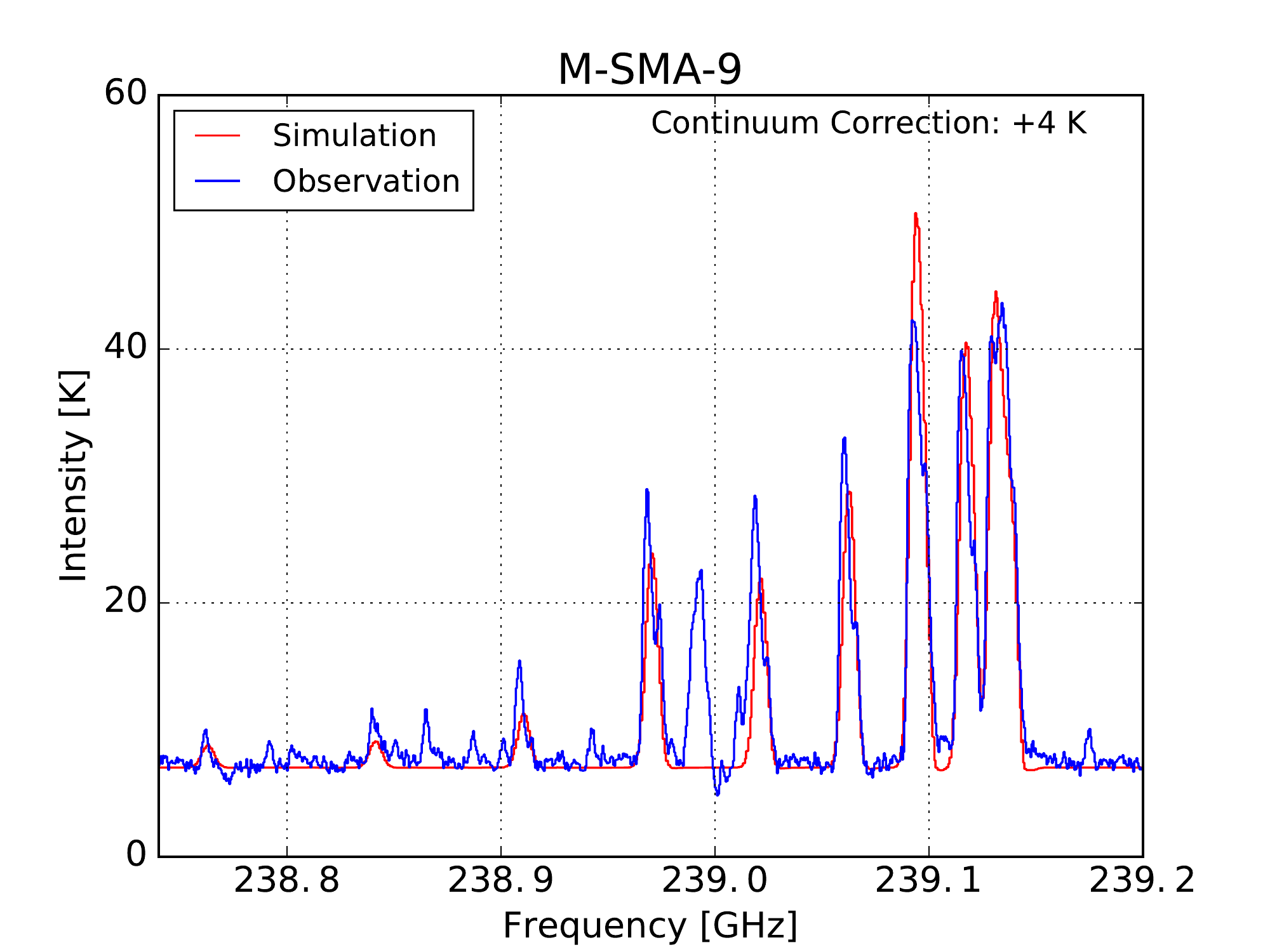}
    \includegraphics[scale=0.36]{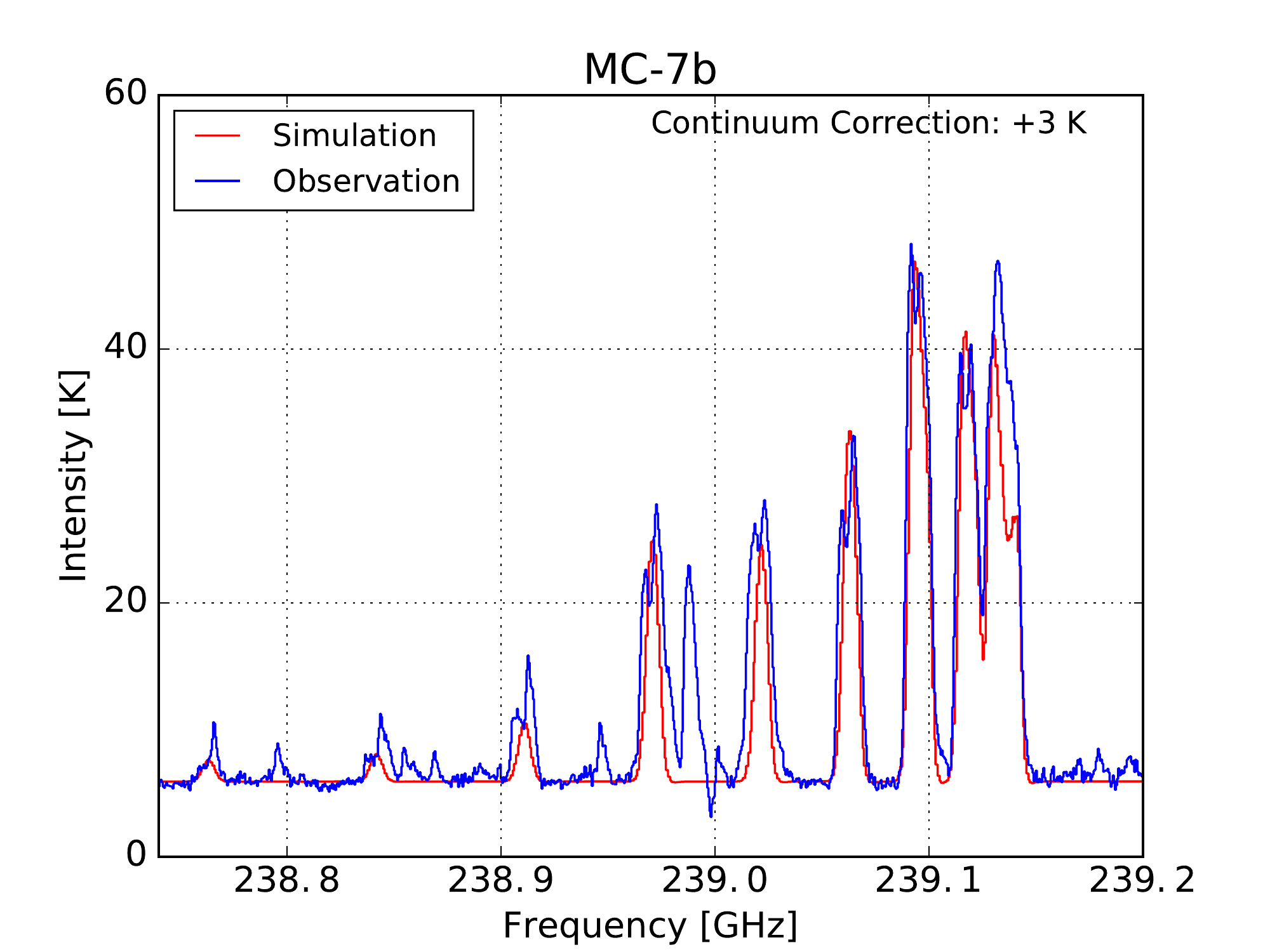}
    \includegraphics[scale=0.36]{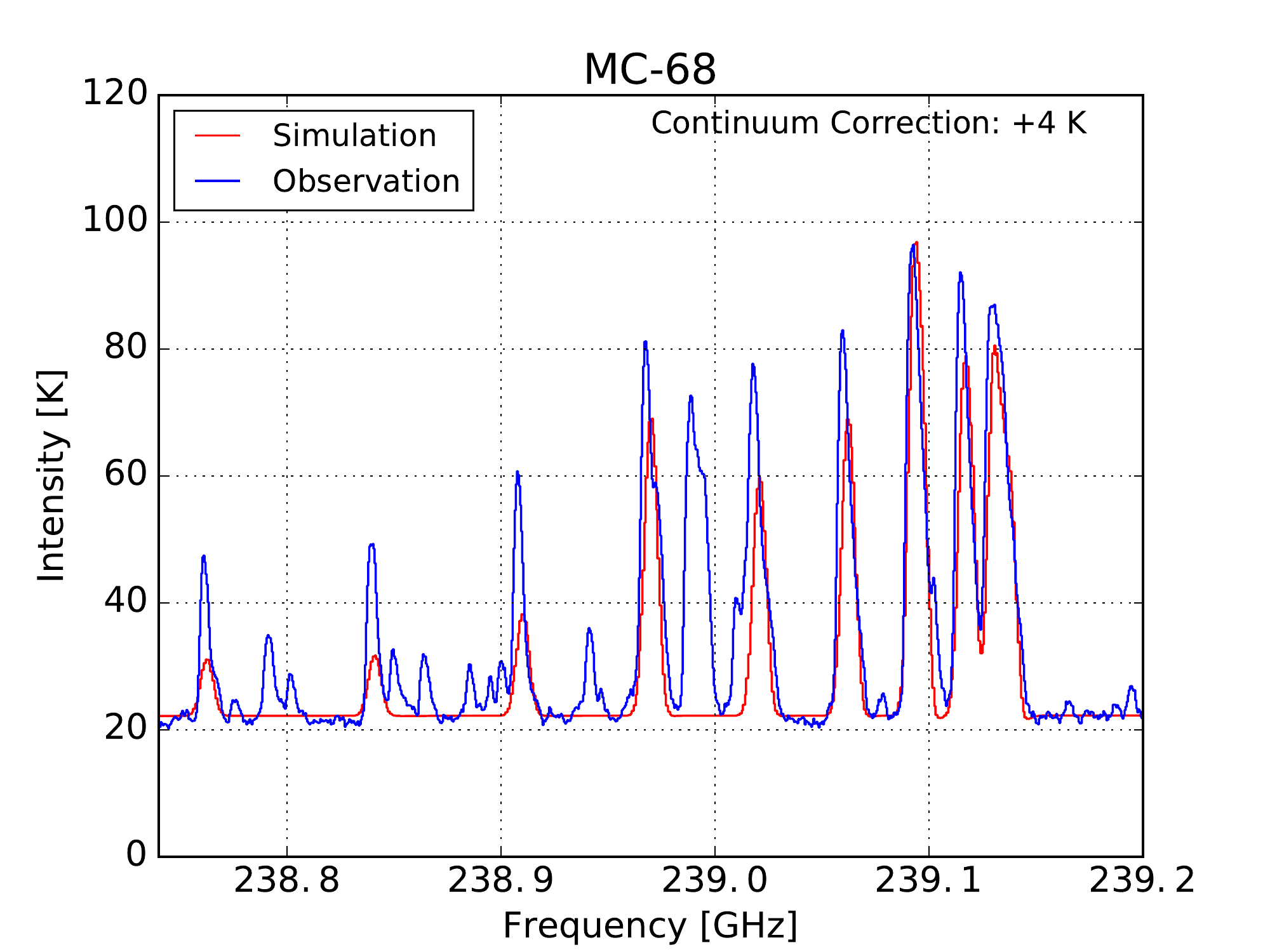}
    \includegraphics[scale=0.36]{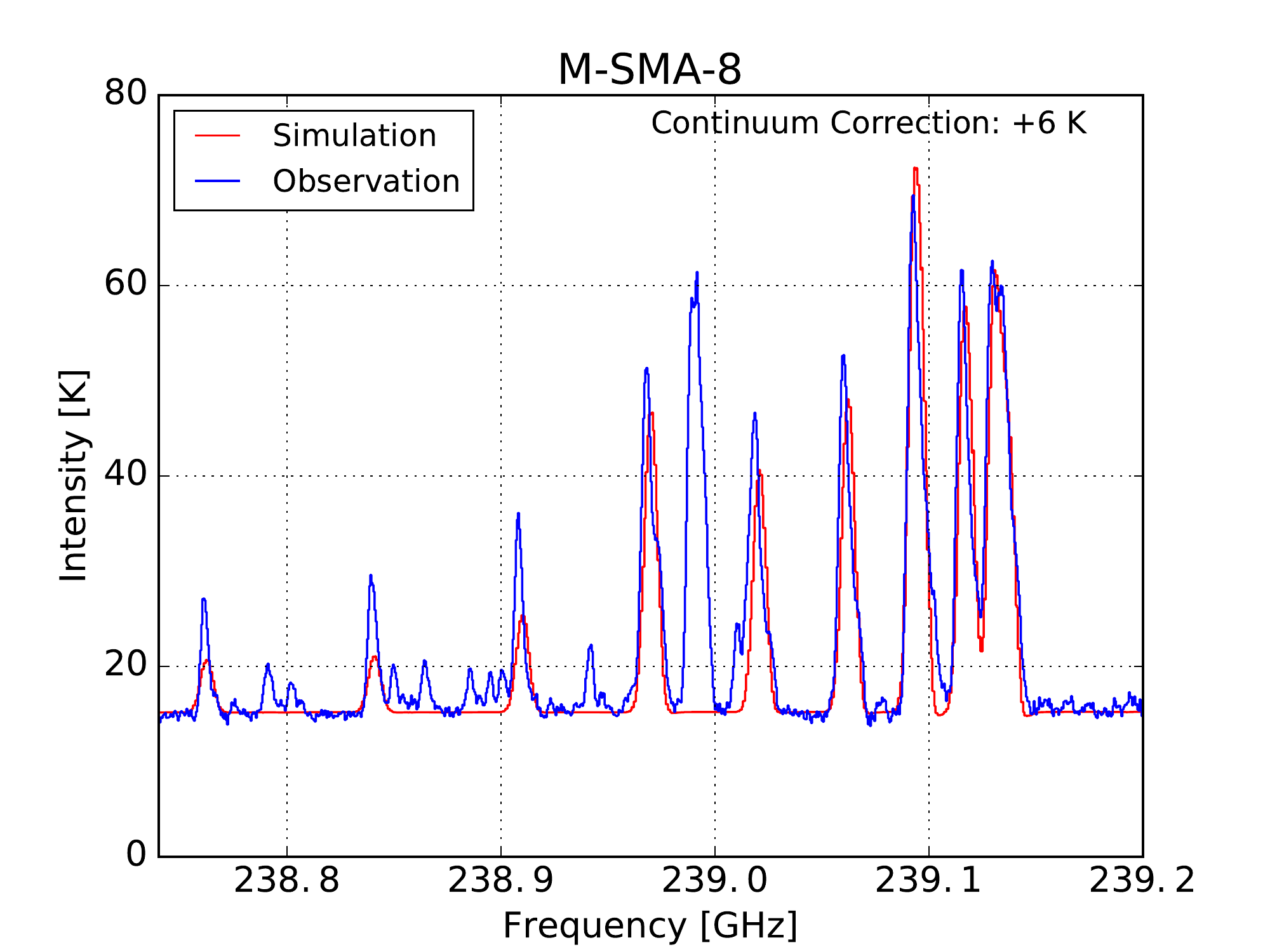}
    \includegraphics[scale=0.36]{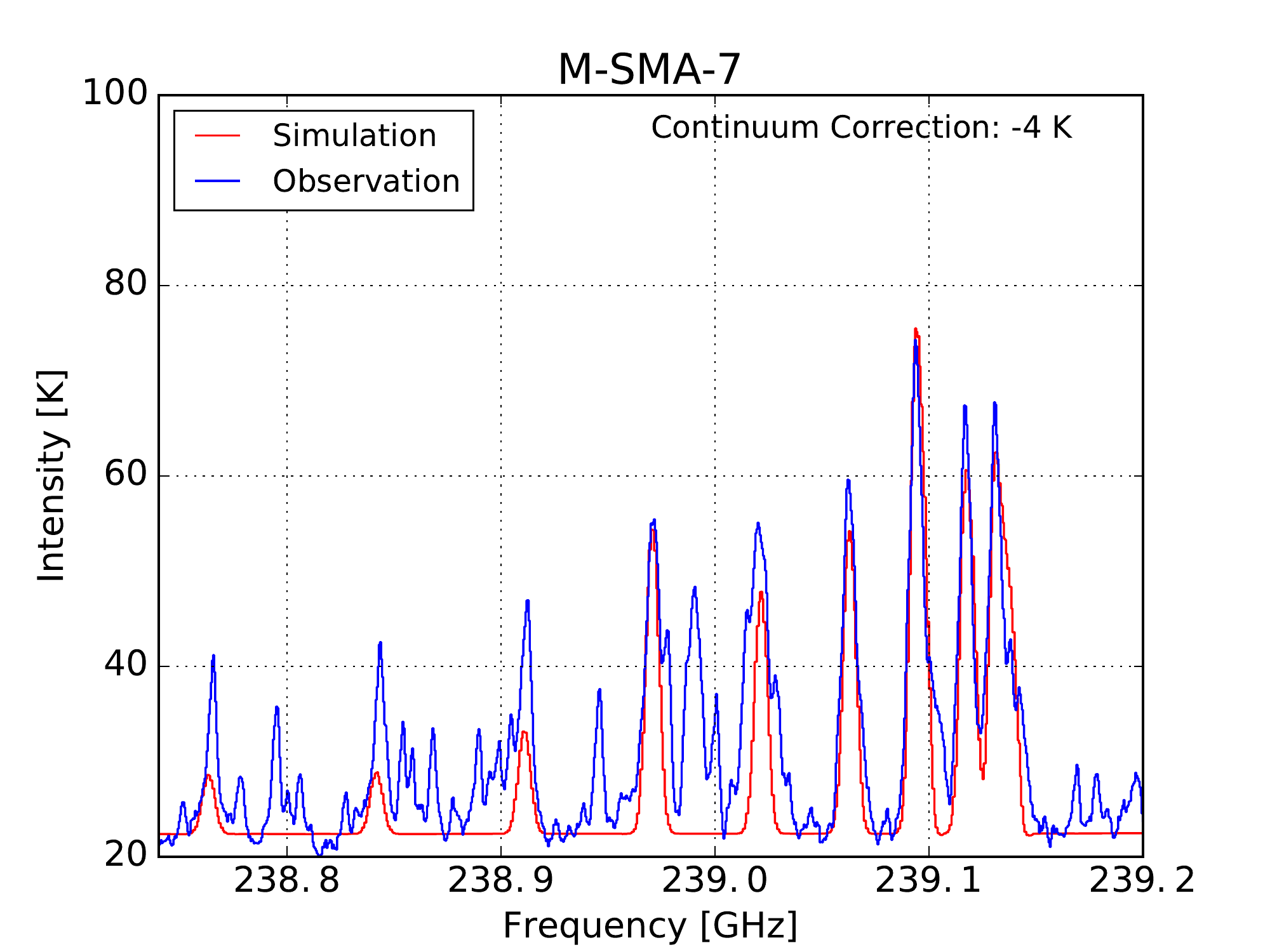}
    \includegraphics[scale=0.36]{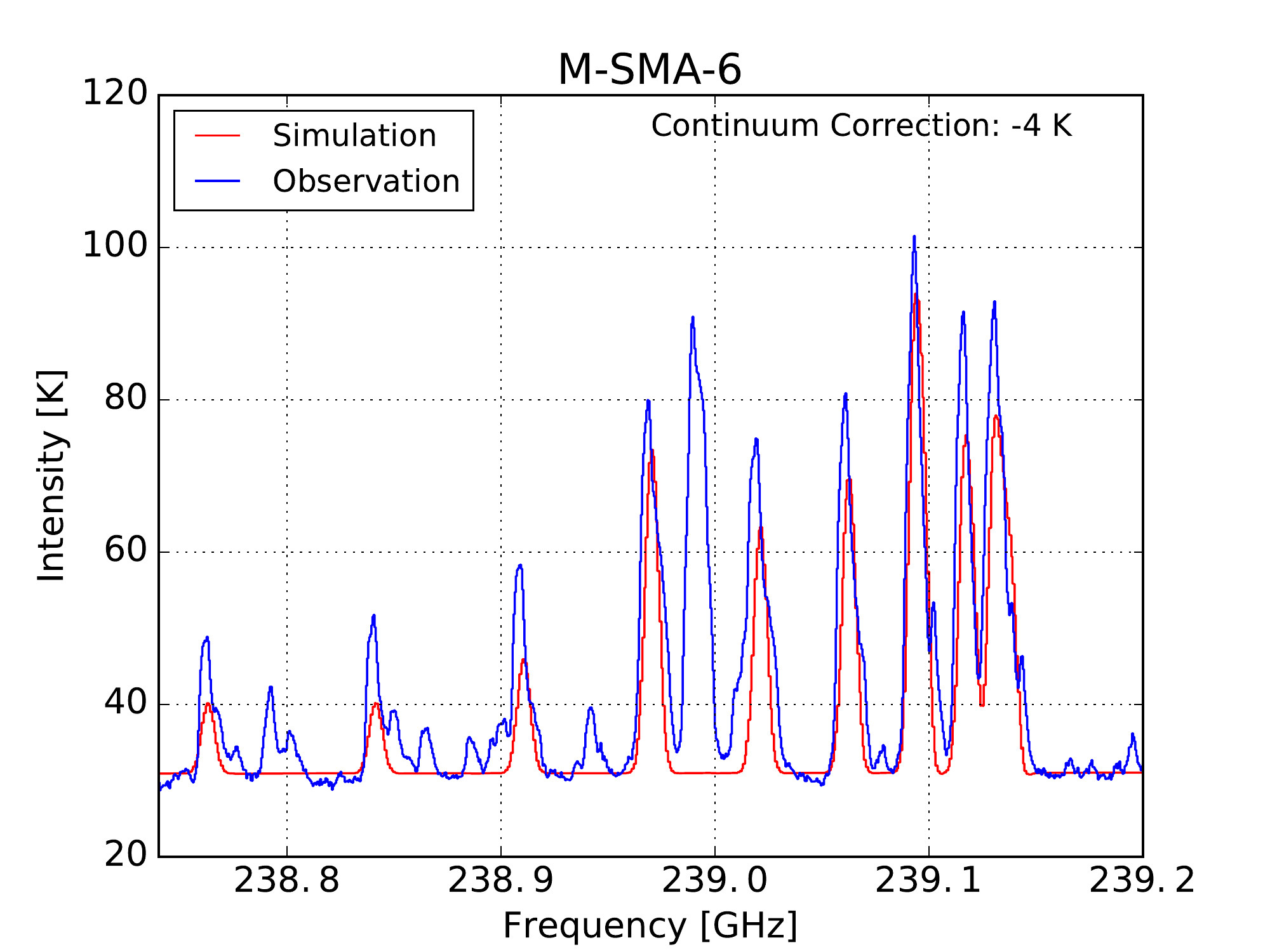}
    \includegraphics[scale=0.36]{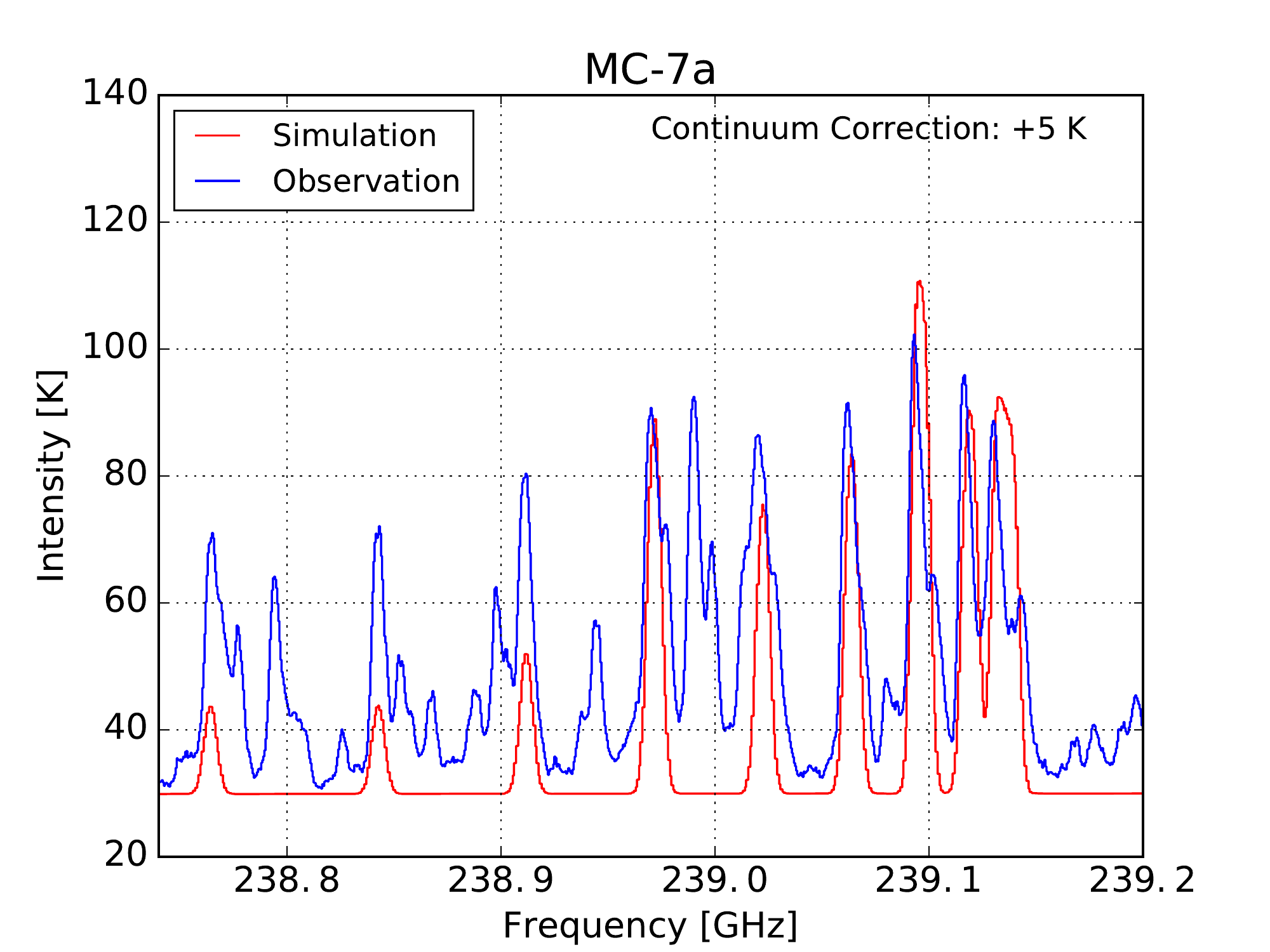}
\caption{Spectra of the observational (in blue) and simulated (in red) data extracted from the position of different cores and molecular centers for the \chcn\ $J$=13--12 transition. The spectra are arranged according to the position of the corresponding cores and molecular centers along a south-north direction, starting from the most southern core. The name of the core and molecular center is shown for each panel. The continuum correction necessary to match the continuum level of the observation and simulation is indicated in the top-right side of each panel.}
\label{fig:spectra_J=13-12_1}
\end{figure*}

\clearpage
\begin{figure*}
\centering
    \includegraphics[scale=0.36]{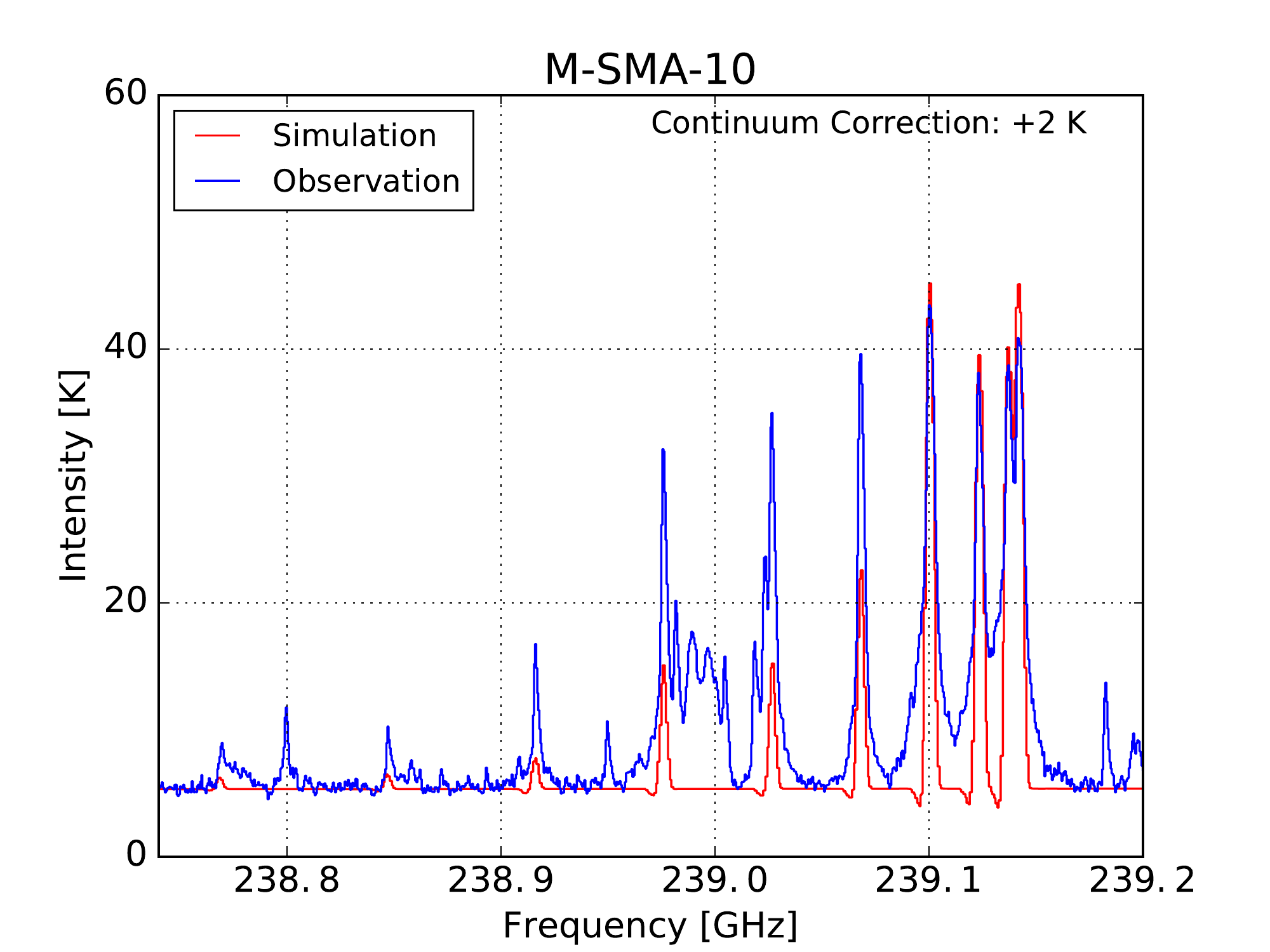}
    \includegraphics[scale=0.36]{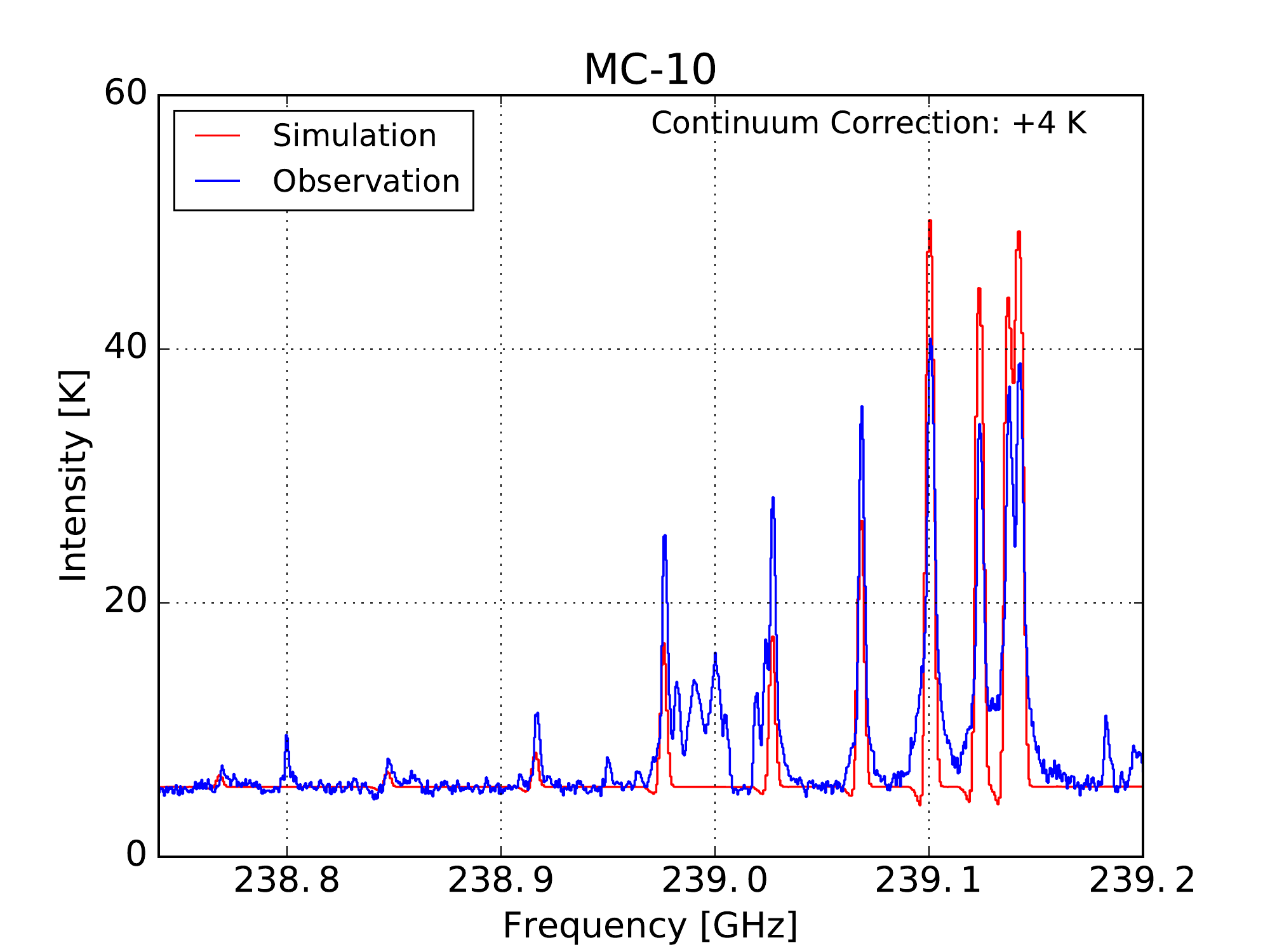}
    \includegraphics[scale=0.36]{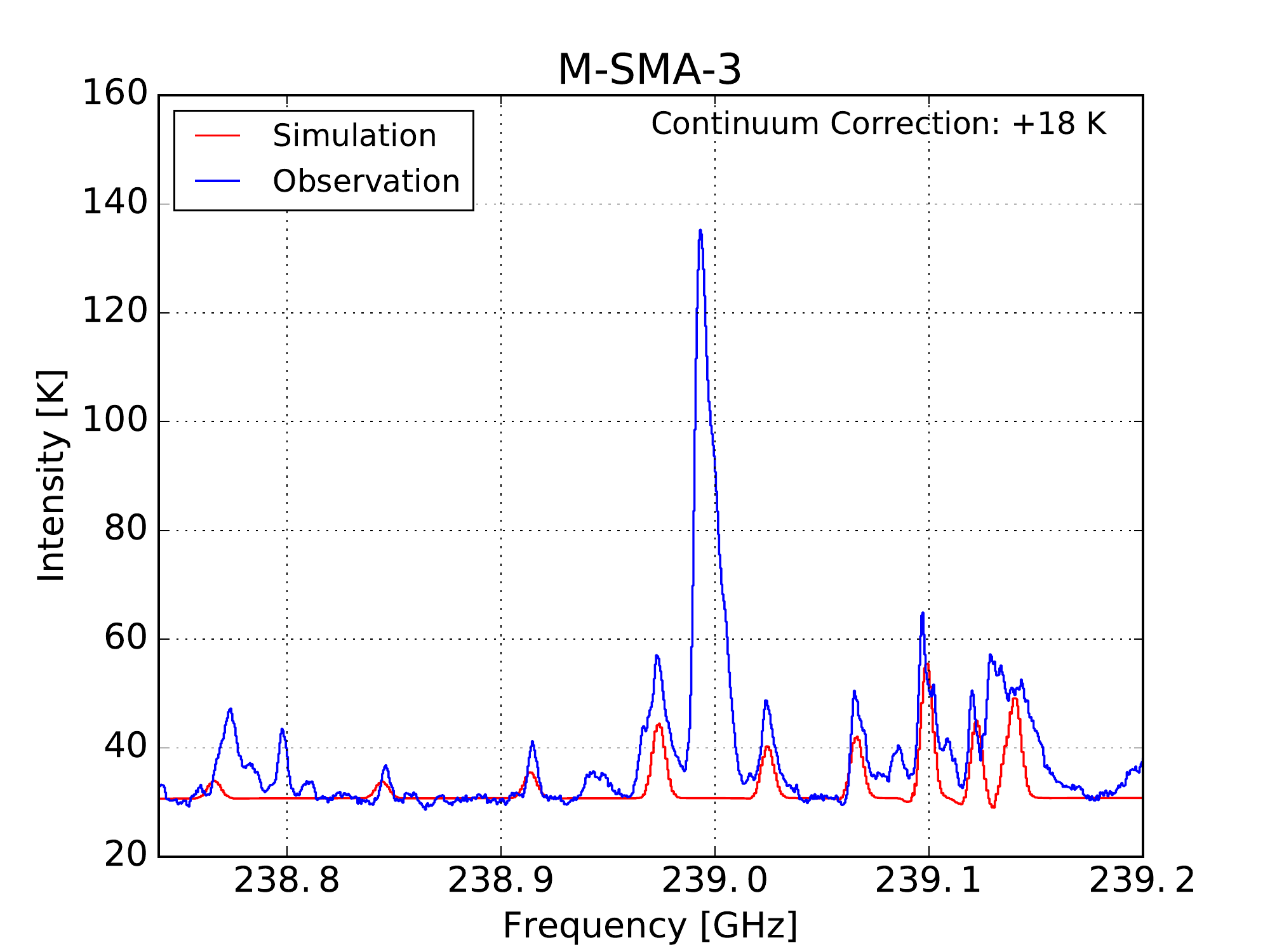}
    \includegraphics[scale=0.36]{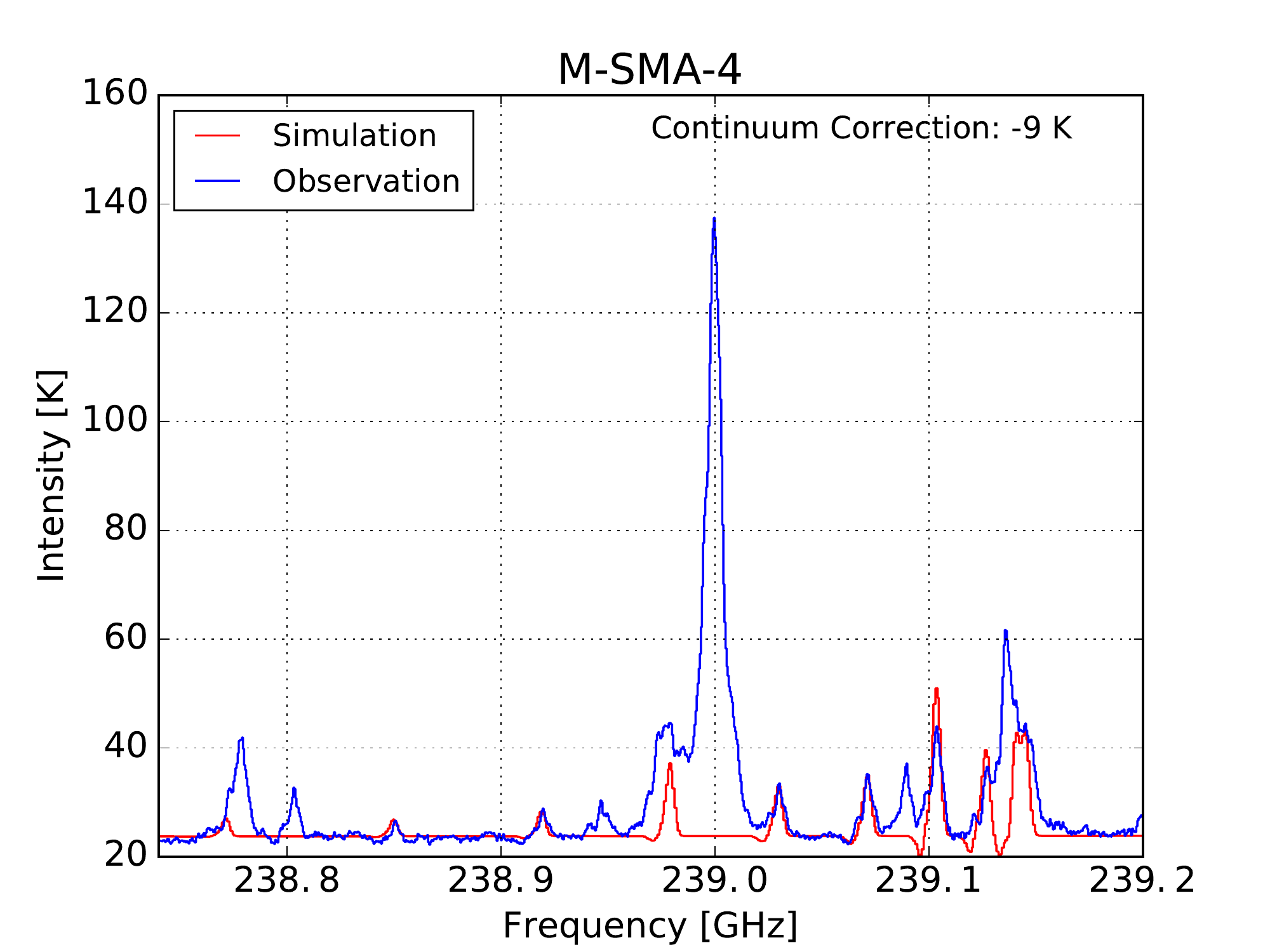}
    \includegraphics[scale=0.36]{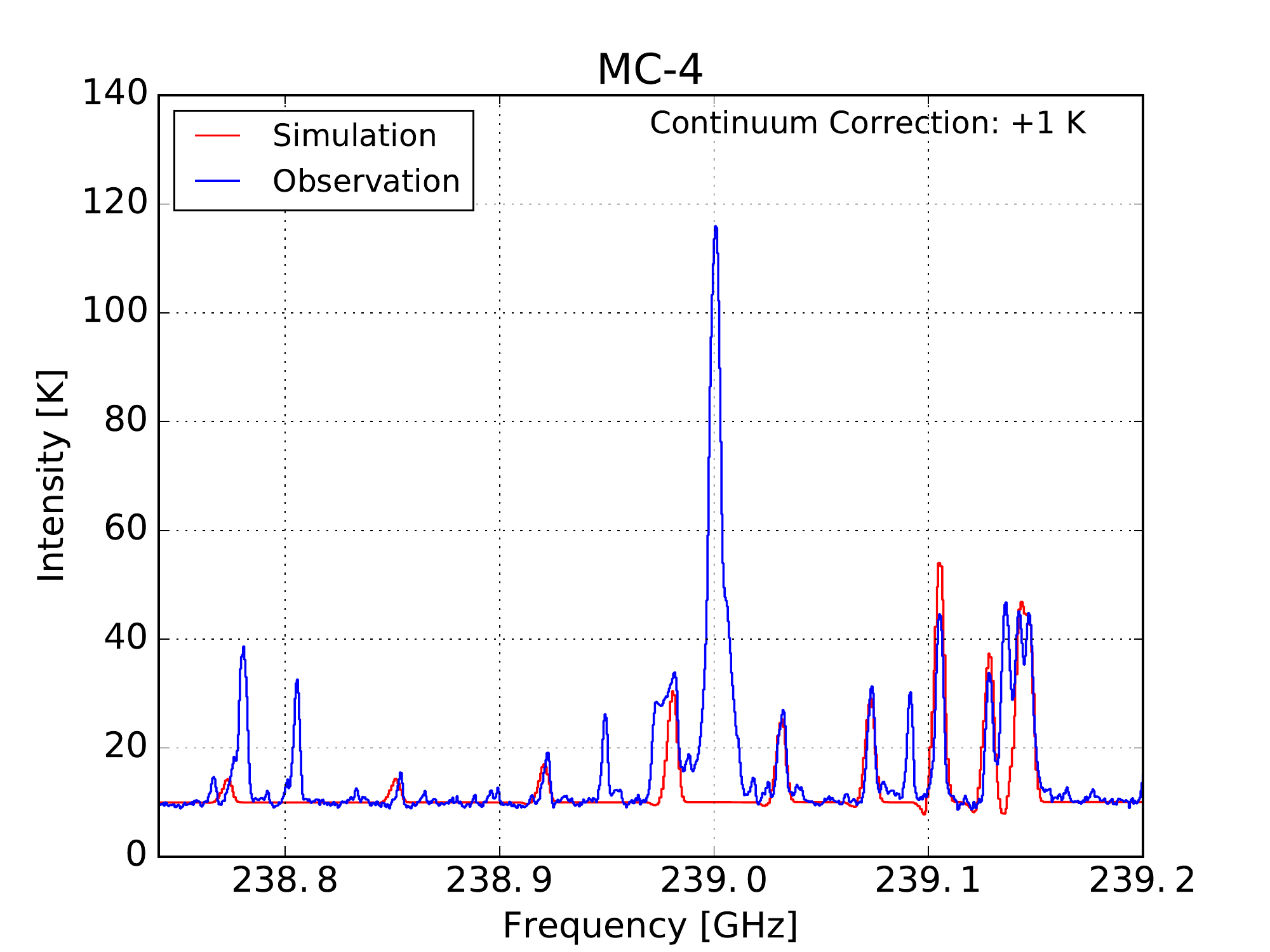}
    \includegraphics[scale=0.36]{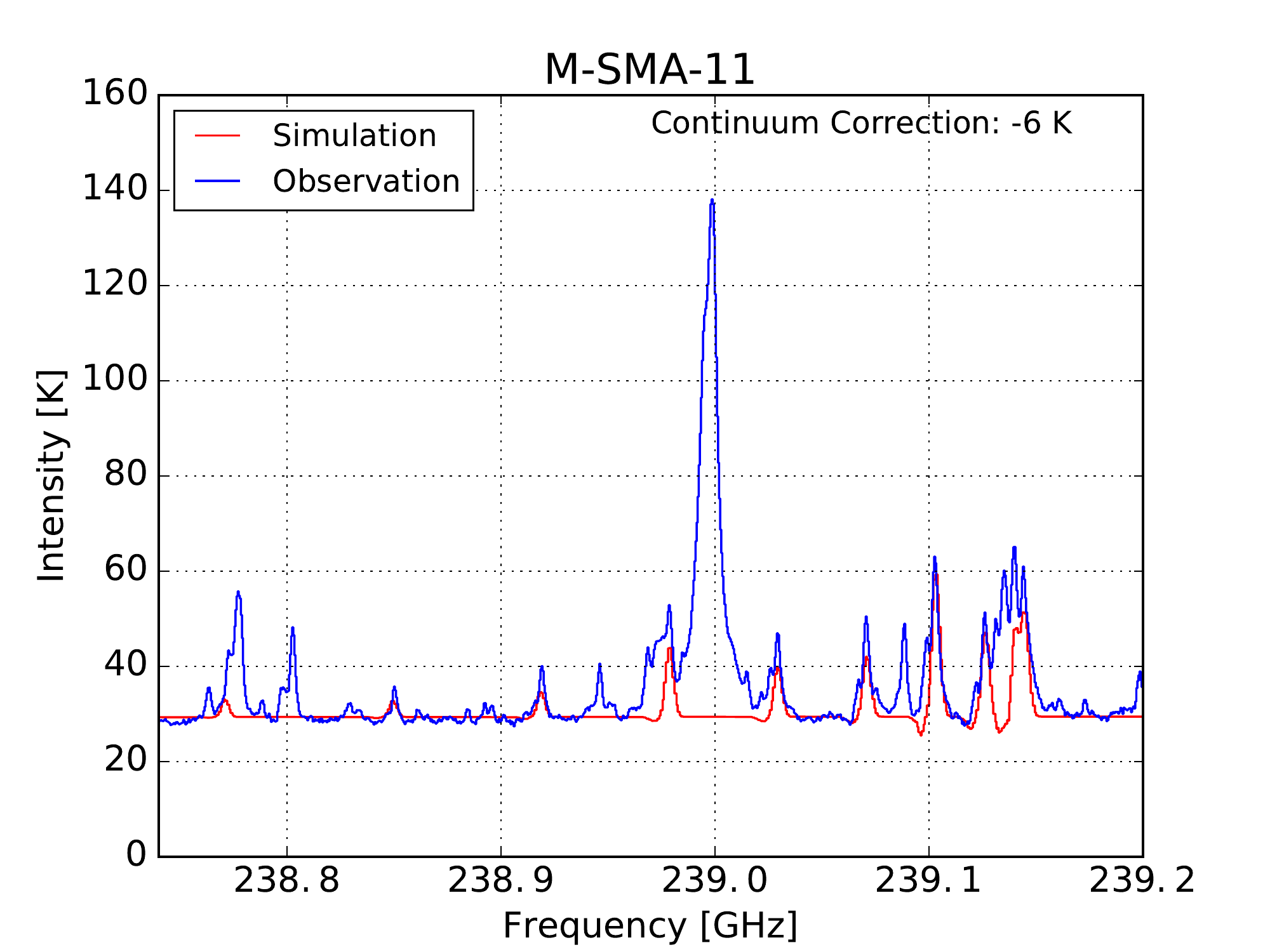}
    \includegraphics[scale=0.36]{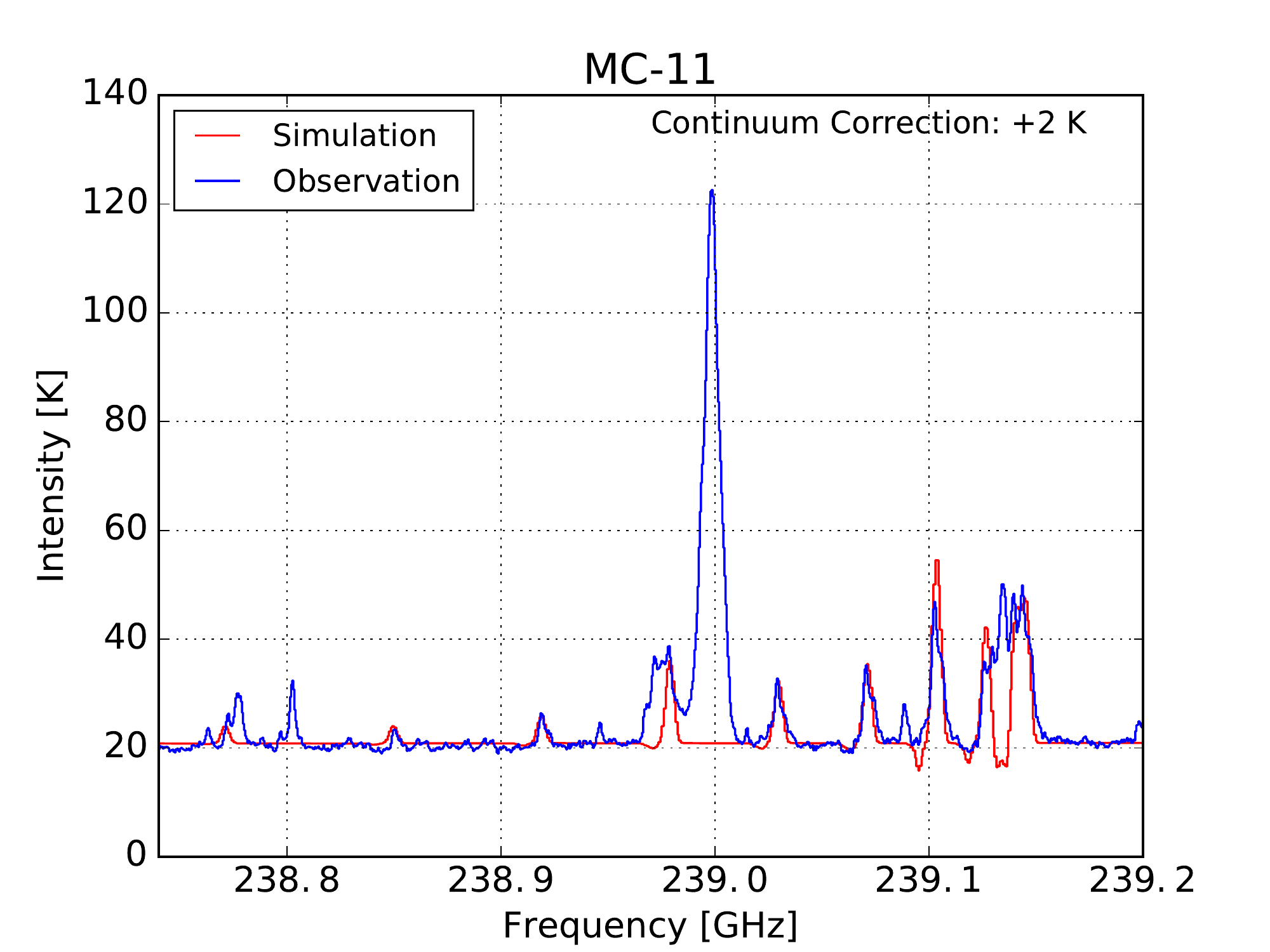}
    \includegraphics[scale=0.36]{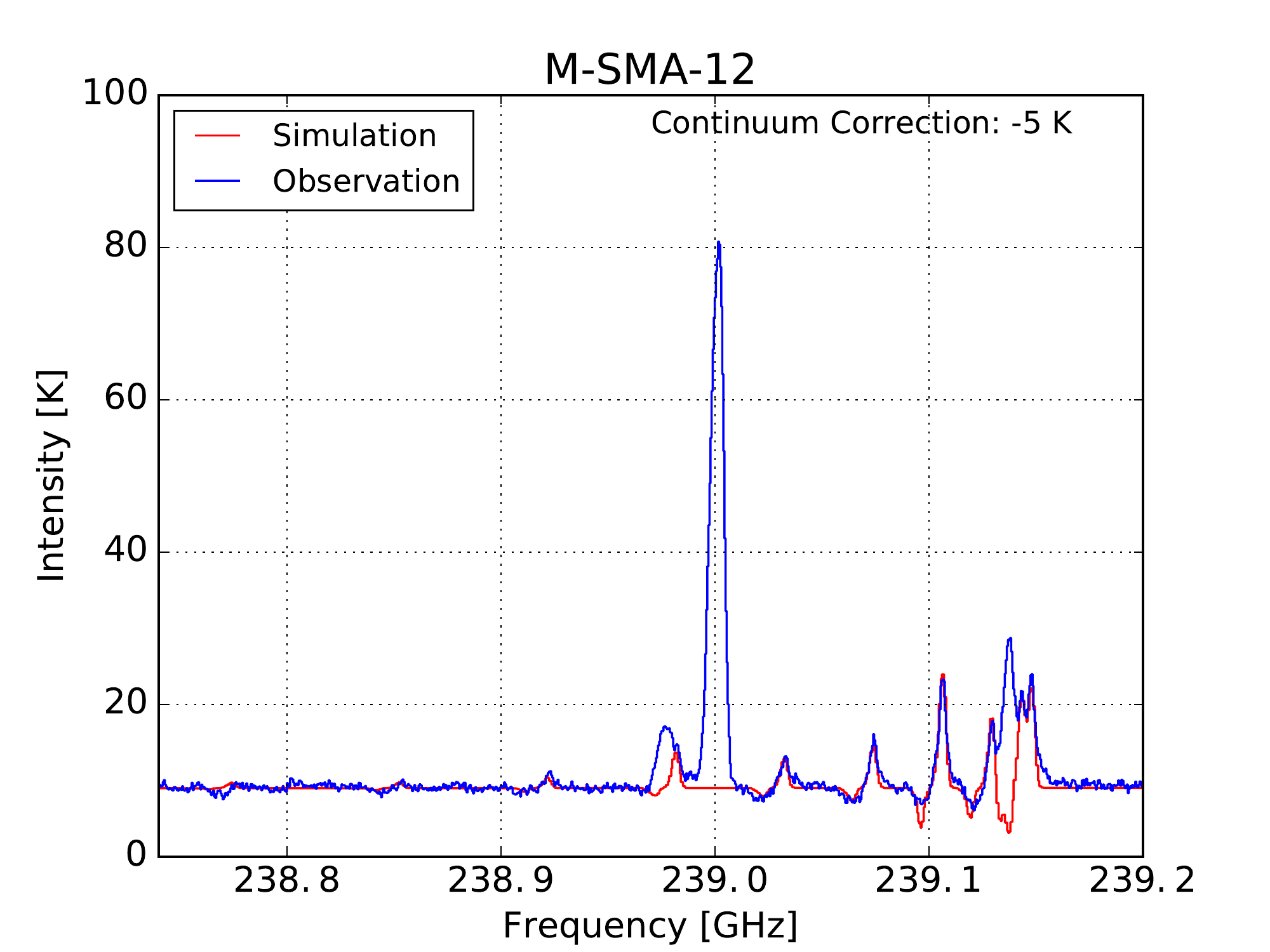}
\caption{Continuation of Fig.~\ref{fig:spectra_J=13-12_1} for the remaining cores and molecular centers.}
\label{fig:spectra_J=13-12_2}
\end{figure*}

\clearpage
\begin{figure*}
\centering
    \includegraphics[scale=0.36]{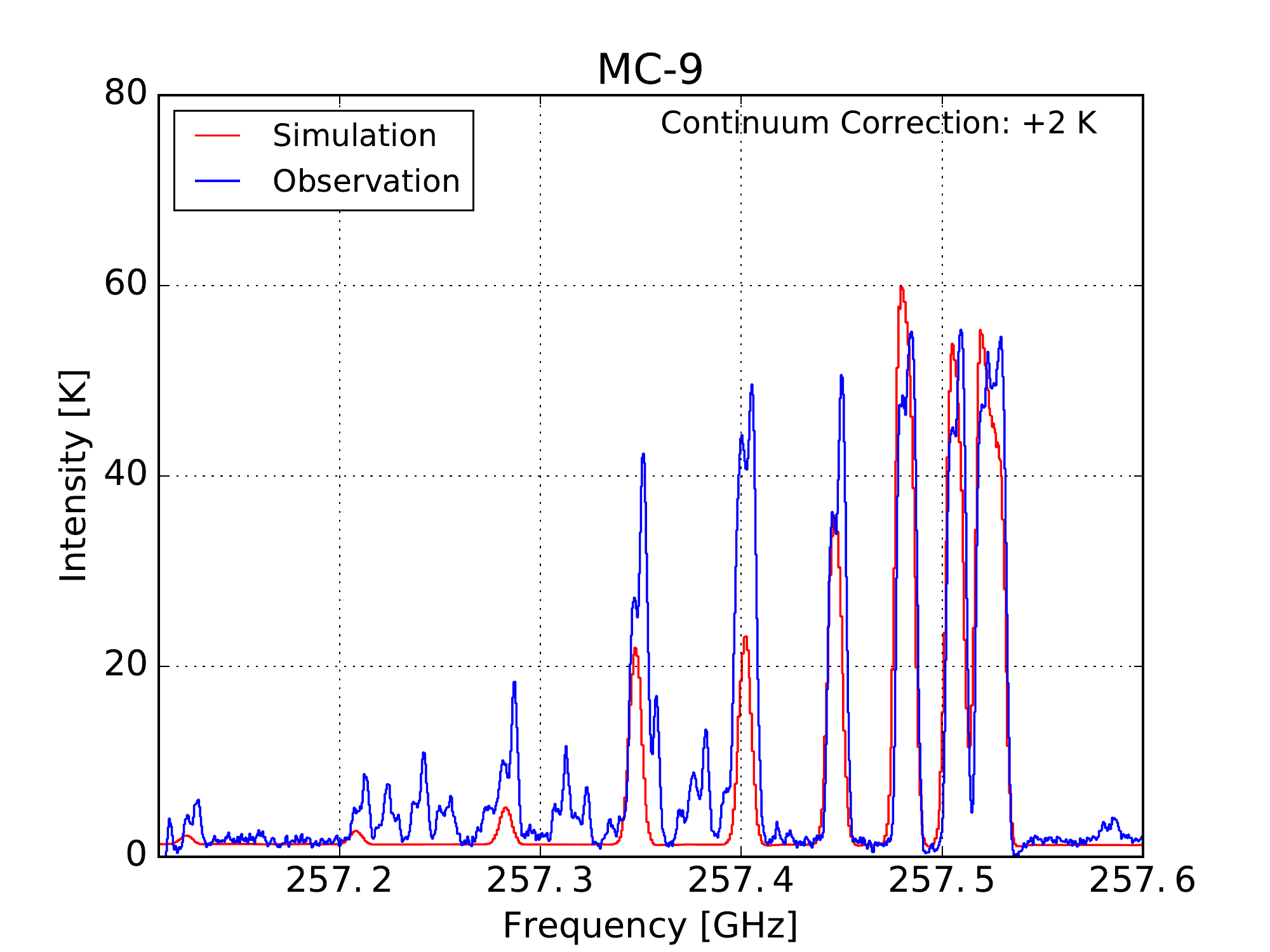}
    \includegraphics[scale=0.36]{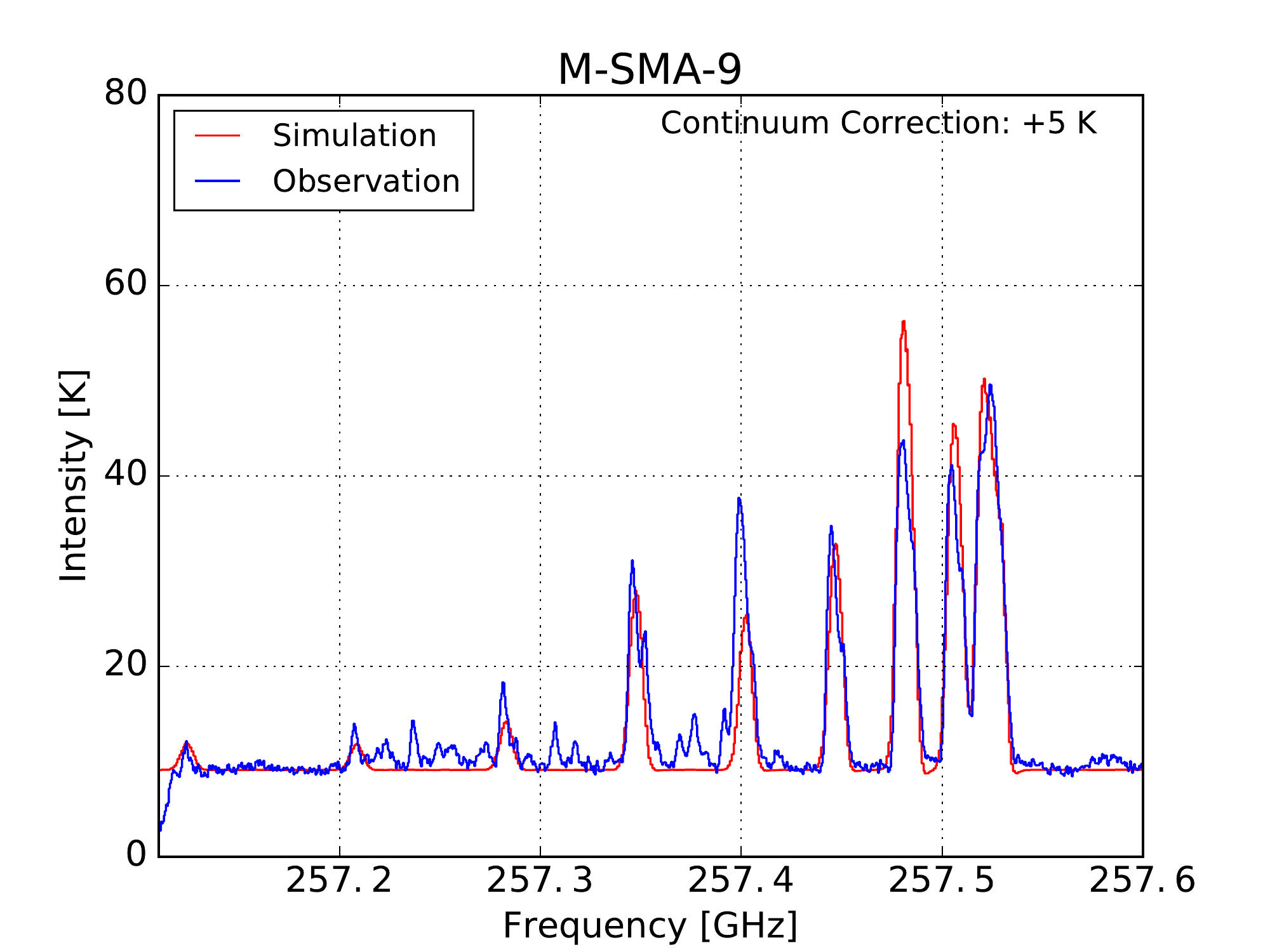}
    \includegraphics[scale=0.36]{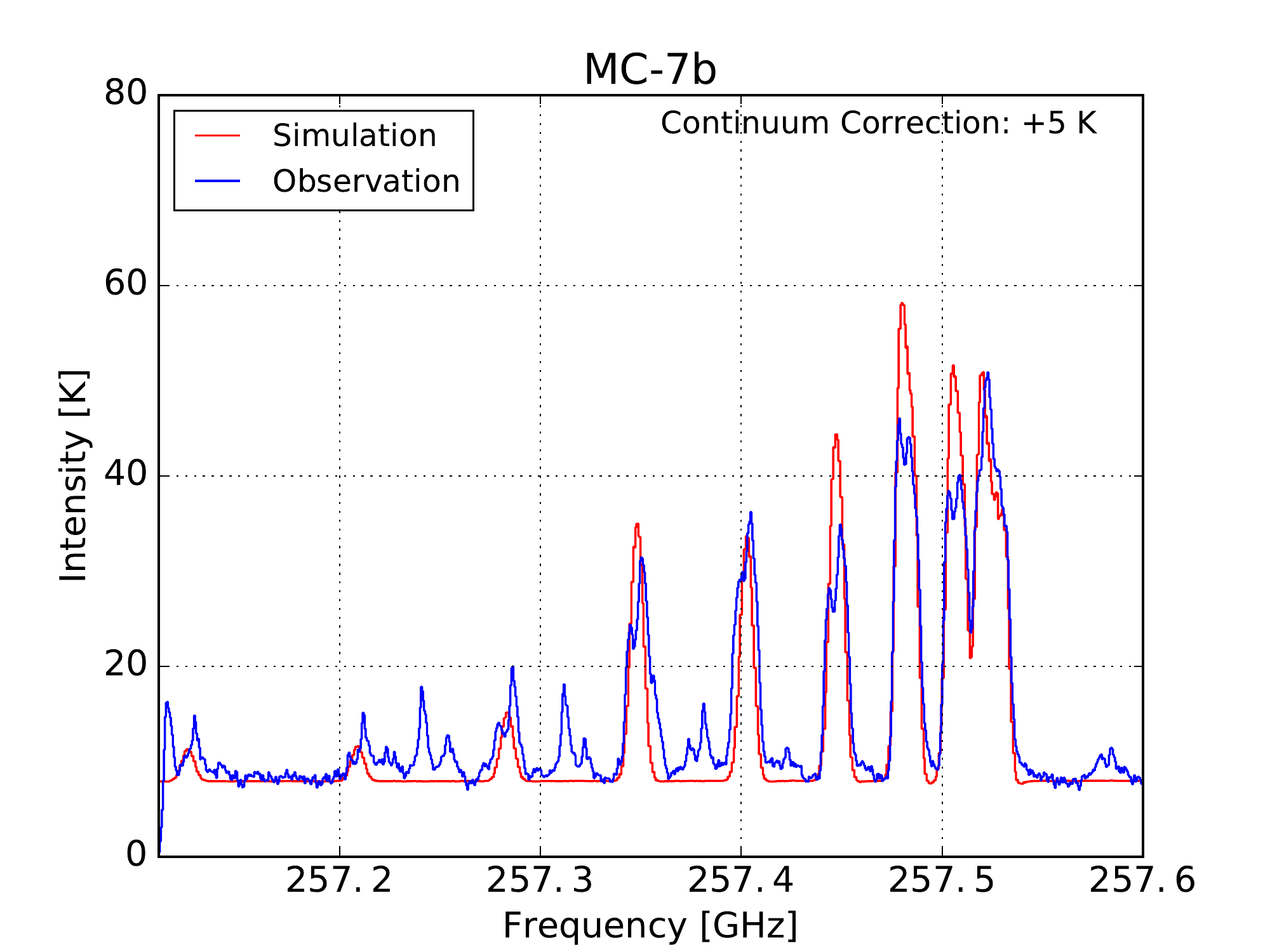}
    \includegraphics[scale=0.36]{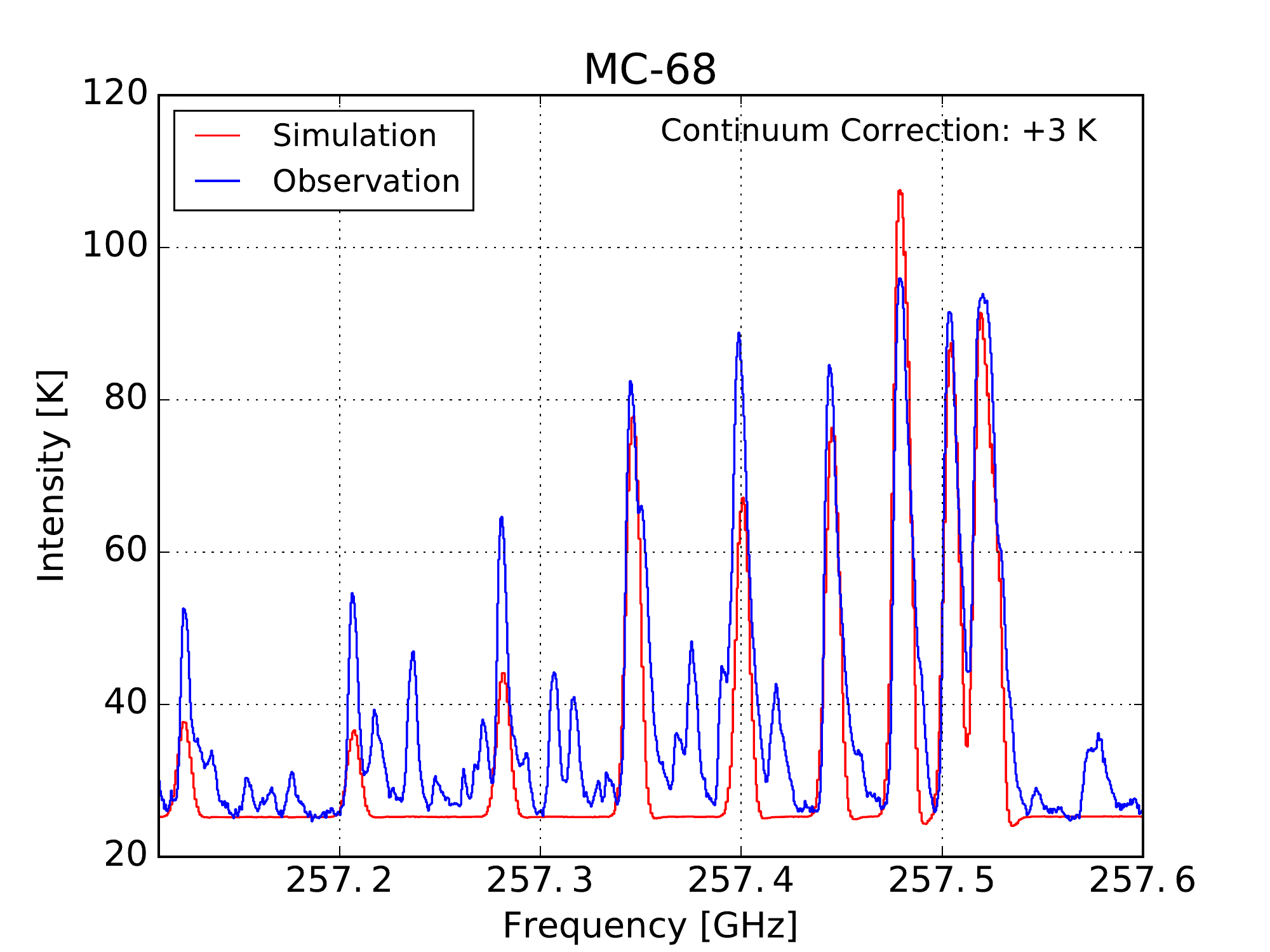}
    \includegraphics[scale=0.36]{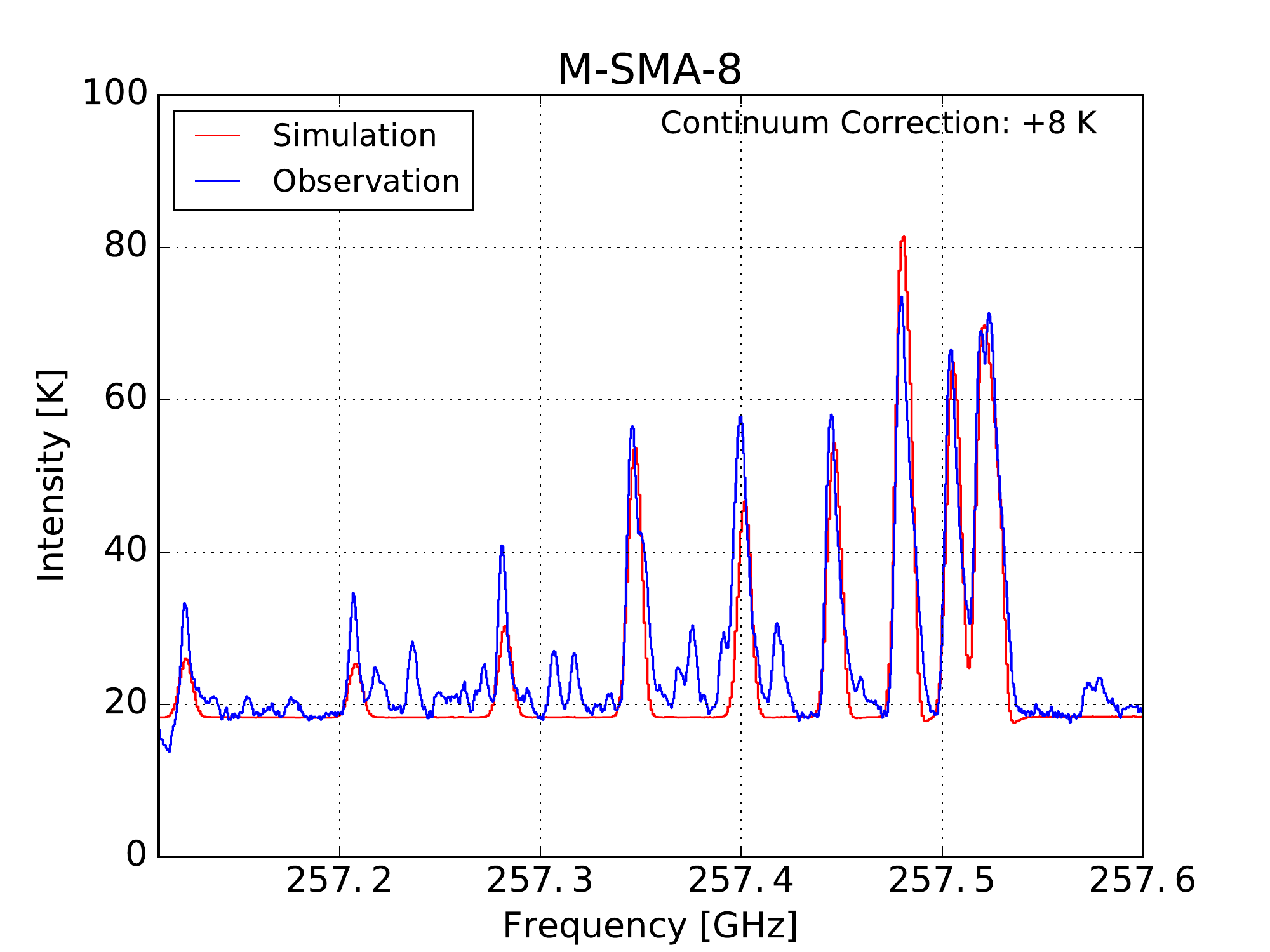}
    \includegraphics[scale=0.36]{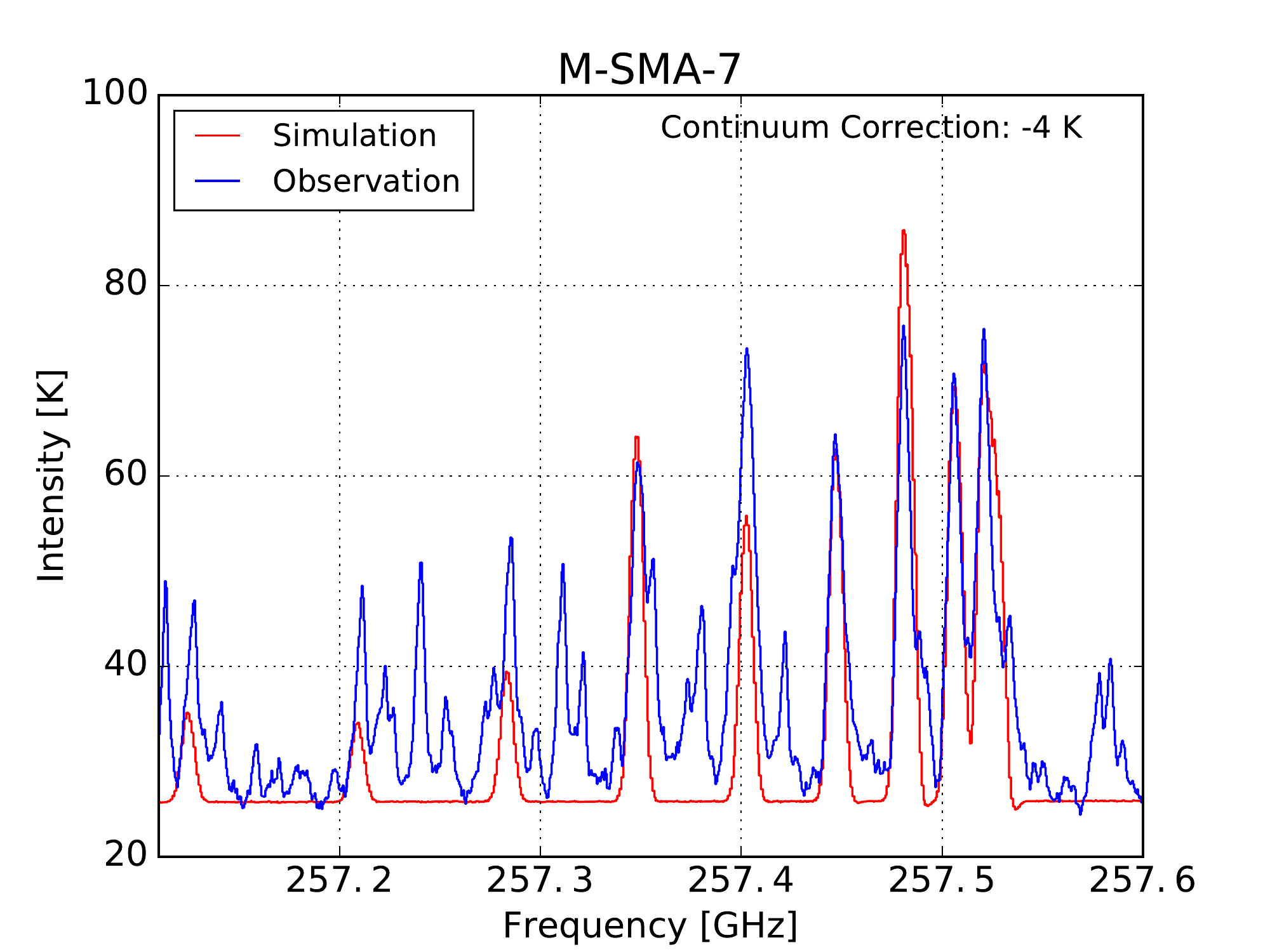}
    \includegraphics[scale=0.36]{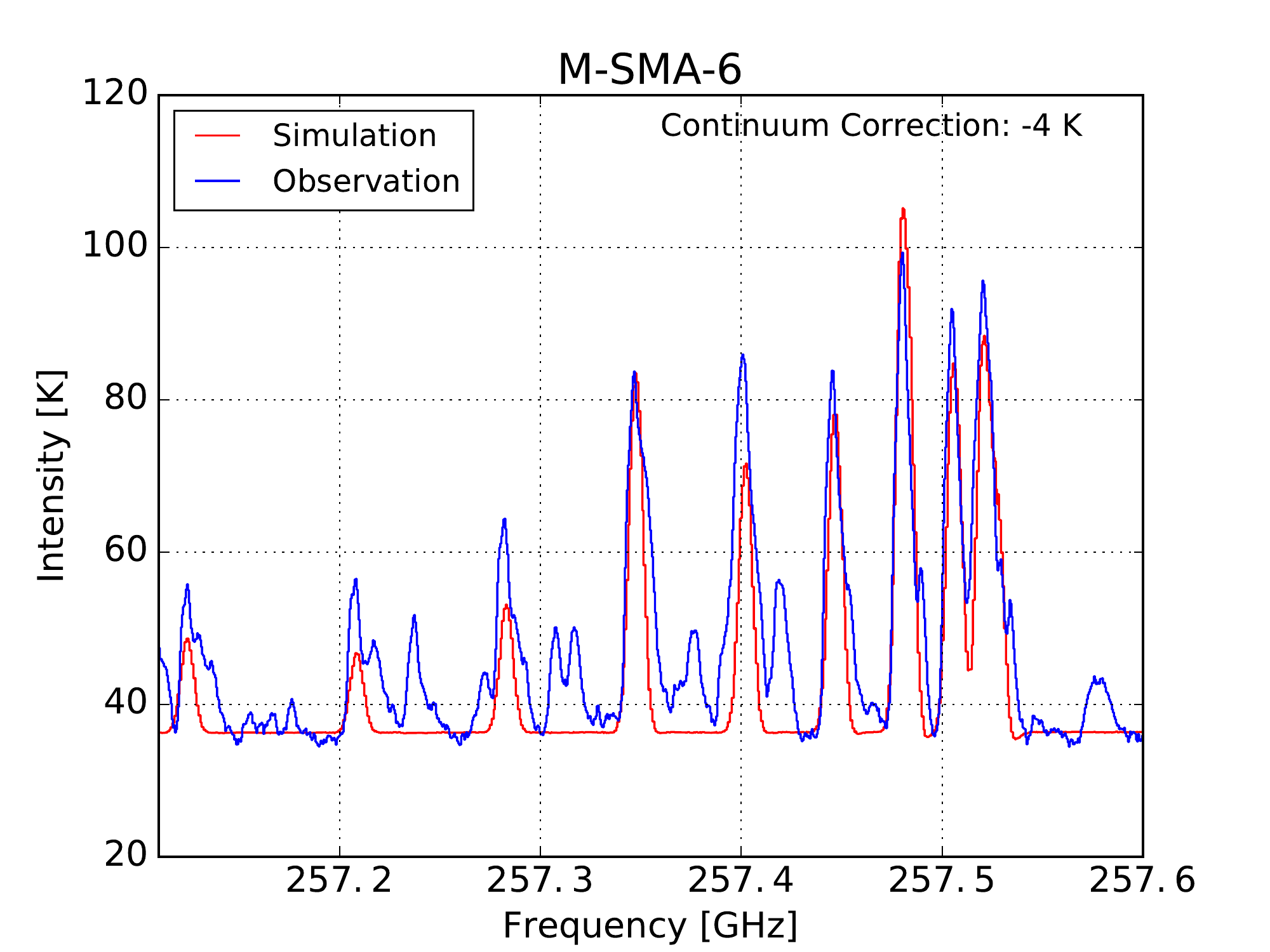}
    \includegraphics[scale=0.36]{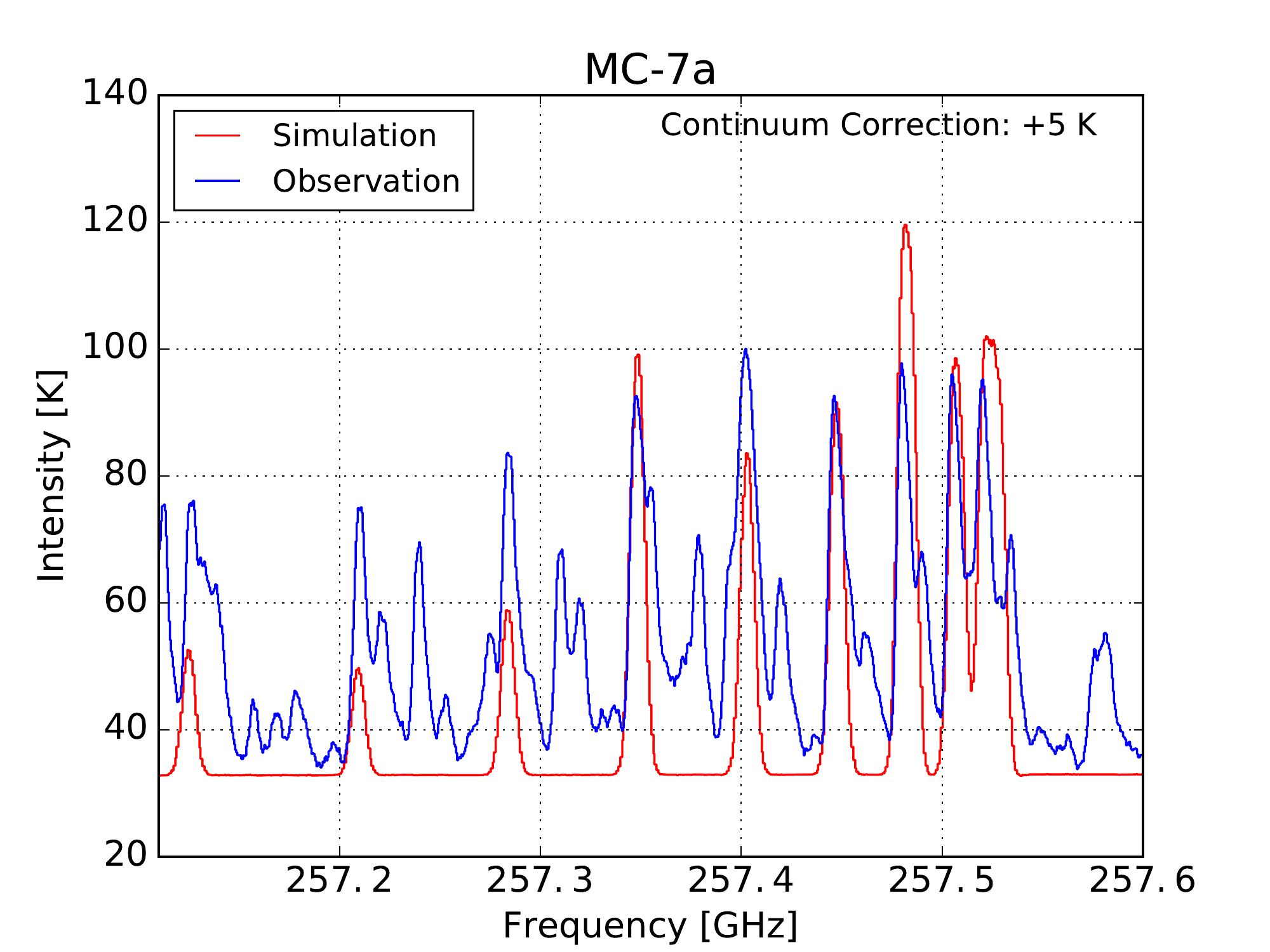}
\caption{Spectra of the observational (in blue) and simulated (in red) data extracted from the position of different cores and molecular centers for the \chcn\ $J$=14--13 transition. The spectra are arranged according to the position of the corresponding cores and molecular centers along a south-north direction, starting from the most southern core. The name of the core and molecular center is shown for each panel. The continuum correction necessary to match the continuum level of the observation and simulation is indicated in the top-right side of each panel.}
\label{fig:spectra_J=14-13_1}
\end{figure*}

\clearpage
\begin{figure*}
\centering
    \includegraphics[scale=0.36]{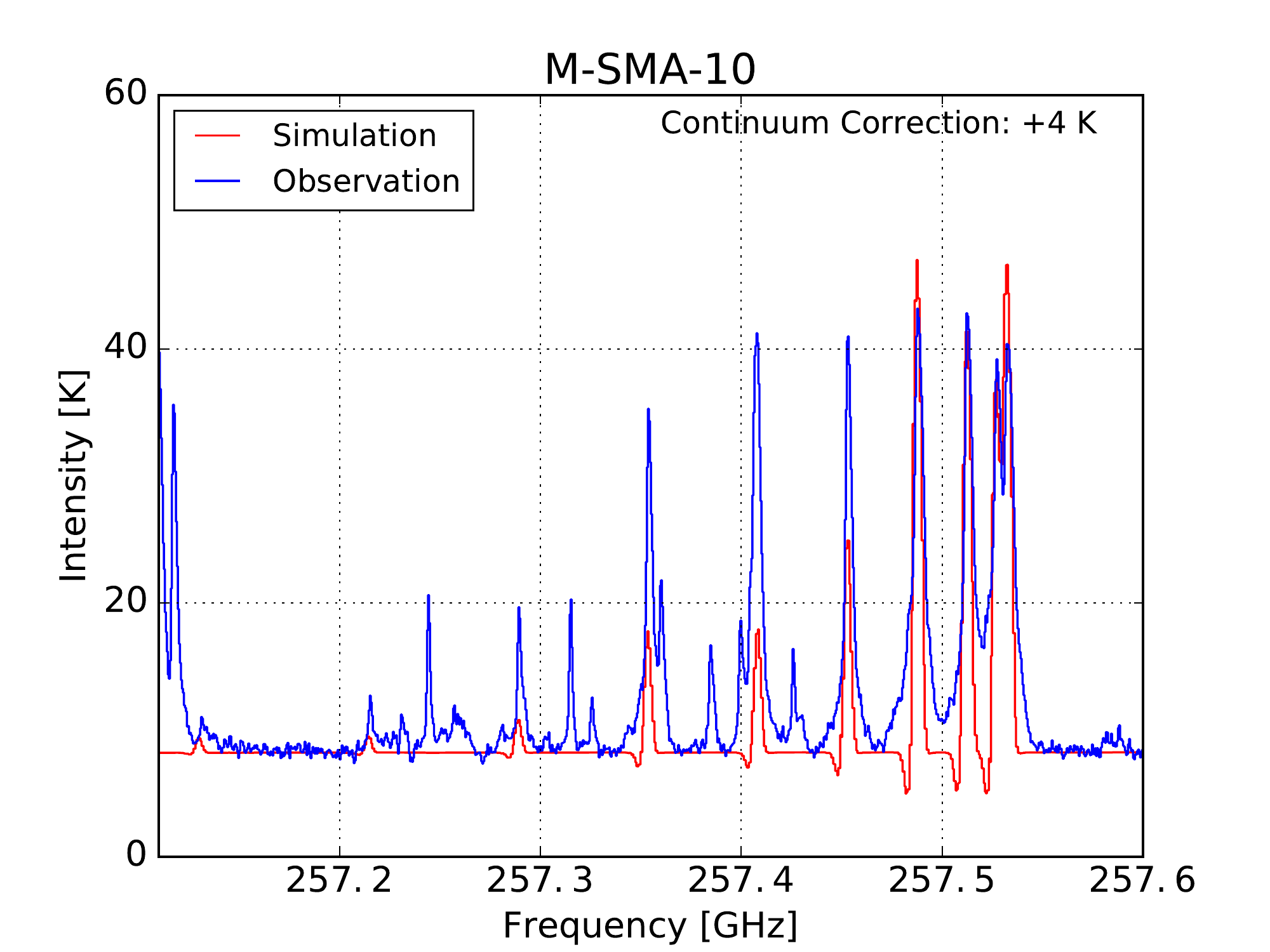}
    \includegraphics[scale=0.36]{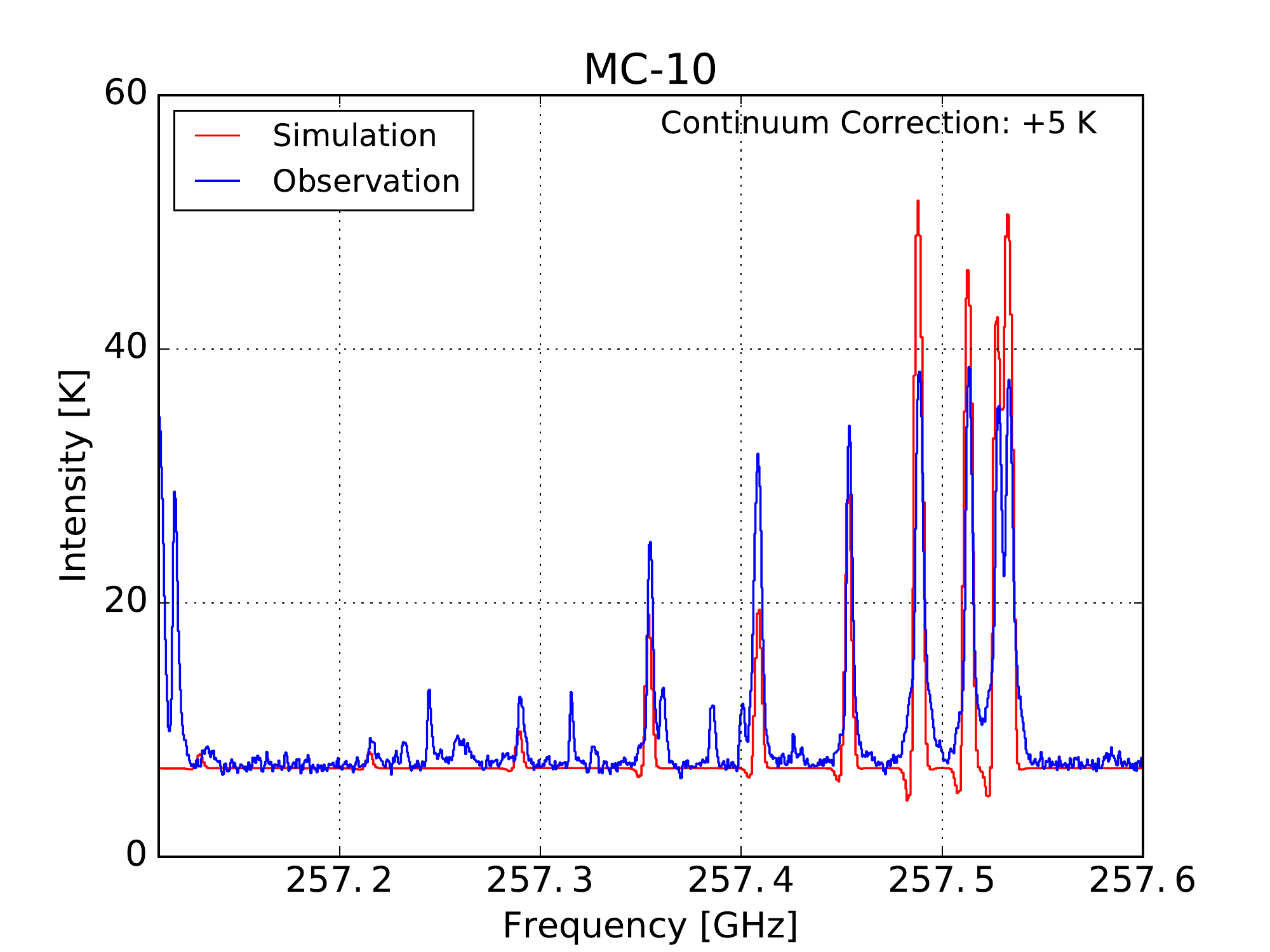}
    \includegraphics[scale=0.36]{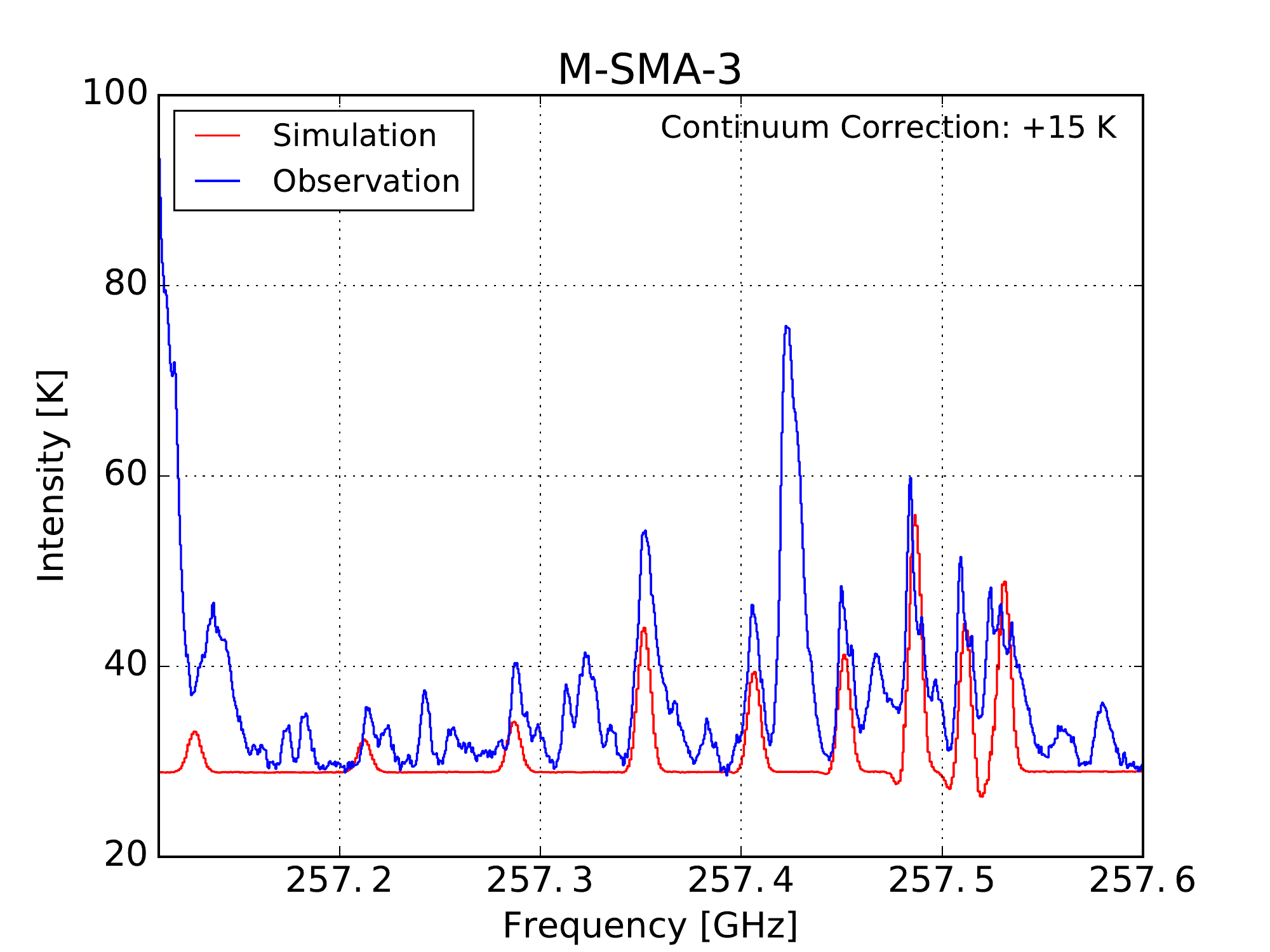}
    \includegraphics[scale=0.36]{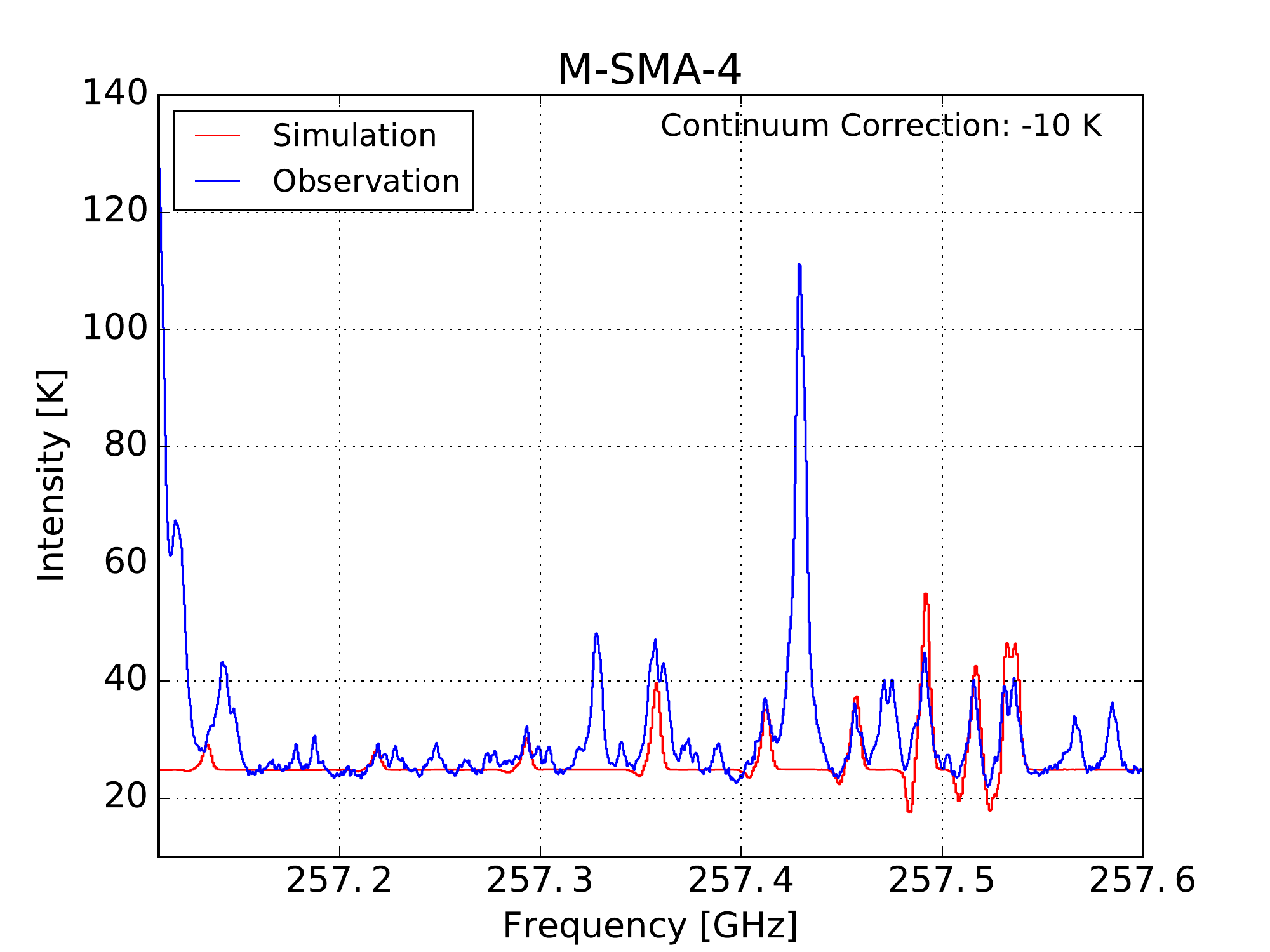}
    \includegraphics[scale=0.36]{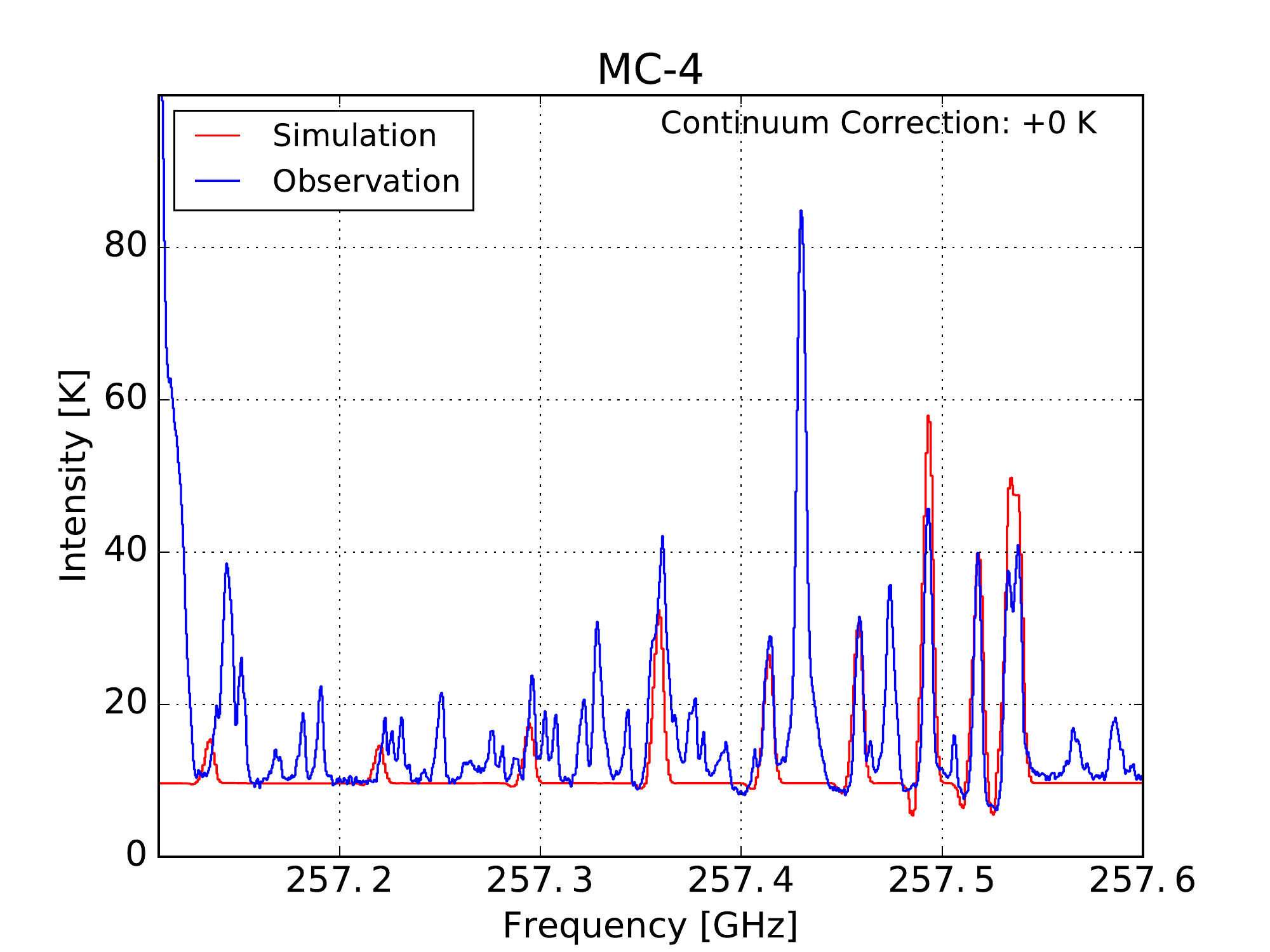}
    \includegraphics[scale=0.36]{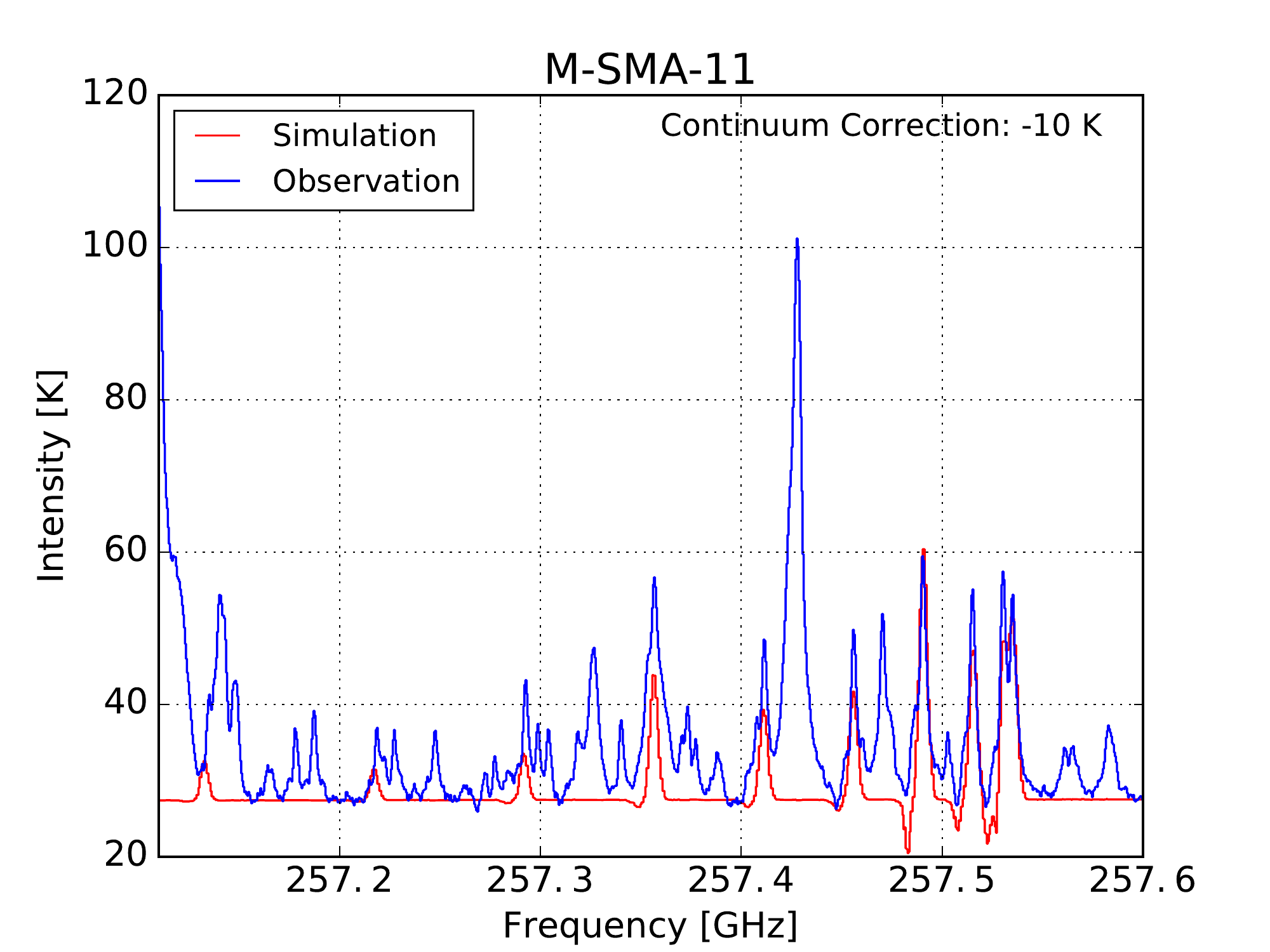}
    \includegraphics[scale=0.36]{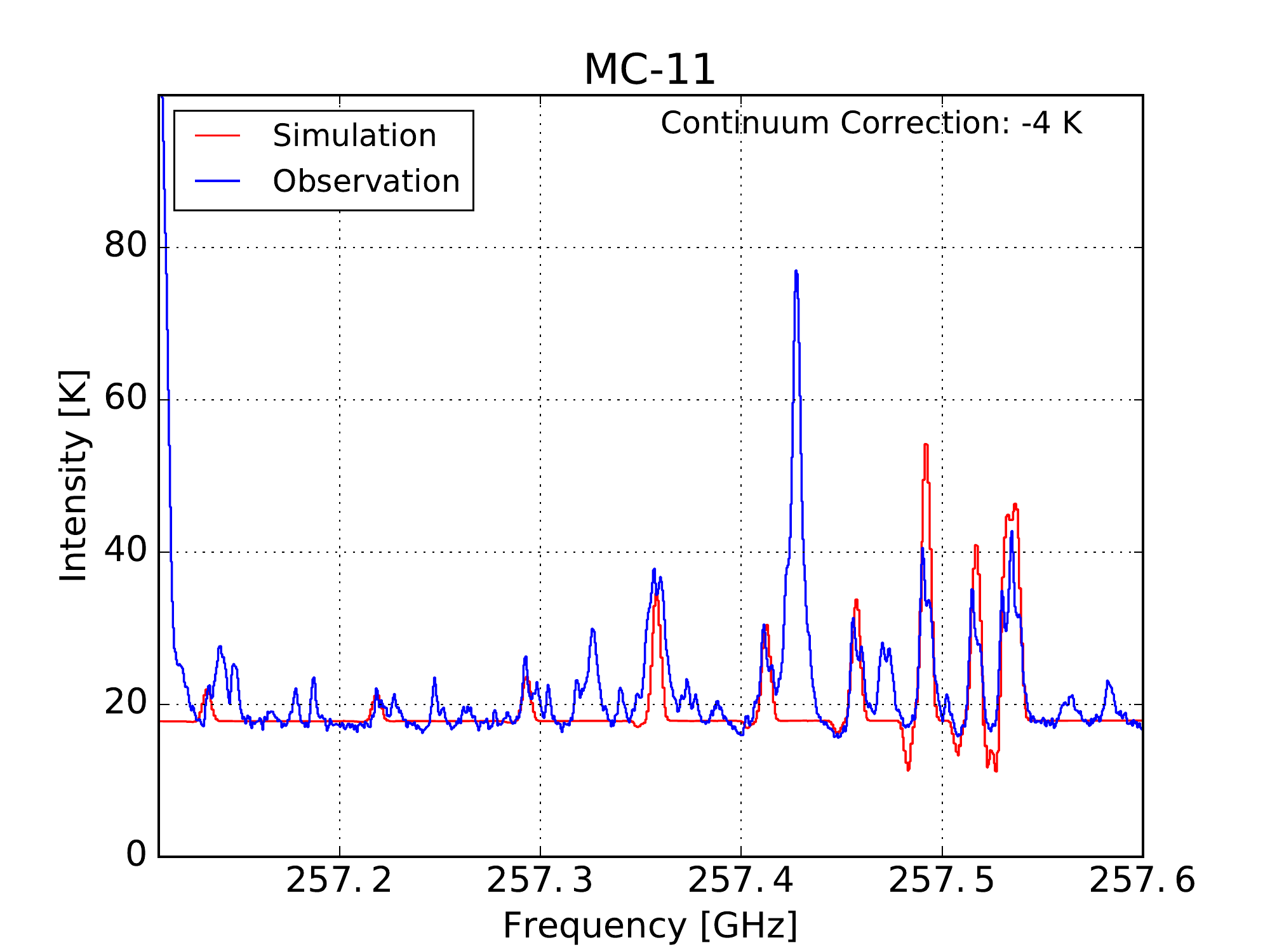}
    \includegraphics[scale=0.36]{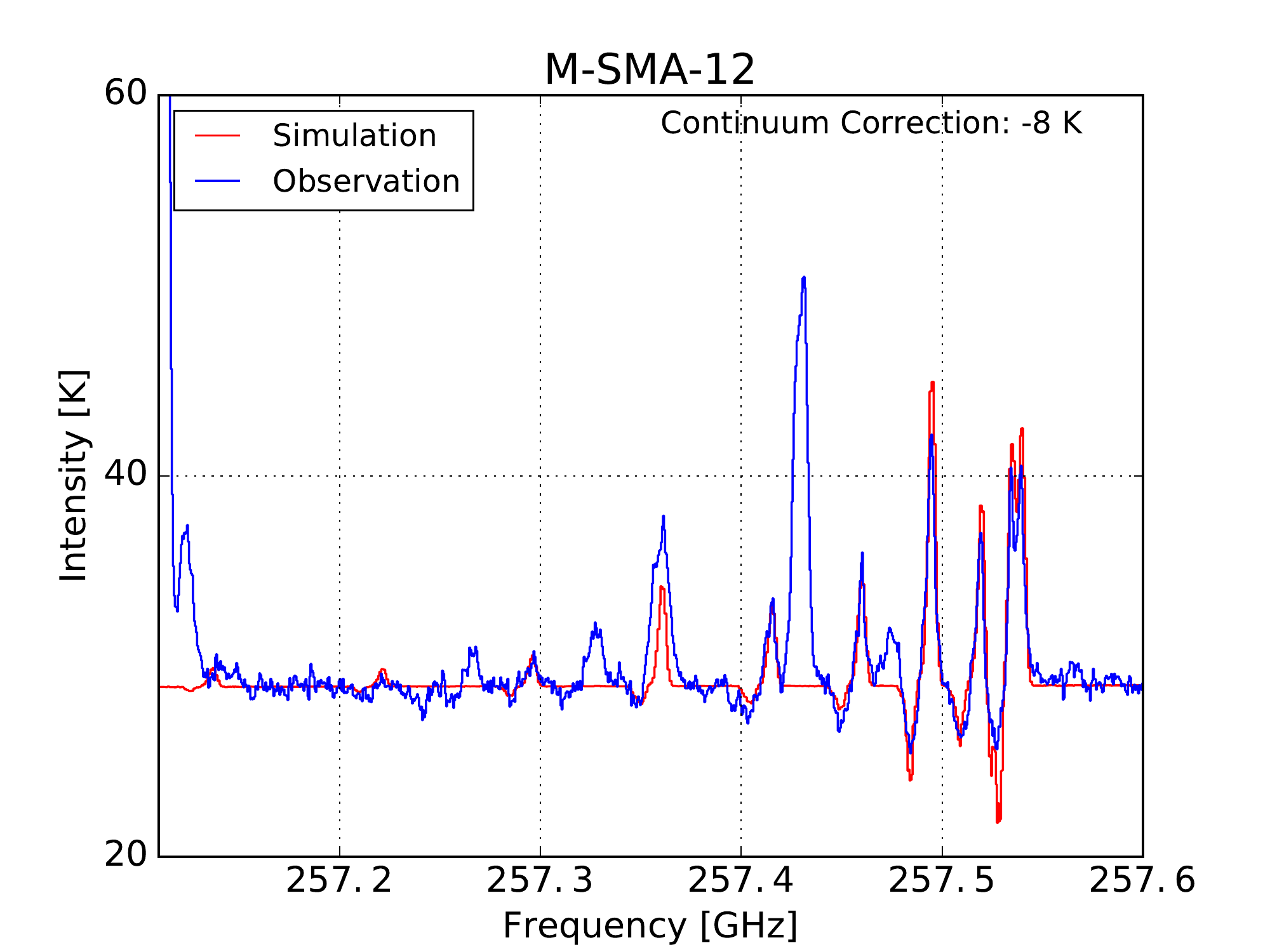}
\caption{Continuation of Fig.~\ref{fig:spectra_J=14-13_1} for the remaining cores and molecular centers.}
\label{fig:spectra_J=14-13_2}
\end{figure*}

\clearpage
\begin{figure*}
\centering
    \includegraphics[scale=0.36]{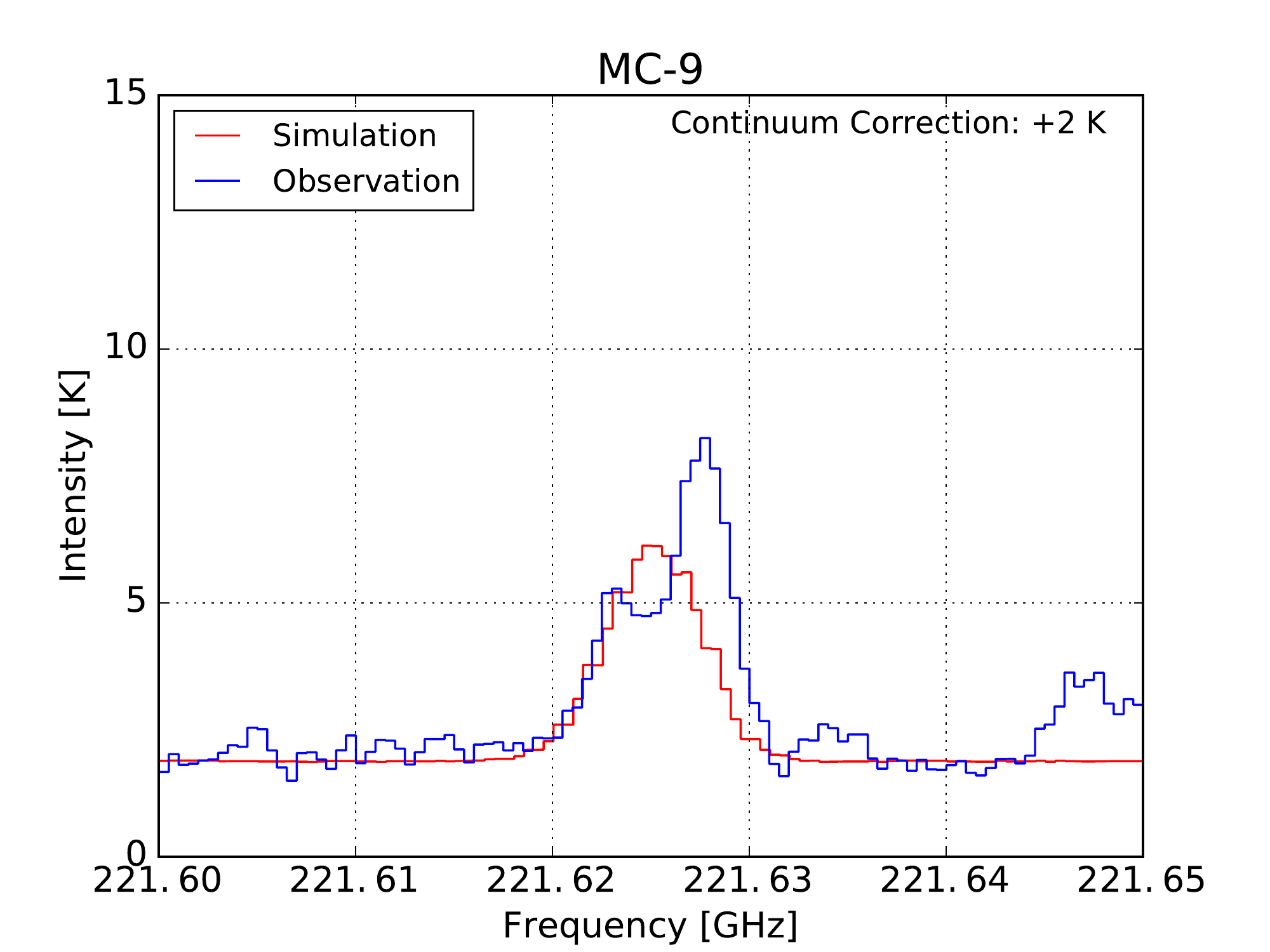}
    \includegraphics[scale=0.36]{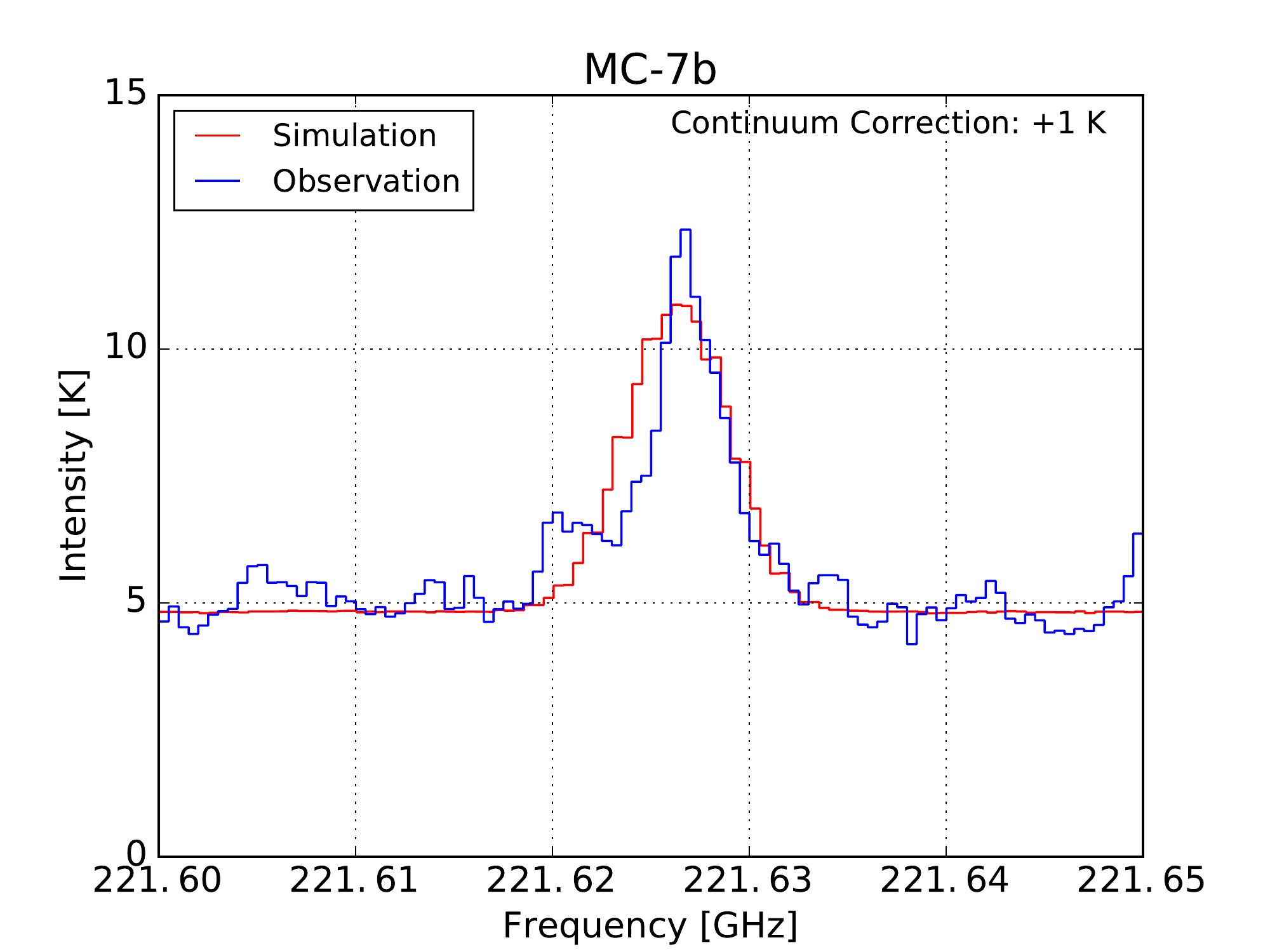}
    \includegraphics[scale=0.36]{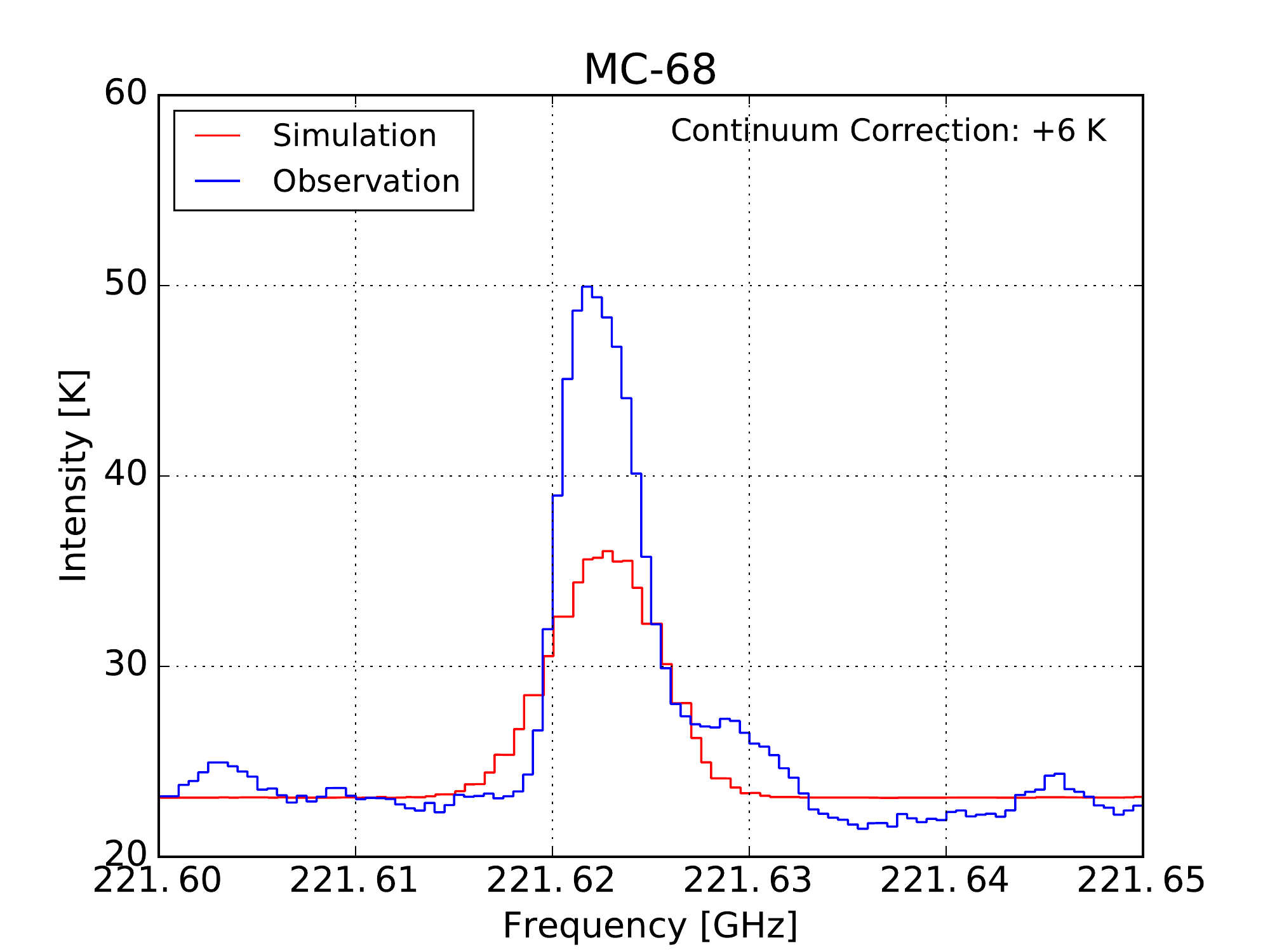}
    \includegraphics[scale=0.36]{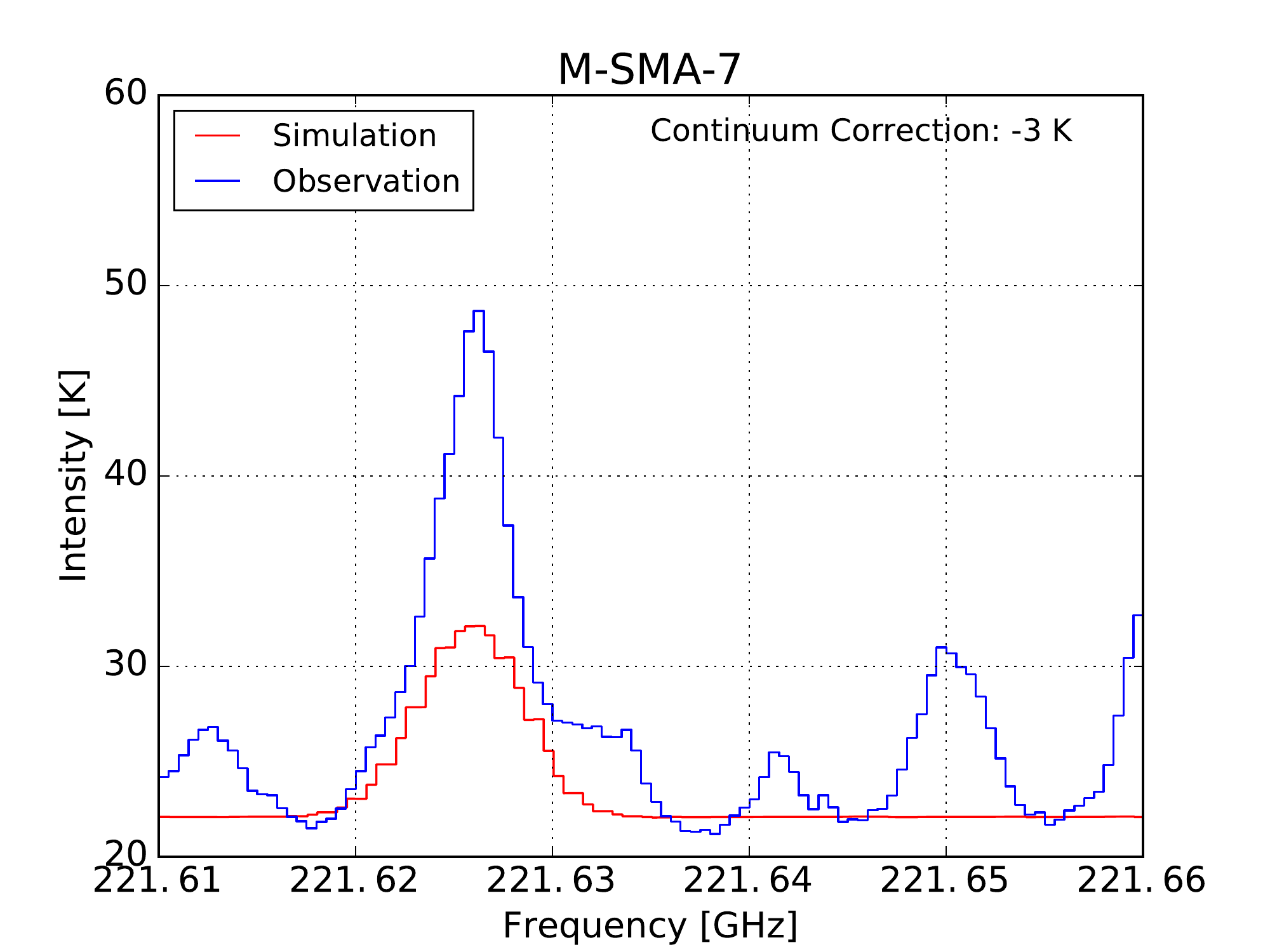}
    \includegraphics[scale=0.36]{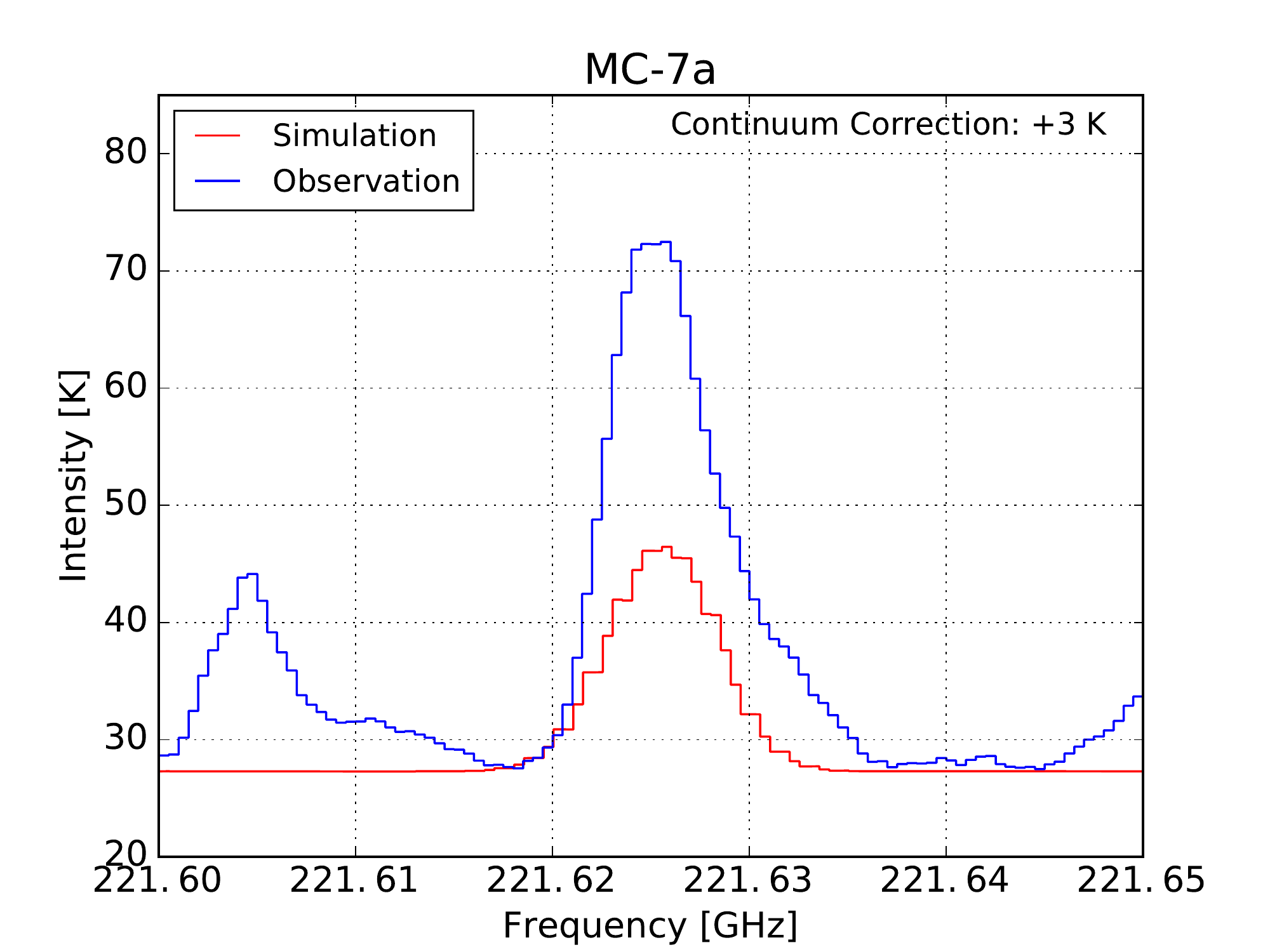}
    \includegraphics[scale=0.36]{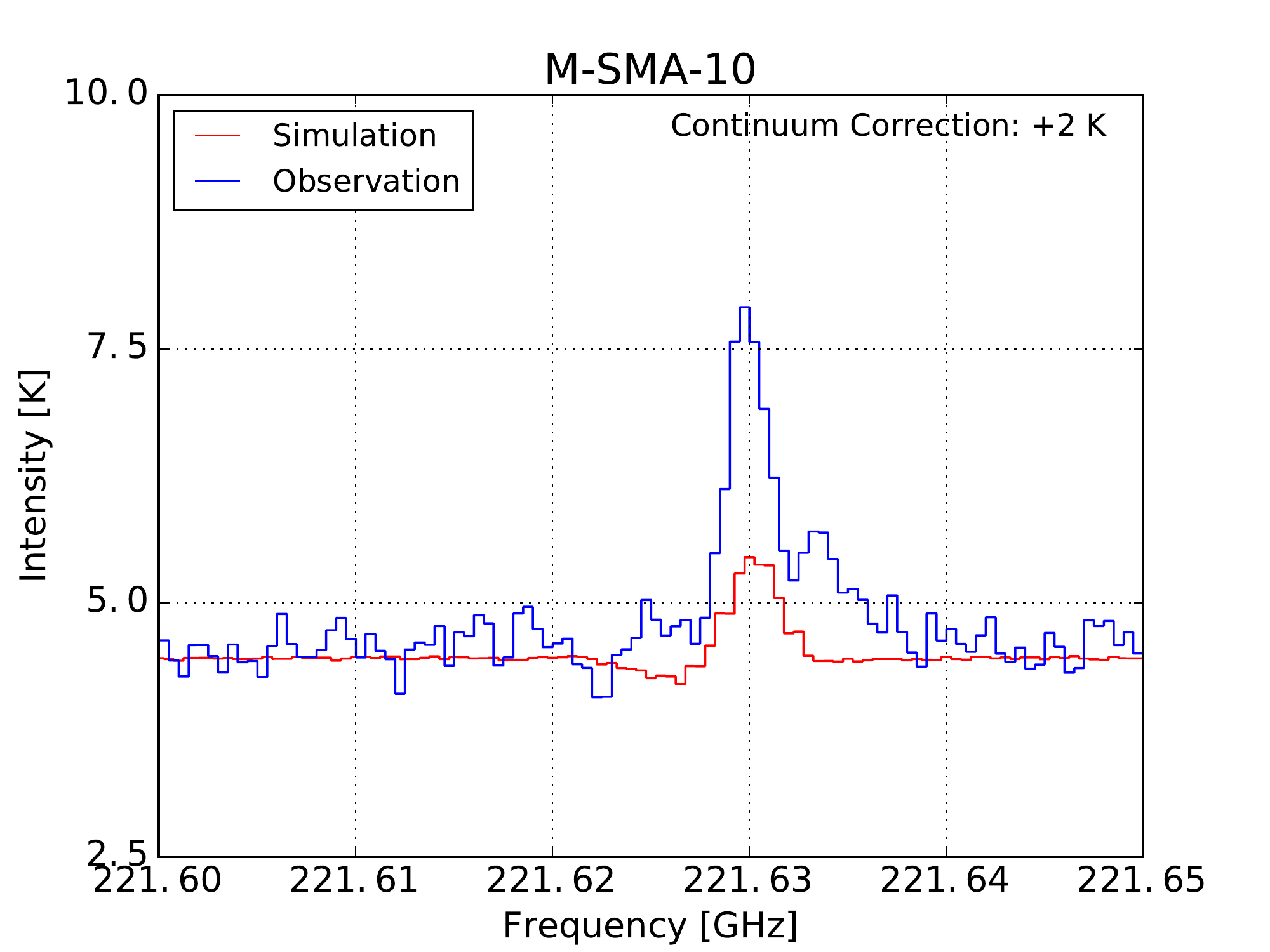}
\caption{Spectra of the observational (in blue) and simulated (in red) data extracted from the position of different cores and molecular centers for the vibrationally excited transition \chcn\  $\nu_8=1$, $J$=12-11, $K=1-(-1)$. The spectra are arranged according to the position of the corresponding cores and molecular centers along a south-north direction, starting from the most southern core. The name of the core and molecular center is shown for each panel. The continuum correction necessary to match the continuum level of the observation and simulation is indicated in the top-right side of each panel.}
\label{fig:spectra_vib_appendix_1}
\end{figure*}

\clearpage
\begin{figure*}
\centering
    \includegraphics[scale=0.36]{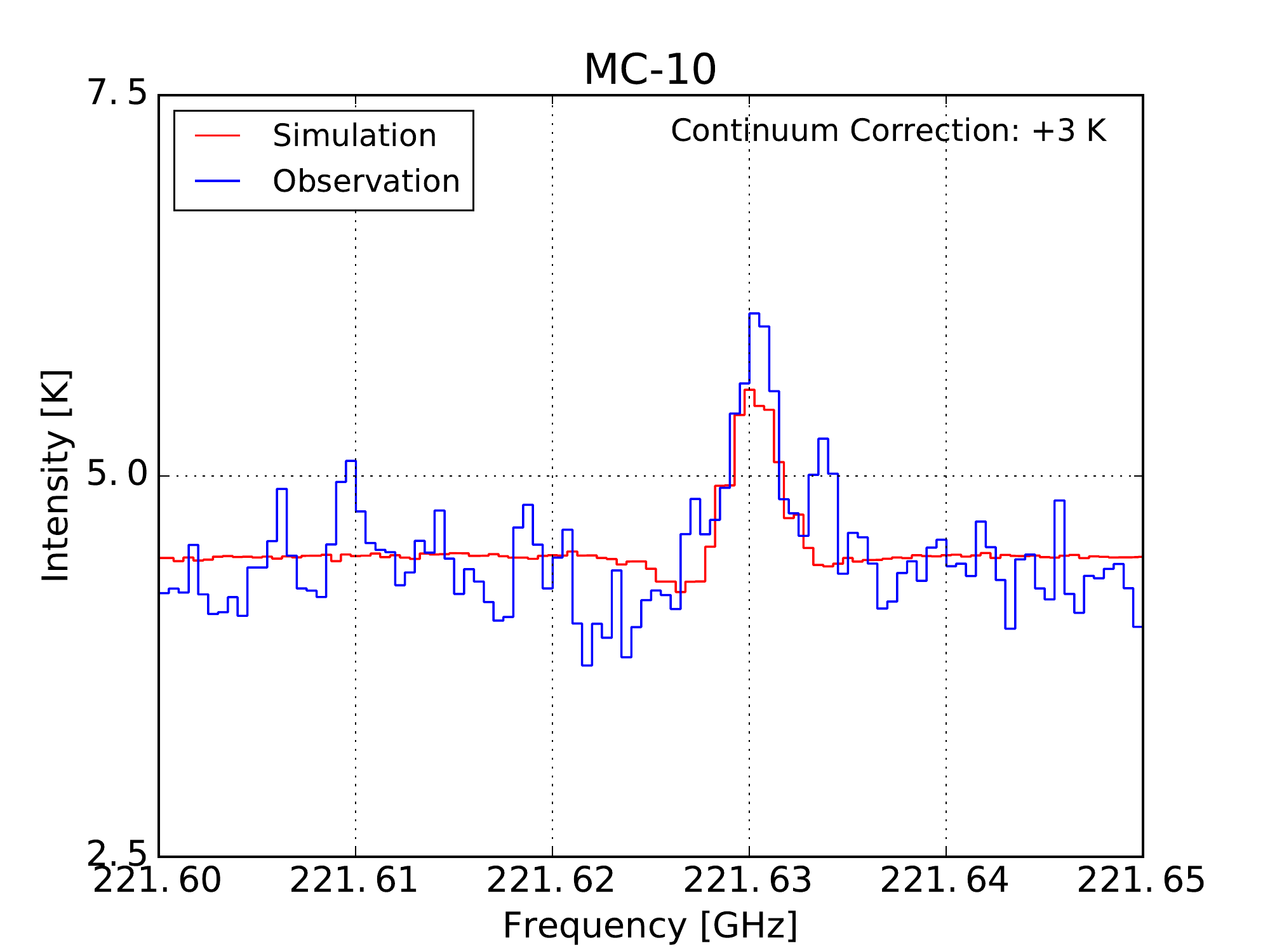}
    \includegraphics[scale=0.36]{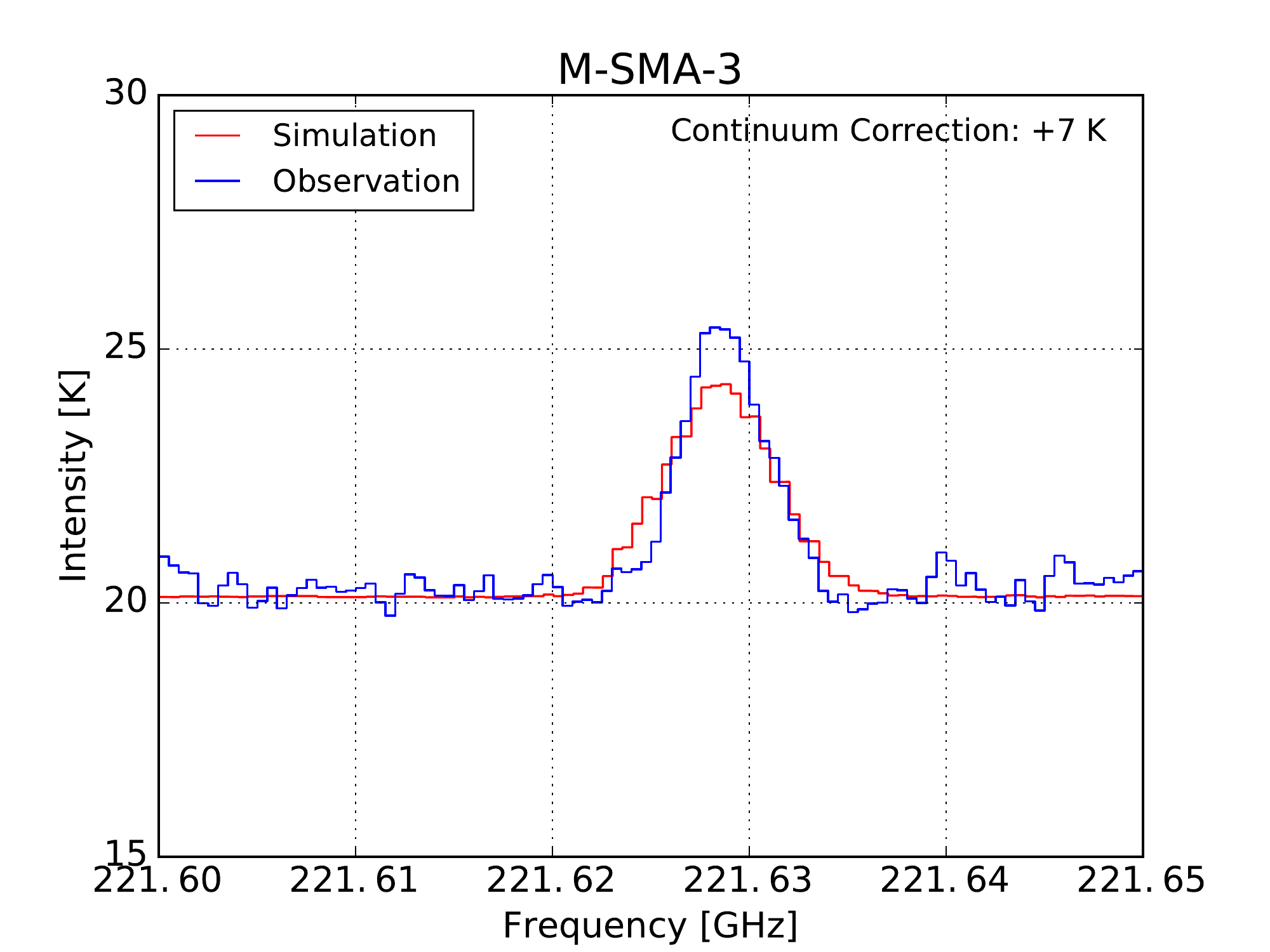}
    \includegraphics[scale=0.36]{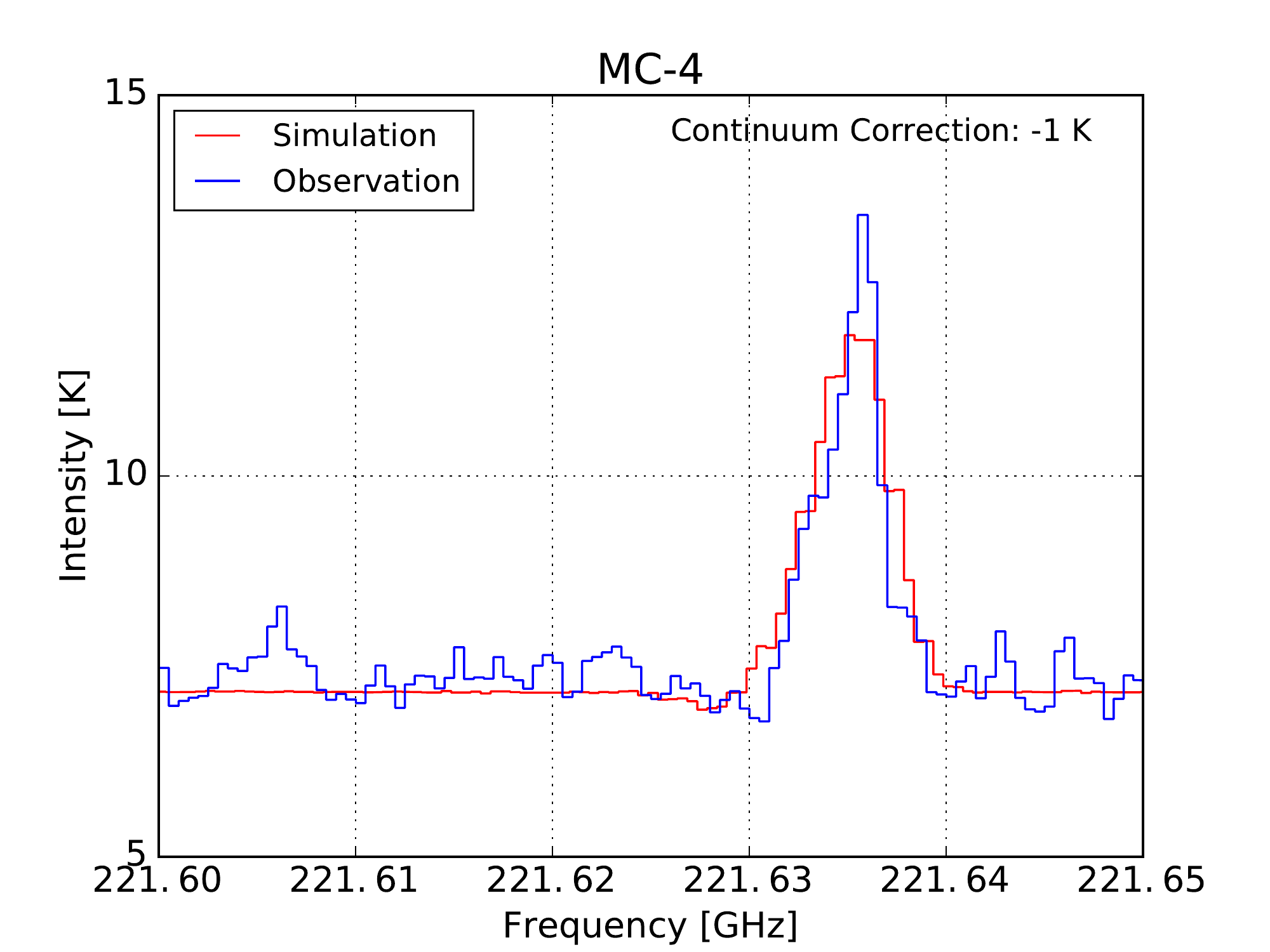}
    \includegraphics[scale=0.36]{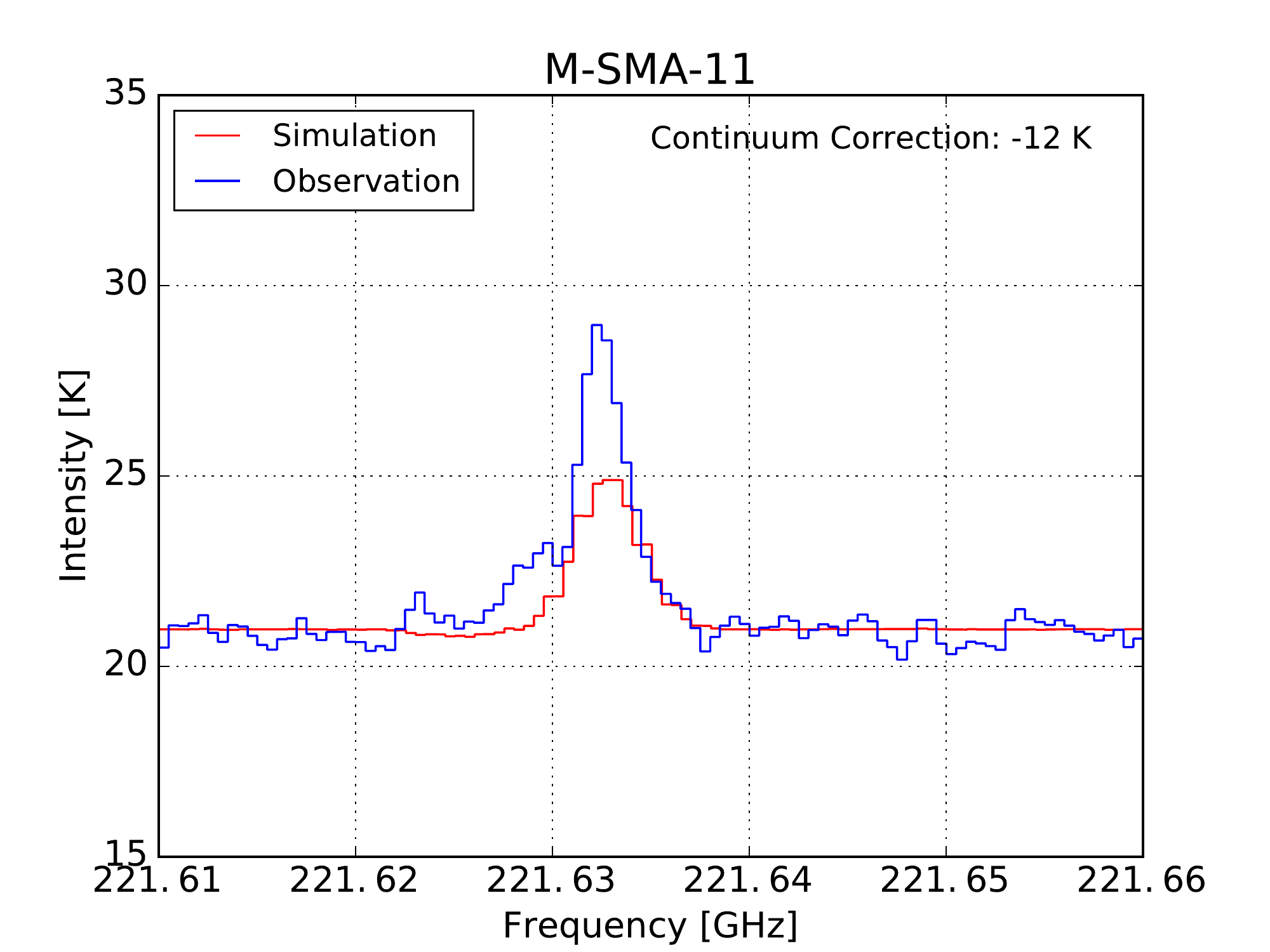}
    \includegraphics[scale=0.36]{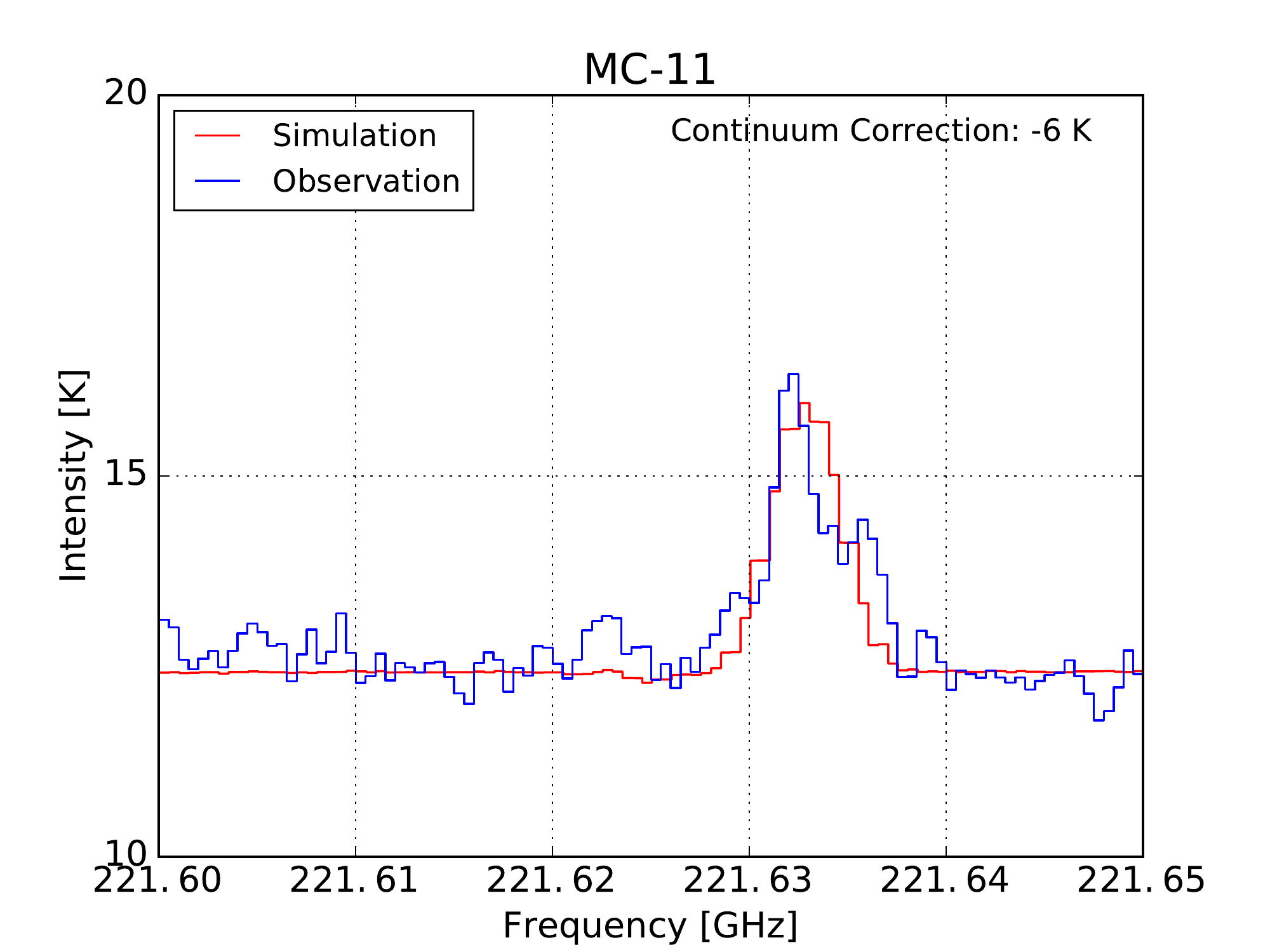}
    \includegraphics[scale=0.36]{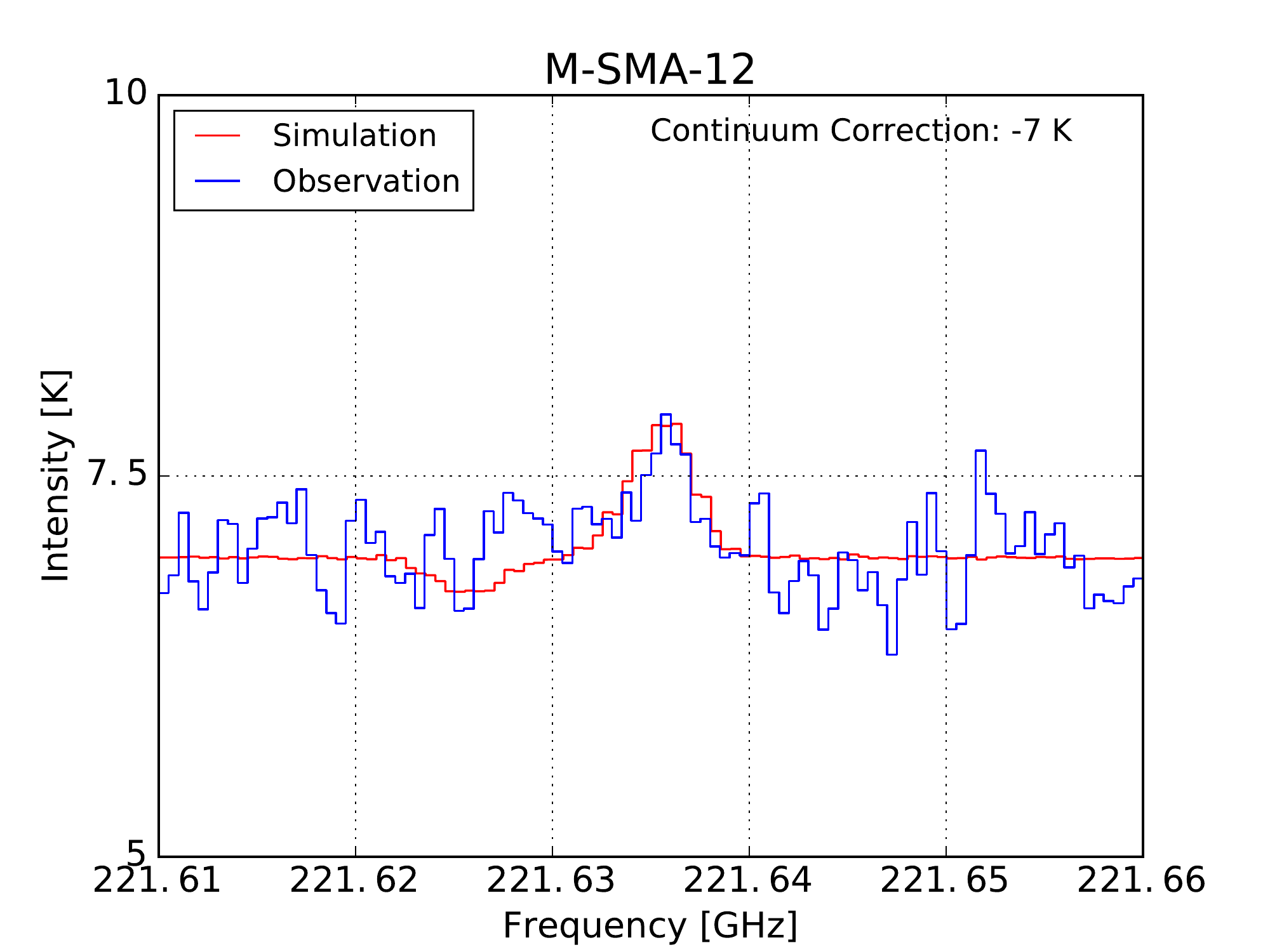}
\caption{Continuation of Fig.~\ref{fig:spectra_vib_appendix_1} for the remaining cores and molecular centers.}
\label{fig:spectra_vib_appendix_2}
\end{figure*}

\clearpage
\begin{figure*}
\centering
   \includegraphics[scale=0.36]{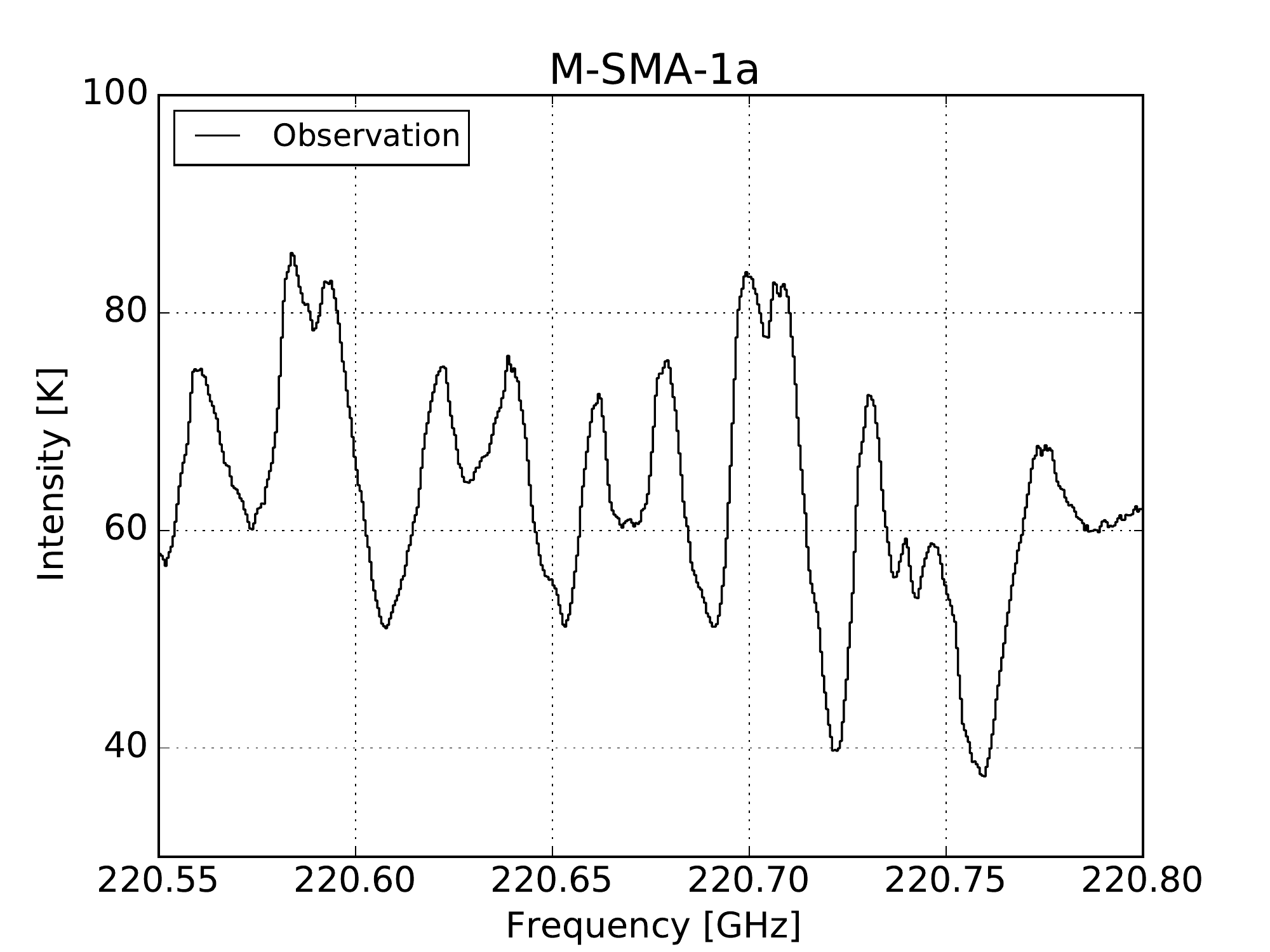}
   \includegraphics[scale=0.36]{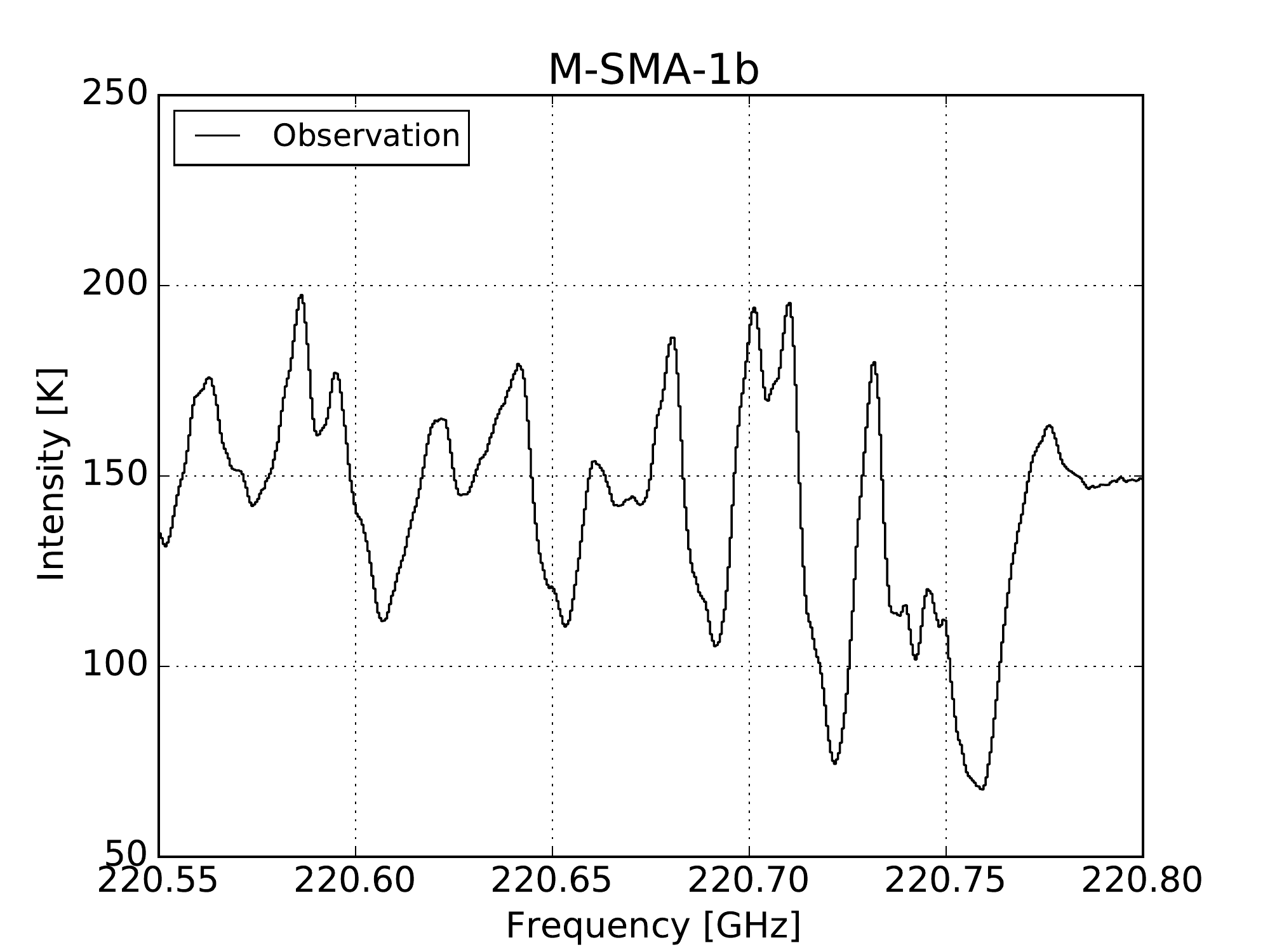}
   \includegraphics[scale=0.36]{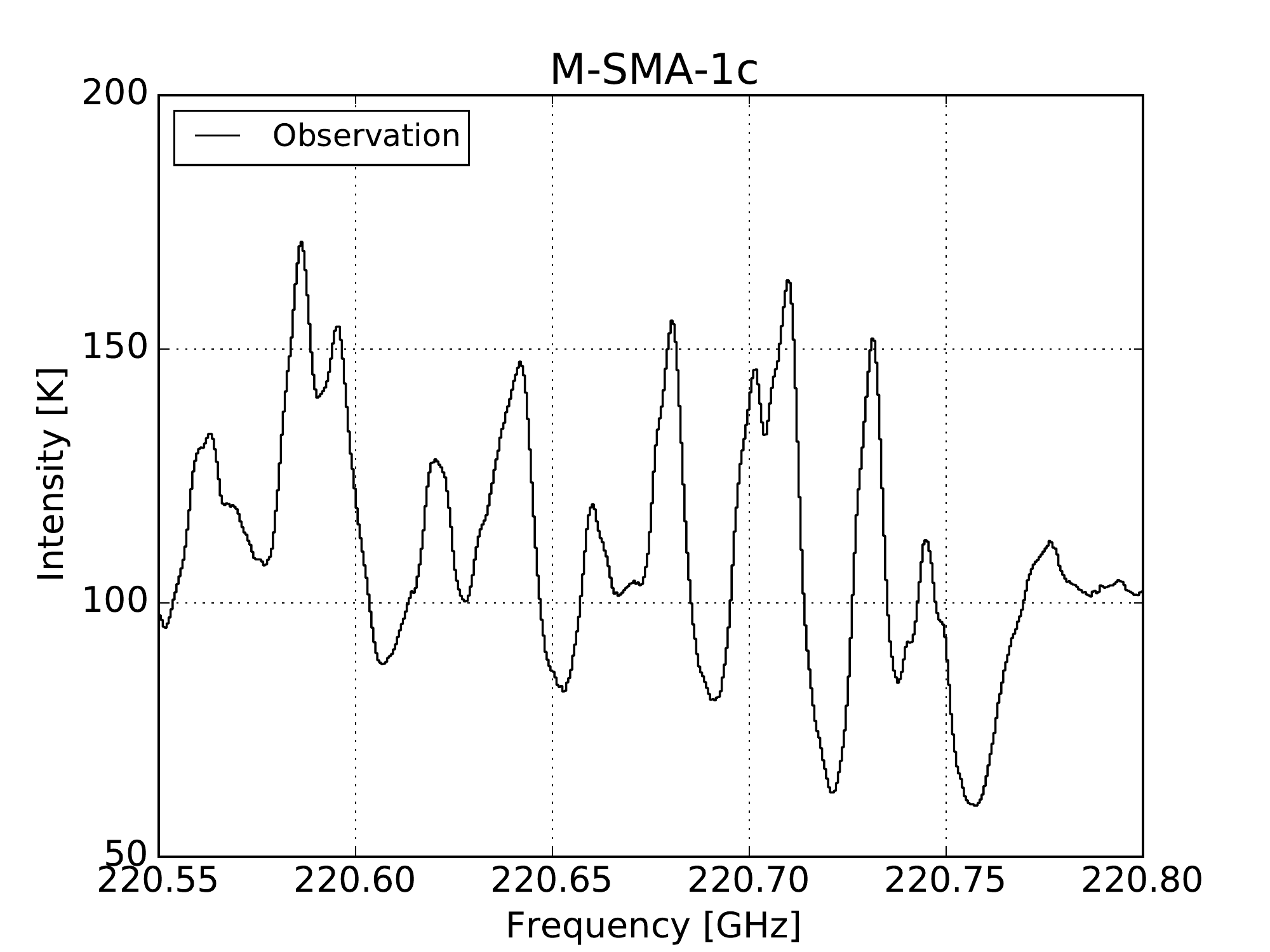}
   \includegraphics[scale=0.36]{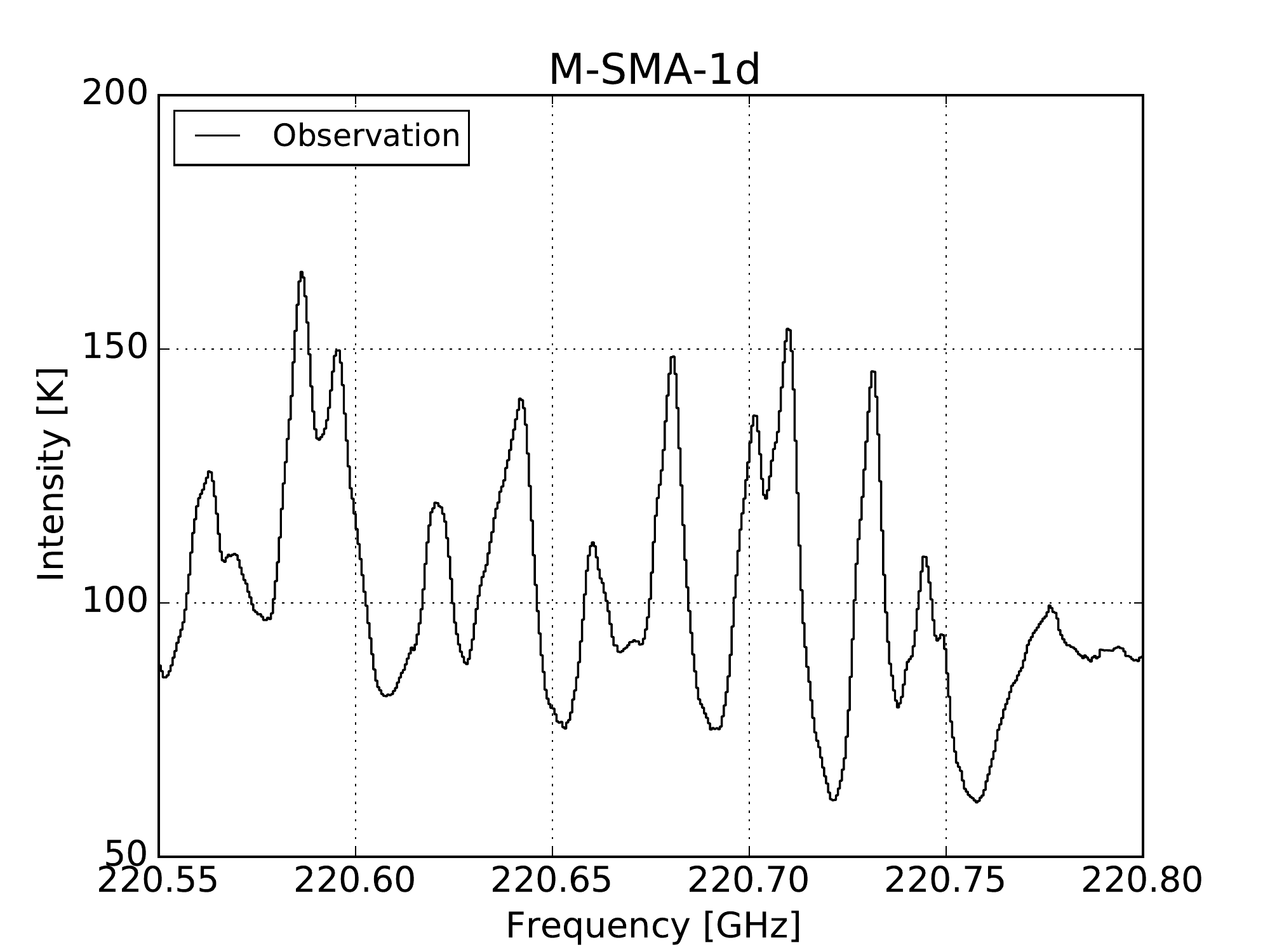}
   \includegraphics[scale=0.36]{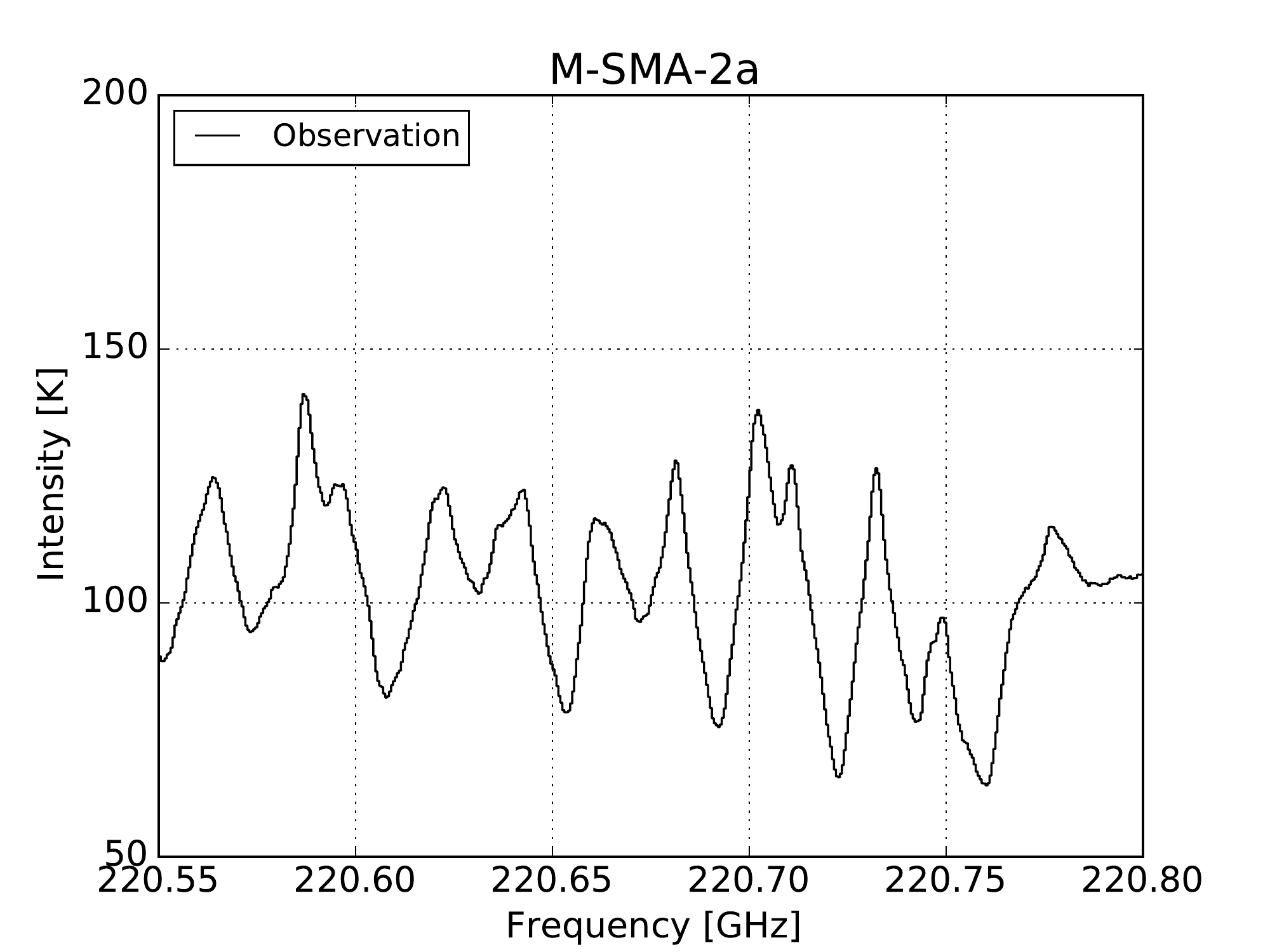}

\caption{Observational spectra (in black) of the ground state \chcn\ $J$=12--11 transition towards the cores located in the central region of \SgrB(M). The continuum level for core M-SMA-1a is about \unit[60]{K}, for core M-SMA-1b it is about \unit[150]{K}, and it is about \unit[100]{K} for the remaining cores. A number of absorption and emission spectral line features are visible in all the spectra.}
\label{fig:center_spectra}
\end{figure*}

\clearpage   
\begin{figure*}
\centering
   \includegraphics[scale=0.36]{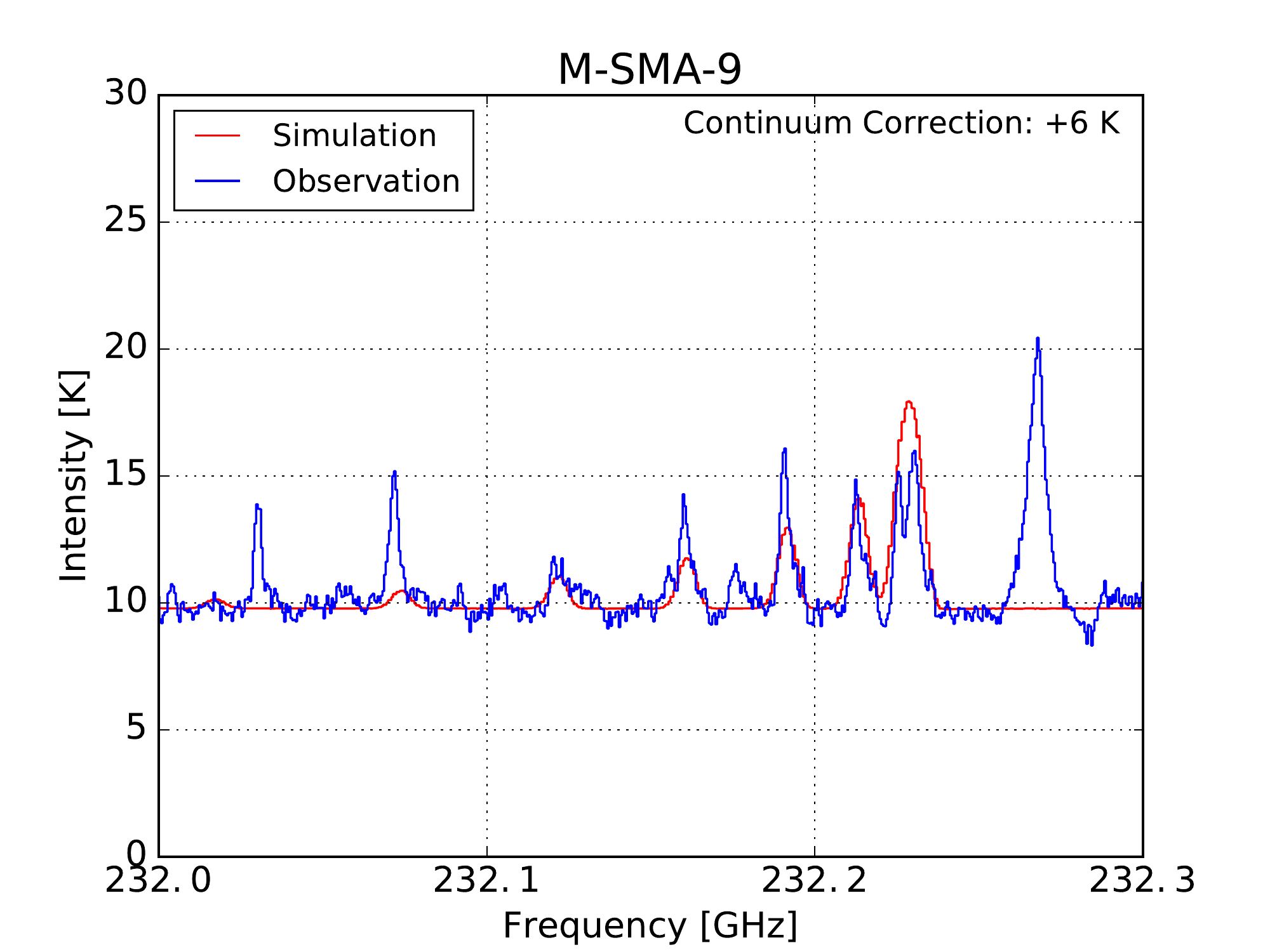}
   \includegraphics[scale=0.36]{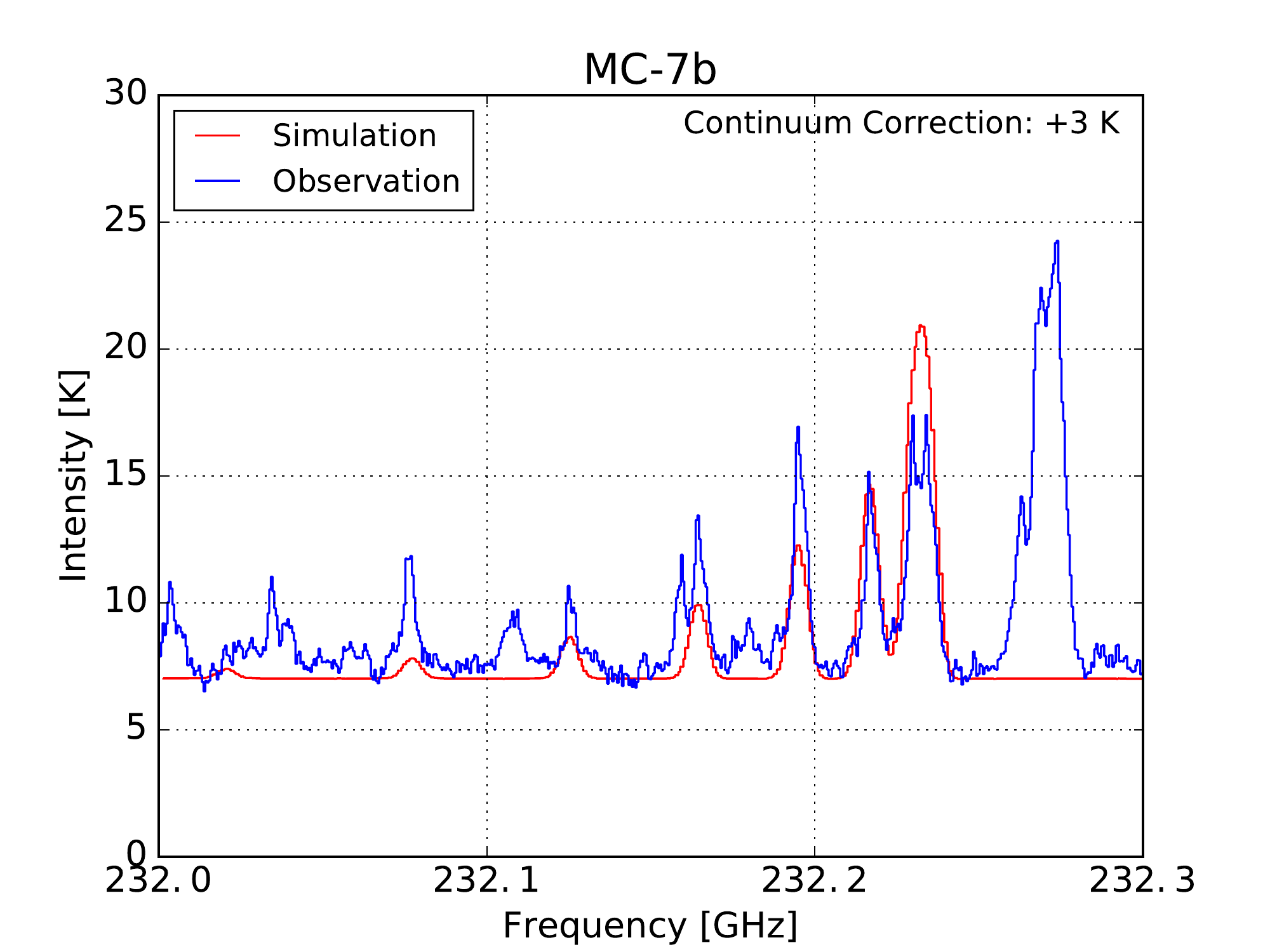}
   \includegraphics[scale=0.36]{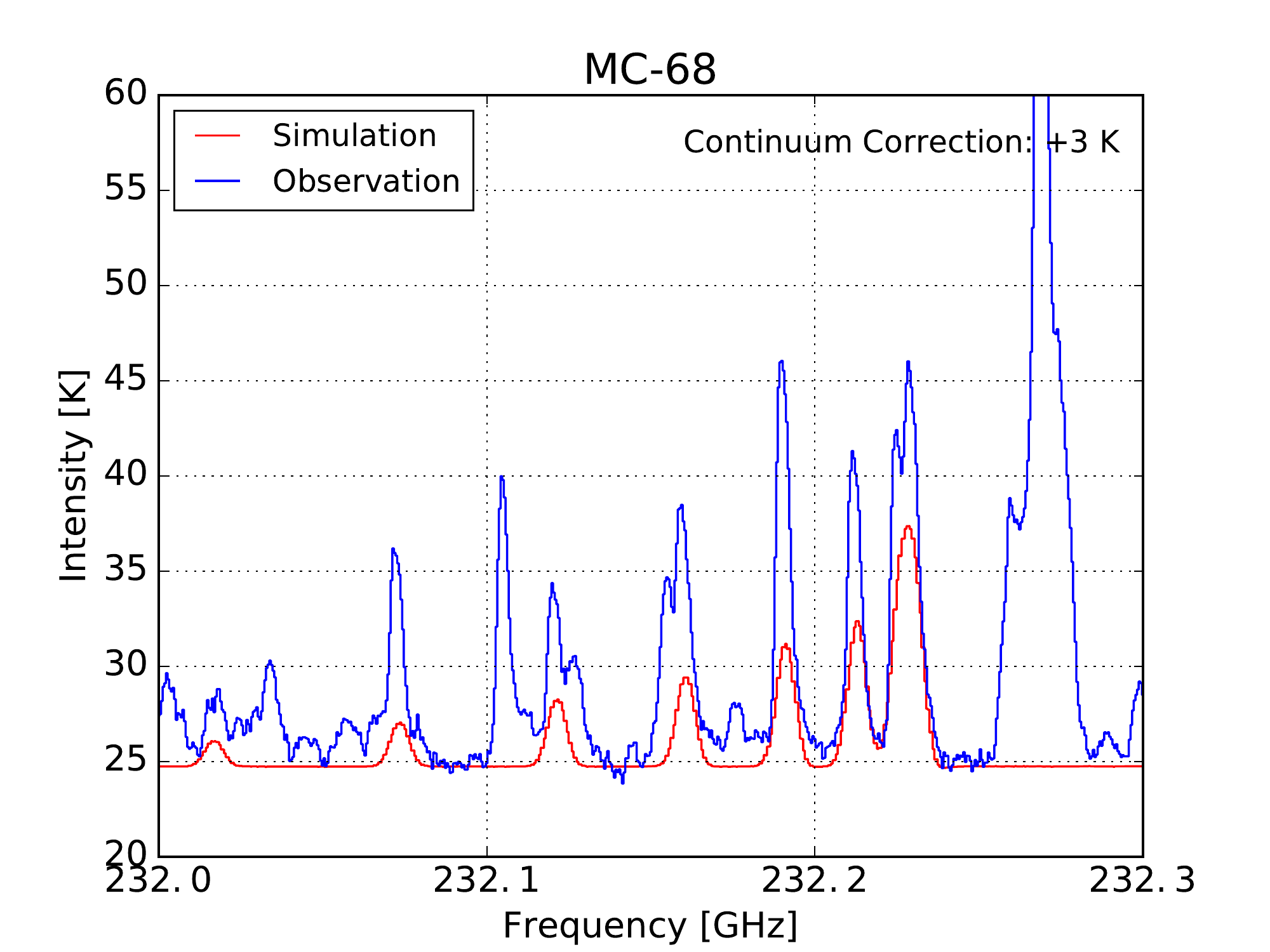}
   \includegraphics[scale=0.36]{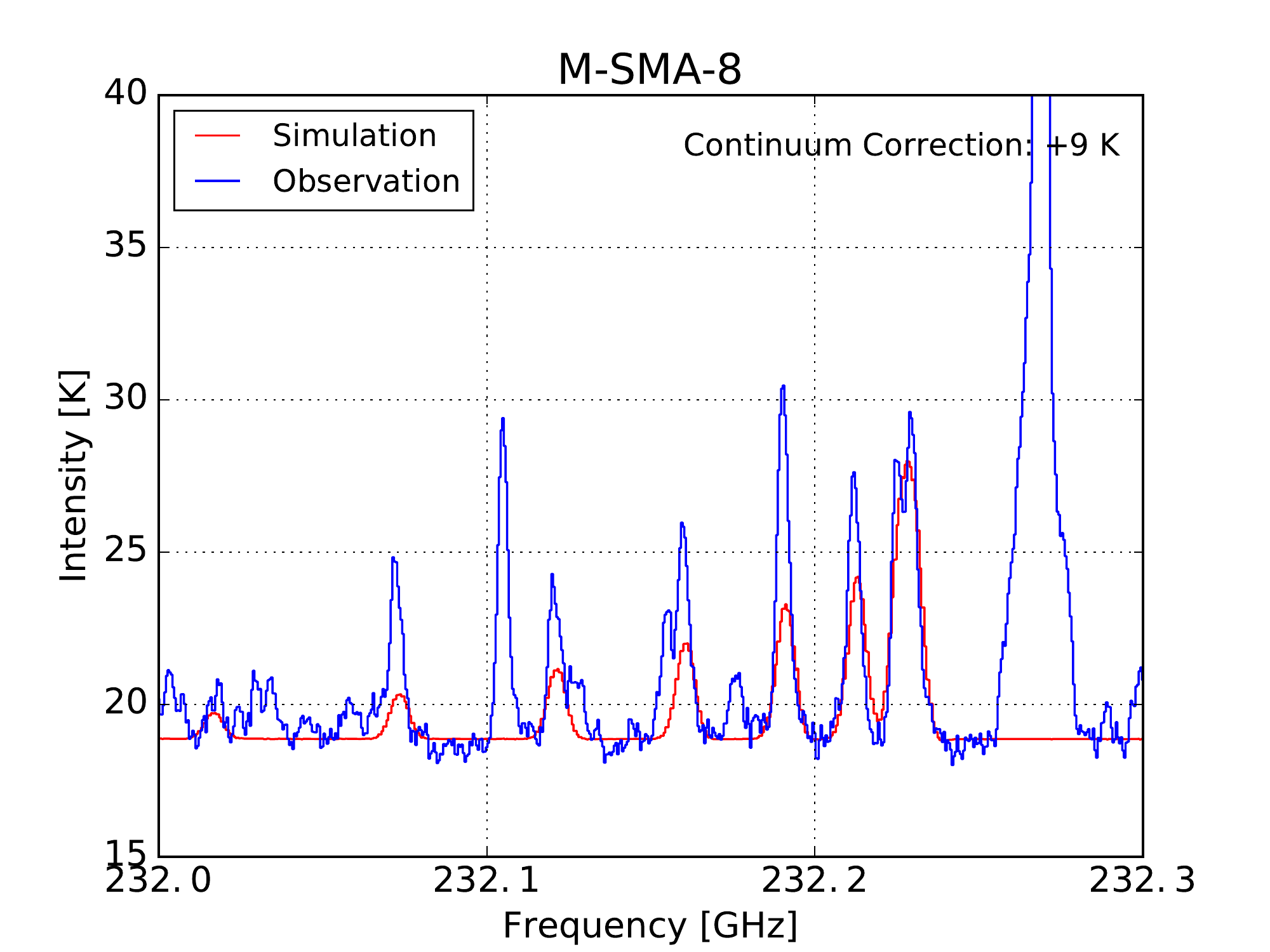}
   \includegraphics[scale=0.36]{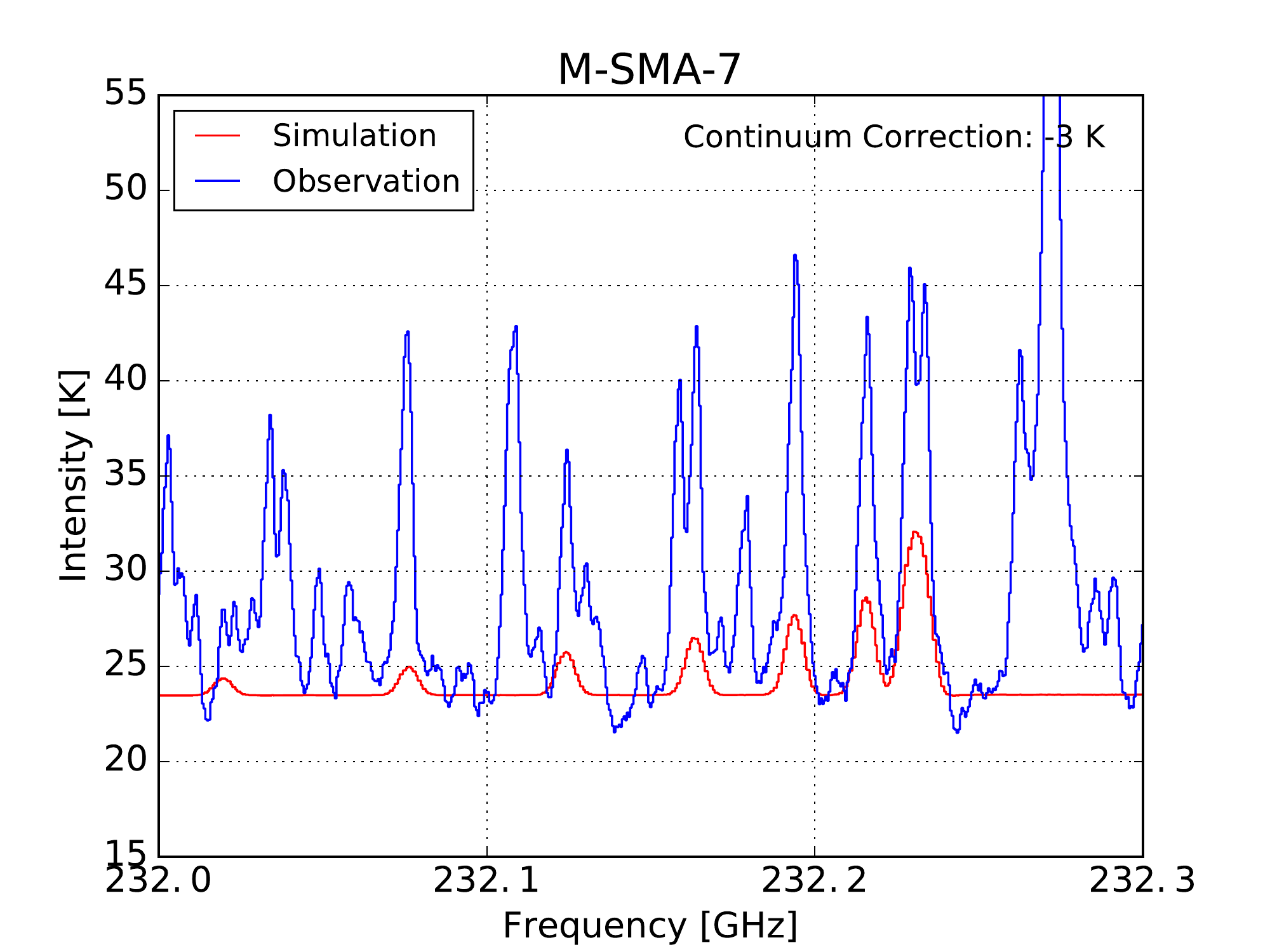}
   \includegraphics[scale=0.36]{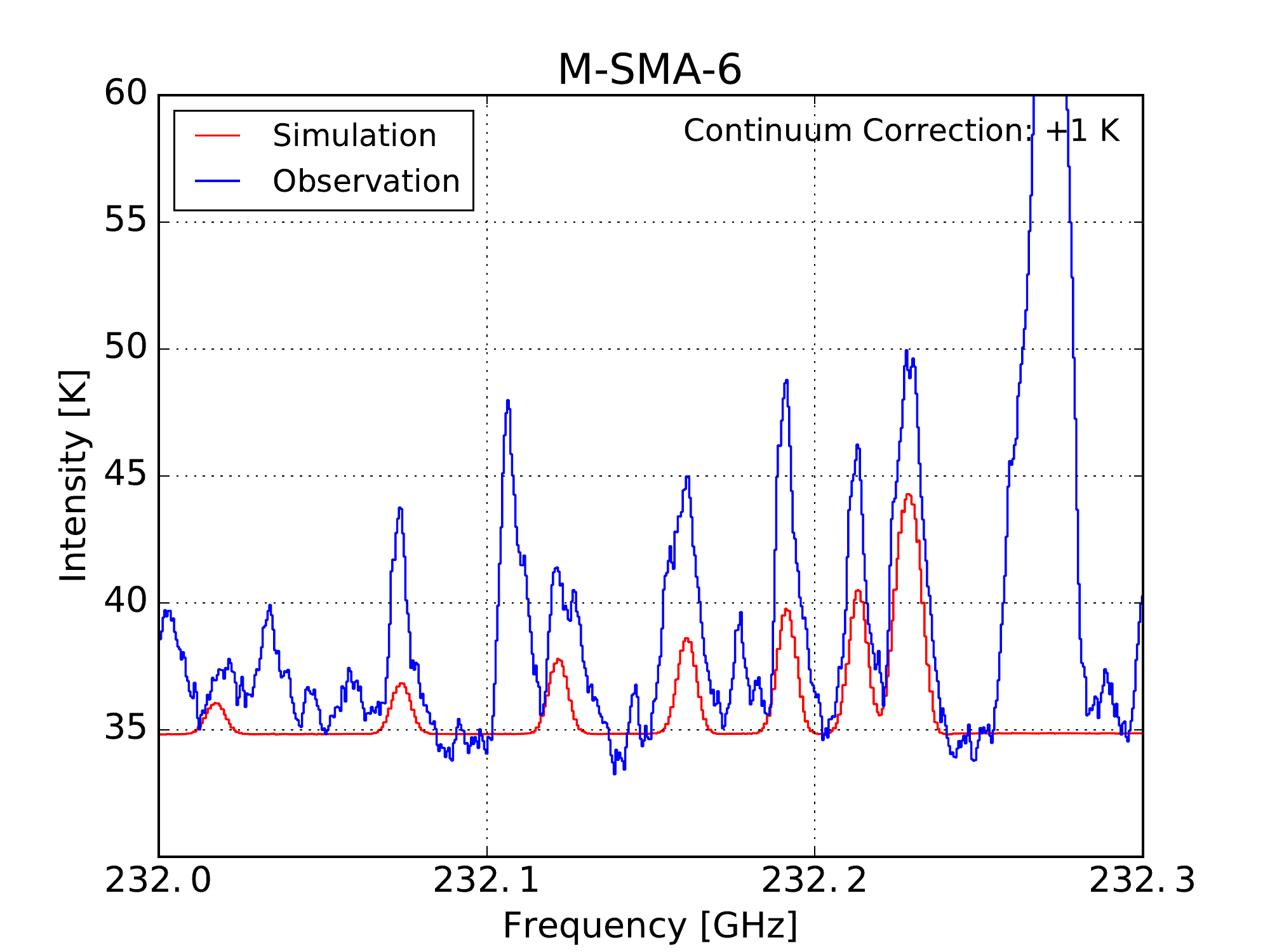}
   \includegraphics[scale=0.36]{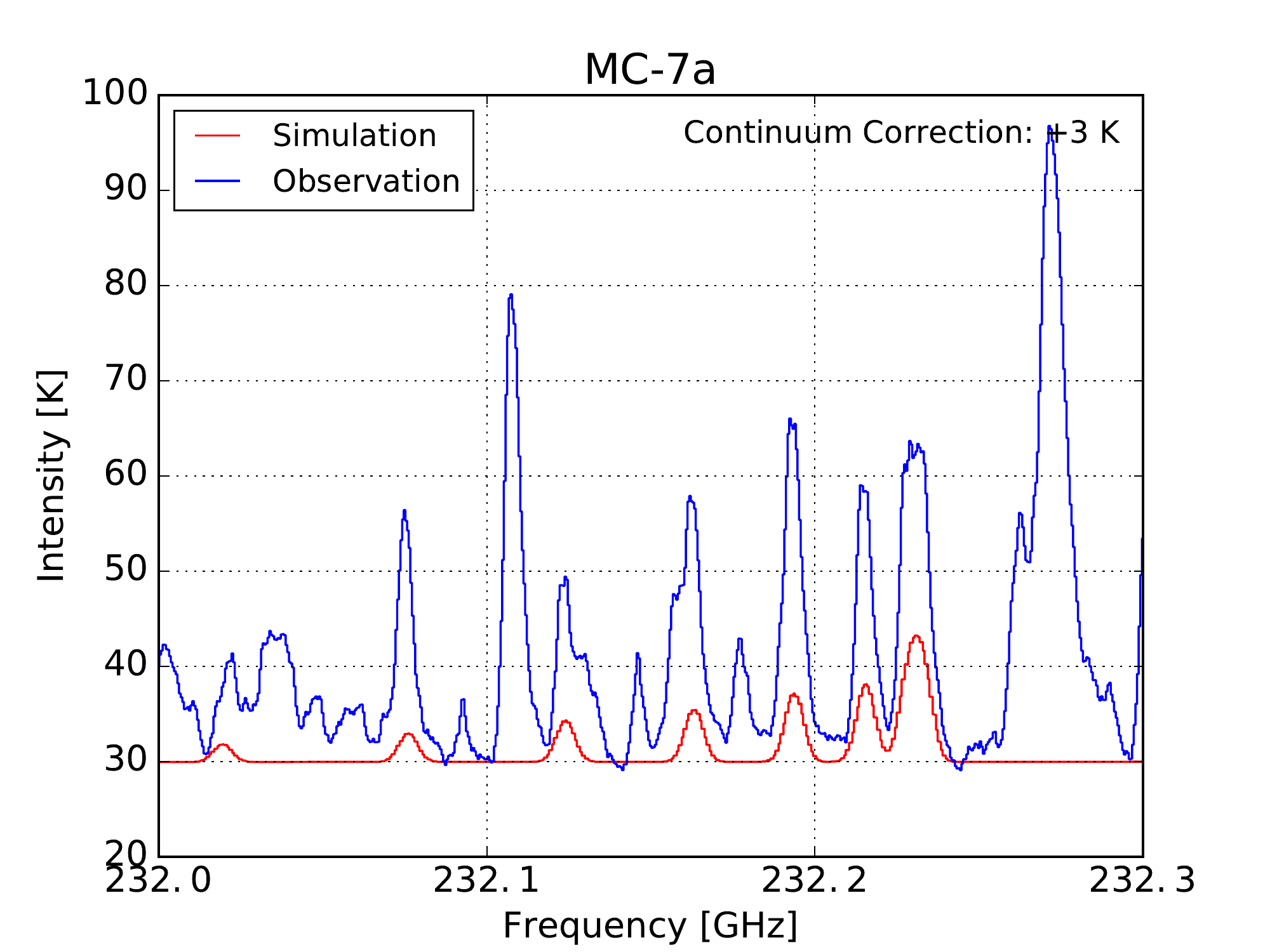}
   \includegraphics[scale=0.36]{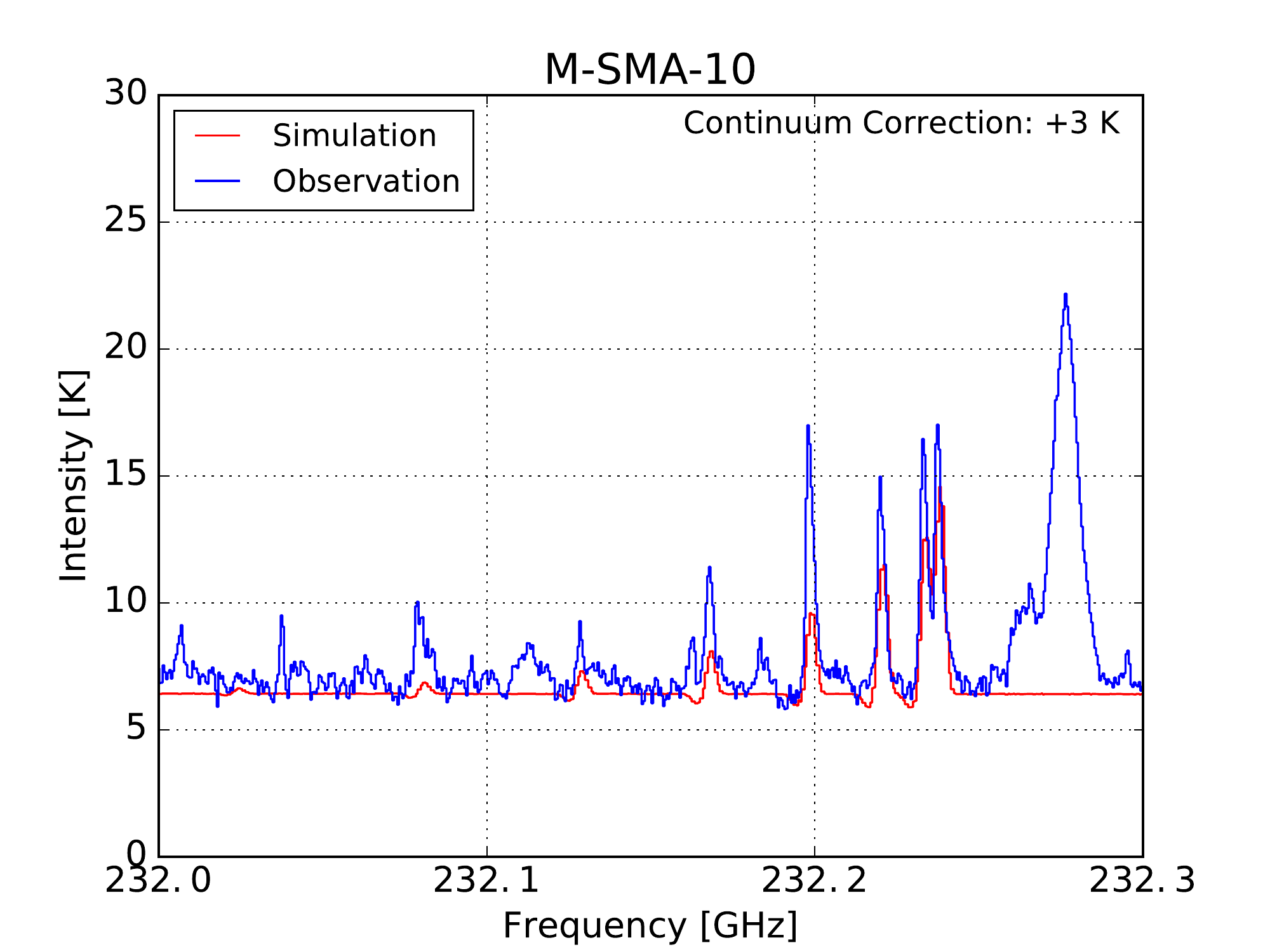}
\caption{Spectra of the observational (in blue) and simulated (in red) data extracted from the position of different cores and molecular centers for the \isochcn\ $J$=13--12 transition. The spectra are arranged according to the position of the corresponding cores and molecular centers along a south-north direction, starting from the most southern core. The name of the core and molecular center is shown for each panel. The continuum correction necessary to match the continuum level of the observation and simulation is indicated in the top-right side of each panel.}
\label{fig:spectra13CH3CN_1}
\end{figure*}

\clearpage
\begin{figure*}
\centering
    \includegraphics[scale=0.36]{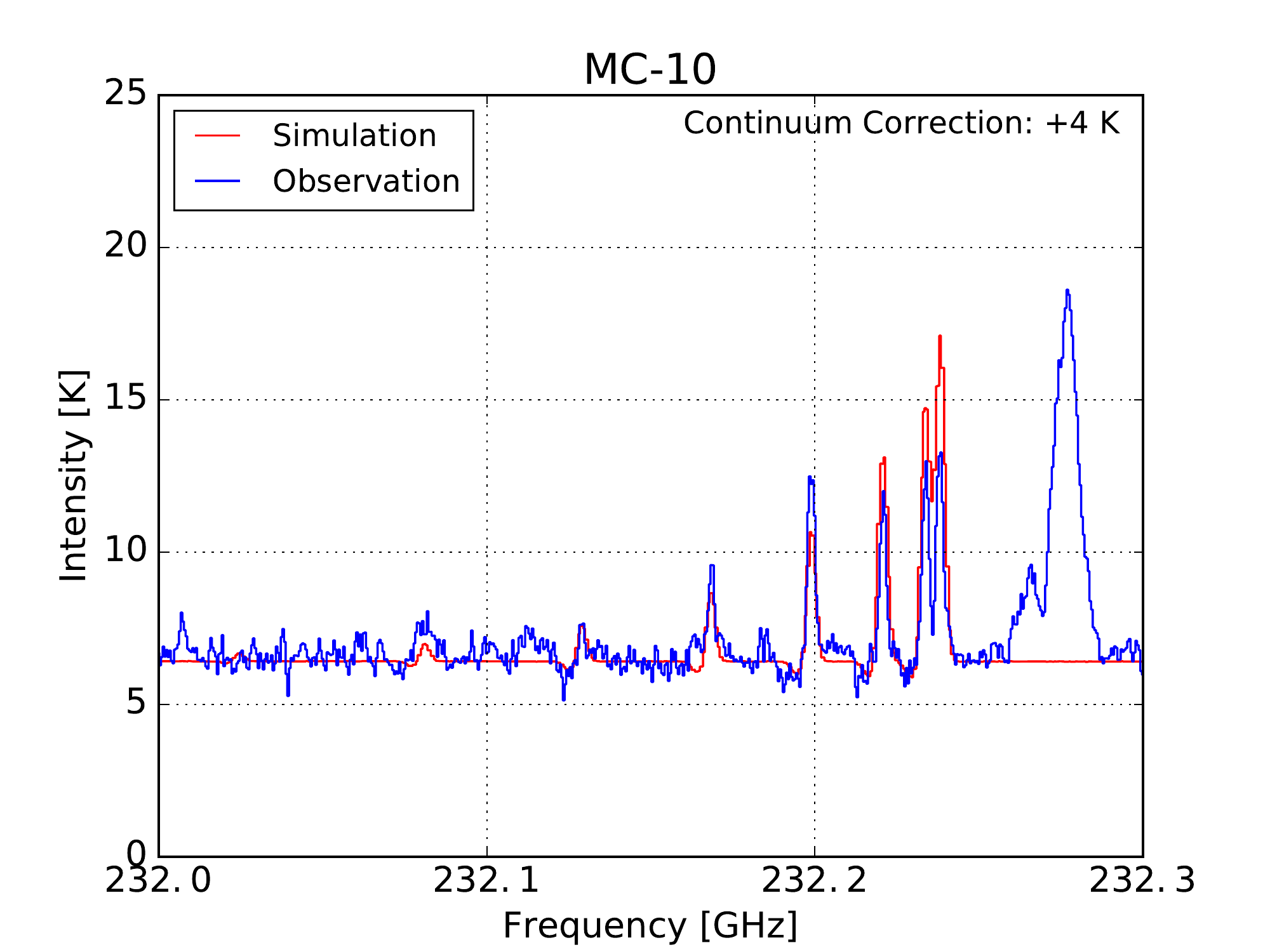}
    \includegraphics[scale=0.36]{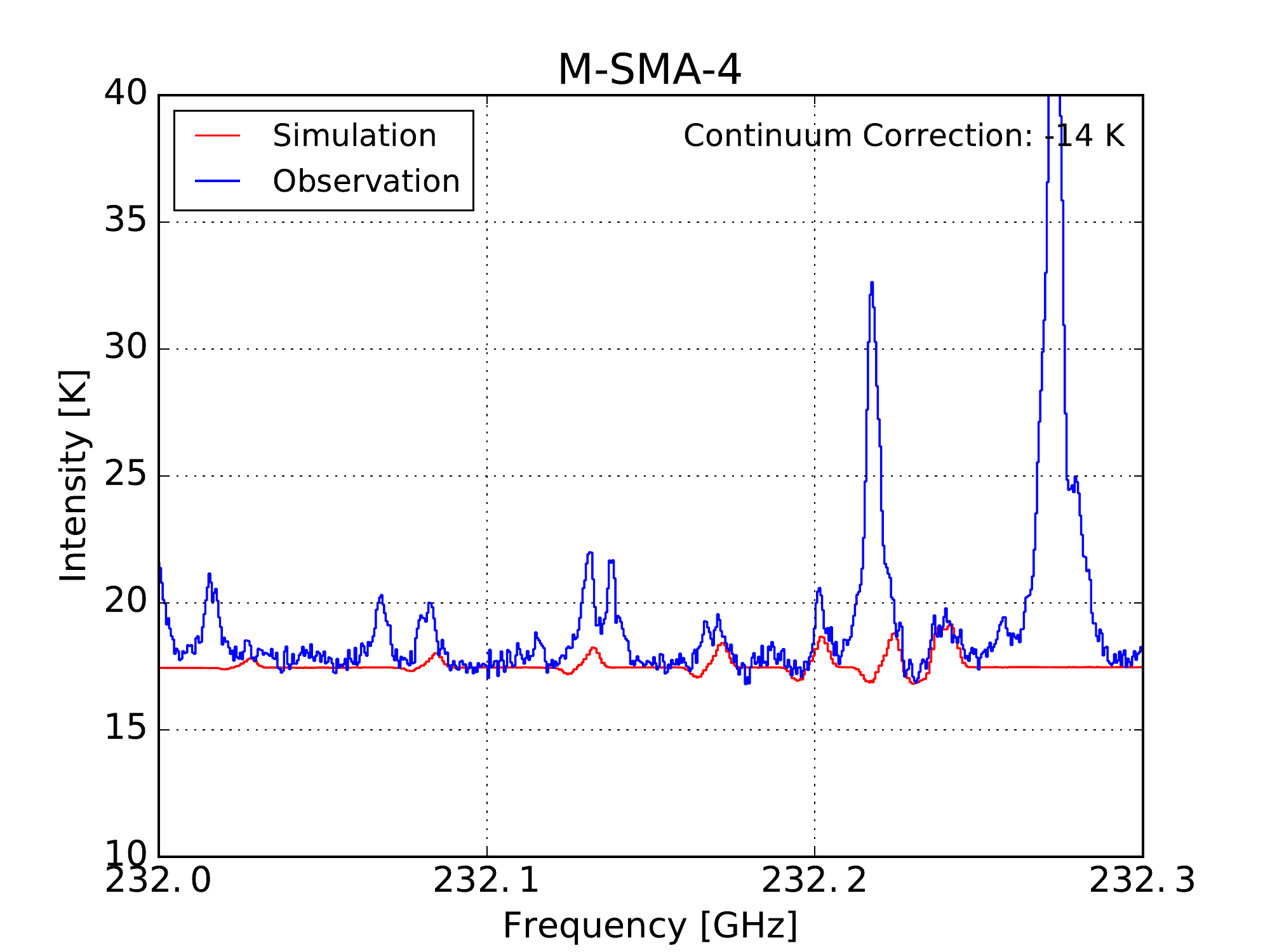}
    \includegraphics[scale=0.36]{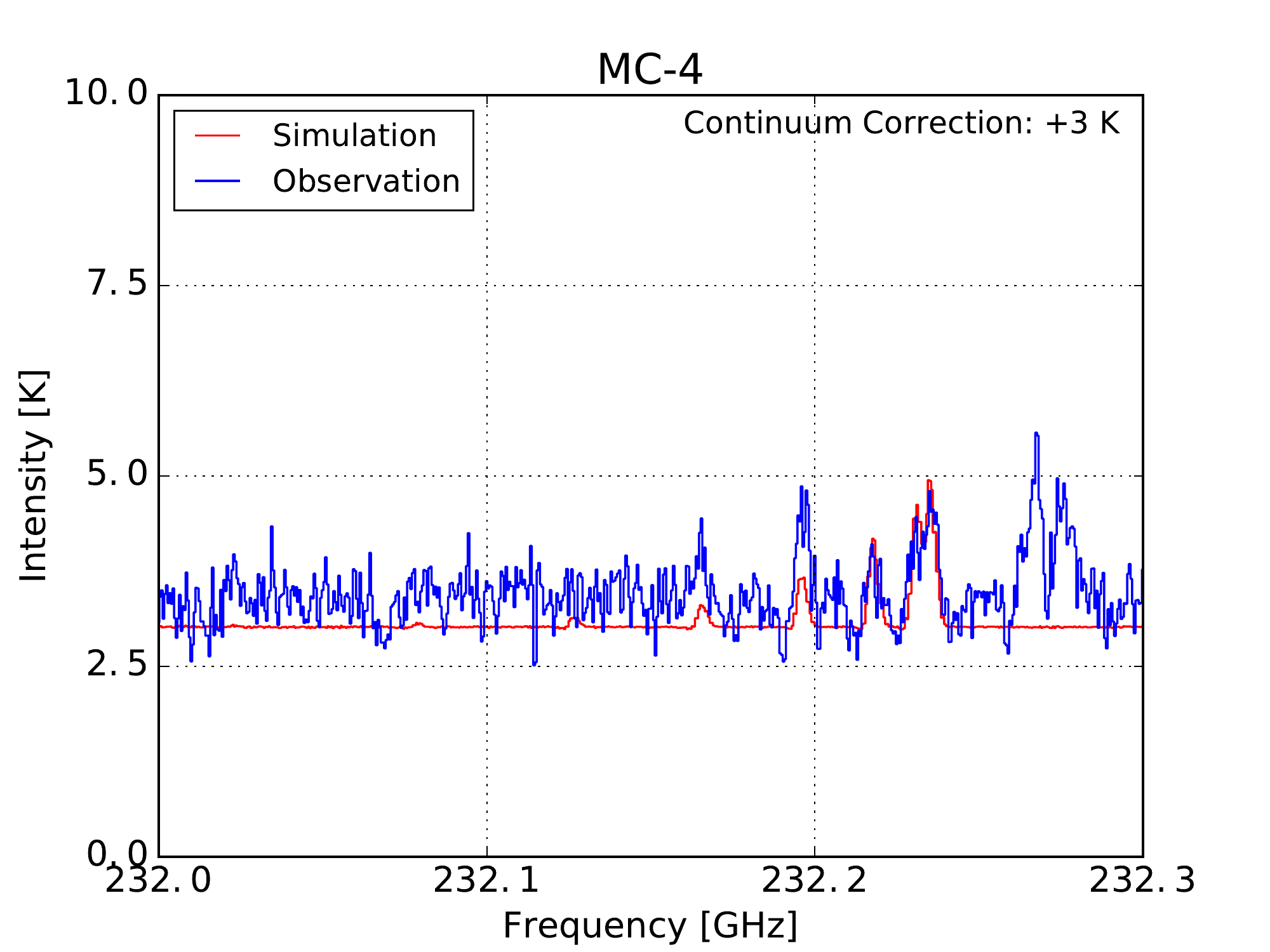}
    \includegraphics[scale=0.36]{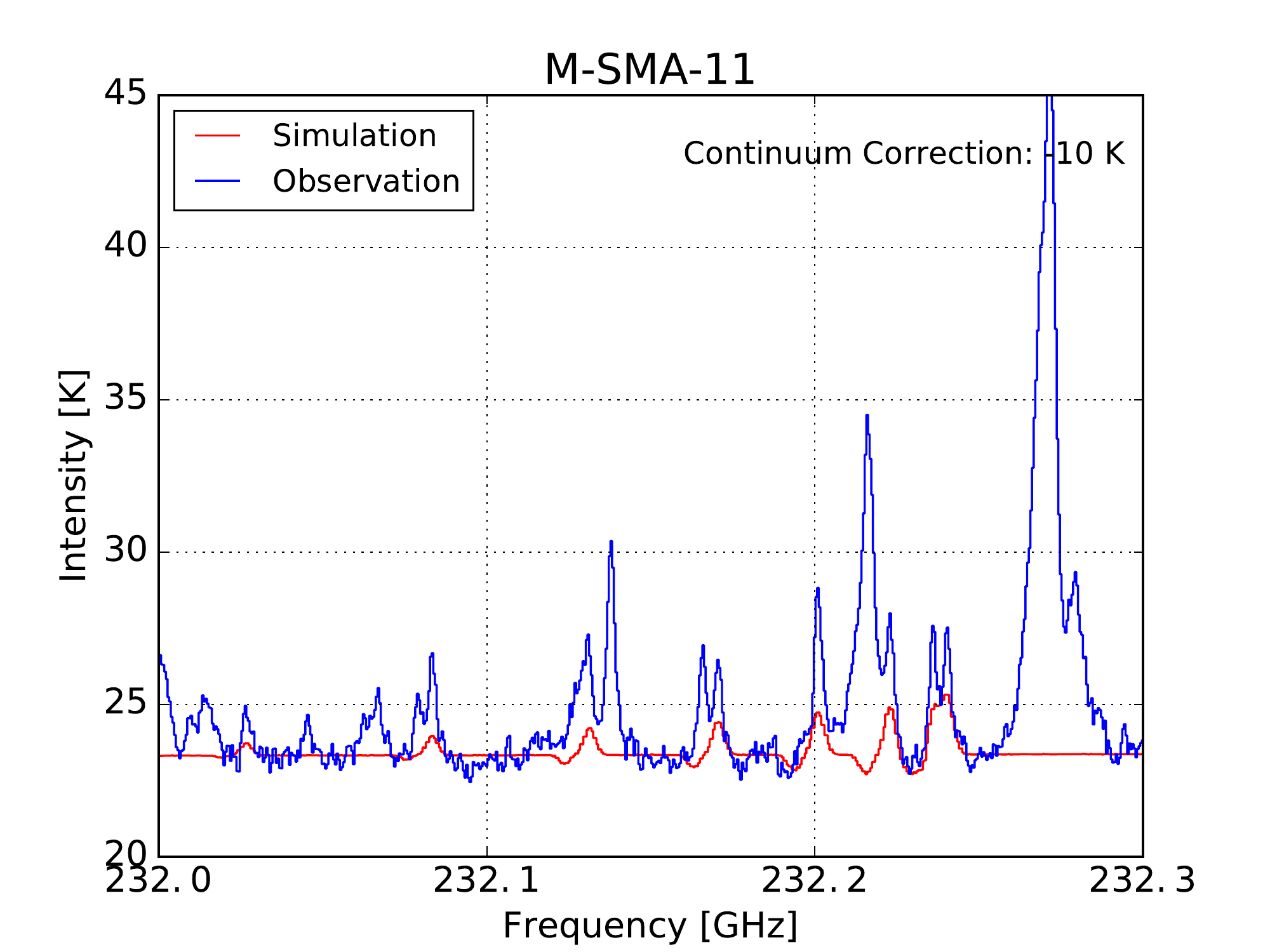}
    \includegraphics[scale=0.36]{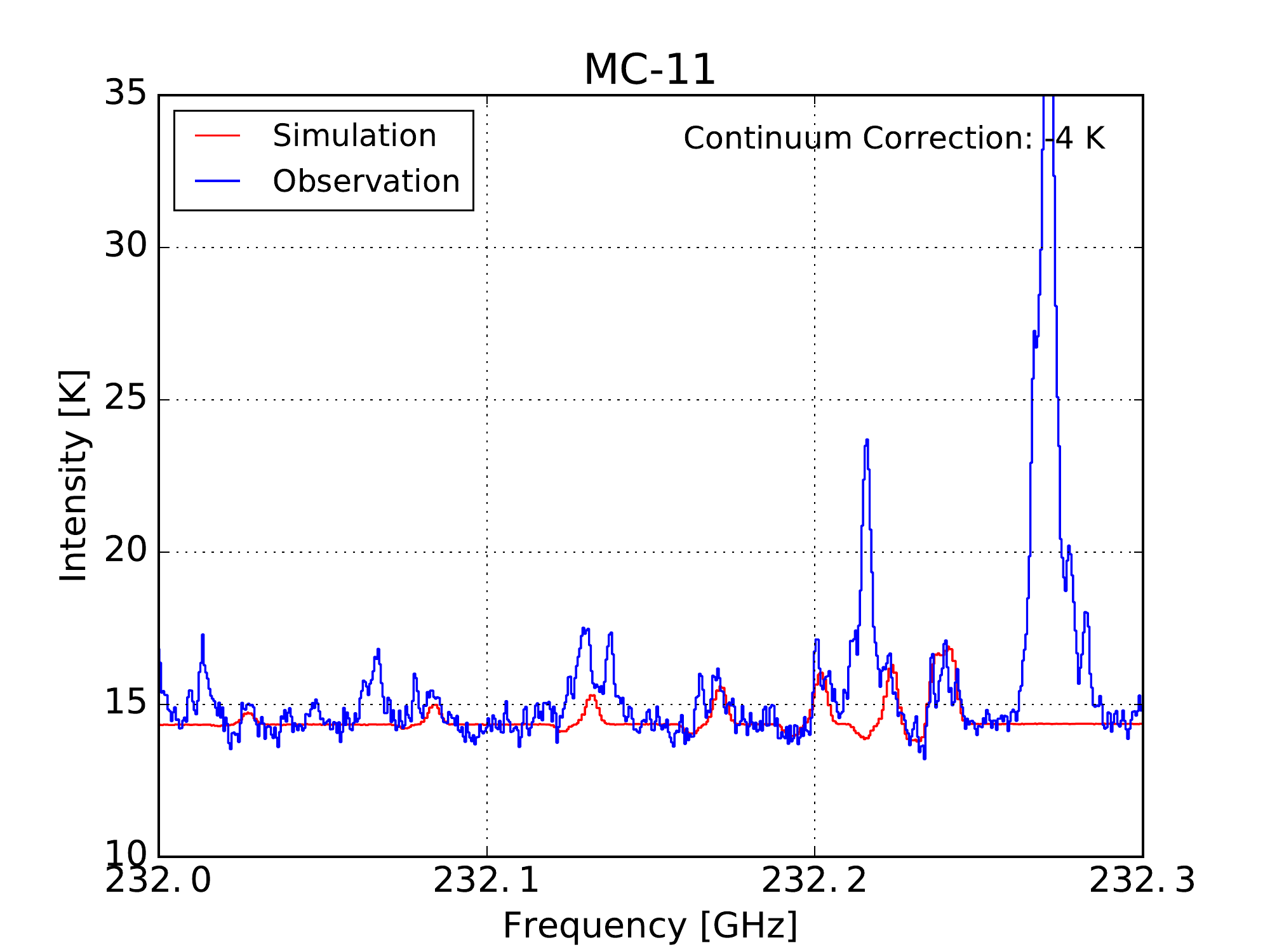}
    \includegraphics[scale=0.36]{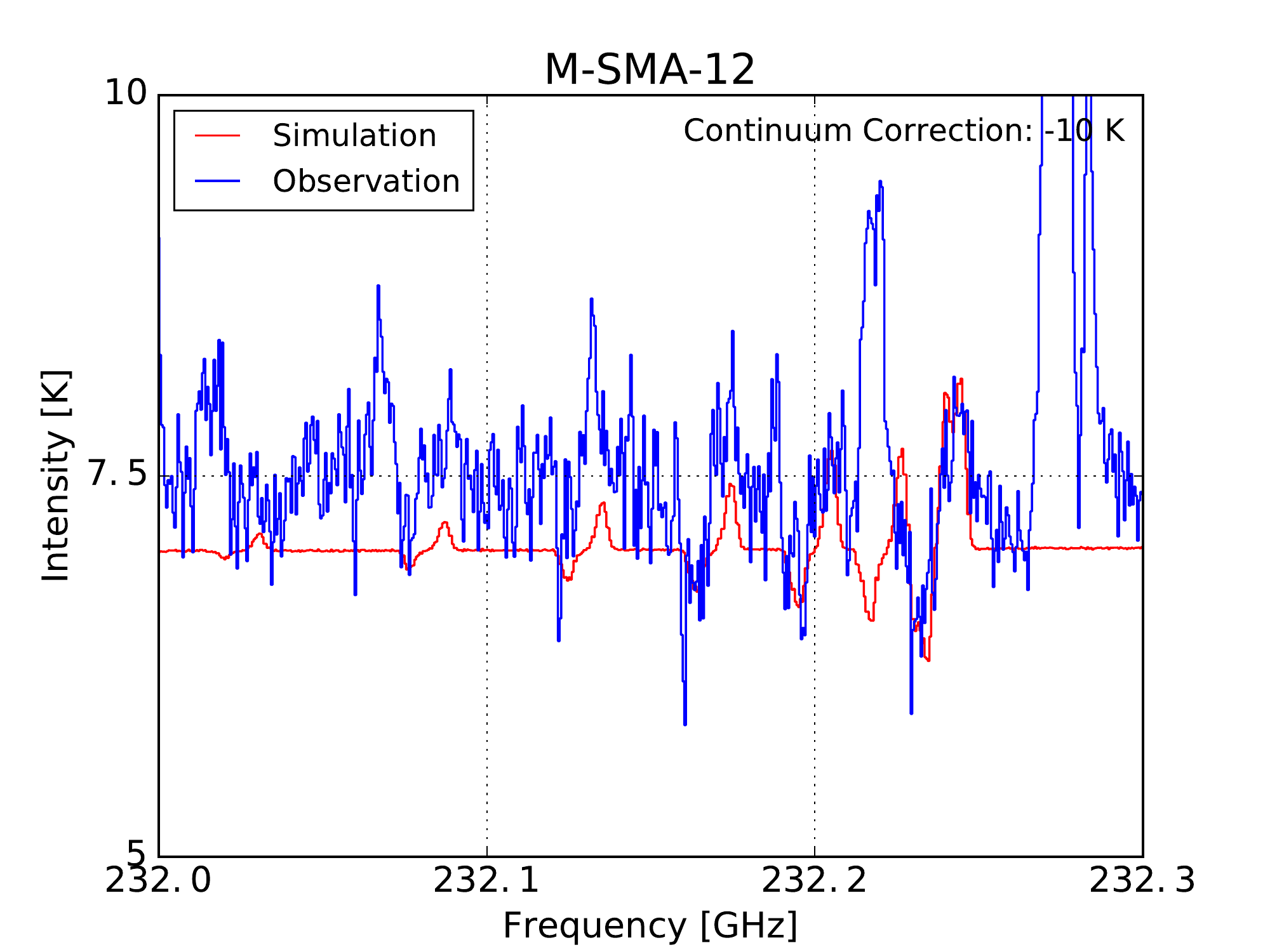}
\caption{Continuation of Fig.~\ref{fig:spectra13CH3CN_2} for the remaining cores and molecular centers.}
\label{fig:spectra13CH3CN_2}
\end{figure*}   

\end{appendix}
\end{document}